\documentclass[a4paper,12pt,justification=centering]{article}
\usepackage{tikz, pgfplots}
\usepackage{tikz-cd} 
\usepackage{latexsym}
\usepackage{rotating}
\usepackage{amsmath,amssymb,amsxtra,amscd,amsthm}
\usepackage[mathscr]{eucal}
\usepackage{graphicx}
\usepackage[justification=centering]{caption}
\usepackage{subfig}
\usepackage{datetime}
\usepackage{pdfsync}
\usepackage{thumbpdf}
\usepackage{color}
\ifx\pdfoutput\undefined

\usepackage{amsmath}
\usepackage{amssymb}
\usepackage{pgfplots}
\usepackage{graphicx} % needed for figures
\usepackage{caption}
\usepackage{mathrsfs}
\usepackage{subcaption}
\usepackage{color}
\usepackage[nottoc,notlot,notlof]{tocbibind}
\usepackage{verbatim}
\usepackage{tensor}
\usepackage{xcolor}
\usepackage{titlesec}

\usepackage{tcolorbox}
\tcbuselibrary{skins}
\usepackage{lipsum}

\usepackage{fancybox}

\newtcolorbox{myenv}{ams align}

\graphicspath {{images/}}
\usepackage{wrapfig}

\usetikzlibrary{shapes.geometric, arrows}

\usetikzlibrary{calc,decorations.markings}

\usetikzlibrary{arrows,decorations.markings}

\usetikzlibrary{calc,decorations.markings}

\tikzstyle{arrow} = [thick,->,>=stealth]

\usepackage{amsmath, amssymb, old-arrows}

\usepackage
[dvips,
 verbose=false,
 bookmarks=true,
 colorlinks=true,
 linktopage=true,
 linkcolor=webred,
 filecolor=webbrown,
 citecolor=webgreen,
 pagecolor=webblue,
 urlcolor=webblue,
 pdftitle={},
 pdfauthor={},
 pdfsubject={},
 pdfkeywords={},
 bookmarksopen=false,
 pdfpagemode=None,
 pdfview=FitH,
 pdfstartview=FitH,
 extension=pdf]{hyperref}
\else
\usepackage
[pdftex,
 verbose=false,
 bookmarks=true,
 colorlinks=true,
 linkcolor=webred,
 filecolor=webbrown,
 citecolor=webgreen,
 pagecolor=webblue,
 urlcolor=webblue,
 pdftitle={},
 pdfauthor={},
 pdfsubject={},
 pdfkeywords={},
 bookmarksopen=false,
 pdfpagemode=None,
 pdfview=FitH,
 pdfstartview=FitH,
 extension=pdf]{hyperref}
\fi
\definecolor{webred}{rgb}{.8,0,0}
\definecolor{webbrown}{rgb}{.6,0,0}
\definecolor{webgreen}{rgb}{0,0.5,0}
\definecolor{webdkgreen}{rgb}{0,0.3,0}
\definecolor{webblue}{rgb}{0,0,0.5}

\numberwithin{equation}{section}

\allowdisplaybreaks
\usepackage{bbm}
\usepackage{rotating}
\newcommand{\be}{\begin{eqnarray}}
\newcommand{\beq}{\begin{eqnarray}}
\newcommand{\ee}{\end{eqnarray}}

\def\e#1\e{\begin{equation}#1\end{equation}}
\def\ea#1\ea{\begin{align}#1\end{align}}

\theoremstyle{plain}% default

\newtheorem*{thm*}{Theorem}

\theoremstyle{definition}

\usepackage{hyperref}
\hypersetup{%
    linktocpage={true} % In TOC, links and colors for page numbers only.
}

\usepackage{enumitem}

\usepackage[a4paper,
            bindingoffset=0.2in,
            left=0.7in,
            right=0.9in,
            top=1.3in,
            bottom=1.3in,
            footskip=.25in]{geometry}

\usetikzlibrary{decorations.markings}

\usepackage[many]{tcolorbox}

\usepackage{empheq}

\usepackage[framemethod=tikz]{mdframed}

\definecolor{ocre}{RGB}{243,102,25}
\definecolor{mygray}{RGB}{243,243,244}

\newcommand*\mymathbox[1]{%
\fcolorbox{ocre}{mygray}{\hspace{1em}#1\hspace{1em}}}

\usepackage{enumitem}

\begin{document}

\setlength{\abovedisplayskip}{20pt}
\setlength{\belowdisplayskip}{20pt}

\begin{titlepage}

\setlength{\parindent}{0cm}
\setlength{\baselineskip}{1.5em}
\title{\textbf{Wild wall-crossing and symmetric quivers \\
in 4d and 3d $\mathcal{N}=2$ field theories} \vspace{1cm}}

\author{Daniel Bryan\footnote{\tt{dbryan@fuw.edu.pl}}, \  Piotr Su{\l}kowski\footnote{\tt {psulkows@fuw.edu.pl}}
\\
\\
\small Faculty of Physics, University of Warsaw, ul. Pasteura 5, 02-093 Warsaw, Poland\\
}

\date{}
\maketitle

\begin{abstract}
We reformulate Kontsevich-Soibelman wall-crossing formulae for 4d $\mathcal{N}=2$ class $\mathcal{S}$ theories and corresponding BPS quivers, including those of wild type, as identities for generating series of symmetric quivers that represent dualities of 3d $\mathcal{N}=2$ boundary theories. We identify such symmetric quivers 
for both sides of the wall-crossing formulae. In the finite chamber such a quiver is captured by the symmetrized BPS quiver, whereas on the other side of the wall we find an infinite quiver with an intricate pattern of arrows and loops.
Invoking diagonalization, for $m$-Kronecker quivers we find a wall-crossing type formula involving trees of unlinkings that expresses closed Donaldson-Thomas invariants of the corresponding 4d theories in terms of open Donaldson-Thomas invariants of the 3d theories and invariants of $m$-loop quivers.  
Using this formula, we determine a number of closed Donaldson-Thomas invariants of wild type.

\end{abstract}

\nonumber

\thispagestyle{empty}

\end{titlepage}

\begin{small}
    \tableofcontents
\end{small}

\newpage

\section{Introduction}

Supersymmetric $\mathcal{N}=2$ theories in 4d and 3d provide an interesting theoretical laboratory where various physical phenomena can be understood in a simplified setting. Such theories in 4d and 3d are also intricately related, as discussed e.g. in  \cite{4d3dwallsbraidsCecotti:2011iy}. In this work we consider such relations in the context of wall-crossing, including and with emphasis on its wild instances. We show that wall-crossing phenomena in 4d $\mathcal{N}=2$ theories known from \cite{KontsevichSoibelman2008, KontsevichCoHA:2010px,GMN4d3d,gaiotto2011wallcrossing} correspond to dualities between specific 3d $\mathcal{N}=2$ theories, and find relations between quivers characterizing the two classes of theories in question, i.e. BPS quivers that encode BPS spectra of 4d theories \cite{cecottivafacompleteclassificationhttps://doi.org/10.48550/arxiv.1103.5832,Alim:2011ae,alim2011n2new}, and symmetric quivers that encode the structure of 3d theories and their BPS states \cite{Knotquivercorrespondence2Kucharski:2017poe,Knotquivercorrespondence1Kucharski:2017ogk,Ekholm:2018eee,multiskeinsEkholm:2019lmb}. We also express wall-crossing formulae associated to 4d BPS quivers as identities between (finite and infinite) symmetric quivers, and then use the formalism of symmetric quivers to determine 4d BPS (wild) degeneracies.

% As discussed also in \cite{}, the relation between 4d and 3d quivers involves symmetrization (or doubling) of the former ones (i.e. adding an arrow in opposite direction for each arrow in a BPS quiver). 

Recall that wall-crossing describes the jumping of BPS degeneracies across special loci in the moduli spaces called walls of marginal stability which hereby divide this moduli space into chambers. 
Over the last decades wall-crossing has become important in the study of 4d $\mathcal{N}=2$ field theories, in particular in the determination of the BPS spectrum throughout the moduli space, for example by finding the spectrum at weak coupling from that at strong coupling \cite{Alim:2011ae,alim2011n2new,Alim:2023doi,Seiberg_1994,Argyres_1995, FramedBPSGMNhttps://doi.org/10.48550/arxiv.1006.0146,spectralnetworksGaiotto:2012rg,  Alim:2024ezg}. This has also led to significant progress in enumerative geometry given that the BPS degeneracies can be interpreted as Donaldson-Thomas (DT) invariants of the BPS quivers or associated to derived categories of coherent sheaves on the Calabi-Yau 3-folds used to construct these theories \cite{joyce2010theory, DonaldsonThomasOriginalpaper:1996kp,DonaldsonThomasdefinitionhttps://doi.org/10.48550/arxiv.math/9806111,Bridgelandconifoldhttps://doi.org/10.48550/arxiv.1703.02776, 2018Bridge}. In particular, the notion of quivers that describe BPS states in string theory was developed for D-brane systems in \cite{QuiverBackgroundDouglasMoorehttps://doi.org/10.48550/arxiv.hep-th/9603167,Diaconescu_1998BPSquivers,BPSorbifoldquiverFiol_2000,QuiversnoncompactCYDouglas_2005} and then used in the classification of 4d $\mathcal{N}=2$ theories \cite{cecottivafacompleteclassificationhttps://doi.org/10.48550/arxiv.1103.5832}. These quivers nicely encode the BPS spectrum \cite{cecottivafacompleteclassificationhttps://doi.org/10.48550/arxiv.1103.5832,Alim:2011ae,alim2011n2new} in the different chambers in terms of stable quiver representations, or alternatively, using quiver mutations as a spectrum generator. The notion of the BPS algebra originally introduced by Harvey and Moore \cite{Harvey_1998, Harvey_1996} is captured by the quiver Yangian \cite{Li:2023zub} or Cohomological Hall Algebra (CoHA) \cite{KontsevichCoHA:2010px, Galakhov_2019BPSalgebrasandframedBPSstates} in simple examples. There have been recent developments in categorifying wall-crossing in terms of these algebras \cite{categorialwallcrossingGaiotto:2023dvs} also for framed BPS states \cite{framedcategorialwallcrossingGaiotto:2024fso}. 

Importantly, Kontsevich and Soibelman \cite{KontsevichSoibelman2008} 
derived wall-crossing formulae, which state that the product of certain symplectic operators associated to BPS states in a given chamber is invariant upon crossing the wall of marginal stability. They also derived the motivic versions of wall-crossing formulae, which correspond to quantum dilogarithm identities with quantum torus valued arguments. The exponents of the operators in these formulas are identified with (ordinary or refined, i.e. involving spins of the states) BPS degeneracies \cite{RefinedMotivicQuantumDimofte_2009}. An interpretation of these formulae in terms of dual open topological string partition functions was also given by Cecotti and Vafa in \cite{Cecotti:2009ufwallcrossingtopologicalstrings}. In well known examples such as Argyres-Doulgas or pure $SU(2)$ Seiberg-Witten theory the BPS spectrum is well understood, however, in other theories refined BPS invariants are difficult to determine. These include pure $SU(3)$ which has chambers with wild BPS spectra \cite{wildwallGalakhov:2013oja, Mainiero:2016xajwildalgebra}, meaning that there exists a dense cone of BPS rays within which all charges support BPS states but with unknown degeneracy. Several approaches have been developed to determine these BPS degeneracies in 4d $\mathcal{N}=2$ theories in the different chambers, such as the spectral networks \cite{spectralnetworksGaiotto:2012rg, wildwallGalakhov:2013oja} and the attractor flow tree formulae \cite{D4Manschot:2010xp,quiver2Manschot:2014fua,AlexandrovPiolinehttps://doi.org/10.48550/arxiv.1804.06928, Bousseau:2022snmnewattractorflowscatteringdiagram}. In this paper we find a new approach that enables to determine such degeneracies, by relating them to 3d $\mathcal{N}=2$ theories and symmetric quivers (without potential).  

The symmetric quivers just mentioned play the main role in this work. Their relation to 3d $\mathcal{N}=2$ theories has been recently found and analyzed in the context of the knots-quivers correspondence and its generalizations \cite{Knotquivercorrespondence2Kucharski:2017poe,Knotquivercorrespondence1Kucharski:2017ogk}. Physically, such symmetric quivers describe 3d $\mathcal{N}=2$ theories $T[L]$ arising from M5-brane on a lagrangian 3-manifold $L$ embedded in a Calabi-Yau three-fold \cite{Ooguri_2000}. In such a setup, open BPS states are described by holomorphic M2 brane disks (ending on the M5-brane), whose linking is encoded in the adjacency matrix of a symmetric quiver $Q$ of a dual quiver theory $T[Q]$ whose spectrum describes BPS vortices in a 3d $\mathcal{N}=2$ theory \cite{Ekholm:2018eee,multiskeinsEkholm:2019lmb}. Motivic generating series of these symmetric quivers take form of Nahm sums \cite{KontsevichCoHA:2010px,Efimov_2012,reineke2011degeneratecohomologicalhallalgebra}, which are identified with the vortex partition functions. In particular, for Calabi-Yau being the resolved conifold and the lagrangian brane $L$ arising from a knot conormal upon the geometric transition, the nodes of a symmetric quiver represent the generators of HOMFLY-PT homology, while the motivic generating series of a quiver encodes colored HOMFLY-PT polynomials, which is the statement of the knots-quivers correspondence. Analogous statements hold for other Calabi-Yau manifolds and choices of $L$, see e.g. \cite{Panfil_2019,Kimura_2021}. On the other hand, analogous 3d $\mathcal{N}=2$ theories arise on the boundary of the 4d theory in 4d-3d coupled systems \cite{4d3dwallsbraidsCecotti:2011iy}, which underlies the relations between 3d and 4d theories and corresponding quivers presented in this work.

%on the boundary of the 4d theory. These can be considered analogous to the 4d-3d coupled systems of \cite{4d3dwallsbraidsCecotti:2011iy}. They can be constructed by considering an M5 brane worldvolume theory on a lagrangian 3-manifold $L$ embedded in a CY 3-fold \cite{PhysicsandGeometryEkholm:2018eee, multicoverskeins}. The open BPS states are described by holomorphic M2 brane disks ending on the M5 brane. The linking of these disks is encoded in the adjacency matrix of a symmetric quiver $Q$ of a dual quiver theory $T[Q]$ where the spectrum now describes BPS vorticies in a 3d $\mathcal{N}=2$ theory. These symmetric quiver CoHAs have a generating series \cite{kontsevich2008stability,CoHAKS,Efimov_2012,reineke2011degeneratecohomologicalhallalgebra} taking the form of a Nahm sum which physically correspond to the vortex partition functions. Recently particularly interesting examples were discovered in which the Lagrangian $L$ corresponded to a knot conormal $L_{K}$. In these cases the symmetric quiver generating series corresponds to the generating series of colored HOMEFLY-PT polynomials   of the knot $K$. This is also known as the knots-quivers correspondence \cite{Knotquivercorrespondence1Kucharski:2017ogk, Knotquivercorrespondence2Kucharski:2017poe}.

\begin{figure}
\begin{center}
\includegraphics[width=0.5\textwidth]{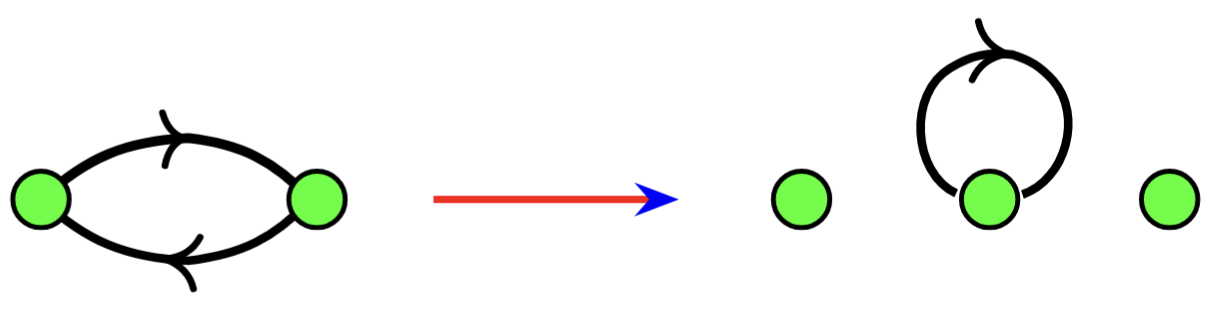}  
\caption{Unlinking of a symmetric quiver arising from the pentagon identity.} \label{fig-unlinking}
\end{center}
\end{figure}

Further, recall that symmetric quivers can be subject to the operations of linking and unlinking, which amount respectively to adding or removing one pair of arrows between two nodes, and at the same time adding one extra node to a quiver (connected by arrows to other nodes in some specific way). Once the generating parameter associated to a new node is appropriately identified, the motivic generating series of the initial and final quiver are the same. These operations have a nice interpretation in terms of multicover skeins and represent creation of a new holomorphic disk upon crossing and gluing of boundaries of other disks \cite{multiskeinsEkholm:2019lmb}. In particular, for a quiver consisting of just two nodes connected by a pair of (oppositely oriented) arrows, their unlinking removes this pair of arrows and produces a new node with a loop, see fig. \ref{fig-unlinking}. This operation is a manifestation of pentagon identity (\ref{pentagon}) and our results can be regarded as its highly non-trivial generalizations. Applications and generalizations of (un)linking were analyzed in \cite{UnlinkingKucharski:2023jds,vangarderen2024}, and its role in the context of the relation between 4d and 3d $\mathcal{N}=2$ theories is also discussed in \cite{kllnps}.

% This symmetric wall-crossing can also be understood in terms of special skein relations. One can define the framed skein module of $L$ from links of the Lagrangian boundaries of the holomorphic disks modulo the skein relation \cite{multiskeinsEkholm:2019lmb}. This skein relation arises from disk boundaries crossing and gluing to form a new disk in addition to the already existing basic disks. One can see that the total curve count remains invariant when considering multiple covers of these disks and can hence be called \textit{multicover skeins}. This unlinking of disks and formation of the new basic disk is translated to the unlinking of arrows in the symmetric BPS quiver to form a new quiver $Q_{U}$ with a new node \cite{UnlinkingKucharski:2023jds} (with new arrows to the other nodes) describing a dual 3d $\mathcal{N}=2$ theory. The new quiver has the same generating series after a suitable specialisation of variables. In the simplest example of unlinking 2 nodes with a single pair of arrows one obtains the pentagon identity and can consider this the standard wall-crossing in the $A_{2}$ model. However, the boundstate doesn't disappear but is mapped to the new basis state (the new node) under the duality.        

Let us summarize the main results of this paper. Consider a BPS quiver, whose nodes $i$ represent charges $\gamma_i$ of states in 4d $\mathcal{N}=2$ theory. The Dirac pairing $\langle \gamma_i,\gamma_j\rangle$ is equal to the number of arrows between nodes $i$ and $j$. To the nodes we assign operators $X_{\gamma_i}$, which satisfy quantum torus algebra  
\begin{align}
X_{\gamma_{i}}X_{\gamma_{j}} = (-q^{\frac{1}{2}})^{ \langle \gamma_{i}, \gamma_{j}\rangle} X_{\gamma_{i}+\gamma_{j}} = q^{\langle \gamma_{i}, \gamma_{j}\rangle} X_{\gamma_{j}}X_{\gamma_{i}}.      \label{quantumtorus}
\end{align}
The BPS spectrum is then captured by the Kontsevich-Soibelman operator, which is the same in each chamber and takes form of a product of quantum dilogarithms with arguments being elements of the quantum torus algebra. The quantum dilogarithm function is defined by $\Phi(x) = \sum_{k \geq 0} \frac{q^{\frac{k}{2}}}{(q,q)_{k}} x^{k}$. It can be shown that in a 4d $\mathcal{N}=2$ theory there is a chamber, in which BPS states are in one-to-one correspondence with nodes of a BPS quiver. %In particular, there is a chamber in which BPS are in one-to-one correspondence with the nodes of a BPS quiver.

We now assign a symmetric quiver to a BPS quiver, by rewriting the Kontsevich-Soibelman operator in terms of the following representation of the quantum torus algebra (\ref{quantumtorus})  \cite{multiskeinsEkholm:2019lmb}
\begin{align}
X_{\gamma_i}\equiv X_{i} = (-1)^{C_{ii}} (q^{\frac{1}{2}})^{C_{ii}-1} \hat{x}_{i} \hat{y}^{C_{ii}}_{i} \prod_{j < i} \hat{y}^{C_{ij}}_{j}, \label{Xixi}
\end{align}
where $\hat{x}_{i}$ and $\hat{y}_{j}$ satisfy the algebra $\hat{y}_{i} \hat{x}_{j} =  q^{\delta_{i,j}} \hat{x}_{j} \hat{y}_{i}$, $C_{ij}=\langle \gamma_{i}, \gamma_{j}\rangle$ for $i>j$, and $C_{ii}$ can take any values. Now, the Kontsevich-Soibleman operator takes form
\begin{align} \label{eq:symmetricquivernoncommutative-intro}
 \mathbb{P}^{Q} = \Phi(X_{m}) \cdot \Phi(X_{m-1}) \cdots  \Phi(X_{1}),
\end{align}
with each $X_i$ assigned to one node of a BPS quiver. Acting with this operator on the constant 1, which amounts to normal ordering :$\ $: and moving $\hat{y}_{i}$ operators to the right (using their commutation relations with $\hat{x}_{j}$) and then removing them from the expression, while identifying $\hat{x}_{i}$ with generating parameters $x_i$, produces the expression of the form of the Nahm sum
\begin{align}
  P_{Q}(\mathbf{x},q) =\  : \mathbb{P}^{Q}: \ = %\sum_{\mathbf{d}} (-q^{\frac{1}{2}})^{\mathbf{d} \cdot C \cdot \mathbf{d}} \frac{\mathbf{x^{\mathbf{d}}}}{(q;q)_{\mathbf{d}}} =  
  \sum_{d_{1}, ... , d_{m} \geq 0} (-q^{\frac{1}{2}})^{\sum^{m}_{i,j = 1} C_{ij}d_{i} d_{j}} \prod^{m}_{i = 1} \frac{x_{i}^{d_{i}}}{(q;q)_{d_{i}}},    \label{PQ}
\end{align}
where $\mathbf{x}=(x_1,\ldots,x_m)$ and $(x ; q )_{n} = \prod^{n-1}_{k = 1} (1- x q^{k})$ is the $q$-Pochammer symbol. This way we obtain the form of a motivic generating series assigned to a symmetric quiver $Q$ consisting of $m$ nodes with $C_{ij}=C_{ji}$ arrows between the nodes $i$ and $j$. In case all $X_i$ operators are independent, this symmetric quiver is given by doubling (symmetrizing) the original BPS quiver, i.e. adding an arrow in opposite direction to every arrow in this BPS quiver. If $C_{ii}\neq 0$ in (\ref{Xixi}), the symmetric quiver would also have the corresponding number of loops (i.e. arrows from a node to itself), so in fact we can assign an infinite family of symmetric quivers (with different values of $C_{ii})$ to a given BPS quiver. Nonetheless, in this work we simply set $C_{ii}=0$. Furthermore, if there are non-trivial relations between $X_i$ (which happens e.g. for $SU(2)$ theory with matter) then apart from the doubling, the normal ordering process generates additional loops in the resulting symmetric quiver.

We stress that the above relation between BPS quivers in 4d and their symmetrized (doubled) analogs in 3d arises in consequence of the realization of quantum torus algebra  (\ref{quantumtorus}) in terms of $\hat{x}_i$ and $\hat{y}_j$ (\ref{Xixi}). A geometric interpretation and derivation of this relation for a series of $A_n$ BPS quivers is presented in detail in a parallel work \cite{kllnps}. While understanding such a geometric picture, also for other types of BPS quivers, is of high interest, in this work we focus on implications of the above relations for (wild) wall-crossing phenomena.
% It is desirable to generalize it to other classes of quivers that we consider in the present work.  

Indeed, once we know how to assign a symmetric quiver to a BPS quiver, we undertake our main task: we rewrite various wall-crossing identities in terms of symmetric quivers. While original such identities describe wall-crossing phenomena in 4d $\mathcal{N}=2$ theories, we interpret the relation between corresponding symmetric quivers as encoding dualities between 3d $\mathcal{N}=2$ theories. We primarily focus on a series of quantum dilogarithm identities assigned to $m$-Kronecker quivers (i.e. quivers consisting of two nodes with $m$ arrows in the same direction between them) in the form presented by Reineke in \cite{reineke2023wildquantumdilogarithmidentities}
\begin{align}
  \Phi_{(1,0)}  \Phi_{(0,1)}     =    \Phi_{(0,1)}   \Phi_{\sigma (0,1)}  \Phi_{\sigma^{2} (0,1)}  \Phi_{\sigma^{3} (0,1)}  \cdots \prod^{\rightarrow}_{\mu_{-} < a/b < \mu_{+}} \Phi_{a,b}^{P_{a,b}}  \cdots \Phi_{\sigma^{-3} (1,0)}   \Phi_{\sigma^{-2} (1,0)}  \Phi_{\sigma^{-1}(1,0)}  \Phi_{(1,0)}   \label{Reineke-ids}
\end{align}
where $\sigma_{a,b} = (b, mb-a)$, $\Phi_{(a,b)}$ are quantum dilogarithms with specific arguments involving elements of quantum torus algebra (\ref{quantumtorus}), and $\Phi_{a,b}^{P_{a,b}}$ denotes a product (over a variable $k$) of several quantum dilogarithms raised to powers $c^{a,b}_k$, which are identified with Donaldson-Thomas invariants or equivalently (upon suitable identification of parameters) with BPS numbers $\Omega^{4d}_{k}(\gamma)$ in 4d $\mathcal{N}=2$ theories. The left hand side of these identities is of the form (\ref{eq:symmetricquivernoncommutative-intro}) with just two quantum dilogarithms. We now take advantage of the identification (\ref{Xixi}) and write both sides of the identities (\ref{Reineke-ids}) in terms of $\hat{x}_i$ and $\hat{y}_j$. Normal ordering such expressions gives $q$-series of the Nahm sum form (\ref{PQ}), from which we can read off the form of symmetric quivers associated to both sides of (\ref{Reineke-ids}). On the left-hand side we get a symmetric quiver, which is a doubling of $m$-Kronecker quiver and has $m$ pairs of arrows (in opposite directions) between the two nodes. As the original BPS quiver ($m$-Kronecker quiver) typically characterizes the strong coupling chamber of 4d $\mathcal{N}=2$ theory, we denote the corresponding symmetric quiver by $Q^s$. Then, from the right hand side of (\ref{Reineke-ids}), which (at least partially) encodes the spectrum in the weak coupling chamber in the original 4d theory, we typically (for $m>1$) get an intricate, infinite symmetric quiver that we denote $Q^w$, which represents the dual 3d $\mathcal{N}=2$ theory. In particular, two specific nodes of $Q^w$ that correspond to the nodes of $Q^s$ are not connected by arrows, so this operation of assigning $Q^w$ to $Q^s$ can be interpreted as unlinking, generalizing the unlinking operation from \cite{multiskeinsEkholm:2019lmb}.

Note that each dilogarithm on either side of (\ref{Reineke-ids}) on one hand corresponds to one BPS state in appropriate chamber, and on the other hand it gives rise to one node of the corresponding symmetric quiver. Therefore the nodes of symmetric quivers $Q^s$ and $Q^w$ are in one-to-one correspondence with BPS states in the relevant chamber of 4d theory.

Our results corresponding to special cases of (\ref{Reineke-ids}) are as follows. For $m=1$ the underlying BPS quiver of the form $\bullet\rightarrow\bullet$ is also referred to as $A_2$ quiver, and it characterizes the spectrum of $A_2$ Argyres-Douglas theory. The identity (\ref{Reineke-ids}) reduces in this case to the well-known pentagon identity \cite{FADDEEV_1994} for the quantum dilogarithm 
\begin{align}
\Phi(X_{\gamma_{2}})\Phi(X_{\gamma_{1}}) = \Phi(X_{\gamma_{1}}) \Phi(X_{\gamma_{1}+\gamma_{2}}) \Phi(X_{\gamma_{2}})  \label{pentagon}
\end{align}
and its two sides represent respectively two chambers of Argyres-Douglas theory, with two or three BPS states. The corresponding symmetric quivers that arise from the identification (\ref{Xixi}) and then normal ordering of the pentagon identity are related by the unlinking operation mentioned earlier, see fig. \ref{fig-unlinking}. Consequences and interpretation of this case are discussed at length in \cite{kllnps}.
% As the two nodes in the quiver on the left get unlinked on the right hand side, this operation is also referred to as unlinking. It was introduced in \cite{multiskeinsEkholm:2019lmb}

\begin{figure}
\begin{center}
\includegraphics[width=0.5\textwidth]{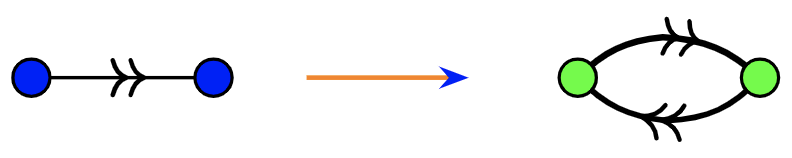}  
\end{center}
\caption{BPS quiver for pure Seiberg-Witten theory (left) and corresponding symmetric quiver $Q^s$ with two pairs of oppositely oriented arrows between the nodes (right) that arises from the left hand side of the identify (\ref{Reineke-ids}) for $m=2$.} 
\label{fig-quiverSU2strong}
\end{figure}

\begin{figure}
\begin{center}
\includegraphics[width=\textwidth]{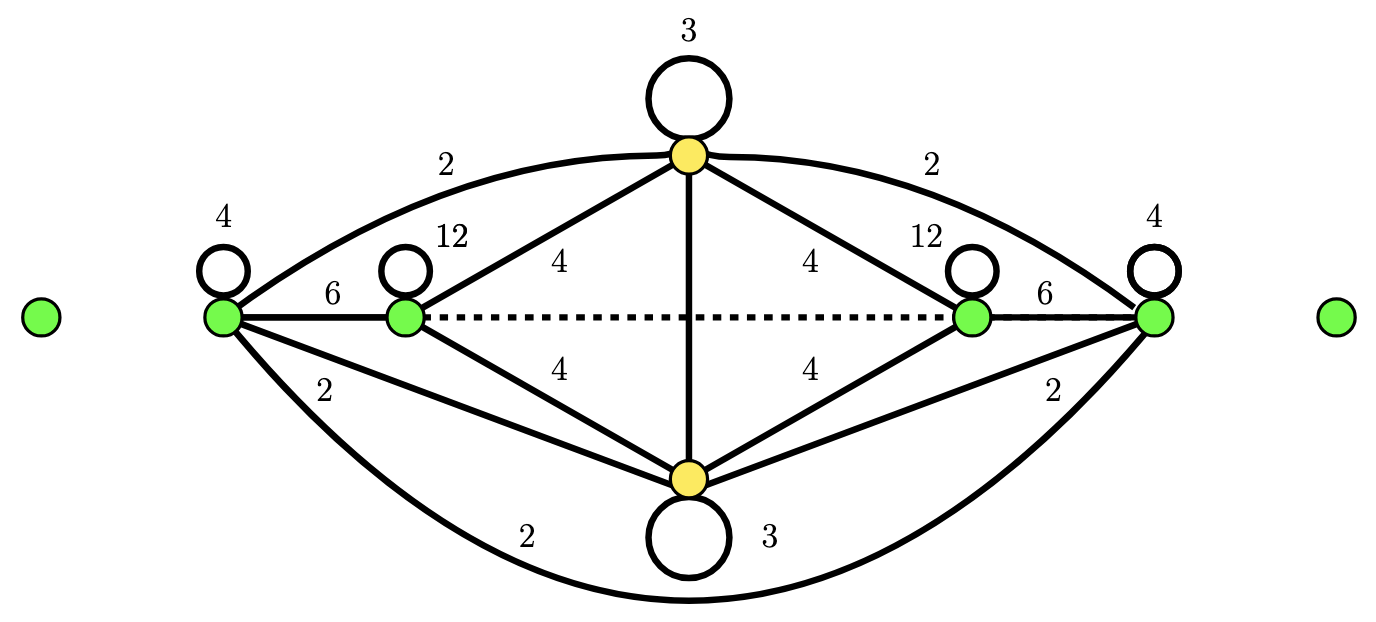}  
\end{center}
\caption{Symmetric quiver $Q^w$ with infinite number of nodes, arising from the right hand side of the identify (\ref{Reineke-ids}) for $m=2$. Two yellow nodes originate from two W-bosons, and an infinite series of green nodes (represented by the dashed segment) originates from dyons in the weak chamber of 4d $\mathcal{N}=2, SU(2)$ theory. The numbers of loops at various nodes, and the numbers of pairs of arrows between the nodes, are also typed.} \label{fig-quiverSU2weak}
\end{figure}

For $m=2$, the identity (\ref{Reineke-ids}) characterizes two chambers of pure $SU(2)$ Seiberg-Witten theory. There are two BPS states in the strongly coupled chamber, corresponding to two nodes of the BPS quiver, and an infinite number of states in the weakly coupled chamber. The symmetric quiver $Q^s$ corresponding to the left hand side of the identity (\ref{Reineke-ids}), which takes form of a doubling of the BPS quiver, is shown in fig. \ref{fig-quiverSU2strong}. In what follows we determine an infinite symmetric quiver $Q^w$, see fig. \ref{fig-quiverSU2weak}, corresponding to the weakly coupled chamber and the right hand side of (\ref{Reineke-ids}). For $SU(2)$ theory all BPS numbers in both chambers are known, so we get an explicit form of the quiver $Q^w$. % and interpret it in terms of 3d $\mathcal{N}=2$ theory.

Yet more interesting results arise for $m$-Kronecker quivers for $m \geq 3$, see fig. \ref{fig-quiver-m}. They are also important in physics, e.g. the quiver for $m=3$ is a subquiver of a BPS quiver for pure $SU(3)$ Seiberg-Witten theory. Such quivers and corresponding wall-crossing are referred to as wild -- the BPS walls for such quivers form a dense set within some cone \cite{reineke2023wildquantumdilogarithmidentities, akagi2024explicitformslowerdegrees, burcroff2024brokenlinescompatiblepairs, burcroff2024scatteringdiagramstightgradings, astruc2024motivescentralslopekronecker, reineke2024expanderrepresentationsquivers,franzen2019cohomologicalhallalgebrakronecker}. There exist walls of marginal stability in these theories such that when crossing them the spectrum jumps to an infinite wild spectrum (including the dense regions) also known as an $m$-cohort. Descriptions of their Donaldson-Thomas invariants or BPS invariants $\Omega(\gamma,y,u)$ have been derived using spectral networks \cite{wildwallGalakhov:2013oja,Galakhov_2015spectralspin} or the attractor flow tree formula \cite{D4Manschot:2010xp,quiver2Manschot:2014fua,AlexandrovPiolinehttps://doi.org/10.48550/arxiv.1804.06928, Bousseau:2022snmnewattractorflowscatteringdiagram}, however they are not known in general. In particular, this is the case with the identity (\ref{Reineke-ids}): for $m\geq 3$ only its overall structure is known \cite{reineke2023wildquantumdilogarithmidentities}, while the invariants $c_k^{a,b}$ in (\ref{Reineke-ids}) are in general unknown -- only the invariants associated to the central wall have a complete description and have been computed. It is therefore important to fined means to determine all other invariants $c_k^{a,b}$. In our construction of the symmetric quiver $Q^{w}$ these $m$-Kronecker DT invariants correspond to an initially unknown number of nodes which we then solve for. In this paper we show that after reformulating the identity (\ref{Reineke-ids}) in terms of symmetric quivers this number of nodes can be determined by diagonalizing \cite{Jankowski:2022qdpnew} both sides of the identity.

We determine $c_k^{a,b}$ or equivalently BPS numbers $\Omega^{4d}_{k}(\gamma)$ of (wild) quivers by expressing them in terms of motivic Donaldson-Thomas invariants of corresponding symmetric quivers, which in turn we write in terms of invariants of $m$-loop quivers (known e.g. from \cite{reineke2011degeneratecohomologicalhallalgebra}) by employing the operation of quiver diagonalization introduced in \cite{Jankowski:2022qdpnew}. Diagonalization operation replaces a quiver $Q$ by another quiver $Q_{\infty}$ without arrows other than loops, i.e. a collection of $m$-loop quivers. Diagonalizations $D_s$ and $D_w$ of quivers $Q^s$ and $Q^w$ produce respectively diagonal quivers $Q^s_{\infty}$ and $Q^w_{\infty}$, according to the following commutative diagram
\begin{equation}  \label{picture:comparingdiagonalisations}
\begin{tikzcd} 
Q^{s} \arrow{r}{\varphi} \arrow[swap]{d}{D_s} & Q^{w} \arrow{d}{D_w} \\%
Q_{\infty}^{s} \arrow{r}{\varphi_\infty} & Q_{\infty}^{w}
\end{tikzcd}
\end{equation}
where $\varphi$ and $\varphi_{\infty}$ denote the operations that relate symmetric and diagonal quivers corresponding to two sides of original the wall-crossing identity. It follows that analogous diagram holds for Donaldson-Thomas invariants DT$(Q)$ of corresponding quivers $Q$ 
\begin{equation} 
\begin{tikzcd} \label{picture:comparingdiagonalisationsDT}
\text{DT}(Q_{\infty}^{s})  \arrow{r}{\sigma} \arrow[swap]{d}{\Sigma_s} & \text{DT}(Q_{\infty}^{w}) \arrow{d}{\Sigma_w} \\%
\text{DT}(Q^{s}) \arrow{r}{\text{Id}} & \text{DT}(Q^{w})
\end{tikzcd}
\end{equation}
where the maps involve identifying variables and summing over contributions for particular spin and charges. More precisely, Donaldson-Thomas invariants of the two symmetric quivers $Q^s$ and $Q^w$ must be the same after identification of parameters, hence the map between them is just the identity Id, while $\sigma$ is the map permuting the invariants of diagonalized quivers (again after a suitable identification of variables). $\Sigma_{s}$ and $\Sigma_{w}$ represent the summation over Donaldson-Thomas invariants of the diagonalized quivers to obtain those of $Q^{s}$ and $Q^{w}$. 

Expressing wall-crossing identities for BPS quivers in terms of symmetric quivers and invoking their diagonalization, ultimately enables us to write down an explicit formula that involves BPS invariants of all these quivers. In particular, this formula can be used to determine (wild) BPS invariants of original BPS quivers. It takes form 
\begin{align} \label{eq:3d4dDTrelation}
\Omega^{3d}_{\mathbf{d},\tilde{k}} = \sum_{\mathbf{T}_{\mathbf{d}, \tilde{k}}, \ C^{loop}_{\mathbf{T}}} \prod^{j, \ q_{max}}_{p,q =1} % \prod^{r^{max}_{pq}}_{r_{pq} = r^{min}_{pq}} 
\Big(\Omega^{4d}_{k}(\gamma)_{i_{p}} - r^{\mathbf{T}}_{pq}\Big) \  \Omega^{3d}_{C^{loop}_{\mathbf{T}}}, 
\end{align}  
where $\Omega^{3d}_{\mathbf{d},\tilde{k}}$ are the 3d BPS degeneracies of the symmetrized $m$-Kronecker quiver $Q^s$ with (vortex) charge $\mathbf{d}$ and spin $\tilde{k}$, and $\Omega^{4d}_{k}(\gamma)_{i_{p}}$ are the 4d (wild) BPS dyon degeneracies $\Omega^{4d}_{k}(\gamma)$ with charge $\gamma$ and spin $k$ represented by the node $i_{p}$ on the infinite quiver $Q^{w}$. Here $p$ labels the end point of the particular tree $\mathbf{T}_{\mathbf{d}, \tilde{k}}$ of unlinkings for the sum in (\ref{eq:3d4dDTrelation}). Furthermore, $\Omega^{3d}_{C^{loop}_{\mathbf{T}}}$ with $m = C^{loop}_{\mathbf{T}}$ are BPS degeneracies of the $m$-loop quivers that build $Q^{w}_{\infty}$ (after the diagonalization) and arise at the ends of the trees of unlinkings $\mathbf{T}_{\mathbf{d}, \tilde{k}}$. We sum over all the trees defined in such a way that they contribute to a particular $\Omega^{3d}_{\mathbf{d},\tilde{k}}$ (meaning a specific final identification in the unlinking sequence) as well as all the possible number of loops that arise on the node produced in the final step in this tree. The complete derivation of this tree formula is given in section \ref{sec:wildDT} where equation (\ref{eq:3d4dDTrelation}) is written out in full detail as equation (\ref{eq:detailed3d4dDTrelation}) and then as (\ref{eq:simplified3d4drelation}).

The formula (\ref{eq:3d4dDTrelation}) and its solution in some specific examples are among the main results of this paper. As the 3d BPS invariants count open states describing M2 branes ending on M5 branes, the formula (\ref{eq:3d4dDTrelation}) can be thought of as an open-closed relation. 
As this formula involves trees of successive unlinkings, we expect it should match the split attractor flow formula of Manschot and Pioline \cite{D4Manschot:2010xp,AlexandrovPiolinehttps://doi.org/10.48550/arxiv.1804.06928}, which was recently applied to wild Kronecker quivers in \cite{Bousseau:2022snmnewattractorflowscatteringdiagram}. Furthermore, this suggests a general relation between the unlinking of symmetric quivers and split flow lines, as well as cluster scattering diagrams. There has been recent work on deriving a formulation of wild Donaldson-Thomas invariants in terms of polynomials with binomial coefficients \cite{akagi2024explicitformslowerdegrees}, and it would be interesting to see whether this can be matched with our result in terms of $m$-loop quivers and trees of unlinkings formulae. Finally, we conjecture that (\ref{eq:3d4dDTrelation}) can be generalized to all 4d $\mathcal{N}=2$ class $\mathcal{S}$ theories with BPS spectrum encoded in a Kontsevich-Soibelman motivic wall-crossing formula \cite{KontsevichSoibelman2008}. 

The plan of the paper is as follows. In section \ref{sec:4dN=2classStheories} we introduce the notation and review basic properties of 4d $\mathcal{N}=2$ class $\mathcal{S}$ theories, and in section \ref{sec:3dN=2boundarytheories} we present analogous introduction to 3d $\mathcal{N}=2$ theories. In section \ref{sec:4dN=2wall-crossingand3dsymmetricquivers} we discuss in general how to determine 3d symmetric quivers associated to 4d BPS quivers. In section \ref{sec:pureSU2} we reinterpret wall-crossing formula for pure $SU(2)$ theory in terms of symmetric quivers and corresponding 3d $\mathcal{N}=2$ theories. In section \ref{sec:wildDT} we interpret wild wall-crossing formulae, for $m$-Kronecker quivers with $m\geq 3$, in terms of symmetric quivers and 3d theories, and determine wild refined BPS invariants. In section \ref{sec-summary} we summarize our work and list possible future directions. In appendix \ref{Appendix:normal ordering} we present details of computation of operator ordering 
 that determines a symmetric quiver. In appendix \ref{appendix:diagonalize} we provide details of diagonalization of symmetric quivers that enable to determine refined BPS invariants. 

%***********************************
%***********************************
%***********************************

\section{$4d$ $\mathcal{N}=2$ class $\mathcal{S}$ theories} \label{sec:4dN=2classStheories}

In this section we summarize basic facts about 4d $\mathcal{N}=2$ class $\mathcal{S}$ theories, in which the BPS states we are counting arise. These theories can be constructed by a twisted compactification of a 6d (2,0) theory on a punctured Riemann surface $\mathcal{C}$ \cite{WKBGMNGaiotto:2009hg,wildwallGalakhov:2013oja}.
The Seiberg-Witten curve $\Sigma$ is a $K$-sheeted branched cover of $\mathcal{C}$. These theories can be labeled as $\mathcal{S}[\mathcal{C},\mathfrak{g},D]$ where $\mathfrak{g}$ is an ADE Lie algebra, where we specifically let $\mathfrak{g}= A_{K-1}$, and $D$ is a set of defect operators at the punctures. These theories include $SU(2)$ Seiberg-Witten and wild quiver theories that we consider in this paper. 

\subsection{BPS structures and BPS quivers}

The data associated to these 4d $\mathcal{N}=2$ theories can be arranged into a BPS structure \cite{WKBGMNGaiotto:2009hg,Alim:2023doi, Alim:2024ezg}. Let $\mathcal{B}$ be the Coulomb branch, i.e. the moduli space of vaccum expectation values of scalar fields in these theories. This is an $r$-dimensional complex manifold, where $r$ is the rank of the gauge group, on which a local coordinate takes form $(u_{1}, \ldots , u_{r})$. We say that there is a local system $\Gamma$ on $\mathcal{B}$ with fiber $ \Gamma_u \cong \mathbb{Z}^{2r}$ when considering gauge charges. Physically, the charge lattice is $\Gamma_u$ and contains both electric and magnetic charges. Often the local sections are written as $\gamma \in \Gamma$ informally. Furthermore, $\Gamma$ has a symplectic pairing
\begin{equation}\label{pairing}
\langle -, - \rangle \colon \Gamma \times \Gamma \to \underline{\mathbb{Z}},
\end{equation}
and $\underline{\mathbb{Z}}$ is the constant sheaf that has fiber $\mathbb{Z}$. Locally, on the moduli space, it is possible to split the charge lattice into electric and magnetic charge sublattices 
$\Gamma = \Gamma^e \oplus \Gamma^m$, with a magnetic basis $\{ \alpha_1,\dots,\alpha_r\}$ for $\Gamma^m$ and an electric basis for $\{ \beta^1,\dots,\beta^r\}$ for $\Gamma^e$, so that
\begin{equation}
\langle \alpha_I , \beta^J \rangle =\delta_I^J \,, \quad  \langle \alpha_I , \alpha_J \rangle = 0 = \langle \beta^I , \beta^J \rangle,\qquad I, J = 1, \dots, r.
\end{equation}
Therefore, denoting the electric-magnetic charges as $(p,q)$, all $\gamma \in \Gamma$ take form
\begin{equation}
\gamma=\sum_{I=1}^r p^I \alpha_I + q_I \beta^I,
\end{equation}
and the electric-magnetic Dirac pairing between the charges $\gamma_{i}, \gamma_{j} \in \Gamma$ is
\begin{equation}
\langle \gamma_{i}, \gamma_{j}\rangle = \sum_{I=1}^{r} (p_{i})^I (q_{j})_I - (p_{j})^I (q_{i})_I\,.
\end{equation}
To define central charges of the BPS states we assume the existence of a holomorphic map
\begin{equation}
Z \colon \mathcal{B} \to \mathrm{Hom}(\Gamma, \mathbb{C})\,
\end{equation}
such that the central charge takes the form of the holomorphic function $ Z_\gamma(u) := Z(u) \cdot \gamma $. The mass of the BPS state is encapsulated in the function
\begin{equation}
M \colon \mathcal{B} \to \mathrm{Map}(\Gamma, \mathbb{R})\,,
\end{equation}
and, of course, the BPS bound must be satisfied for all $\gamma \in \Gamma$
and $u \in \mathcal{B}$:
\begin{equation}\label{eq:bpsbound}
M_\gamma(u):= M(u)\cdot \gamma \geq |Z_\gamma (u)| .
\end{equation}
The BPS states are the states that saturate this bound so that $M_\gamma(u)= |Z_\gamma (u)|$. The charge lattice $\Gamma$ can be considered a sublattice of $H_{1}(\Sigma, \mathbb{Z})$, and one can define a meromorphic 1-form $\lambda$ on $\Sigma$ such that the central charge and mass can be written as      
\begin{equation}\label{eq:bpsbound}
Z_\gamma (u) = \int_{\gamma} \lambda_{u} , \  \  \  \   M_\gamma (u) = \int_{\gamma}  |\lambda_{u}| .
\end{equation}

Once the central charge is introduced, we define the BPS particles and antiparticles respectively as the states with central charges in the upper and lower half of the $Z$ plane. CPT invariance implies that every particle with charge $\gamma$ has a corresponding antiparticle with charge $-\gamma$. Including flavor charges, we can consider a charge lattice $\Gamma$ of rank $2r+f$ and find a minimal basis of hypermultiplets $\gamma_{i}$ such that the charge of any BPS particle can be written as a positive integral linear combination of these basis charges
\begin{equation}
\gamma =\sum_{i=1}^{2r+f}n_{i}\gamma_{i}, \qquad n_{i}\in \mathbb{Z}^{+}. \label{nsum}
\end{equation}
One can then use this minimal basis to construct a BPS quiver by assigning a node to each charge $\gamma_i$, and for $\langle \gamma_{i}, \gamma_{j}\rangle > 0$ introducing $\langle \gamma_{i}, \gamma_{j}\rangle$ arrows from the node corresponding to charge $\gamma_{j}$ to the node corresponding to $\gamma_{i}$  \cite{cecottivafacompleteclassificationhttps://doi.org/10.48550/arxiv.1103.5832,alim2011n2new,Alim:2024ezg,Lecturenotescecotti2010trieste}. Such BPS quivers are of our main interest. Recall that the BPS algebra originally defined in terms of tensor products of Hilbert spaces by Harvey and Moore \cite{Harvey_1996, Harvey_1998} is now encapsulated in the quiver BPS algebra as a Yangian \cite{Li:2023zub} or CoHA \cite{KontsevichCoHA:2010px, categorialwallcrossingGaiotto:2023dvs, framedcategorialwallcrossingGaiotto:2024fso}. 

%\subsection{4d $\mathcal{N} = 2$ BPS quivers}

%Quivers have long been used to characterize supersymmetric theories. Originally they arose when describing D-brane worldvolume theories \cite{QuiverBackgroundDouglasMoorehttps://doi.org/10.48550/arxiv.hep-th/9603167,Diaconescu_1998BPSquivers,BPSorbifoldquiverFiol_2000,QuiversnoncompactCYDouglas_2005}, for example in type IIA string theory on ALE spaces. More recently it was realized that they can be emplyed to classify 4d $\mathcal{N}=2$ theories and characterize their BPS spectra \cite{cecottivafacompleteclassificationhttps://doi.org/10.48550/arxiv.1103.5832,N=2BPSquivershttps://doi.org/10.48550/arxiv.1112.3984,Lecturenotescecotti2010trieste}. 

%**********

\subsection{Wall-crossing} \label{sec:wallcrossing}

At a generic point in $\mathcal{B}$ the BPS spectrum is invariant under continuous deformation of the moduli, however it can jump discontinuously at special loci called walls of marginal stability when bound states are either formed or decay into constituents, which is referred to as wall-crossing \cite{alim2011n2new, FramedBPSGMNhttps://doi.org/10.48550/arxiv.1006.0146, WKBGMNGaiotto:2009hg,GMN4d3d}. The wall of marginal stability $\text{$MS_{\gamma_{i}, \gamma_{j}}$}$ for bound states of $\gamma_{i}$ and $\gamma_{j}$ occurs when the central charges of these states align, meaning that $\text{Im}[ Z_{\gamma_{i}}(u)\bar{Z}_{\gamma_{j}}(u)] = 0$ and $|Z_{\gamma_{i}}(u) + Z_{\gamma_{j}}(u) | = |Z_{\gamma_{i}}(u)| + |Z_{\gamma_{j}}(u)|$, so that the change of ordering of the phases induces a jump in the BPS spectrum. The number of bound states on one side of the wall (often referred to at the weak coupling chamber) may be finite or infinite, depending on the theory and the electro-magnetic pairing  $\langle \gamma_{i}, \gamma_{j}\rangle$ between the charges. Determining the BPS spectrum in the chamber with bound states is in general a challenge. The methods to determine it include conducting a series of quiver mutations that involve redefining the upper half-plane in $Z$ \cite{alim2011n2new}, the spectral networks that use the data of BPS solitons interpolating between defects \cite{spectralnetworksGaiotto:2012rg}, or the flow tree formula that represents the process of successively crossing walls starting from the attractor chamber and using the wall-crossing formula for split flows to build the spectrum \cite{D4Manschot:2010xp,quiver2Manschot:2014fua,AlexandrovPiolinehttps://doi.org/10.48550/arxiv.1804.06928, Bousseau:2022snmnewattractorflowscatteringdiagram}.     

% \subsubsection{BPS and Donaldson-Thomas invariants}

Recall that the BPS Hilbert space has a grading of the form $\mathcal{H}_{BPS} = \bigoplus_{\gamma \in H_{1}(\Sigma_{u}, \mathbb{Z})} \mathcal{H}_{(\gamma,u)}$. It is useful to encode degeneracies of BPS states in the index called the second helicity supertrace, which is piecewise constant but jumps at walls of marginal stability \cite{FramedBPSGMNhttps://doi.org/10.48550/arxiv.1006.0146, wildwallGalakhov:2013oja} 
\begin{align} \label{eq:4dDTindex}
 \Omega(\gamma,u) = -\frac{1}{2} \text{Tr}_{\mathcal{H}_{(\gamma,u)}} (2J_{3})^{2}(-1)^{2J_{3}},
\end{align}
where $J_{3}$ can be considered an angular momentum. This index can be refined to include the spin degeneracies by instead considering a finite dimensional representation $\mathfrak{h}$ of $\mathfrak{so}(3) \oplus \mathfrak{su}(3)_{R}$, the Clifford vacuum, after factoring out a half-hypermultiplet representation.  With the $R$-symmetry generator $I_{3}$, one then obtains the protected spin character 
\begin{align} \label{eq:4drefinedDTindex}
 \Omega(\gamma,u,y) =  \text{Tr}_{\mathfrak{h}_{\gamma}} y^{2J_{3}}(-y)^{2I_{3}}
 =  y^{(\mathbf{\gamma}, \mathbf{\gamma})-1} \sum_{k} \Omega^{4d}_{k}(\gamma)\, y^{k}.
\end{align}
The coefficients of its expansion in $y$ encode BPS invariants $\Omega^{4d}_{k}(\gamma)$, and in order to match the conventions with the quiver representation theory discussed below we introduce an overall shift in powers of $y$, which involves the Euler form $(\mathbf{\gamma}, \mathbf{\gamma})$ defined below in (\ref{Euler}). The protected spin character recovers the helicity supertrace in the $y \rightarrow -1$ limit. From a mathematical perspective, these BPS invariants are conjectured to correspond to (motivic) Donaldson-Thomas (DT) invariants counting stable objects in the derived category of coherent sheaves on Calabi-Yau 3-folds used to construct the theory. 
% For a theory with a BPS quiver description, the DT invariants are defined in terms of dimensions of intersection homology of moduli spaces of stable representations of the quiver \cite{reineke2023wildquantumdilogarithmidentities}.

% \subsubsection{Kontsevich-Soibelman wall-crossing formulae}

Let us also recall the formulae that encapsulate how the above indices change upon wall-crossing. In unrefined case, these formulae involve a symplectomorphism $\mathcal{K}_{\gamma}$ acting on functions $x_{\gamma}$ on a complex torus $T_{u}$, which can be considered a fibre of $T = \Gamma^{*} \otimes_{\mathbb{Z}} \mathbb{C}^{\times}$ \cite{KontsevichSoibelman2008}. The functions obey $x_{\gamma_{1}}x_{\gamma_{2}} = x_{\gamma_{1}+\gamma_{2}}$ and the operators also have an interesting physical interpretation, for example, in the context of GMN \cite{FramedBPSGMNhttps://doi.org/10.48550/arxiv.1006.0146} where 4d $\mathcal{N} =2$ BPS states, now called vanilla, bind to a line operator or large core charge to form framed BPS states. At the BPS walls $W_{\gamma}$ at which these boundstates of charge $\gamma$ can form (in the unrefined case) one acts with $\mathcal{K}_{\gamma}$ to obtain the generating function on the other side of the BPS wall. If one considers two basis charges $\gamma_{1}, \gamma_{2}$ and a wall of marginal stability $MS_{\gamma_{1}, \gamma_{2}}$ then all the BPS walls intersect on the wall of marginal stability and one can consider a clockwise loop in the moduli space crossing all the BPS walls on both sides of $MS_{\gamma_{1}, \gamma_{2}}$. Every time one crosses a BPS wall one must act with the symplectomorphism to the power of the BPS invariant $\Omega(\gamma,u)$ from (\ref{eq:4dDTindex}). These products take the form:
\begin{align}
   A : = \prod^{ \curvearrowleft}_{ \substack{\gamma = n \gamma_{1}+ m \gamma_{2} \\ n,m > 0}}  \mathcal{K}^{\Omega(\gamma,u)}_{\gamma}.
   \end{align}
One must take this product on both sides of $MS_{\gamma_{1}, \gamma_{2}}$ and invert the order of the operators. This then amounts to the well known Kontsevich-Soibelman wall-crossing formula. 

\begin{figure}
\begin{center}
\includegraphics[width=0.2\textwidth]{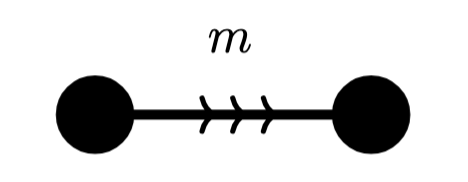}  
\end{center}
\caption{$m$-Kronecker quiver with two nodes connected by $m$ arrows.} 
\label{fig-quiver-m}
\end{figure}

Of our main interest are motivic versions of wall-crossing formulae that also include spin contributions of refined BPS degeneracies (\ref{eq:4drefinedDTindex}). 
They were originally derived by Kontsevich and Soibelmann \cite{KontsevichSoibelman2008} and can be considered as two ways of writing the character of the CoHA of the BPS quiver in question. They can be also derived physically, using framed BPS states by including all spin contributions on both sides of $MS_{\gamma_{1},\gamma_{2}}$. These formulae involve non-commutative variables $X_{\gamma}$ for $\gamma \in \Gamma$ as elements of a quantum torus algebra, such that
\begin{align} \label{eq:4dquantumtorus}
 X_{\gamma_{1}}X_{\gamma_{2}} = (-q^{\frac{1}{2}})^{ \langle \gamma_{1}, \gamma_{2}\rangle} X_{\gamma_{1}+\gamma_{2}} = q^{\langle \gamma_{1}, \gamma_{2}\rangle} X_{\gamma_{2}}X_{\gamma_{1}},   \end{align}
where $q^{\frac{1}{2}} = -y$ from the protected spin character 
 \cite{DimofteGukovSoibelman}. They satisfy commutation relations
\begin{align}
\bigl[X_{\gamma_{1}}, X_{\gamma_{2}} \bigr] = ((-q^{\frac{1}{2}})^{ \langle \gamma_{1}, \gamma_{2}\rangle}-(-q^{-\frac{1}{2}})^{ \langle \gamma_{1}, \gamma_{2}\rangle}) X_{\gamma_{1}+\gamma_{2}}.
\end{align}
The motivic wall-crossing formulae then take form of quantum dilogarithm identities. One series of such identities, which are of our main interest and correspond to Kronecker quivers shown in fig. \ref{fig-quiver-m}, can be written following Reineke in the form \cite{reineke2023wildquantumdilogarithmidentities}
\begin{equation} \label{eq:motivicwallcrossingformula}
  \Phi_{(1,0)}  \Phi_{(0,1)}     =    \Phi_{(0,1)}   \Phi_{\sigma (0,1)}  \Phi_{\sigma^{2} (0,1)}  \Phi_{\sigma^{3} (0,1)}  \cdots \prod^{\rightarrow}_{\mu_{-} < a/b < \mu_{+}} \Phi_{a,b}^{P_{a,b}}  \   \  \cdots \Phi_{\sigma^{-3} (1,0)}   \Phi_{\sigma^{-2} (1,0)}  \Phi_{\sigma^{-1}(1,0)}  \Phi_{(1,0)}.
\end{equation}
In this setup we write a general charge as $\gamma = a \gamma_{2}+b \gamma_{1}$, denote the electric-magnetic product by $m =\langle \gamma_{2}, \gamma_{1}\rangle $, introduce the operator $\sigma(a,b) = (b, mb-a)$, and $\mu_{\pm} = \frac{m \pm \sqrt{m^{2}-4}}{2}$ are solutions of the quadratic equation $x^{2}-mx+1=0$. We also introduce 
\begin{align}
\Phi_{a,b} = \Phi( (-1) ^{mab}q^{ \frac{a^{2} +  b^{2} -2mab-1 }{2} } x^{a}y^{b})^{(-1)^{a^{2}+b^{2}-mab-1}} ,
\end{align}
with $x = X_{\gamma_{2}}$ and $y = X_{\gamma_{1}}$, which is a redefinition of a quantum dilogarithm
\begin{align}
\Phi(x) = \sum_{k \geq 0} \frac{q^{\frac{k}{2}}}{(q,q)_{k}} x^{k},\qquad \textrm{for}\ (\alpha , q )_{k} = \prod^{k-1}_{i = 1} (1- \alpha q^{i})  \label{qdilog}
\end{align}
In particular $\Phi_{(1,0)} = \Phi(x)$ and $\Phi_{(0,1)} = \Phi(y)$. Furthermore
\begin{align} \label{phi}
\Phi^{P_{a,b}}_{a,b} = \prod_{k}\Phi( (-1) ^{mab}q^{ \frac{a^{2} +  b^{2} -2mab-1+k }{2} } x^{a}y^{b})^{(-1)^{(a^{2}+b^{2}-mab-1+k)} c^{a,b}_{k} },
\end{align}
where a polynomial\footnote{Taking the positivity condition on the DT invariants into account this becomes a polynomial in $\mathbb{N}[q]$.}
\begin{align} \label{eq:generatingfunction}
P_{a,b}(q^{\frac{1}{2}}) = \sum_{k} c^{a,b}_{k} \ (-q^{\frac{1}{2}})^{k} \in \mathbb{Q} [q^{\pm \frac{1}{2}}]
\end{align}
has coefficients that are identified with physical refined BPS degeneracies counting the states with charge vector $(a,b)$ in the expansion (\ref{eq:4drefinedDTindex})
\begin{align}
    c^{a,b}_{k} \equiv \Omega^{4d}_{k}(\gamma), \qquad \textrm{for}\ \gamma= a \gamma_{2}+b \gamma_{1},
\end{align}
so that, upon the identification $y\equiv -q^{\frac12}$, we have
% captures refined BPS degeneracies from (\ref{eq:4drefinedDTindex}) and its coefficients count the states with charge vector $(a,b)$. We now denote the refined BPS degeneracies in the expansion (\ref{eq:4drefinedDTindex}) by $\Omega^{4d}_{k}(\gamma= a \gamma_{1}+b \gamma_{2}) = c^{a,b}_{k}$. However, we should note that  there is a shift in the powers of $y\equiv -q^{\frac12}$ variables relative to the definition of $\Omega(\gamma,u,y)$. We can write out the expansion  
\begin{align} \label{eq:4drefinedDTindexexpansion}
 \Omega(\gamma,u, -q^{\frac{1}{2}}) =  (-q^{\frac{1}{2}})^{(\mathbf{\gamma}, \mathbf{\gamma})-1} \sum_{k} \Omega^{4d}_{k}(\gamma)\ (-q^{\frac{1}{2}})^{k} = (-q^{\frac{1}{2}})^{(\mathbf{\gamma}, \mathbf{\gamma})-1}P_{a,b}(q^{\frac{1}{2}}),
 \end{align}
 %for example in \cite{DimofteGukovSoibelman} where the prefactor $(-q^{\frac{1}{2}})^{(\mathbf{d}, \mathbf{d})-1}$ in (\ref{eq:DTpoly}) is not incorporated into the definition of the dilogarithm in (\ref{phi}) and a basis of purely electric and magnetic charges i.e. $\gamma_{1}$ and $\gamma_{1}+\gamma_{2}$ is used rather than the monopole and dyon that we are using. In this alternative labelling $k$ would be shifted and can be negative. 
where we emphasize that $(\gamma, \gamma)$ is not the Dirac pairing but the Euler form in (\ref{Euler}).  
Note the shift in the label $k$ (arising from the prefactor $(-q^{\frac{1}{2}})^{(\mathbf{\gamma}, \mathbf{\gamma})-1}$) does not affect the values of the expansion coefficients. Geometrically, this polynomial captures dimensions in intersection homology 
%The polynomial for $c^{a,b}_{k}$ is part of a general geometric property of the DT invariants that in full can be written out as  
\cite{meinhardt2016donaldsonthomasinvariantsversusintersection,reineke2023wildquantumdilogarithmidentities}   
\begin{align} \label{eq:DTpoly}
 \text{DT}^{\mu}_{\mathbf{d}}(q) = (-q^{\frac{1}{2}})^{(\mathbf{d}, \mathbf{d})-1} \sum_{k \geq 0} \text{dim IH}^{k}(M^{\mu -sst}_{\mathbf{d}}(Q), \mathbb{Q)}(-q^{\frac{1}{2}})^{k} = (-q^{\frac{1}{2}})^{(\mathbf{d}, \mathbf{d})-1} P_{\mathbf{d}}(q),
 \end{align}
where $\mathbf{d} =(a,b)$ is the dimension vector of the quiver representation, $\mu$ denotes the stability of the DT invariants, $M^{\mu -sst}_{\mathbf{d}}(Q)$ is the moduli space of representations of the quiver $Q$ with the dimension vector $\mathbf{d}$ that are also $\mu$-semistable (sst),
and $\text{IH}^{k}(M^{\mu -sst}_{\mathbf{d}}(Q), \mathbb{Q)}$ is the intersection homology of this moduli space graded by $k$. We define the set of nodes $Q_{0} \in \{i_{1}, \ldots , i_{l} \}$ in $Q$ and use the ordering $i > j$ if $i_{i} \rightarrow i_{j}$. For $\mathbf{d}, \mathbf{e} \in \mathbb{Z}Q_{0}$,
the Euler form of $Q$ is now defined as  
\begin{align}   \label{Euler}
(\mathbf{d}, \mathbf{e}) = \sum_{i \in Q_{0}} d_{i}e_{i} - \sum_{\alpha: i \rightarrow j} d_{i}e_{j} .
\end{align}

%*******

% \subsubsection{$A_{2}$ and pure $SU(2)$}

The left-hand side of the wall-crossing identity (\ref{eq:motivicwallcrossingformula}) represents BPS states encoded in the $m$-Kronecker quiver shown in fig. \ref{fig-quiver-m}, which consists of two nodes connected by $m=\langle \gamma_2,\gamma_1 \rangle$ arrows. Let us summarize basic properties of such systems for various values of $m$. The simplest example occurs for $m = 1$, which represents $A_2$ quiver that characerizes Argyres-Douglas theory. In this case (\ref{eq:motivicwallcrossingformula}) 
reduces to pentagon identity
\begin{align}
\Phi(X_{\gamma_{2}})\Phi(X_{\gamma_{1}}) = \Phi(X_{\gamma_{1}}) \Phi(X_{\gamma_{1}+\gamma_{2}}) \Phi(X_{\gamma_{2}}),
\end{align}
whose right-hand side is finite and 
apart from quantum dilogarithms representing the initial electric and magnetic monopole, it also includes an additional factor that represents their bound state (a dyon).

The case $m=2$, with BPS quiver shown in fig. \ref{fig-quiverSU2strong} (left), corresponds to pure $SU(2)$ Seiberg-Witten theory whose spectrum, upon crossing $MS_{\gamma_{1}, \gamma_{2}}$, jumps to infinitely many hypermultiplet dyons with charges $n\gamma_{1}+ (n+1)\gamma_{2}$ and $(n+1)\gamma_{1}+ n\gamma_{2}$ as well as a W-boson $\gamma_{1}+\gamma_{2}$. In this case (\ref{eq:motivicwallcrossingformula}) reduces to 
\begin{align}
\begin{split}   \label{m2identity}
\Phi(X_{\gamma_{2}}) \Phi(X_{\gamma_{1}}) = & \  
 \Phi(X_{\gamma_{1}})\Phi(X_{\gamma_{1}+(\gamma_{1}+ \gamma_{2})}) \Phi(X_{\gamma_{1}+2(\gamma_{1}+ \gamma_{2})})\cdots
  \Phi(X_{\gamma_{1}+n(\gamma_{1}+ \gamma_{2})}) \cdots  \\  
& \times  \Phi(q^{\frac{1}{2}} X_{(\gamma_{1}+ \gamma_{2})})^{-1}  \Phi(q^{-\frac{1}{2}}X_{(\gamma_{1}+ \gamma_{2})})^{-1} \times \\ 
& \cdots
\Phi(X_{\gamma_{2}+n(\gamma_{1}+ \gamma_{2})})\cdots \Phi(X_{\gamma_{2}+2(\gamma_{1}+ \gamma_{2})})
 \Phi(X_{\gamma_{2}+(\gamma_{1}+ \gamma_{2})}) \Phi(X_{\gamma_{2}}).
\end{split}
\end{align}

%\subsubsection{Wild wall-crossing and pure $SU(3)$}

For $m \geq 3$ the wall-crossing is referred to as wild, as it is much more complicated and the motivic wall-crossing formula (\ref{eq:motivicwallcrossingformula}) must be used in full generality. In the weak coupling chamber there is again an infinite spectrum, but now for all the ratios of the charges in the range $\mu_{-} < \frac{a}{b} < \mu_{+}$ every charge $(a,b)$ has a non-vanishing degeneracy, meaning that inside this cone the BPS walls are dense. We denote this region by $\mathcal{C}$ and we define the regions on either side of $\mathcal{C}$ as $\mathcal{C}^{-}$ for $\frac{a}{b} < \mu_{-}$ and $\mathcal{C}^{+}$ for
$\frac{a}{b} > \mu_{+}$.
This dense region makes finding these BPS degeneracies particularly difficult and methods such as quiver mutations cannot be used effectively. Physically, these wild spectra can be realized in pure $SU(3)$ $\mathcal{N}=2$ theory, where all the $m$-Kronecker quivers can be realized as subquivers of that of the full theory after mutation \cite{wildwallGalakhov:2013oja}, meaning there are walls $MS_{\gamma_{i}, \gamma_{j}}$ for all $m \geq 3$. These are called $m$-cohorts or $m$-herds when realized in spectral networks. There have also been works on wild $m$-Kronecker BPS indices from the perspective of quiver quantum mechanics \cite{Kim:2015fbaJK,Kim:2015oxawildJK,Cordova:2015zrawildJK,Beaujard:2019pknwildJK} where the index is computed using Jeffery-Kirwan residues arising from localisation. Interestingly, in \cite{Cordova:2015qkatreewildJK} a formula for the degeneracies of a particular class of charges was given in terms of trees derived from the contour integral. It would be interesting to relate these to our tree formula derived in sec. \ref{sec:wildDT}. 

The aim of our paper is to develop a combinatorial model for BPS numbers $c^{a,b}_{k}$, by relating them to symmetric quivers and 3d $\mathcal{N}=2$ theories that we review next.

%**************************
%**************************
%**************************

\section{$3d$ $\mathcal{N}=2$ boundary theories and symmetric quivers} \label{sec:3dN=2boundarytheories}  

Of our interest are also 3d $\mathcal{N}=2$ theories that can be considered as boundary theories of the 4d $\mathcal{N}=2$ class $\mathcal{S}$ theories or as a component of 4d-3d coupled system. They can be constructed by considering a toric CY 3-fold $X$ in an M-theory setup and wrapping an M5 brane on a Lagrangian submanifold $L$ and $\mathbb{R}^{2} \times S^{1}$ \cite{Knotquivercorrespondence1Kucharski:2017ogk,multiskeinsEkholm:2019lmb,Ekholm:2018eee} 
\begin{align*}
\text{space}  \  \   \  \   \  \   \  \  \    \mathbb{R}^{4} \times S^{1} \times X 
\end{align*}
\vspace{-1.75cm}
\begin{align*}
\hspace{-0.55cm}
 \text{\phantom{....................''..}} \cup  \ \text{\phantom{aa.........}} \cup 
\end{align*}
\vspace{-1.7cm}
\begin{align*}
\hspace{-0.8cm}
\text{M5 brane}  \  \   \  \   \  \   \  \  \  \mathbb{R}^{2} \times S^{1} \times L 
\end{align*}
One then gets an effective 3d $\mathcal{N}=2$ theory $T[L]$ living on $\mathbb{R}^{2} \times S^{1}$ in which the spectrum of M2 branes with boundary on the M5 brane becomes that of BPS vorticies. The vorticies are localised at the origin of the $\mathbb{R}^{2}$ so that they can be considered to rotate around the $S^{1}$. The K-theoretic vortex partition function coincides with the the open topological string partition function counting holomorphic M2 brane disks with boundary on a Lagrangian M5 brane, and can be also written as a motivic generating series (\ref{eq:symmetricquiver}) for some symmetric quiver $Q$ that encodes a dual abelian Cherns-Simons matter theory $T[Q]$. This is a gauge theory with a $U(1)$ at every node with a charged chiral multiplet, with (shifted) effective Chern-Simons couplings coinciding with the elements $C_{ij}$ of the adjacency matrix of the quiver $Q$, and with Fayet-Illiopolis (FI) parameters mapped onto the generating variables in the quiver generating series. The charges of the vorticies correspond to the entries of the dimension vector of the quiver. There is a network of dualities of such 3d $\mathcal{N}=2$ theories, which have the same generating series however correspond to different symmetric quivers. 
% Later in this work we relate these symmetric quivers to 4d $\mathcal{N}=2$ BPS quivers, and use this network of dualities to derive the 4d BPS invariants $\Omega(\gamma,y,u)$ in terms of the open counts of these BPS vorticies. 

% \subsubsection{Relation to 4d $\mathcal{N}=2$ theory}

Note that the 4d $\mathcal{N}=2$ theories discussed in the previous section can also be constructed using an M5 brane worldvolume theory, however compactified on the Seiberg-Witten curve $\Sigma$ rather than the Lagangian 3-manifold $L$. As with the 3d theory the cycles $\gamma_{i} \in H_{1}(\Sigma)$ can be considered M2 brane boundaries for basic disks. It was shown in GMN \cite{spectralnetworksGaiotto:2012rg, WKBGMNGaiotto:2009hg} using spectral networks that bound states arise from new curves formed by connecting these disks when they intersect. This was later used in \cite{quiverspectralnetworkGabella:2017hpz} to construct 4d $\mathcal{N}=2$ BPS quivers including symmetric examples. One can look at the 3d $\mathcal{N}=2$ theory as a boundary theory by considering BPS particles of $T[\Sigma]$ in the 4d spacetime $\mathbb{R}^{3} \times \mathbb{R}_{>0}$ that when moved to the boundary become BPS vorticies on a 3d theory
$T[M]$ such that $\Sigma = \partial M$.

There has also been recent work relating protected operator algebras in 4d $\mathcal{N}=2$ SCFTs to 3d $\mathcal{N}=4$ chiral boundary algebras \cite{Dedushenko:2023cvd} using the cigar reduction. Followups on how to construct the 3d theories using 4d $\mathcal{N}=1$ Maruyoshi-Song Lagrangians and the IR formula for the Schur index respectively include \cite{ArabiArdehali:2024ysy,Gaiotto:2024ioj4d3d}. Further work building families of these theories is described in \cite{ArabiArdehali:2024vli,Kim:2024dxu, Go:2025ixu4d3d}. In this case a 3d $\mathcal{N}=2$ boundary theory flows in the infrared to an $\mathcal{N}=4$ theory. The Schur indices correspond to the characters of these algebras and take the form of a Nahm sum \cite{Gang:2024loaNahmsum3d4d}, as has also been found for 3d $\mathcal{N}=2$ half indices \cite{Chung:2023qth}. The relation between 4d Argyres-Douglas theories and 3d $\mathcal{N}=2$ theories including their Schur indicies is also presented in detail in \cite{kllnps}.

%**********
%**********

\subsection{Symmetric quivers and Nahm sums}

To proceed we need to introduce the notation and review some properties of symmetric quivers  \cite{KontsevichSoibelman2008,KontsevichCoHA:2010px,Knotquivercorrespondence2Kucharski:2017poe, Knotquivercorrespondence1Kucharski:2017ogk,Efimov_2012,reineke2011degeneratecohomologicalhallalgebra, Jankowski:2022qdpnew, Jankowski:2021flt}. In general, a quiver $Q$ is a pair $(Q_{0},Q_{1})$ where $Q_{0}$ is the set of vertices (or nodes) and $Q_{1}$ is the set of arrows between vertices. The number of arrows between vertices $i$ and $j$ is denoted $C_{ij}$, which can be regarded as an element of the adjacency matrix of a quiver $C$ of size $|Q_{0}|$. A quiver is symmetric if $C_{ij} = C_{ji}$, i.e. the number of arrows from $i$ to $j$ is equal to the number of arrows from $j$ to $i$. 

Define the dimension vector $\mathbf{d} \in \mathbb{N}^{|Q_{0}|}$ of $Q$ as a vector in an integral lattice with the vertices $Q_{0}$ as a basis. A quiver representation with $\mathbf{d} = (d_{0}, \ldots  , d_{|Q_{0}|})$ is an assignment of a complex vector space at every node $i$ having dimension $d_{i}$ together with the linear maps $\gamma_{ij}: \mathbb{C}^{d_{i}} \rightarrow \mathbb{C}^{d_{j}}$ assigned to arrows between $i$ and $j$. We then look at the moduli spaces of stable representations of these quivers. They are characterized by motivic Donaldson-Thomas invariants, which are identified with intersection homology Betti numbers 
for the moduli space of semi-simple representations having dimension vector $\mathbf{d}$ or the equivalent for the Chow Betti numbers of simple representations, and physically are counts of 3d $\mathcal{N}=2$ BPS vortices  \cite{multiskeinsEkholm:2019lmb, Jankowski:2022qdpnew}. These invariants are encoded in a motivic generating series that takes form of a Nahm sum, with generating parameters $x_{i}$  associated to vertices $i \in Q_{0}$
\begin{align} \label{eq:symmetricquiver}
  P_{Q}(\mathbf{x},q) = \sum_{\mathbf{d}} (-q^{\frac{1}{2}})^{\mathbf{d} \cdot C \cdot \mathbf{d}} \frac{\mathbf{x^{\mathbf{d}}}}{(q;q)_{\mathbf{d}}} =  \sum_{d_{1}, \ldots , d_{|Q_{0}|} \geq 0} (-q^{\frac{1}{2}})^{\sum^{|Q_{0}|}_{i,j = 1} C_{ij}d_{i} d_{j}} \prod^{|Q_{0}|}_{i = 1} \frac{x_{i}^{d_{i}}}{(q;q)_{d_{i}}}.
\end{align}
This motivic generating series can be factorized into quantum dilogarithms with exponents identified with motivic DT invariants $\Omega_{\mathbf{d},s}$, such that $(-1)^{s+1}\Omega_{\mathbf{d},s}$ are positive integers 
\begin{align}
  P_{Q}(\mathbf{x},q) = \prod_{\mathbf{d},s} (\mathbf{x}^{\mathbf{d}} q^{\frac{s}{2}} ;q)^{\Omega_{\mathbf{d},s}}_{\infty} = \prod_{\mathbf{d} \in \mathbb{N}^{|Q_{0}|} \setminus \mathbf{0}} \prod_{s \in \mathbb{Z}} \prod_{k \geq 0} (1- (x^{d_{1}}_{1} \cdots  x^{d_{|Q_{0}|}}_{|Q_{0}|}) q^{k + \frac{s}{2}})^{\Omega_{(d_{1}, \ldots , d_{|Q_{0}|}),s}}.
\end{align}
One can also write down a motivic generating series of DT invariants for symmetric quivers
\begin{align} \label{eq:3dsymmetricDTinvariants}
\Omega(\mathbf{x},q) = \sum_{\mathbf{d}} \Omega_{\mathbf{d}}(q) \mathbf{x}^{\mathbf{d}} =  \sum_{\mathbf{d} \in \mathbb{N}^{|Q_{0}|} \setminus \mathbf{0}}\sum_{s \in \mathbb{Z}}\Omega_{d_{1}, \ldots , d_{|Q_{0}|},s}  \mathbf{x}^{\mathbf{d}} q^{\frac{s}{2}}.
\end{align}
% Later in this paper we will develop a combinatorial model expressing the 4d $\mathcal{N}=2$ BPS invariants $\Omega(\gamma, u, y)$ in terms of these DT of symmetric quivers counting vorticies in the 3d $\mathcal{N}=2$ boundary theory.    

%**********
%**********

\subsection{Quiver generating series from quantum torus algebra}

Quiver generating series (\ref{eq:symmetricquiver}) is annihilated by operators of the form
\begin{align}
\hat{A}_{i} (\mathbf{x}, \mathbf{y}) P_{Q} (\mathbf{x}, q) = 0,\qquad \hat{A}_{i} (\mathbf{x}, \mathbf{y}) =  1 - \hat{y}_{i} - \hat{x}_{i}(-q^{\frac{1}{2}} \hat{y}_{i})^{C_{ii}} \prod_{j \neq i} \hat{y}_{j}^{C_{ij}}.
\end{align}
They are referred to as quantum quiver A-polynomials and reduce to classical quiver A-polynomials in $q=1$ limit \cite{multiskeinsEkholm:2019lmb,Larraguivel:2020sxk}. The operators $\hat{x}_{i}$ and $\hat{y}_{i}$ are defined by the action on the functions $f=f(x_{1}, \ldots , x_{m}, y_{1}, \ldots , y_{m})$
\begin{align}
\hat{x}_{i} f = x_{i} f, \qquad
\hat{y}_{i} f(x_{1}, \ldots , x_{m}, y_{1}, \ldots , y_{m}) =   f(x_{1}, \ldots , q x_{i}, \ldots , x_{m}, y_{1}, \ldots , y_{m})
\end{align}
and they generate a new quantum torus algebra
\begin{align} \label{eq:3dquantumtorus}
\hat{y}_{i} \hat{x}_{j} =  q^{\delta_{i,j}} \hat{x}_{j} \hat{y}_{i},   \qquad \hat{x}_{i} \hat{x}_{j} =   \hat{x}_{j} \hat{x}_{i},  \qquad \hat{y}_{i} \hat{y}_{j} =   \hat{y}_{j} \hat{y}_{i}.
\end{align}
% which we will later explain can be related to the quantum torus algebra already discussed, which arises in the argument of the quantum dilogarithim the context of 4d $\mathcal{N}=2$ wall-crossing.
%
%\subsubsection{Symmetric quivers and Nahm sums from non-commutative variables}
%
One can write the symmetric quiver generating series (\ref{eq:symmetricquiver}) using this quantum torus algebra as follows \cite{multiskeinsEkholm:2019lmb}. Consider a symmetric quiver $Q$ and add an extra node, labeled as $0$, with $l$ loops and $v_{i}$ pairs of oppositely oriented arrows between the node $0$ and nodes $i\in Q$. Call this new quiver as $Q'$. Its generating series can be written as
\begin{align}
\begin{split}
P_{Q'}(x_{0},\mathbf{x},q) & =  \sum_{d_{0}, \mathbf{d}} (-q^{\frac{1}{2}})^{\mathbf{d} \cdot C \cdot \mathbf{d} + 2 d_{0} \mathbf{v} \cdot \mathbf{d} + l d^{2}_{0}} \frac{\mathbf{x^{\mathbf{d}}} x^{d_{0}}_{0}}{(q;q)_{\mathbf{d}}(q;q)_{d_{0}}} =  \\
& = \sum_{d_{0}} (-q^{\frac{1}{2}})^{l d^{2}_{0}} \frac{ x^{d_{0}}_{0}}{(q;q)_{d_{0}}} \sum_{\mathbf{d}}(-q^{\frac{1}{2}})^{\mathbf{d} \cdot C \cdot \mathbf{d}} \frac{(-q^{\frac{1}{2}})^{ 2 d_{0} \mathbf{v} \cdot \mathbf{d}} \mathbf{x}^{\mathbf{d}}}{(q;q)_{\mathbf{d}}} = \\ 
& = \Biggl [ \sum_{d_{0} \geq 0} \frac{(-q^{\frac{1}{2}})^{l d^{2}_{0}}}{(q;q)_{d_{0}}} \hat{x}^{d_{0}}_{0} \biggl( \prod_{i} \hat{y}^{v_{i}}_{i} \biggr)^{d_{0}} \Biggr] P_{Q} (\mathbf{x},q) = \\
& = \Phi \biggl( (-1)^{l} (q^{\frac{1}{2}})^{l-1} \hat{x}_{0} \hat{y}^{l}_{0} \prod_{i} \hat{y}^{v_{i}}_{i} \biggr) P_{Q} (\mathbf{x},q),
\end{split}
\end{align}
where $\Phi$ is the quantum dilogarithm (\ref{qdilog}) and we used quantum torus algebra to write $(\hat{x}_{0} \hat{y}^{k}_{0})^{n} = \hat{x}^{n}_{0} \hat{y}^{n k}_{0} q^{(n^{2}-n) k/2}$. This can be  interpreted as adding a node to the quiver $Q$ by action of an operator. Assigning the series $P_{\varnothing}(q) = 1$ to an empty quiver $Q = \varnothing$, we can now construct the generating series of any symmetric quiver (\ref{eq:symmetricquiver}) iteratively, starting from $\varnothing$ and successively adding nodes with appropriate linking numbers upon the action of quantum dilogarithms 
\begin{align} \label{eq:symmetricquivernoncommutative}
 \mathbb{P}^{Q} = \Phi(X_{m}) \cdot \Phi(X_{m-1}) \cdots  \Phi(X_{1})
 \end{align}
with operator arguments 
\begin{align} \label{eq:quantumtorusnoncommutative}
X_{i} = (-1)^{C_{ii}} (q^{\frac{1}{2}})^{C_{ii}-1} \hat{x}_{i} \hat{y}^{C_{ii}}_{i} \prod_{j < i} \hat{y}^{C_{ij}}_{j}.
\end{align}
Amusingly, these operators satisfy quantum torus algebra 
\begin{align}  \label{algebra-XY}
X_{i} X_{j} =  q^{A_{ij}} X_{j} X_{i},\qquad A_{ij} =
 \begin{cases}
 C_{ij}  &  i > j    \\ 
  0        &  i =j  \  \  \\
 -C_{ij} &   i < j
 \end{cases} 
\end{align}
analogous to the one we found for 4d theories (\ref{eq:4dquantumtorus}). The series (\ref{eq:symmetricquiver}) arises now as
\begin{align} \label{eq:symmetricquivernormalordering} 
P_{Q}(\mathbf{x},q) = \ :\mathbb{P}^{Q}: \ =  \  :\Phi(X_{m}) \cdot \Phi(X_{m-1}) \cdots  \Phi(X_{1}):
\end{align}
where $:\, :$ is the normal ordering the quantum torus algebra generators, which amounts to moving every $\hat{y}_{i}$ to the right using (\ref{eq:3dquantumtorus}) and then removing it (in result of acting on the constant function 1). Note that it is possible to change the order of the nodes of the quiver $Q$ so that the number of dilogarithms stays the same but the variables $X_{i}$ are redefined with the quantum torus algebra and the product (\ref{eq:symmetricquivernoncommutative}) is subsequently reordered.

%**********************
%**********************
%**********************

\subsection{Linking and unlinking from multicover skein relations} \label{sec:linkingandunlinking}

As we already recalled, 3d $\mathcal{N}=2$ theories $T[Q]$ are dual to theories $T[L]$ constructed from M5 branes wrapping lagrangians $L$ in toric Calabi-Yau 3-folds $X$. M2 branes with boundary on the M5 brane can be reinterpreted as generalised holomorphic curves corresponding to open topological strings with boundary on $L$. This means that the quiver generating series also encodes open Gromov-Witten invariants. This has been directly shown for strip geometries with Aganagic-Vafa branes in \cite{Panfil_2019}. More precisely, the quiver generating series do not exactly count holomorphic curves but generalized versions of them that correspond to combinations of multiple covers of basic disks \cite{multiskeinsEkholm:2019lmb}. This requires us to introduce a Morse function $f$ on $L$ and a 4-chain $C$ with boundary $ \partial C = 2 \cdot L$. The  4-chain is compatible with the Morse function on its boundary. The purpose of the Morse function is to define linking numbers between the boundaries of basic holomorphic disks by constructing a bounding chain for these boundaries in $L$. The self linking of the disks corresponds to their intersections with the 4-chain. Looking back at the symmetric quivers we can match the adjacency matrix with these linking numbers so that every node corresponds to a basic disk and the number of pairs of arrows in the quiver corresponds to the linking numbers of the disk boundaries. The self-linking  is the number of loops in the quiver. It is interesting to see what happens to the quiver when we unlink these disks.

For actual holomorphic curves the curve count is not invariant under deformations in the moduli space because when the disk boundaries cross, the bound states disappear in standard wall-crossing. However, for generalized holomorphic curves this works differently because we are now considering multiple covers. Now when the disk boundaries cross one can consider the glued basic disks as a new basic disk in place of the bound state. Before the crossing of the boundaries this glued disk could be thought of as the bound state. One can represent the holomorphic curve boundaries as elements in the \textit{framed skein module} of $L$. This can be thought of as a module of 1-dimensional defects in Cherns-Simons theory on $L$ (modulo the skein relation and isotopy). Now the unlinking of disk boundaries can be thought of as a \textit{multicover skein relation} of this framed skein module of $L$.

It should be noted that a similar construction involving multi-disks and linking numbers has been introduced in \cite{Iacovino:2009xf,iacovino2009open}. 
Interestingly in \cite{iacovino2018kontsevichsoibelmanwallcrossingformula} an interpretation of the motivic Kontsevich-Soibelman wall-crossing formula is given in terms of trees of these multi-disk relations although not in the context of symmetric quivers. In section \ref{sec:wildDT} we derive a formula for $m$-Kronecker DT invoving trees of unlinkings and it would be interesting to determine the relation between these formulations.  

There is a general pattern of how these disks can be unlinked under these multicover skeins, and, when one introduces redundant pairs of nodes (by cancelling pairs of dilogarithms corresponding to 0 and 1-loop quivers), also how such disks can be linked \cite{multiskeinsEkholm:2019lmb,UnlinkingKucharski:2023jds}. This can be encoded in the quiver $Q$ by adding an extra node with a particular specialization of variables for this new node such that the generating series and the motivic DT invariants are preserved. More precisely, unlinking of two nodes $a$ and $b$ in $Q$ yields a new quiver $Q^{u}$ with an extra node $n$, $Q^{u}_{0} = Q_{0} \cup \{n \}$, and with the number of arrows
\begin{align} \label{eq:unlinking}
    C^{u}_{ab} = C_{ab} - 1, \quad C^{u}_{nn} = C_{aa} + 2C_{ab} + C_{bb} -1 , \quad C^{u}_{in} = C_{ai} + C_{bi} - \delta_{ai} - \delta_{bi},   
\end{align}
and $C^{u}_{ij} = C_{ij}$ otherwise. Specializing the generating parameters of $Q^u$ as
\begin{align} \label{unlinking-identification}
x_{n}^{u}(x_{1}, \ldots , x_{|Q_{0}|}, q) = q^{-\frac{1}{2}}x_{a}x_{b}, \qquad
 & x_{i}^{u}(x_{1}, \ldots , x_{|Q_{0}|}, q) = x_{i}  \qquad  \forall\, i \neq n  
\end{align}
guarantees that generating series of $Q$ and $Q^u$ are the same. 

There are analogous rules for the linking of nodes $a$ and $b$, which also produces a new quiver $Q^{l}$ with $Q^{l}_{0} = Q_{0} \cup \{n \}$ from $Q$. In this case
\begin{align} \label{eq:linking}
C^{l}_{ab} = C_{ab} + 1, \quad C^{l}_{nn} = C_{aa} + 2C_{ab} + C_{bb} , \quad C^{l}_{in} = C_{ai} + C_{bi},  
\end{align}
and $C^{l}_{ij} = C_{ij}$ otherwise. Adjusting the generating parameters of $Q^u$ as follows
 \begin{align}
x_{n}^{l}(x_{1}, \ldots , x_{|Q_{0}|}, q) = x_{a}x_{b},  \qquad x_{i}^{l}(x_{1}, \ldots , x_{|Q_{0}|}, q) = x_{i}  \quad   \forall \, i \neq n   
 \end{align}
 also makes the generating series of $Q$ and $Q^u$ equal.

%*************
%*************
%*************

\section{$4d$ wall-crossing and $3d$ symmetric quivers} \label{sec:4dN=2wall-crossingand3dsymmetricquivers}

The main result of this work is reinterpretation of the Kontsevich-Soibelman wall-crossing identities (\ref{eq:motivicwallcrossingformula}) of the 4d $\mathcal{N}=2$ class $\mathcal{S}$ theories in the realm of 3d boundary theories discussed in section \ref{sec:3dN=2boundarytheories}. We achieve this by writing both sides of these identities in the form (\ref{eq:symmetricquivernormalordering}) and using the representation of the quantum torus algebra (\ref{eq:quantumtorusnoncommutative}). The two sides of the wall of marginal stability in the 4d $\mathcal{N}=2$ theory get then an interpretation in 3d as two dual theories with identical vortex partition functions, taking form of two  symmetric quiver generating series (\ref{eq:symmetricquiver}) with suitable identification of parameters. In the Argyres-Douglas $A_{2}$ model the simple jump from 2 to 3 BPS states according to pentagon identity, when reinterpreted in terms of symmetric quivers, becomes the unlinking operation (\ref{eq:unlinking}) that removes a pair of arrows from two nodes and creates a new node with a single loop.  %However, for pure $SU(2)$ the 3d quiver for the weak coupling region already has infinitely many nodes corresponding to the spectrum of Seiberg-Witten theory. 
In what follows we generalize this process to $SU(2)$ theory and wild $m$-Kronecker quivers, and determine the adjacency matrix of an infinite symmetric quiver on the other side of the wall of marginal stability.  
% by first doubling the linear Kronecker quiver and then applying the Kontsevich-Soibelman wall-crossing formula. By identifying generators of the two quantum torus algebras and applying the normal ordering operation we determine the adjacency matrix of the infinite symmetric quiver on the other side of the wall of marginal stability.   
%
% The unlinking (\ref{eq:unlinking}) operation removes one pair of arrows between two nodes and introduces one extra node. 
For $m$-Kronecker quivers, our results thus can be regarded as a generalization of the unlinking, such that $m$ pairs of arrows between two nodes (in a symmetrized $m$-Kronecker quiver) are removed at once, at the same time generating
%obtained from symmetrising the wild wall-crossing in the 4d $\mathcal{N}=2$ class $\mathcal{S}$ theories is a generalisation of this transformation to the case of 2 nodes and $m$ pairs of arrows which removes these pairs of arrows in a single step with the addition of 
infinitely many new nodes. In the language of multi-cover skeins we have infinitely many bound states of multiple covers of two basic disks becoming infinitely many basic disks together with a new pattern of bound states of these new basic disks.

%*****************

\subsection{Doubling (symmetrizing) 4d $\mathcal{N}=2$ BPS quivers} \label{sec:doubledquivers}

We focus first on expressions involving products of a finite number of quantum dilogarithms, which are in one-to-one correspondence with the nodes of a 4d BPS quiver. Such expressions appear on the left-hand side of the wall-crossing identities, such as (\ref{eq:motivicwallcrossingformula}). We show that representing the operator arguments of these quantum dilogarithms in the form (\ref{eq:quantumtorusnoncommutative}) enables us to identify a symmetric quiver, which includes a doubling or symmetrization of the 4d BPS quiver, i.e. for each arrow in the BPS quiver there is a new arrow in the opposite direction in the resulting quiver. In fact, the values of $C_{ii}$ could be arbitrary and we choose to set them to zero, so that (\ref{eq:quantumtorusnoncommutative}) takes form
\begin{align}  \label{eq:quantumtorusnoncommutative2}
X_{\gamma_{i}} = X_{i} =  (q^{\frac{1}{2}})^{-1} \hat{x}_{i} \prod_{j < i} \hat{y}^{\langle \gamma_{i}, \gamma_{j}\rangle}_{j},
\end{align}
with $\langle \gamma_{i}, \gamma_{j}\rangle$ given by the numbers of arrows in the original BPS quiver. For other values of $C_{ii}$, the symmetric quivers that we find in what follows would have extra loops. Furthermore, if quantum torus generators $X_i$ satisfy extra relations, the resulting symmetric quiver has additional loops and identifications between generating parameters. 

\begin{figure}
\begin{center}
\includegraphics[width=0.9\textwidth]{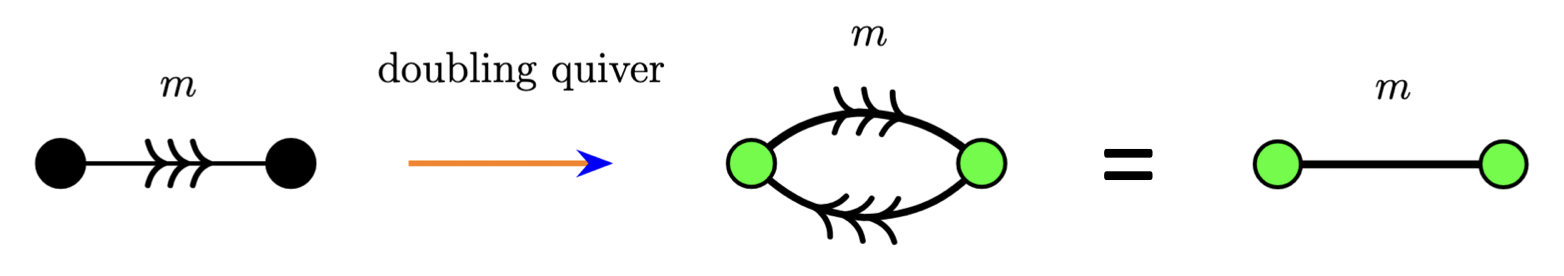}  
\end{center}
\caption{Doubling $m$-Kronecker quiver (left) produces a symmetric quiver (middle). In what follows we often represent a set of arrows between two nodes by a thick segment with the number of arrows typed next to it (right).} 
\label{fig-double}
\end{figure}

Consider first (\ref{eq:motivicwallcrossingformula}) with two quantum dilogarithms on the left-hand side, $\Phi(x)\Phi(y)$, representing $m$-Kronecker, and express it in the form (\ref{eq:symmetricquivernoncommutative})
%At this point we must symmetrise or double the $m$-Kronecker quiver by replacing every arrow by a pair of arrows. This can be thought of as the symmetric quiver on the side of $MS$ in the 4d $\mathcal{N}=2$ theory with just 2 BPS states which should for $m$-arrow kronecker quivers coincide, on the 3d boundary theory, with the symmetric quivers with $m$ pairs of arrows. This doubling of the Kronecker quivers can be done by considering the product of dilogarithms generating the 4d spectrum on that side of the wall which for Kronecker quivers is
%\begin{align}
%\Phi(x)\Phi(y)
%\end{align}
\begin{align}
 \mathbb{P}^{Q} =  \Phi(X_{2} ) \Phi(X_{1} )   \label{PPQ}
\end{align}
with the operator arguments (\ref{eq:quantumtorusnoncommutative2}) in this case taking form
\begin{align} \label{eq:quantumtorusident}
x = X_{2} =(q^{\frac{1}{2}})^{-1} \hat{x}_{2} \hat{y}_{1}^{m}, \qquad
y  = X_{1} =(q^{\frac{1}{2}})^{-1} \hat{x}_{1},
\end{align}
so that their algebra (\ref{algebra-XY}) indeed reproduces that of the $m$-Kronecker quiver (\ref{eq:4dquantumtorus}). Normal ordering (\ref{PPQ}), i.e. moving $\hat{y}_1$ to the right using the relation (\ref{eq:3dquantumtorus}), we find
\begin{align}
P_{Q}(\mathbf{x}, q) & = \ : \Phi( (q^{\frac{1}{2}})^{-1} \hat{x}_{2} \hat{y}_{1}^{m} )     \Phi(  (q^{\frac{1}{2}})^{-1} \hat{x}_{1}   ) : \ 
= \sum_{k \geq 0} \frac{(q^{\frac{1}{2}})^{k}}{(q,q)_{k}} (  (q^{\frac{1}{2}})^{-1} \hat{x}_{2} \hat{y}_{1}^{m} )^{k} \sum_{l \geq 0} \frac{(q^{\frac{1}{2}})^{l}}{(q,q)_{ l }} (  (q^{\frac{1}{2}})^{-1} \hat{x}_{1} )^{l}  \nonumber \\
& =  \sum_{k,l \geq 0} \frac{(-q^{\frac{1}{2}})^{2mkl}}{(q,q)_{k} (q,q)_{ l }  } (x_{2})^{k} ( x_{1} )^{l}.  \label{PPQ-ordered}
\end{align}
This final expression indeed takes form of a motivic generating series (\ref{eq:symmetricquiver}) for a symmetric quiver with two nodes and $m$ pairs of arrows, with adjacency matrix
\begin{align}  \label{C0mm0}
C= 
\begin{pmatrix}
0  & m \\
m & 0
\end{pmatrix}.
\end{align}
This means that we have doubled or symmetrized the $m$-Kronecker quiver, see fig. \ref{fig-double}. As explained in section \ref{sec:3dN=2boundarytheories}, we can assign 3d $\mathcal{N}=2$ theory to such a symmetric quiver, which is a boundary theory for the 4d theory described by the original BPS $m$-Kronecker quiver.

\begin{figure}
\begin{center}
\includegraphics[width=0.9\textwidth]{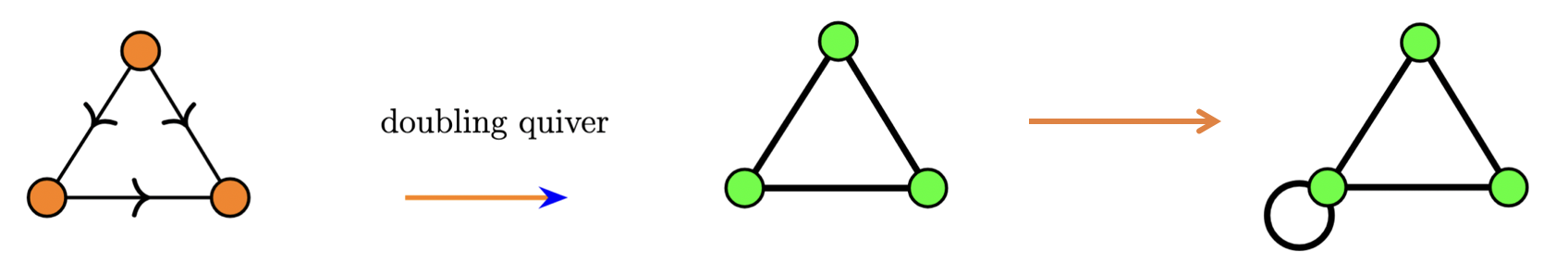}  
\end{center}
\caption{BPS quiver for $SU(2)$ theory with $N_f=1$ (left), symmetric quiver arising from doubling (middle), and the resulting symmetric quiver with extra loop arising from the relation $\gamma_{2} = \gamma_{1}+\gamma_{3}$ between charges, which encodes the corresponding 3d $\mathcal{N}=2$ theory (right).} 
\label{fig-quiverSU2Nf1}
\end{figure}

As another example we consider $SU(2)$ theory with $N_{f} = 1$. This theory has a strong coupling chamber with three BPS states with charges
$\gamma_{3} = (1,-1), \  \gamma_{2} = (1,0),   \  \gamma_{1} = (0,1) $. The BPS quiver is shown in fig. \ref{fig-quiverSU2Nf1} (left). We have $\langle  \gamma_{2}  ,   \gamma_{1} \rangle = 1,   \    \langle  \gamma_{3}  ,   \gamma_{2} \rangle = 1,     \  \langle  \gamma_{3}  ,   \gamma_{1} \rangle = 1$, so we can identify the quantum torus algebra as
\begin{align}
X_{\gamma_{1}} =  (q^{\frac{1}{2}})^{-1}\hat{x}_{1} , \quad  X_{\gamma_{2}} =  (q^{\frac{1}{2}})^{-1}\hat{x}_{2} \hat{y}_{1}, \quad  X_{\gamma_{3}} =  (q^{\frac{1}{2}})^{-1}\hat{x}_{3} \hat{y}_{2}  \hat{y}_{1}.
\end{align}
Now writing down (\ref{eq:symmetricquivernoncommutative}) and normal ordering we would get
\begin{align}
\begin{split}
: \Phi(X_{\gamma_{3}})   \Phi(X_{\gamma_{2}})\Phi(X_{\gamma_{1}}) : \ &  =  \ : \sum^{\infty}_{n,m,l=0} \frac{1}{(q;q)_{n} (q;q)_{m} (q;q)_{l} } ( \hat{x}_{3} \hat{y}_{2}  \hat{y}_{1} )^{n}( \hat{x}_{2} \hat{y}_{1} )^{m} ( \hat{x}_{1})^{l} :\ = \\
& = \sum^{\infty}_{n,m,l=0} \frac{(-q^{\frac{1}{2}})^{2nm+2nl+2ml}}{(q;q)_{n} (q;q)_{m} (q;q)_{l} }  x_{3}^{n} x_{2}^{m} x_{1}^{l},
\end{split}
\end{align}
which encodes a symmetric quiver shown in fig. \ref{fig-quiverSU2Nf1} (middle). However, in this case we also have to implement the relation for the 4d $\mathcal{N}=2$ charges $\gamma_{2} = \gamma_{1}+\gamma_{3}$, which in terms of the quantum torus algebra reads
\begin{align}
X_{\gamma_{2}} = X_{\gamma_{1}+\gamma_{3}} = (-q^{\frac{1}{2}}) X_{\gamma_{1}}X_{\gamma_{3}}    
\end{align}
and thus imposes a relation for the generators of the 3d quantum torus algebra
$\hat{x}_{2} (-\hat{y}_{2}) = - q^{-\frac{1}{2}}\hat{x}_{1}\hat{x}_{3} \hat{y}_{2} $.\footnote{Here we can insert $-q^{\frac{1}{2}}\hat{y}_{2}$ operator into the definition of $X_{\gamma_{2}}$ by hand to give a consistent identification for both algebras.} Taking this into account we get
\begin{align}
\begin{split}
P_{Q}(\mathbf{x}, q) & = \ : \sum^{\infty}_{n,m,l=0} \frac{1}{(q;q)_{n} (q;q)_{m} (q;q)_{l} } ( \hat{x}_{3} \hat{y}_{2}  \hat{y}_{1} )^{n}(-\hat{x}_{1}\hat{x}_{3} \hat{y}_{2}\hat{y}_{1} )^{m} ( \hat{x}_{1})^{l} : \ = \\ 
& = \sum^{\infty}_{n,m,l=0} \frac{(-q^{\frac{1}{2}})^{m^{2}+2nm+2nl+2ml}}{(q;q)_{n} (q;q)_{m} (q;q)_{l} }  x_{3}^{n} (q^{-\frac{1}{2}}x_{1}x_{3})^{m} x_{1}^{l} = \\
& = \sum^{\infty}_{n,m,l=0} \frac{(-q^{\frac{1}{2}})^{m^{2}+2nm+2nl+2ml}}{(q;q)_{n} (q;q)_{m} (q;q)_{l} } q^{-\frac{m}{2}} x_{1}^{m+l} x_{3}^{n+m}.
\end{split}
\end{align}
This expression encodes a symmetric quiver shown in fig. \ref{fig-quiverSU2Nf1} (right), with adjacency matrix
\begin{align}
C= 
\begin{pmatrix}
0  & 1 & 1 \\
1  & 1 & 1 \\
1  & 1 & 0 \\
\end{pmatrix}.
\end{align}
Therefore, the relation $\gamma_{2} = \gamma_{1}+\gamma_{3}$ gives rise to a loop on the second node and also imposes a specific identification of generating parameters $x_i$ assigned to the nodes of this quiver.

%*********************
%*********************

\subsection{Infinite symmetric quiver $Q^w$ for the dual 3d theory} \label{sec:infinitequivers}

We now reinterpret wall-crossing identities for 4d $\mathcal{N}=2$ theories in 3d terms. To be specific, we focus on wall-crossing formulae (\ref{eq:motivicwallcrossingformula}). We have already shown that two quantum dilogarithms on the left-hand side, associated to $m$-Kronecker quiver, represent strong coupling spectrum in 4d, and after using the representation (\ref{eq:quantumtorusident}) and normal ordering give rise to the motivic generating series (\ref{PPQ-ordered}) for a quiver that we denote $Q^{s}$, with $m$ pairs of arrows and the adjacency matrix (\ref{C0mm0}), which encodes some specific 3d $\mathcal{N}=2$ theory. We now focus on the right-hand side of this identity, which from 4d perspective represents weak coupling spectrum on the other side of the wall of marginal stability. Invoking normal ordering (\ref{eq:symmetricquivernormalordering}) we show that it also can be reinterpreted in terms of a symmetric quiver that we denote $Q^w$ that encodes a dual 3d $\mathcal{N}=2$ theory, so that the identity (\ref{eq:motivicwallcrossingformula}) yields equality of generating series of symmetric quivers $Q^s$ and $Q^w$ and a duality between corresponding 3d theories
\begin{align}
\begin{split}
& P_{Q^{s}}(\mathbf{x}, q) = \   :\Phi_{(1,0)}  \Phi_{(0,1)}: \     = \ P_{Q^{w}} (\mathbf{x}^{w}(\mathbf{x}), q) \  = \\
& =\  : \Phi_{(0,1)}   \Phi_{\sigma (0,1)}  \Phi_{\sigma^{2} (0,1)}  \Phi_{\sigma^{3} (0,1)}  \cdots \prod^{\rightarrow}_{\mu_{-} < a/b < \mu_{+}} \Phi_{a,b}^{P_{a,b}}  \   \  \cdots \Phi_{\sigma^{-3} (1,0)}   \Phi_{\sigma^{-2} (1,0)}  \Phi_{\sigma^{-1}(1,0)}  \Phi_{(1,0)} :
\end{split}
\end{align}
The quiver $Q^w$ has infinitely many nodes, one node for every dilogarithm on the right-hand side of (\ref{eq:motivicwallcrossingformula}), with an intricate pattern of loops and linkings. In particular, it turns out that two nodes originally linked by $m$ pairs of arrows in the quiver $Q^s$ are now unlinked both from each other and also from all the new nodes generated by the wall-crossing. This hereby represents a generalization of the unlinking related to pentagon identity, which removes one pair of arrows at a time between two nodes. We stress that quiver generating series  (\ref{eq:symmetricquiver}) and therefore the motivic DT invariants (\ref{eq:3dsymmetricDTinvariants}) for both quivers $Q^{s}$ and $Q^{w}$ are the same, after specializing the generating parameters in a way that follows from the normal ordering.

%************
%************

% \subsubsection{Adjacency matrix and diagram of a quiver $Q^w$}

The computation of the infinite adjacency matrix of the quiver $Q^{w}$ is conducted in appendix \ref{Appendix:normal ordering}. We just write it here in a compact form, and to this end let us introduce some notation. First, note that the product on the right-hand side of the identity (\ref{eq:motivicwallcrossingformula}) can be split into three factors: a product on the left-hand side (LHS), a product of factors in the (possibly dense) cone $\mathcal{C}$ in the middle, and a product on the right-hand side (RHS). Denote the number of the nodes on the LHS (counted inwards from the left) by $n$, on the RHS (also counting inwards, now from the right) by $n'$. Define
\begin{align} \label{LHSandRHS}
\begin{split}
\textrm{on the LHS of infinite product:} & \quad \   \   a =  \sigma^{n} (0,1)_{1} =  \sigma^{n,m} _{1}, \  \  \  b =  \sigma^{n} (0,1)_{2} =\sigma^{n,m}_{2}, \\
\textrm{on the RHS of infinite product:} & \quad \   \   a =  \sigma^{n'} (0,1)_{2} =  \sigma^{n',m} _{2}, \  \  \  b =  \sigma^{n'} (0,1)_{1} =\sigma^{n',m}_{1},
\end{split}
\end{align}
where $1$ and $2$  represent the first and second component of the vector $  \sigma^{n} (0,1)$. Further, we label the nodes inside the dense cone by $i$ starting with the lowest ratio of $(a_{i},b_{i})$ and moving in the direction of increasing ratio. In contrast to the LHS and RHS in (\ref{LHSandRHS}), inside $\mathcal{C}$ all (integer) charges of general form $(a,b)$ are allowed. We denote the contributions to the numbers of loops in these three regions respectively by $r_{n}, \ r_{n'}, \ l_{i}$. They depend on whether the power of the dilogarithm is positive or negative. We find for the general form
\begin{align}
r_{n} = 
\begin{cases}
0     & \   \   \   (\sigma^{n,m} _{1})^{2}+  (\sigma^{n,m} _{2}) ^{2}-m \sigma^{n,m} _{1} \sigma^{n,m} _{2} -1  \   \   \in   \  2\mathbb{Z} \\
1     & \   \   \   (\sigma^{n,m} _{1})^{2}+  (\sigma^{n,m} _{2}) ^{2}-m \sigma^{n,m} _{1} \sigma^{n,m} _{2} -1  \   \   \in   \  2\mathbb{Z} + 1
\end{cases}   
\end{align}
and this also holds for $n'$.\footnote{This is because this expression is symmetric when permuting 1 and 2 which one must carry out when moving from $n$ to $n'$.} However, we should note that in practice that because we start from (0,1) and (1,0) and there is invariance under the dihedral symmetry $\sigma_{a,b}$ we can determine $r_{n} = 0$ for all $n$. Therefore from now on we can assume this in the rest of the paper.

Inside the cone we have a slightly different situation as $(a,b)$ can take any value within the appropriate range and therefore following the same argument as the dilogarithm can indeed have negative powers. Furthermore (before positivity conditions are considered) the power of the dilogarithm depends on the spin number $k$
\begin{align}
l_{i} = 
\begin{cases}
0     & \   \   \ a^{2}+b^{2}-mab-1+k \   \   \in   \  2\mathbb{Z} \\
1     & \   \   \ a^{2}+b^{2}-mab-1+k \   \   \in   \  2\mathbb{Z} + 1
\end{cases}   
\end{align}
We stress that the form of the adjacency matrix of $Q^w$, and specifically the number of various nodes, depends on the Donaldson-Thomas invariants $c^{a,b}_{k}$ that we only determine later by comparing to the quiver in the strong coupling chamber $Q^s$ and by imposing the wall-crossing identity (\ref{eq:motivicwallcrossingformula}). So, with the above notation, the quiver matrix of $Q^w$ takes the form:

\newpage

\begin{scriptsize}
\begin{align*}
 & C =\  \  \ \phantom{aaaaaaaaaaaa} \text{Number of entries in cone:} \  \ \  \  \   c^{a_{1},b_{1}}_{k} \  \  \  \   \  \  \  \   \  \   \  \  \  \  \  \    c^{a_{2},b_{2}}_{\tilde{k}} \  \  \   \  \ \phantom{aaaaaaaaaaa..} 1  \phantom{aaaaaaaaaaaaaaaaaaa.} 1 \\ \nonumber 
 & 
 \hspace{-1.3cm}
 \left( \begin{array}{ccccccccccccccccccc}
 ... &  ...  &  ... &  ...  &... & ...  & ...  & ... &  ... &  ... & ... & ... & ...  \\
  ...  & \footnotesize{m  \sigma^{n,m} _{1}  \sigma^{n,m} _{2} + r_{n}} & ... & m  \sigma^{n_{1},m} _{1}  \sigma^{n_{2},m} _{2} &... & m  \sigma^{n,m} _{1} b_{1} & ... & m  \sigma^{n,m} _{1} b_{2} & ... & m  \sigma^{n,m} _{1}   \sigma^{n',m} _{1} & ... & ... & ...   \\
 ...  & ...  & ... &... &...   & ... & ... & ... & ... & ...  & ... & ... & ...  \\
 ...   & m  \sigma^{n_{1},m} _{1}  \sigma^{n_{2},m} _{2} & ... & m  \sigma^{n,m} _{1}  \sigma^{n,m} _{2} + r_{n} &...   & m  \sigma^{n,m} _{1} b_{1} & ... & m  \sigma^{n,m} _{1} b_{2} &  ... &  m  \sigma^{n,m} _{1}   \sigma^{n',m} _{1} & ... & m  \sigma^{n,m} _{1}   \sigma^{n',m} _{1} & ...   \\
 ...  & ...  & ... & ... & \textcolor{red}{...} & \textcolor{red}{...} & \textcolor{red}{...} & \textcolor{red}{...} & \textcolor{red}{...} & ...  & ...   & ... & ...  \\
 ...  & m  \sigma^{n,m} _{1} b_{1} & ... & m  \sigma^{n,m} _{1} b_{1} & \textcolor{red}{...} &  ma_{1}b_{1} + l_{1} &   ...  & ma_{1}b_{2} & \textcolor{red}{...} & ma_{1} \sigma^{n',m} _{1} & ...   & ma_{1} \sigma^{n',m} _{1} & ...  \\
 ...  & ...  & ... & ... & \textcolor{red}{...} & ... &  ...  & ... & \textcolor{red}{...} & ... & ... & ...  & ...    \\
  ...  & m  \sigma^{n,m} _{1} b_{2} & ... & m  \sigma^{n,m} _{1} b_{2} & \textcolor{red}{...} & ma_{1}b_{2} &  ...  & ma_{2}b_{2} + l_{2} & \textcolor{red}{...} & ma_{2} \sigma^{n',m} _{1} & ... & ma_{2} \sigma^{n',m} _{1} & ...   \\
...  & ...  & ... &...  & \textcolor{red}{...} & \textcolor{red}{...} & \textcolor{red}{...} & \textcolor{red}{...} & \textcolor{red}{...} & ... & ... & ... &  ...  \\
 ... & m  \sigma^{n,m} _{1}   \sigma^{n',m} _{1} & ... & m  \sigma^{n,m} _{1}   \sigma^{n',m} _{1} & ... & ma_{1} \sigma^{n',m} _{1} &  ...  & ma_{2} \sigma^{n',m} _{1} & ... & m  \sigma^{n',m} _{1}  \sigma^{n',m} _{2} + r_{n'} &  ... &  m  \sigma^{n'_{2},m} _{2}  \sigma^{n'_{1},m} _{1} & ...  \\
 ...  & ...  &  ...  & ...   &...& ... & ... &  ... & ... & ... &  ...  & ... & ...  \\
 ...  & ...  & ...  & m  \sigma^{n,m} _{1}   \sigma^{n',m} _{1} & ... & ma_{1} \sigma^{n',m} _{1} &  ...  & ma_{2} \sigma^{n',m} _{1} & ... &    m  \sigma^{n'_{2},m} _{2}  \sigma^{n'_{1},m} _{1} & ... & m  \sigma^{n',m} _{1}  \sigma^{n',m} _{2} + r_{n'}& ... \\
 ...  & ...  & ... & ... &... & ... & ... &  ...  & ... & ... & ... &  ... & ...   \\
 \end{array} \right),
\end{align*}

\end{scriptsize}

Let us also zoom in to the central region of the cone where the Donaldson-Thomas invariants are unknown and look more closely at the structure of loops and linkings. The factor of $(-1)^{a^{2}+b^{2}-mab-1+k}$ generates an alternating pattern of loops as one moves from $ k \rightarrow k+1 $. This means that if the number of loops at spin $k$ is $mab$ then the number becomes $mab+1$ at spin $k+1$. The reverse also holds if one starts with $mab+1$. However, at this point we note that the positivity conditions on the DT invariants in Reineke \cite{reineke2023wildquantumdilogarithmidentities} imply that for any $c^{a,b}_{k}$ we have $c^{a,b}_{k+1} = 0$. This means that on the diagram of $Q^{w}$ we could start by drawing an alternating pattern of loops as $k \rightarrow k+1$ but we will later find (when deriving the DT invariants in the next section) that the number of nodes at $k+1$ must vanish. 

Therefore, in this section we can use the adjacency matrix inside the dense cone $\mathcal{C}$ and the quiver diagram by only including the nodes (that actually exist) for spin $k \rightarrow k +2$.   
We now depict this by looking at the central region of the adjacency matrix $C$. We define a new matrix $A$ which contains just the central region of the adjacency matrix and the linkings to the outer region of the adjacency matrix.

\newpage

\begin{small}

\begin{align*}
 & A =  \ \phantom{aaaaaaaaa} \text{Number of entries:} \\  
 & \ \  \  \ \phantom{aaaaaaaa} c^{a_{1},b_{1}}_{k} \phantom{aaaaaaaaaaaaaaaaa..} c^{a_{1},b_{1}}_{k+2} \  \  \  \   \  \  \  \   \  \   \  \  \  \  \  \ \phantom{aaaaaaaaaaaaa} c^{a_{2},b_{2}}_{\tilde{k}} \phantom{aaaaaaaaaaaaaaaa} c^{a_{2},b_{2}}_{\tilde{k}+2} \\ \nonumber 
 & 
 \hspace{-1cm}
 \left( \begin{array}{ccccccccccccccccccc}
 ... &  ...  &  ... &  ...  &... & ...  & ...  & ... &  ... &  ... & ... & ... & ...  \\
 ... & m  \sigma^{n,m} _{1} b_{1} &  ... & m  \sigma^{n,m} _{1} b_{1} & m  \sigma^{n,m} _{1} b_{1} & ...  & m  \sigma^{n,m} _{1} b_{1} & ... & m  \sigma^{n,m} _{1} b_{2} &  ... & m  \sigma^{n,m} _{1} b_{2} & m  \sigma^{n,m} _{1} b_{2} & ...  \\
 ... &  ...  &  ... &  ...  &... & ...  & ...  & ... &  ... &  ... & ... & ... & ...  \\
  ...  & ma_{1}b_{1} & ... & ma_{1}b_{1} & ma_{1}b_{1} & ... & ma_{1}b_{1} & ... & ma_{1}b_{2} & ... & ma_{1}b_{2} & ma_{1}b_{2} & ...   \\
 ...  & ...  & ... & ... & ... & ... & ... & ... & ... & ...  & ... & ... & ...  \\
 ...   & ma_{1}b_{1} & ... & ma_{1}b_{1} & ma_{1}b_{1} & ...  & ma_{1}b_{1} & ... & ma_{1}b_{2} &  ...  & ma_{1}b_{2} & ma_{1}b_{2} & ...  \\
 ...  & ma_{1}b_{1} & ... & ma_{1}b_{1} & ma_{1}b_{1} & ... & ma_{1}b_{1} & ... & ma_{1}b_{2} & ...  & ma_{1}b_{2} & ma_{1}b_{2} & ...  \\
 ...  & ... & ... & ... & ... & ... & ... & ... & ... & ... & ...   & ... & ...  \\
...  & ma_{1}b_{1} & ... & ma_{1}b_{1} &  ma_{1}b_{1} & ... & ma_{1}b_{1} & ... & ma_{1}b_{2} & ... & ma_{1}b_{2} & ma_{1}b_{2} & ...    \\
  ...  & ... & ... & ... & ... & ... &  ...  & ...  & ... & ... & ... & ... & ...   \\
...  & ma_{1}b_{2} & ... & ma_{1}b_{2} & ma_{1}b_{2} & ... & ma_{1}b_{2} & ... &  ma_{2}b_{2}+1 & ... & ma_{2}b_{2} & ma_{2}b_{2} &  ...  \\
 ... & ...  & ... & ... & ... & ... &  ...  & ... & ... & ... & ... & ...  & ...  \\
 ...  & ma_{1}b_{2} &  ...  & ma_{1}b_{2} & ma_{1}b_{2} & ... & ma_{1}b_{2} &  ... & ma_{2}b_{2} & ... & ma_{2}b_{2}+1 & ma_{2}b_{2} & ...  \\
 ...  & ma_{1}b_{2} & ...  & ma_{1}b_{2} & ma_{1}b_{2} & ...  & ma_{1}b_{2} & ... & ma_{2}b_{2} &  ... & ma_{2}b_{2} & ma_{2}b_{2}+1 & ... \\
 ...  & ...  & ... & ... &... & ... & ... &  ...  & ... & ... & ... &  ... & ...   \\
...  & ma_{1} \sigma^{n',m} _{1} & ... & ma_{1} \sigma^{n',m} _{1} & ma_{1} \sigma^{n',m} _{1} & ... & ma_{1} \sigma^{n',m} _{1} & ... & ma_{2} \sigma^{n',m} _{1} & ... & ma_{2} \sigma^{n',m} _{1} & ma_{2} \sigma^{n',m} _{1}& ...   \\
...  & ...  & ... & ... &... & ... & ... &  ...  & ... & ... & ... &  ... & ...   \\
 \end{array} \right),
\end{align*}

\end{small}

Based on these matrices, in fig. \ref{infinitequiver1} we draw the general form of the quiver $Q^w$. The green nodes represent the dilogarithms outside the cone and the yellow nodes represent those inside the cone (their number is given by $c^{a,b}_{k}$ for each $a,b$ and spin $k$). We have put the yellow nodes with the same charges along a long row and have drawn two yellow nodes as the ends of a shorter row representing the set of $c^{a,b}_{k}$ nodes with spin $k$ (we also implement the $k \rightarrow k+2$ alternating pattern). For this diagram we have explicitly drawn $l_{i} =0$ for the first long row of nodes and $l_{i} =1$ for the second, but any combination is possible. 

%Then we have drawn another 2 nodes representing the end nodes of the adjacent set of $c^{a,b}_{k+1}$ where we have now increased the spin to $k+1$ such that .

%************ 

\begin{figure}[h!] 
	
	\begin{center}

 \vspace{-4.5cm}
 
		\begin{tikzpicture}

  \hspace{-5cm}

		%loops
		
		\path [draw=black,line width=0.8mm] (17.72, 3) circle (0.5cm);
		
		\path [draw=black,line width=0.8mm] (12.28, 3) circle (0.5cm);

         \path [draw=black,line width=0.8mm] (9.0, 3.7) circle (0.5cm);

         \path [draw=black,line width=0.8mm] (20.5, 3.7) circle (0.5cm);

         \path [draw=black,line width=0.8mm] (21.0, -2.3) circle (0.5cm);

         \path [draw=black,line width=0.8mm] (24.0, -2.3) circle (0.5cm);

         \path [draw=black,line width=0.8mm] (8.9, -2.3) circle (0.5cm);

         \path [draw=black,line width=0.8mm] (6.0, -2.3) circle (0.5cm);

		\path [draw=black,line width=0.8mm] (15.62,-12.75) circle (0.3cm);
		
		%\path [draw=black,line width=0.8mm] (15.62, 9.75) circle (0.3cm);
		
		\path [draw=black,line width=0.8mm] (15.62,-10.5) circle (0.3cm);
		
		\path [draw=black,line width=0.8mm] (15.62, 8) circle (0.3cm);
		
		\path [draw=black,line width=0.8mm] (15.62, 6.5) circle (0.3cm);

		%paths
		
		\path [draw=black,line width=0.8mm] (15, 6.3)--(17, 3) ;
		
		\path [draw=black,line width=0.8mm] (15, 6.5) to (13, 3) ;
		
		\path [draw=black,line width=0.8mm] (15,-10.5) [bend right= 15] to [bend left = -15] (17, 3) ;

		\path [draw=black,line width=0.8mm] (15,-10.5)--(13, 3) ;
		
		\path [draw=black,line width=0.8mm] (15,-12.5)--(15, -10.75);

         \path [draw=black,line width=0.8mm] (15,-12.5)--(15, 6.75);
         
		\path [draw=black,line width=0.8mm] (15,-4.5)--(15,-12.75);
		
		\path [draw=black,line width=0.8mm] (15, 4.5)--(15, 8.3);

		\draw[thick,line width=0.8mm]  (17, 3) [bend right=-15] to [bend left = 25] (15,-12.75);
		
		\draw[thick,line width=0.8mm]   (17.1, 3) [bend right= 15] to (15.12, 8) (16.98, 2.5) ;
		
		%paths between nodes on either side of the identity
		
		\draw[thick,line width=0.8mm] (15, -10.65) to [bend right= -65] (15, 7.9) ;
		
		\draw[dashed,line width=0.6mm] (14.9, 6.5) [bend right= 77] to [bend right= 77] (15, -12.5) ;
		
		%\draw[dashed,line width=0.8mm] (15, -4.5) parabola bend (15, 7.9) (8, 3);
		
		%\draw[dashed,line width=0.8mm] (15, 7.5)  parabola bend (15, -10.65) (8, -4);
		
		%paths between central nodes at infinity
		
		\path [draw=black,line width=0.8mm] (13, 3)--(17, 3);
		
		\path [draw=black,line width=0.8mm] (15, -12.75) [bend right= -5] to (12.85, 3);
		
		\path [draw=black,line width=0.8mm] (15, 8.3)--(12.85, 3);

		%nodes

\draw[thick,line width=0.3mm]  (6.0, -3.2) [bend right= 45] to [bend left = -25] (9, 3);

  \draw[thick,line width=0.3mm]  (6.0, -3) [bend right= 45] to [bend left = -45]( 8.9,-3);

  \draw[thick, dashed , line width=0.3mm]  (6.0, -3.2) [bend right= 20] to [bend left = -10]( 13, 3);

  \draw[thick, dashed , line width=0.3mm]  (6.0, -3.2) [bend right= 0] to [bend left = -0]( 17, 3);  
\draw[thick, dashed , line width=0.3mm]  (6.0, -3.2) [bend right= 0] to [bend left = -0]( 20.5, 3);  
\draw[thick, dashed , line width=0.3mm]  (6.0, -3.2) [bend right= 25] to [bend left = -25]( 21, -3.2);

\draw[thick, dashed , line width=0.3mm]  (6.0, -3.2) [bend right= 25] to [bend left = -25]( 23.9, -3.2); 

\draw[thick, dashed , line width=0.3mm]  (6.0, -3.2) [bend right= 25] to [bend left = -25]( 23.9, -3.2); 

\draw[thick , line width=0.3mm]  (6.0, -3.2) [bend right= 65] to [bend left = -65]( 15, -12.75); 

\draw[thick , line width=0.3mm]  (6.0, -3.2) [bend right= 65] to [bend left = -65]( 15, -10.5); 

\draw[thick , line width=0.3mm]  (6.0, -3.28) [bend right= -85] to [bend left = 85]( 15, 8); 

\draw[thick , dashed,  line width=0.3mm]  (9.0, -3) [bend right= -45] to [bend left = -45]( 17, 3);

\draw[thick , dashed,  line width=0.3mm]  (9.0, -3) [bend right= -35] to [bend left = -35]( 20.5, 3);

\draw[thick , dashed,  line width=0.3mm]  (9.0, -3) [bend right= -35] to [bend left = -35]( 12.8, 3);

\draw[thick , dashed,  line width=0.3mm]  (9.0, -3) [bend right= 100] to [bend left = 100]( 15, 8);

\draw[ dashed,  line width=0.3mm]  (9.0, -3) [bend right= 100] to [bend left = 100]( 15, 6.5);

\draw[ dashed,  line width=0.3mm]  (9.0, -3) [bend right= 100] to [bend left = 100]( 9, 3);

\draw[ dashed,  line width=0.3mm]  (9.0, -3) [bend right= -20] to [bend left = -20]( 21, -3);

\draw[ dashed,  line width=0.3mm]  (9.0, -3) [bend right= -80] to [bend left = -80]( 23.9, -3);

\draw[ thick,  line width=0.3mm]  (15, -10.7) [bend right= -60] to [bend left = -60]( 24, -3);

\draw[ thick,  line width=0.3mm]  (15, -12.75) [bend right= -70] to [bend left = -70]( 24, -3);

\draw[ thick, dashed,  line width=0.3mm]  (15, -12.75) [bend right= -70] to [bend left = -70]( 21, -3);

\draw[ thick, dashed,  line width=0.3mm]  (15, -10.7) [bend right= -70] to [bend left = -70]( 21, -3);

\draw[ thick,  line width=0.3mm]  (21, -3.1) [bend right= -40] to [bend left = -40]( 24, -3.05);

\draw[ thick, dashed,  line width=0.3mm]  (8.9, -3.1) [bend right= -70] to [bend left = -70]( 15, -12.75);

\draw[ thick, dashed,  line width=0.3mm]  (8.9, -3.1) [bend right= -70] to [bend left = -70]( 15, -10.5);

\draw[ thick,  line width=0.3mm]  (9, 3) [bend right= -30] to [bend left = -30]( 13, 3);

\draw[ thick,  line width=0.3mm]  (17, 2.9) [bend right= -50] to [bend left = -50]( 20.5, 2.9);

\draw[ thick,  line width=0.3mm]  (21, -3.2) [bend right= 20] to [bend left =  20]( 17.09, 3);

\draw[ thick,  line width=0.3mm]  (21, -3.2) [bend right= 20] to [bend left =  20]( 17.09, 3);

\draw[ thick,  line width=0.3mm]  (21, -3.2) [bend right= 40] to [bend left =  40]( 20.6, 3);

\draw[ thick,  line width=0.3mm]  (24, -3.1) [bend right= 25] to [bend left =  25]( 20.5, 3.1);

\draw[ thick, dashed,  line width=0.3mm]  (24, -3.2) [bend right= 0] to [bend left =  0]( 16.9, 2.9);

\draw[ thick, dashed,  line width=0.3mm]  (24, -3.2) [bend right= 0] to [bend left =  0]( 13, 2.9);

\draw[ thick, dashed,  line width=0.3mm]  (24, -3.2) [bend right= 0] to [bend left =  0]( 9, 2.9);

\draw[ thick,  line width=0.3mm]  (24, -3.2) [bend right= -100] to [bend left =  -100]( 15, 8);

\draw[ thick, dashed,  line width=0.3mm]  (24, -3.2) [bend right= -95] to [bend left =  -95]( 15, 6.5);

\draw[ thick, dashed,  line width=0.3mm]  (5.9, -3.2) [bend right= 85] to [bend left =  85]( 15, 6.5);

\draw[ thick, dashed,  line width=0.3mm]  (9,  3) [bend right= 40] to (5 ,-3 )  to [bend left =  -40]( 15, -12.75);

\draw[ thick, dashed,  line width=0.3mm]  (9,  3) [bend right= 0]  to [bend left =  -0](15, -10.5);

% \draw[ thick, dashed,  line width=0.3mm]  (13,  3) [bend right= 5]  to [bend left =  -5](15, -10.5);

%\draw[ thick, dashed,  line width=0.3mm]  (13,  3) [bend right= 20]  to [bend left =  -20](15, -12.75);

\draw[ thick, dashed,  line width=0.3mm]  (20.5,  3.2) [bend right= 0]  to [bend left =  -0](15, -10.5);

\draw[ thick, dashed,  line width=0.3mm]  (20.5,  3) [bend right= -40] to (25.1 ,-3 )  to [bend left =  40](15, -12.75);

\draw[ thick, dashed,  line width=0.3mm]  (20.5,  2.9) [bend right= 0] to (20.5 ,2.9 ) to [bend left =  -0](15, 6.4);

\draw[ thick, dashed,  line width=0.3mm]  (20.5,  3) [bend right= 60] to  (24 , 6 )  to [bend left =  -60](15, 8);

\draw[ thick, dashed,  line width=0.3mm]  (9,  3) [bend right= -60] to  (5.5 , 6 ) to [bend left =  60](15, 8);

\draw[ thick, dashed,  line width=0.05mm]  (20.5,  2.9) [bend right= 45] to (20.5 ,2.9 )  to [bend left =  -45](9, 2.9);

\draw[ thick, dashed,  line width=0.05mm]  (20.5,  2.9) [bend right= 45] to (20.5 ,2.9 )  to [bend left =  -45](13, 2.9);

\draw[ thick, dashed,  line width=0.05mm]  (15, -12.75) [bend right= -40] to (5,-3) to [bend left = 55]( 15, 8);

%\draw[ thick, dashed,  line width=0.05mm]  (15, -10.5) [bend right= 50] to (25,-3) to [bend left = -50](15, 6.5);

\draw[ thick, dashed,  line width=0.05mm]  (9, 3) [bend right= 60] to (17,3) to [bend left = -60](15, 6.5);

 \path [draw=black,fill=green,line width=0.4mm] (15, -15) circle (0.3cm);
		
		\path [draw=black,fill=green,line width=0.4mm] (15, -12.75) circle (0.3cm);
		
		\path [draw=black,fill=green,line width=0.4mm] (15, -10.5) circle (0.3cm);

		\path [draw=black,fill=green,line width=0.4mm] (15, 8) circle (0.3cm);
		
		\path [draw=black,fill=green,line width=0.4mm] (15, 9.75) circle (0.3cm);
		
		\path [draw=black,fill=green,line width=0.4mm] (15, 6.5) circle (0.3cm);

\path [draw=black,fill=yellow,line width=0.4mm] (17, 3) circle (0.3cm);
		
		\path [draw=black,fill=yellow,line width=0.4mm] (13, 3) circle (0.3cm);

        \path [draw=black,fill=yellow,line width=0.4mm] (9, 3) circle (0.3cm);

        \path [draw=black,fill=yellow,line width=0.4mm] (20.5, 3) circle (0.3cm); 

        \path [draw=black,fill=yellow,line width=0.4mm] (21, -3) circle (0.3cm);	

        \path [draw=black,fill=yellow,line width=0.4mm] (8.9, -3) circle (0.3cm); 

        \path [draw=black,fill=yellow,line width=0.4mm] (6.0, -3) circle (0.3cm);

        \path [draw=black,fill=yellow,line width=0.4mm] (24, -3) circle (0.3cm);

      \path [draw=red,fill=black,line width=0.4mm] (15, 8.9) circle (0.02cm);

      \path [draw= red,fill=black,line width=0.4mm] (15, 8.7) circle (0.02cm);	

      \path [draw= red,fill=black,line width=0.4mm] (15, 9.1) circle (0.02cm);

      \path [draw= red ,fill=black,line width=0.4mm] (15, -13.7) circle (0.02cm);

      \path [draw= red ,fill=black,line width=0.4mm] (15, -13.9) circle (0.02cm);

      \path [draw= red ,fill=black,line width=0.4mm] (15, -14.1) circle (0.02cm);

      \path [draw= red ,fill=black,line width=0.4mm] (11.2, 5.3) circle (0.02cm);

      \path [draw= red ,fill=black,line width=0.4mm] (11.2, 5.1) circle (0.02cm);

      \path [draw= red ,fill=black,line width=0.4mm] (11.2, 4.9) circle (0.02cm);

      \path [draw= red ,fill=black,line width=0.4mm] (18.5, 5.3) circle (0.02cm);

      \path [draw= red ,fill=black,line width=0.4mm] (18.5, 5.1) circle (0.02cm);

      \path [draw= red ,fill=black,line width=0.4mm] (18.5, 4.9) circle (0.02cm);

      \path [draw= red ,fill=black,line width=0.4mm] (18.5, -6.3) circle (0.02cm);

      \path [draw= red ,fill=black,line width=0.4mm] (18.5, -6.1) circle (0.02cm);

      \path [draw= red ,fill=black,line width=0.4mm] (18.5, -5.9) circle (0.02cm);

      \path [draw= red ,fill=black,line width=0.4mm] (9.5, -6.3) circle (0.02cm);

      \path [draw= red ,fill=black,line width=0.4mm] (9.5, -6.1) circle (0.02cm);

      \path [draw= red ,fill=black,line width=0.4mm] (9.5, -5.9) circle (0.02cm);

       \path [draw= red ,fill=black,line width=0.4mm] (9.2, 0.3) circle (0.02cm);

      \path [draw= red ,fill=black,line width=0.4mm] (9.2, 0.1) circle (0.02cm);

      \path [draw= red ,fill=black,line width=0.4mm] (9.2, 0.5) circle (0.02cm);

      \path [draw= red ,fill=black,line width=0.4mm] (20.6, 0.3) circle (0.02cm);

      \path [draw= red ,fill=black,line width=0.4mm] (20.6, 0.1) circle (0.02cm);

      \path [draw= red ,fill=black,line width=0.4mm] (20.6, 0.5) circle (0.02cm);

      \path [draw= blue ,fill=black,line width=0.4mm] (18.8, 2.8) circle (0.02cm);

      \path [draw= blue ,fill=black,line width=0.4mm] (19.0, 2.8) circle (0.02cm);

      \path [draw= blue ,fill=black,line width=0.4mm] (19.2, 2.8) circle (0.02cm);

      \path [draw= blue ,fill=black,line width=0.4mm] (9.9, 2.9) circle (0.02cm);

      \path [draw= blue ,fill=black,line width=0.4mm] (10.1, 2.9) circle (0.02cm);

      \path [draw= blue ,fill=black,line width=0.4mm] (10.3, 2.9) circle (0.02cm);

  \path [draw= blue ,fill=black,line width=0.4mm] (5.0, 2.9) circle (0.02cm);

      \path [draw= blue ,fill=black,line width=0.4mm] (5.2, 2.9) circle (0.02cm);

      \path [draw= blue ,fill=black,line width=0.4mm] (5.4, 2.9) circle (0.02cm);

       \path [draw= blue ,fill=black,line width=0.4mm] (22.5, 2.8) circle (0.02cm);

      \path [draw= blue ,fill=black,line width=0.4mm] (22.7, 2.8) circle (0.02cm);

      \path [draw= blue ,fill=black,line width=0.4mm] (22.9, 2.8) circle (0.02cm);

      \path [draw= blue ,fill=black,line width=0.4mm] (22.2, -3.35) circle (0.02cm);

      \path [draw= blue ,fill=black,line width=0.4mm] (22.4, -3.35) circle (0.02cm);

      \path [draw= blue ,fill=black,line width=0.4mm] (22.6, -3.35) circle (0.02cm);

      \node at (22.0, -2.9){\scalebox{0.85}{$c^{a_{2},b_{2}}_{k+2}$}};

      \path [draw= blue ,fill=black,line width=0.4mm] (7.2, -3.38) circle (0.02cm);

      \path [draw= blue ,fill=black,line width=0.4mm] (7.4, -3.38) circle (0.02cm);

      \path [draw= blue ,fill=black,line width=0.4mm] (7.6, -3.38) circle (0.02cm); 

      \node at (7.6, -3.0){\scalebox{0.85}{$c^{a_{2},b_{2}}_{k}$}};

		\node at (18.8, 3.3){\scalebox{0.85}{$c^{a_{1},b_{1}}_{k+2}$}};

        \node at (10.3, 3.4){\scalebox{0.85}{$c^{a_{1},b_{1}}_{k}$}};

        \node at (21.7, 0.7){\scalebox{0.75}{$ma_{1}b_{2}$}};

        \node at (15.93, 3.35){\scalebox{0.75}{$ma_{1}b_{1}$}};

        \node at (20.33, 4.55){\scalebox{0.75}{$ma_{1}b_{1}$}};

        \node at (7.95, 3.75){\scalebox{0.75}{$ma_{1}b_{1}$}};

        \node at (6.95, 9.05){\scalebox{0.75}{$ m  \sigma^{n_{1},m} _{1} b_{1} $}};

        \node at (22.35, 9.05){\scalebox{0.75}{$ m  \sigma^{n_{1},m} _{1} b_{1} $}};

        \node at (22.75, 6.03){\scalebox{0.75}{$ m  \sigma^{n_{1},m} _{1} b_{2} $}};

         \node at (17.34, 5.83){\scalebox{0.75}{$ m  \sigma^{n_{2},m} _{1} b_{1} $}};

        \node at (8.15, 7.05){\scalebox{0.75}{$ m  \sigma^{n_{1},m} _{1} b_{2} $}};

        \node at (11.05, 6.35){\scalebox{0.75}{$ m  \sigma^{n_{2},m} _{1} b_{2} $}};

        \node at (20.3, 6.65){\scalebox{0.75}{$ m  \sigma^{n_{2},m} _{1} b_{2} $}};

        \node at (22.0, -11.55){\scalebox{0.75}{$ma_{2} \sigma^{n'_{1},m} _{1}$}};

        \node at (7.95, -11.55){\scalebox{0.75}{$ma_{2} \sigma^{n'_{1},m} _{1}$}};

        \node at (7.15, -6.45){\scalebox{0.75}{$ma_{2} \sigma^{n'_{2},m} _{1}$}};

        \node at (6.2, -0.05){\scalebox{0.75}{$ma_{1} \sigma^{n'_{1},m} _{1}$}};

        \node at (7.35, -4.85){\scalebox{0.75}{$ma_{2} \sigma^{n'_{1},m} _{1}$}};

        \node at (12.85, -7.85){\scalebox{0.75}{$m  \sigma^{n_{1},m} _{1}   \sigma^{n'_{2},m} _{1}$}};

        \node at (17.85, -8.85){\scalebox{0.75}{$m a _{1} \sigma^{n'_{2},m} _{1}$}};

        \node at (16.65, -13.45){\scalebox{0.75}{$m  \sigma^{n'_{1},m} _{1}  \sigma^{n'_{1},m} _{2} + r_{n'_{1}}$}};

        \node at (17.25, -11.34){\scalebox{0.75}{$m  \sigma^{n'_{2},m} _{1}  \sigma^{n'_{2},m} _{2} + r_{n'_{2}}$}};

        \node at (17.35, 6.84){\scalebox{0.75}{$m  \sigma^{n_{2},m} _{1}  \sigma^{n_{2},m} _{2} + r_{n_{2}}$}};

        \node at (17.35, 8.04){\scalebox{0.75}{$m  \sigma^{n_{1},m} _{1}  \sigma^{n_{1},m} _{2} + r_{n_{1}}$}};

        \node at (13.45, 7.64){\scalebox{0.75}{$m  \sigma^{n_{1},m} _{1}  \sigma^{n_{2},m} _{2} $}};

    \node at (16.0, -2.8){\scalebox{0.75}{$m  \sigma^{n_{2},m} _{1}  \sigma^{n'_{2},m} _{1}$}};
                        
    \node at (12.5, -9.8){\scalebox{0.75}{$m  \sigma^{n_{2},m} _{1}  \sigma^{n'_{1},m} _{1}$}};

    \node at (13.75, -11.3){\scalebox{0.75}{$m  \sigma^{n'_{2},m} _{2}  \sigma^{n'_{1},m} _{1}$}};

    \node at (23.55, -0.0){\scalebox{0.75}{$m a_{1} \sigma^{n'_{1},m} _{1}$}}; 

    \node at (12.05, -2.0){\scalebox{0.75}{$m a_{1} \sigma^{n'_{2},m} _{1}$}};

    \node at (6.52, 5.3){\scalebox{0.75}{$m \sigma^{n_{1},m} _{1} \sigma^{n'_{1},m} _{1}$}};

        \node at (22.8, -6.45){\scalebox{0.75}{$ma_{2} \sigma^{n'_{2},m} _{1}$}};

        \node at (19.83, -6.35){\scalebox{0.75}{$ma_{2}b_{2}$}};

        \node at (6.03, -1.45){\scalebox{0.75}{$ma_{2}b_{2}+1$}}; 

        \node at (23.83, -1.45){\scalebox{0.75}{$ma_{2}b_{2}+1$}};

        \node at (19.83, -4.3){\scalebox{0.75}{$ma_{2}b_{2}$}};

		%\node at (16.9, 2){$2$};
		
		%\node at (16.9, -4){$2$};

		%\node at (15.3, 3.4){$6$};
		
		%\node at (15.3, -5.4){$6$};
		
		%\node at (14.1, 3.5){$2$};
		
		%\node at (14.1, -5.5){$2$};
		
		%\node at (16.25, 4.75){$4$};
		
		%\node at (16.25, -6.75){$4$};
		
		%\node at (16, 1.93){$12$};
		
		%\node at (16, -3.93){$12$};
		
		%\node at (16, -3.93){$12$};
		
		%\node at (15.7, 0.5){$4$};
		
		%\node at (15.7, -2.6){$4$};
		
		%\node at (14.36, 0.5){$4$};
		
		%\node at (14.36, -2.6){$4$};
		
		%\node at (4.5, -1){$2$};
		
		%\node at (7.5, 0.3){$4$};
		
		%\node at (7.5, -2.2){$4$};
		
		%\node at (10.5, -1){$8$};
		
		%\label{infinitequiver1}

		\end{tikzpicture}
		
	\end{center}

\caption{Infinite quiver $Q^w$.} \label{infinitequiver1}
\end{figure}

\clearpage

%***********

% \subsubsection{Zoomed in quiver diagram in dense cone $\mathcal{C}$ with unknown DT}

\begin{figure}[h!]    
\begin{center}

 %\vspace{1.2cm}

		\begin{tikzpicture}

  \hspace{-1.55cm}

   \node at (-0.9, 0.4){\scalebox{0.9}{$c^{a_{1},b_{1}}_{k}$}};

   \node at (5.05, 0.4){\scalebox{0.9}{$c^{a_{1},b_{1}}_{k+2}$}};

   \node at (11.05, 0.4){\scalebox{0.9}{$c^{a_{1},b_{1}}_{k+4}$}};

   \node at (-6.9, 0.4){\scalebox{0.9}{$c^{a_{1},b_{1}}_{k-2}$}};

   \node at (-4.0, -0.1){\scalebox{0.7}{$ma_{1}b_{1}$}};

   \node at (2.0, -0.1){\scalebox{0.7}{$ma_{1}b_{1}$}};

   \node at (8.0, -0.1){\scalebox{0.7}{$ma_{1}b_{1}$}};

   \node at (-4.0, -4.7){\scalebox{0.7}{$ma_{2}b_{2}$}};

   \node at (2.0, -4.7){\scalebox{0.7}{$ma_{2}b_{2}$}};

   \node at (8.0, -4.7){\scalebox{0.7}{$ma_{2}b_{2}$}};

   \node at (-5.6, -2.3){\scalebox{0.7}{$ma_{1}b_{2}$}}; 

   \node at (9.6, -2.3){\scalebox{0.7}{$ma_{1}b_{2}$}};

   \node at (9.0, 1.63){\scalebox{0.7}{$ma_{1}b_{1}$}};

   \node at (7.0, 1.63){\scalebox{0.7}{$ma_{1}b_{1}$}};

   \node at (1.0, 1.63){\scalebox{0.7}{$ma_{1}b_{1}$}};

   \node at (3.0, 1.63){\scalebox{0.7}{$ma_{1}b_{1}$}};

   \node at (-3.0, 1.63){\scalebox{0.7}{$ma_{1}b_{1}$}};

   \node at (-5.0, 1.63){\scalebox{0.7}{$ma_{1}b_{1}$}};

   \node at (9.0, -6.57){\scalebox{0.7}{$ma_{2}b_{2}+1$}};

   \node at (7.0, -6.57){\scalebox{0.7}{$ma_{2}b_{2}+1$}};

   \node at (1.0, -6.57){\scalebox{0.7}{$ma_{2}b_{2}+1$}};

   \node at (3.0, -6.57){\scalebox{0.7}{$ma_{2}b_{2}+1$}};

   \node at (-3.0, -6.57){\scalebox{0.7}{$ma_{2}b_{2}+1$}};

   \node at (-5.0, -6.57){\scalebox{0.7}{$ma_{2}b_{2}+1$}};

   \node at (-0.9, -4.7){\scalebox{0.9}{$c^{a_{2},b_{2}}_{\tilde{k}}$}};

   \node at (5.05, -4.7){\scalebox{0.9}{$c^{a_{2},b_{2}}_{\tilde{k}+2}$}};

   \node at (11.05, -4.7){\scalebox{0.9}{$c^{a_{2},b_{2}}_{\tilde{k}+4}$}};

   \node at (-6.9, -4.7){\scalebox{0.9}{$c^{a_{2},b_{2}}_{\tilde{k}-2}$}};

\path [draw=black, dashed, line width=0.2mm] (-3, -0.15) [bend right= 0]  to [bend left =  -0] (1, -4.7);

\path [draw=black, dashed, line width=0.2mm] (-3, -0.15) [bend right= 0]  to [bend left =  -0] (3, -4.7);

\path [draw=black, dashed, line width=0.2mm] (-3, -0.15) [bend right= 0]  to [bend left =  -0] (7, -4.7);

\path [draw=black, dashed, line width=0.2mm] (-3, -0.15) [bend right= 0]  to [bend left =  -0] (9, -4.7);

\path [draw=black, dashed, line width=0.2mm] (-5, -0.15) [bend right= 0]  to [bend left =  -0] (1, -4.7);

\path [draw=black, dashed, line width=0.2mm] (-5, -0.15) [bend right= 0]  to [bend left =  -0] (3, -4.7);

\path [draw=black, dashed, line width=0.1mm] (-5, -0.15) [bend right= 0]  to [bend left =  -0] (7, -4.7);

\path [draw=black, dashed, line width=0.1mm] (-5, -0.15) [bend right= 0]  to [bend left =  -0] (9, -4.7);

\path [draw=black, dashed, line width=0.1mm] (1, -0.15) [bend right= 0]  to [bend left =  -0] (7, -4.7);

\path [draw=black, dashed, line width=0.1mm] (1, -0.15) [bend right= 0]  to [bend left =  -0] (3, -4.7);

\path [draw=black, dashed, line width=0.1mm] (1, -0.15) [bend right= 0]  to [bend left =  -0] (9, -4.7);

\path [draw=black, dashed, line width=0.1mm] (3, -0.15) [bend right= 0]  to [bend left =  -0] (9, -4.7);

\path [draw=black, dashed, line width=0.1mm] (3, -0.15) [bend right= 0]  to [bend left =  -0] (7, -4.7);

\path [draw=black, dashed, line width=0.1mm] (7, -0.15) [bend right= 0]  to [bend left =  -0] (9, -4.7);

\path [draw=black, dashed, line width=0.1mm] (-5, -0.15) [bend right= 0]  to [bend left =  -0] (-3, -4.7);

\path [draw=black, dashed, line width=0.3mm] (-5, -0.15) [bend right= 0]  to [bend left =  -0] (-5, -4.7);

\path [draw=black, dashed, line width=0.3mm] (9, -0.15) [bend right= 0]  to [bend left =  -0] (9, -4.7);

\path [draw=black, dashed, line width=0.3mm] (7, -0.15) [bend right= 0]  to [bend left =  -0] (7, -4.7);

\path [draw=black, dashed, line width=0.3mm] (1, -0.15) [bend right= 0]  to [bend left =  -0] (1, -4.7);

\path [draw=black, dashed, line width=0.3mm] (3, -0.15) [bend right= 0]  to [bend left =  -0] (3, -4.7);

\path [draw=black, dashed, line width=0.3mm] (-3, -0.15) [bend right= 0]  to [bend left =  -0] (-3, -4.7);

\path [draw=black, dashed, line width=0.1mm] (-3, -4.7) [bend right= -35]  to [bend left =  35] (1, -4.7);

\path [draw=black, dashed, line width=0.1mm] (-3, -4.7) [bend right= -35]  to [bend left =  35] (3, -4.7);

\path [draw=black, dashed, line width=0.1mm] (-3, -4.7) [bend right= -35]  to [bend left =  35] (7, -4.7);

\path [draw=black, dashed, line width=0.1mm] (-3, -4.7) [bend right= -35]  to [bend left =  35] (9, -4.7);

\path [draw=black, dashed, line width=0.1mm] (-5, -4.7) [bend right= -35]  to [bend left =  35] (1, -4.7);

\path [draw=black, dashed, line width=0.1mm] (-5, -4.7) [bend right= -35]  to [bend left =  35] (3, -4.7);

\path [draw=black, dashed, line width=0.1mm] (-5, -4.7) [bend right= -35]  to [bend left =  35] (7, -4.7);

\path [draw=black, dashed, line width=0.1mm] (-5, -4.7) [bend right= -35]  to [bend left =  35] (9, -4.7);

\path [draw=black, dashed, line width=0.1mm] (1, -4.7) [bend right= -35]  to [bend left =  35] (7, -4.7);

\path [draw=black, dashed, line width=0.1mm] (1, -4.7) [bend right= -35]  to [bend left =  35] (3, -4.7);

\path [draw=black, dashed, line width=0.1mm] (1, -4.7) [bend right= -35]  to [bend left =  35] (9, -4.7);

\path [draw=black, dashed, line width=0.1mm] (3, -4.7) [bend right= -35]  to [bend left =  35] (9, -4.7);

\path [draw=black, dashed, line width=0.1mm] (3, -4.7) [bend right= -35]  to [bend left =  35] (7, -4.7);

\path [draw=black, dashed, line width=0.1mm] (7, -4.7) [bend right= -35]  to [bend left =  35] (9, -4.7);

\path [draw=black, dashed, line width=0.1mm] (-5, -4.7) [bend right= -35]  to [bend left =  35] (-3, -4.7);

\path [draw=black, dashed, line width=0.3mm] (-3, -4.7) [bend right= 0]  to [bend left =  -0] (1, -0.15);

\path [draw=black, dashed, line width=0.3mm] (-3, -4.7) [bend right= 0]  to [bend left =  -0] (3, -0.15);

\path [draw=black, dashed, line width=0.3mm] (-3, -4.7) [bend right= 0]  to [bend left =  -0] (7, -0.15);

\path [draw=black, dashed, line width=0.3mm] (-3, -4.7) [bend right= 0]  to [bend left =  -0] (9, -0.15);

\path [draw=black, dashed, line width=0.3mm] (-5, -4.7) [bend right= 0]  to [bend left =  -0] (1, -0.15);

\path [draw=black, dashed, line width=0.3mm] (-5, -4.7) [bend right= 0]  to [bend left =  -0] (3, -0.15);

\path [draw=black, dashed, line width=0.3mm] (-5, -4.7) [bend right= 0]  to [bend left =  -0] (7, -0.15);

\path [draw=black, dashed, line width=0.3mm] (-5, -4.7) [bend right= 0]  to [bend left =  -0] (9, -0.15);

\path [draw=black, dashed, line width=0.3mm] (1, -4.7) [bend right= 0]  to [bend left =  -0] (7, -0.15);

\path [draw=black, dashed, line width=0.3mm] (1, -4.7) [bend right= 0]  to [bend left =  -0] (3, -0.15);

\path [draw=black, dashed, line width=0.3mm] (1, -4.7) [bend right= 0]  to [bend left =  -0] (9, -0.15);

\path [draw=black, dashed, line width=0.3mm] (3, -4.7) [bend right= 0]  to [bend left =  -0] (9, -0.15);

\path [draw=black, dashed, line width=0.3mm] (3, -4.7) [bend right= 0]  to [bend left =  -0] (7, -0.15);

\path [draw=black, dashed, line width=0.3mm] (7, -4.7) [bend right= 0]  to [bend left =  -0] (9, -0.15);

\path [draw=black, dashed, line width=0.3mm] (-5, -4.7) [bend right= 0]  to [bend left =  -0] (-3, -0.15);

\path [draw=black, dashed, line width=0.1mm] (-3, -0.15) [bend right= 30]  to [bend left =  -30] (1, -0.15);

\path [draw=black, dashed, line width=0.1mm] (-3, -0.15) [bend right= 30]  to [bend left =  -30] (3, -0.15);

\path [draw=black, dashed, line width=0.1mm] (-3, -0.15) [bend right= 30]  to [bend left =  -30] (7, -0.15);

\path [draw=black, dashed, line width=0.1mm] (-3, -0.15) [bend right= 30]  to [bend left =  -30] (9, -0.15);

\path [draw=black, dashed, line width=0.1mm] (-5, -0.15) [bend right= 30]  to [bend left =  -30] (1, -0.15);

\path [draw=black, dashed, line width=0.1mm] (-5, -0.15) [bend right= 30]  to [bend left =  -30] (3, -0.15);

\path [draw=black, dashed, line width=0.1mm] (-5, -0.15) [bend right= 30]  to [bend left =  -30] (7, -0.15);

\path [draw=black, dashed, line width=0.1mm] (-5, -0.15) [bend right= 30]  to [bend left =  -30] (9, -0.15);

\path [draw=black, dashed, line width=0.1mm] (1, -0.15) [bend right= 30]  to [bend left =  -30] (7, -0.15);

\path [draw=black, dashed, line width=0.1mm] (1, -0.15) [bend right= 30]  to [bend left =  -30] (3, -0.15);

\path [draw=black, dashed, line width=0.1mm] (1, -0.15) [bend right= 30]  to [bend left =  -30] (9, -0.15);

\path [draw=black, dashed, line width=0.1mm] (3, -0.15) [bend right= 30]  to [bend left =  -30] (9, -0.15);

\path [draw=black, dashed, line width=0.1mm] (3, -0.15) [bend right= 30]  to [bend left =  -30] (7, -0.15);

\path [draw=black, dashed, line width=0.1mm] (7, -0.15) [bend right= 30]  to [bend left =  -30] (9, -0.15);

\path [draw=black, dashed, line width=0.1mm] (-5, -0.15) [bend right= 30]  to [bend left =  -30] (-3, -0.15);

 \path [draw=black,line width=0.8mm] (-5, 0.74) circle (0.5cm);

   \path [draw= black ,fill=yellow,line width=0.4mm] (-5, 0) circle (0.3cm);

  \path [draw= gray ,line width=0.8mm] (-3, 0.74) circle (0.5cm);

   \path [draw= black,fill=yellow,line width=0.4mm] (-3, 0) circle (0.3cm);

   \path [draw= gray ,line width=0.8mm] (1, 0.74) circle (0.5cm);

   \path [draw=black,fill=yellow,line width=0.4mm] (1, 0) circle (0.3cm);

   \path [draw=black,line width=0.8mm] (3, 0.74) circle (0.5cm);

   \path [draw=black,fill=yellow,line width=0.4mm] (3, 0) circle (0.3cm);

   \path [draw= black,line width=0.8mm] (7, 0.74) circle (0.5cm);

   \path [draw=black,fill=yellow,line width=0.4mm] (7, 0) circle (0.3cm);

   \path [draw=gray,line width=0.8mm] (9, 0.74) circle (0.5cm);

   \path [draw=black,fill=yellow,line width=0.4mm] (9, 0) circle (0.3cm);

    \path [draw= black,line width=0.8mm] (-5, -5.74) circle (0.5cm);

   \path [draw=black,fill=yellow,line width=0.4mm] (-5, -5) circle (0.3cm);

  \path [draw=gray,line width=0.8mm] (-3, -5.74) circle (0.5cm);

   \path [draw=black,fill=yellow,line width=0.4mm] (-3, -5) circle (0.3cm);

   \path [draw=gray,line width=0.8mm] (1, -5.74) circle (0.5cm);

   \path [draw=black,fill=yellow,line width=0.4mm] (1, -5) circle (0.3cm);

   \path [draw=black,line width=0.8mm] (3, -5.74) circle (0.5cm);

   \path [draw=black,fill=yellow,line width=0.4mm] (3, -5) circle (0.3cm);

   \path [draw=black,line width=0.8mm] (7, -5.74) circle (0.5cm);

   \path [draw=black,fill=yellow,line width=0.4mm] (7, -5) circle (0.3cm);

   \path [draw=gray,line width=0.8mm] (9, -5.74) circle (0.5cm);

   \path [draw=black,fill=yellow,line width=0.4mm] (9, -5) circle (0.3cm);

   \path [draw= blue ,fill=black,line width=0.4mm] (-1.8, -0.2) circle (0.02cm);

    \path [draw= blue ,fill=black,line width=0.4mm] (-1.5, -0.2) circle (0.02cm);

    \path [draw= blue ,fill=black,line width=0.4mm] (-1.2, -0.2) circle (0.02cm);

    \path [draw= blue ,fill=black,line width=0.4mm] (-0.9, -0.2) circle (0.02cm);

    \path [draw= blue ,fill=black,line width=0.4mm] (-0.6, -0.2) circle (0.02cm); 

    \path [draw= blue ,fill=black,line width=0.4mm] (-0.3, -0.2) circle (0.02cm);

    \path [draw= blue ,fill=black,line width=0.4mm] (-1.8, -5.3) circle (0.02cm);

    \path [draw= blue ,fill=black,line width=0.4mm] (-1.5, -5.3) circle (0.02cm);

    \path [draw= blue ,fill=black,line width=0.4mm] (-1.2, -5.3) circle (0.02cm);

    \path [draw= blue ,fill=black,line width=0.4mm] (-0.9, -5.3) circle (0.02cm);

    \path [draw= blue ,fill=black,line width=0.4mm] (-0.6, -5.3) circle (0.02cm); 

    \path [draw= blue ,fill=black,line width=0.4mm] (-0.3, -5.3) circle (0.02cm);

    \path [draw= blue ,fill=black,line width=0.4mm] (5.73, -5.3) circle (0.02cm);

    \path [draw= blue ,fill=black,line width=0.4mm] (5.43, -5.3) circle (0.02cm);

    \path [draw= blue ,fill=black,line width=0.4mm] (5.13, -5.3) circle (0.02cm);

    \path [draw= blue ,fill=black,line width=0.4mm] (4.83, -5.3) circle (0.02cm);

    \path [draw= blue ,fill=black,line width=0.4mm] (4.53, -5.3) circle (0.02cm); 

    \path [draw= blue ,fill=black,line width=0.4mm] (4.23, -5.3) circle (0.02cm);

    \path [draw= blue ,fill=black,line width=0.4mm] (5.73, -0.2) circle (0.02cm);

    \path [draw= blue ,fill=black,line width=0.4mm] (5.43, -0.2) circle (0.02cm);

    \path [draw= blue ,fill=black,line width=0.4mm] (5.13, -0.2) circle (0.02cm);

    \path [draw= blue ,fill=black,line width=0.4mm] (4.83, -0.2) circle (0.02cm);

    \path [draw= blue ,fill=black,line width=0.4mm] (4.53, -0.2) circle (0.02cm); 

    \path [draw= blue ,fill=black,line width=0.4mm] (4.23, -0.2) circle (0.02cm);

     \path [draw= blue ,fill=black,line width=0.4mm] (11.73, -0.2) circle (0.02cm);

    \path [draw= blue ,fill=black,line width=0.4mm] (11.43, -0.2) circle (0.02cm);

    \path [draw= blue ,fill=black,line width=0.4mm] (11.13, -0.2) circle (0.02cm);

    \path [draw= blue ,fill=black,line width=0.4mm] (10.83, -0.2) circle (0.02cm);

    \path [draw= blue ,fill=black,line width=0.4mm] (10.53, -0.2) circle (0.02cm); 

    \path [draw= blue ,fill=black,line width=0.4mm] (10.23, -0.2) circle (0.02cm);

   \path [draw= blue ,fill=black,line width=0.4mm] (11.73, -5.3) circle (0.02cm);

    \path [draw= blue ,fill=black,line width=0.4mm] (11.43, -5.3) circle (0.02cm);

    \path [draw= blue ,fill=black,line width=0.4mm] (11.13, -5.3) circle (0.02cm);

    \path [draw= blue ,fill=black,line width=0.4mm] (10.83, -5.3) circle (0.02cm);

    \path [draw= blue ,fill=black,line width=0.4mm] (10.53, -5.3) circle (0.02cm); 

    \path [draw= blue ,fill=black,line width=0.4mm] (10.23, -5.3) circle (0.02cm);

    \path [draw= blue ,fill=black,line width=0.4mm] (-7.83, -5.3) circle (0.02cm);

    \path [draw= blue ,fill=black,line width=0.4mm] (-7.43, -5.3) circle (0.02cm);

    \path [draw= blue ,fill=black,line width=0.4mm] (-7.13, -5.3) circle (0.02cm);

    \path [draw= blue ,fill=black,line width=0.4mm] (-6.83, -5.3) circle (0.02cm);

    \path [draw= blue ,fill=black,line width=0.4mm] (-6.53, -5.3) circle (0.02cm); 

    \path [draw= blue ,fill=black,line width=0.4mm] (-6.23, -5.3) circle (0.02cm);

    \path [draw= blue ,fill=black,line width=0.4mm] (-7.83, -0.2) circle (0.02cm);

    \path [draw= blue ,fill=black,line width=0.4mm] (-7.43, -0.2) circle (0.02cm);

    \path [draw= blue ,fill=black,line width=0.4mm] (-7.13, -0.2) circle (0.02cm);

    \path [draw= blue ,fill=black,line width=0.4mm] (-6.83, -0.2) circle (0.02cm);

    \path [draw= blue ,fill=black,line width=0.4mm] (-6.53, -0.2) circle (0.02cm); 

    \path [draw= blue ,fill=black,line width=0.4mm] (-6.23, -0.2) circle (0.02cm);

    \end{tikzpicture}
\end{center}

\caption{Central part of an infinite quiver $Q^w$.} \label{fig-central}

\end{figure}
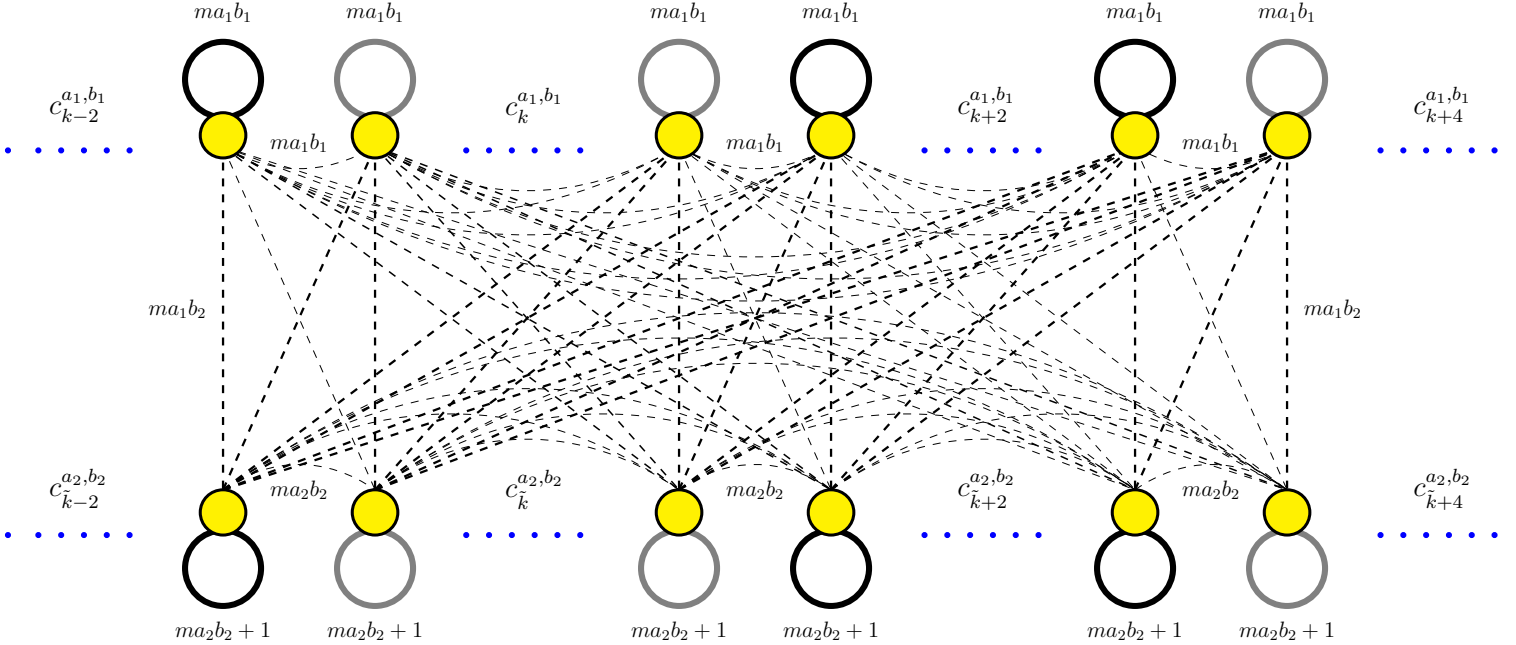

In fig. \ref{fig-central} we draw in more detail a part of the quiver corresponding to the central region inside the cone, where we illustrate the unknown DT invariants $c^{a,b}_{k}$ as the number of nodes and the alternating pattern of allowed spins generated by shifting $k$ to $k+2$.
%These $c^{a,b}_{k}$ will be computed in the next section by diagonalising the quivers on both sides of the wall and comparing the symmetric DT invariants of the $m$-loop quivers to constrain the non symmetric DT invariants of the original $m$-arrow Kronecker quiver.

%\clearpage

%************

% \subsubsection{Identifications of quiver generating parameters}

Furthermore, we write down the identification of parameters. Similarly as with linking and unlinking (\ref{eq:unlinking})-(\ref{eq:linking}), the generating parameters for the symmetric quiver generating series of the two quivers $Q^{s}$ and $Q^{w}$ must be appropriately identified, so that the wall-crossing identity (\ref{eq:motivicwallcrossingformula}) indeed holds. Denoting the generating parameters $\{x_{i} \}_{i \in Q^{s}_{0}}$ for $Q^s$ simply by $x_1$ and $x_2$, the identifications in question are in general of the form $ \{ x^{w}_{i} = x_{i}^{w}(x_{1}, \ldots , x_{|Q^{s}_{0}|}, q) \} = x_{i}^{w}(x_{1}, x_{2}, q) \}$. However, when labeling we will use the charges and spins rather than the node number $i$ remembering that the identification is the same for all nodes representing the same charge and spin. These identifications are also determined by the normal ordering operation carried out in the appendix \ref{Appendix:normal ordering}. They can be obtained simply by normal ordering all the generators within a single term in the expansion of the dilogarithm. Here we list the general results for both the LHS and the RHS of $Q^{w}$, as well as inside the dense cone $\mathcal{C}$. 

While we are working with the Reineke general form from \cite{reineke2023wildquantumdilogarithmidentities} and list identifications for both positive and negative dilogarithm powers we should note that in practice the powers of the dilogarithm on the LHS and RHS are actually positive which can be seen from the invariance of the Euler form $(\mathbf{d}, \mathbf{d})$ under the $\sigma_{a,b} = (b, mb-a)$ symmetry.
This also follows physical expectations that there are only hypermultiplets outside the dense region. However, inside $\mathcal{C}$ both positive and negative powers continue to exist. 

For LHS of the spectrum generator, for positive dilogarithm powers we find
\begin{align} \label{ident1}
x_{\sigma^{n}(0,1)}  = q^{ \frac{ (\sigma^{n,m} _{1}) ^{2} +  (\sigma^{n,m} _{2})^{2}  - m  \sigma^{n,m} _{1}  \sigma^{n,m} _{2}    -\sigma^{n,m} _{1} -  \sigma^{n,m} _{2} }{2} } x_{2}^{\sigma^{n,m} _{1}   } x_{1}^{ \sigma^{n,m} _{2}  }, 
\end{align}
and if there were existing negative powers of the  dilogarithm they would take the following form with a shift in the identification 
\begin{align} \label{ident2}
x_{\sigma^{n}(0,1)}  = 
q^{ \frac{ (\sigma^{n,m} _{1}) ^{2} +  (\sigma^{n,m} _{2})^{2} - m \sigma^{n,m} _{1}  \sigma^{n,m} _{2}    - \sigma^{n,m} _{1} -  \sigma^{n,m} _{2}  -1 }{2} }  x_{2}^{  \sigma^{n,m} _{1}} x_{1} ^{ \sigma^{n,m} _{2} } .
\end{align}
For RHS of the spectrum generator, for existing physical positive and hypothetical negative dilogarithm powers we find
\begin{align} \label{ident3}
x_{\sigma^{-n}(1,0)} = q^{ \frac{ (\sigma^{n',m} _{1}) ^{2} +  (\sigma^{n',m} _{2})^{2}  - m  \sigma^{n',m} _{1}  \sigma^{n',m} _{2}    -\sigma^{n',m} _{1} -  \sigma^{n',m} _{2} }{2} } x_{2}^{\sigma^{n',m} _{2}   } x_{1}^{ \sigma^{n',m} _{1}  }  
\end{align}
and respectively
\begin{align} \label{ident4}
x_{\sigma^{-n}(1,0)} = q^{ \frac{ (\sigma^{n',m} _{1}) ^{2} +  (\sigma^{n',m} _{2})^{2} - m \sigma^{n',m} _{1}  \sigma^{n',m} _{2}    - \sigma^{n',m} _{1} -  \sigma^{n',m} _{2}  -1 }{2} }  x_{2}^{  \sigma^{n',m} _{2}} x_{1} ^{ \sigma^{n',m} _{1} } .
\end{align}
Inside the dense cone $\mathcal{C}$ the identifications also depend on the spin $k$, and both positive and negative powers of dilogarithms must exist.  We find for the postive and negative identifications respectively
\begin{align} \label{ident5}  
x_{(a,b)_{k}} =  \  \  
q^{ \frac{a^{2} +  b^{2}-mab-a-b +k }{2} } x^{a}_{2}  x_{1}^{b} ,  \   \   \   \   q^{ \frac{a^{2} +  b^{2}-mab-a-b -1 +k }{2} } x^{a}_{2}  x_{1}^{b}.
\end{align}

%************
%************

\section{Pure $SU(2)$ Seiberg-Witten theory}  \label{sec:pureSU2}

In this section we illustrate our approach in the case of the pure $SU(2)$ Seiberg-Witten theory. We identify corresponding symmetric quivers and discuss the meaning of the formula (\ref{eq:3d4dDTrelation}). The BPS spectrum in two chambers of this theory is captured by the wall-crossing identity (\ref{m2identity}). Its left-hand side corresponds to the strong coupling chamber and the BPS quiver identified as $m$-Kronecker quiver with $m=2$, which has two nodes and two arrows, see fig. \ref{fig-quiverSU2strong}. These nodes represent two BPS states, a monopole and a dyon, represented by $\gamma_{1}$ and $\gamma_{2}$, such that $\langle  \gamma_{1}, \gamma_{2} \rangle = 2$. In the weak coupling chamber the spectrum is infinite and includes dyons of the form $n\gamma_{1}+(n+1) \gamma_{2}$ and $(n+1)\gamma_{1}+n\gamma_{2}$ together with the W-boson $\gamma_{1}+\gamma_{2}$. In contrast to the $m \geq 3$ examples, the DT invariants or refined BPS indices $\Omega(\gamma,u,y)$ are well known for $m=2$ \cite{RefinedMotivicQuantumDimofte_2009} and take form
\begin{align}
\Omega(n\gamma_{1}+ (n+1) \gamma_{2},u,y) \  = \  \Omega((n+1)\gamma_{1}+ n \gamma_{2},u,y)= 1, \quad  
\Omega(\gamma_{1}+ \gamma_{2},u,y) = y+y^{-1}.    \label{DTSU2}
\end{align}
This means that there is no dense cone of BPS walls where the BPS invariants are unknown. 

From the perspective of the 3d $\mathcal{N}=2$ boundary theory we have the two dual theories $T[Q^{s}]$ and $T[Q^{w}]$ with dual vortex partition functions given by the symmetric quiver generating series of $Q^{s}$ and $Q^{w}$. The symmetric quiver $Q^{w}$ has an infinite number of nodes, one node for every (refined) BPS invariant. Looking back at the identity (\ref{m2identity}) or the invariants listed in (\ref{DTSU2}) we see that the weak coupling spectrum generator has a LHS with infinitely many nodes represented by quantum dilogarithms with positive exponents for each dyon and a RHS with the same pattern but for the other set of dyons. It also has two central nodes for the W-boson with negative exponents due to the sign in the BPS index. As in other examples we can obtain the infinite quiver by normal ordering the quantum torus algebra. However, as we already found the results for general $m$ in section \ref{sec:infinitequivers}, we simply specialize them to $m=2$, obtaining the quiver shown in fig. \ref{fig-quiverSU2weak} and the adjacency matrix:

\begin{align*}
 & C = \\
 \\ \nonumber
 & 
 \hspace{-1cm}
 \left( \begin{array}{cccccccccccccc}
0  & 0  & 0  & ...&... &... & 0 & 0 & ... & ... & ...  & 0 & 0 & 0 \\
0  & 4 &  6  &  ... &  2n_{1} (n_{2}+1)   &... & 2 & 2 &  ...  & 2nn' & ...  & 4 & 2  & 0 \\
0  &  6  & 12  & ... &... &... & 4 &  4 & ... & ... & ...   & 8 & 4 & 0 \\
...  & ...  & ...  & ... &... &...   & ... & ... & ... & ...  & ...   & ... & ... &... \\
...  &  2n_{1} (n_{2}+1)   & ...  & ... & 2n^{2}+2n  &...   & 2n & 2n & ... &  2nn'  & ...   & ... & 2nn' &... \\
...  & ...  & ...  & ... &... &...   & ... & ... & ... & ...  & ...   & ... & ... &... \\
0  & 2  & 4  & ... & 2n & ... &  3 &   2  & ...  & 2n'  & ...   & 4 & 2 & 0  \\
0  & 2  & 4  & ... & 2n &...& 2 &  3  & ...  & 2n'  & ...   &   4  & 2 & 0 \\
...  & ...  & ...  & ... &...  &...& ... &  ...  & ... & ...   & ... & ...& ... & ... \\
...  & 2nn'  & ...  & ... & 2nn'  & ... & 2n' &  2n'  & ... &  2n'^{2}+2n'  & ... & ...& 2n'_{3} (n'_{4}+1) & ... \\
...  & ...  & ...  &  ...  & ...   &...& ... &  ...  & ... & ...   &  ...  & ...& ... & ... \\
0  & 4  & 8  & ... &... &... & 4 &  4  & ...  & ...   & ...  & 12 & 6 & 0 \\
0  & 2  & 4  & ... & 2nn' &... & 2 &  2  & ...   &  2n'_{3} (n'_{4}+1)   & ...   & 6 & 4 & 0 \\
0  & 0  & 0 & ... &... &... & 0 &  0  & ...  & ...  & ...  & 0 & 0 & 0 \\
 \end{array} \right),
\end{align*}

Here $n_{2} >  n_{1}$ and $n'_{4} >  n'_{3}$, and as in the general case, these labels are both ordered from the outside of the identity inwards, and therefore arise on opposite sides of the adjacency matrix. The identifications read 
\begin{align}
x'_{n} = q ^{-n} \hat{x}^{n+1} _{1} \hat{x}^{n}_{2}, \  \   \   \  \   \   x'_{\infty,n} = q^{-\frac{3}{2}}  x_{2} x_{1},  \   \   \   \   \   \  x'_{\infty,n'} =  q^{-\frac{1}{2}}  x_{2} x_{1},   \   \  \   \   \    \  x'_{n'} = q^{-n'} \hat{x}^{n'}_{1} \hat{x}^{n'+1}_{2},
\end{align}
where we label the new variables in the infinite quiver by their node number $n$ or $n'$, and the variables $ x_{\infty,n'}, \,   x_{\infty,n} $ represent the terms at infinity in the central region of the product.

%*************
%*************
%*************

\subsection{BPS invariants for pure $SU(2)$ theory}

Let us illustrate now the formula (\ref{eq:3d4dDTrelation}) and the relation between 4d and 3d BPS numbers in this example. For pure 4d $SU(2)$ theory the refined BPS degeneracies are known so we do not need to determine them. However, we can simply plug them into (\ref{eq:3d4dDTrelation}) and illustrate that the BPS numbers of the symmetric quivers on both sides of the wall are the same. This also imposes constraints on the number of loops on $Q^{w}$. More precisely, we have $\Omega^{4d}_{k}(\gamma)_{i_{p}} =1, r^{\mathbf{T}}_{pq} = 0$, so plugging these values into (\ref{eq:3d4dDTrelation}) we get 
\begin{align}
\Omega^{3d}_{\mathbf{d},\tilde{k}} = \sum_{\mathbf{T}_{\mathbf{d}, \tilde{k}}, \ C^{loop}_{\mathbf{T}}} \Omega^{3d}_{C^{loop}_{\mathbf{T}}}, 
\end{align}
which is the statement that $\Omega^{Q^{s}}_{\mathbf{d},k} = \Omega^{Q^{w}}_{\mathbf{d},k}$ after identification of quiver generating parameters.

In this example we can also explicitly compute the open DT invariants of both quivers and see how they match. For this we first diagonalize the doubled $2$-Kronecker quiver to read off specific DT invariants. We start unlinking two nodes originally linked with each other  
\begin{align} \label{eq:quivers}
C= 
\begin{pmatrix}
0  & 2 \\
2  & 0
\end{pmatrix}, \
\begin{pmatrix}
0  & 1  & 1\\
1  & 0 &  1 \\
1  & 1  & 3 
\end{pmatrix}, \
\begin{pmatrix}
0  & 0  & 1  & 0 \\
0  & 0 &  1  &  0\\
1  & 1  & 3  & 2 \\
0  & 0  & 2  & 1
\end{pmatrix}.
\end{align}
Here we immediately get a 3-loop and a 1-loop quiver on the diagonal. As we have just unlinked 2 nodes we now immediately get an identification of $q^{-\frac{1}{2}}x_{2}x_{1}$. Now we must consider DT invariants for a 3-loop and a 1-loop quiver.
The general form of the refined DT invariants of an $m$-loop quiver can be written as:  
\begin{align}
 \Omega_{m-loop} (x,q) =  \sum_{r=1}^{\infty}  \Omega^{m}_{r} (q)x^{r} = \sum_{r,s} \Omega^{m}_{r,s}x^{r}q^{\frac{s}{2}}.
\end{align}
There is an algorithm that can compute the DT invariants for $m$-loop quivers. Here we can focus on just some of the DT-invariants that constrain the loops on the quivers. At first we can do an order approximation and we know that \cite{Jankowski:2022qdpnew}:
\begin{align}
\Omega_{m-loop} (x,q) =  (-1)^{m-1}q^{\frac{m}{2}}x + O(x^{2}).
\end{align}
We can now substitute in our specific identifications from the unlinking to show that these DT invariants of $Q^{s}$ are given by
\begin{align}
\begin{split}
\Omega_{3-loop} (q ^{-\frac{1}{2}} x _{2} x_{1} ,q) =  q^{\frac{3}{2}}( q ^{-\frac{1}{2}} x_{2} x_{1} ) = q x_{2} x_{1} + \ldots\\ 
\Omega_{1-loop} (q ^{-\frac{1}{2}} x _{2} x_{1} ,q) =  q^{\frac{1}{2}}( q ^{-\frac{1}{2}} x_{2} x_{1} ) =  x_{2} x_{1} + \ldots, 
\end{split}
\end{align}
and this is the only contribution to these charges and spins from the diagonalization as all other steps give different identifications. Now we look at the infinite quiver $Q^{w}$ on the other side of the wall. In this quiver we look for the contribution to this open DT invariant. We see that the only contibution can come from the 2 nodes (representing the vector W-boson in the 4d theory) in the center of the infinite quiver or product as all unlinkings in a diagonalization would produce higher order terms. The quiver has 2 3-loop nodes in this place with identifications of
$q^{-\frac{3}{2}}  x_{2} x_{1}$ and $q^{-\frac{1}{2}}  x_{2} x_{1}$, and DT invariants generated here are
\begin{align}
\begin{split}
\Omega_{3-loop} ( q^{-\frac{3}{2}}  x_{2} x_{1} ,q) & =  q^{\frac{3}{2}}( q^{-\frac{3}{2}}  x_{2} x_{1}) = 
 x_{2} x_{1} + \ldots,  \\ 
\Omega_{3-loop} ( q^{-\frac{1}{2}}  x_{2} x_{1} ,q) & =  q^{\frac{3}{2}}(q^{-\frac{1}{2}}  x_{2} x_{1}) =  q  x_{2} x_{1} + \ldots. 
\end{split}
\end{align}
We see that at the leading order the DT invariants are the same despite different loop numbers. The loops thus constrain the DT invariants in such a way as to give the same answer for both quivers. 

Now we continue the diagonalization and specifically look for nodes with identifications that correspond to those in $Q^{w}$. This means identifications of the generating parameters as $q^{-n}x^{n+1}_{2}x^{n}_{1}$ and $q^{-n}x^{n}_{2}x^{n+1}_{1}$. For this we must start by successively unlinking the third node in the rightmost quiver in equation (\ref{eq:quivers}). We therefore obtain
\begin{align} \label{eq:quivers}
C= 
\begin{pmatrix}
0  & 0  & 1  & 0 \\
0  & 0 &  1  &  0\\
1  & 1  & 3  & 2 \\
0  & 0  & 2  & 1
\end{pmatrix}, \
\begin{pmatrix}
0  & 0  & 0  & 0   & \\
0  & 0 &  1  &  0  &\\
0  & 1  & 3  & 2  &  3\\
0  & 0  & 2  & 1  &  2\\
   &    &  3  &  2   & 4
\end{pmatrix}, \
\begin{pmatrix}
0  & 0  & 0  & 0   &  &  \\
0  & 0 &  1  &  0  &  & \\
0  & 1  & 3  & 2  &  2 &  5 \\
0  & 0  & 2  & 1  &  2  &  4 \\
   &    &  2  &  2   & 4  & \\
   &    &  5   &  4    &     & 12
\end{pmatrix}, \
\begin{pmatrix}
0  & 0  & 0  & 0   &  &  & \\
0  & 0 &  1  &  0  &  & & \\
0  & 1  & 3  & 2  &  2 &  4 & \\
0  & 0  & 2  & 1  &  2  &  4  & \\
   &    &  2  &  2   & 4  &  & \\
   &    &  4   &  4    &     & 12 & \\
   &    &      &       &     &    & 24
\end{pmatrix}
\end{align}
Now we can see a pattern of loops: $ 4, \ 12, \ 24, \ \ldots $,  and we notice that this is the same pattern $2n^{2}+2n$ as occurs in the infinite quiver $Q^{w}$. When we check the identifications that are multiplied in this sequence of unlinkings we obtain $ q^{-\frac{n}{2}} \times (q^{-\frac{1}{2}}  x_{2} x_{1})^{n} \times x_{2} = q^{-n}x^{n+1}_{2}x^{n}_{1}$. This is also the same as in $Q^{w}$, so this explains why the numbers of loops are the same. This is essential to satisfy the condition of equal (open) DT invariants of the two symmetric quivers on both sides of the wall which can be obtained by completing a diagonalization. The final diagonal quiver, as explained in \cite{Jankowski:2022qdpnew}, must contain the loops in its diagonal that we obtain here at low order. One can check using this diagonalization that these are the only contributions (at lowest order in the $m$-loop formula) to the DT of both symmetric quivers. Therefore, if the identification at a particular loop resulting from unlinking in $Q^{s}$ is the same as that immediately existing in $Q^{w}$ before any unlinkings then the number of loops must also be the same.

We can also prove this using a recursion for the number of loops. The new loops with this identification are always generated by unlinking a node $n$ from node 3 so the number of loops on the next node is given by 
\begin{align*}
C_{l,l} = C_{l-1,l-1} +3 +2C_{l-1,3}-1, \  \  \  \  \   C_{l-1,3} = C_{l-2,3}+C_{33} -1 = C_{l-2,3}+2,
\end{align*}
 and we start with $C_{55} =4$ and $C_{53}=3$ so we can now write
\begin{align*}
C_{4+n,4+n} = 4+\sum^{n-1}_{m=1} (2 )+ 2\sum^{n-1}_{m=1}(2(m-1)+3), \  \  \  \  \   C_{4+n,3} = 3+2(n-1).
\end{align*}
We then proceed by taking the sum and indeed we find that $C_{4+n,4+n} = 2n^{2}+2n$ as expected. This then matches the number of loops in $Q^{w}$ in the weak coupling chamber.

%**************
%**************
%**************

\section{Wild Donaldson-Thomas invariants} \label{sec:wildDT}

In this section we finally consider the general case of wild $m$-Kronecker quivers.  
%and the infinite quivers generated by the general motivic wall-crossing formula (\ref{eq:motivicwallcrossingformula}). 
Our goal is to derive the identity (\ref{eq:3d4dDTrelation}) that enables to determine the closed DT invariants $c^{a,b}_{k}$ in the dense region $\mathcal{C}$, which arise in the exponents on the right-hand side of the quantum dilogarithm identity (\ref{Reineke-ids}). To this end we consider symmetric quivers $Q^s$ and $Q^w$ representing the two sides of this identity and conduct their diagonalization, as summarized by commutative diagrams (\ref{picture:comparingdiagonalisations}) and (\ref{picture:comparingdiagonalisationsDT}) presented in the introduction. Recall that quiver diagonalization, developed in \cite{Jankowski:2022qdpnew}, enables to determine the open DT invariants of an arbitrary symmetric quivers; it amounts to repeatedly unlinking the quiver until the resulting (usually infinite) quiver $Q_{\infty}$ has a diagonal adjacency matrix, i.e. infinitely many nodes with no arrows other than loops. This allows one to express the DT invariants of the original symmetric quiver by (the summations over) the DT invariants of the $m$-loop quivers. 
% (after the appropriate identification of variables) and hence also proves that the DT invariants of an arbitrary symmetric quiver are given by the sums over those of $m$-loop quivers. 

In our context, we diagonalize quivers $Q^s$ and $Q^w$, thereby producing $Q^s_{\infty}$ and $Q^w_{\infty}$. As we explained earlier, the invariants $c^{a,b}_{k}$ we are after determine the numbers of certain nodes in $Q^w$ (so in fact we do not fully know the structure of $Q^w$). However, the wall-crossing identity (\ref{Reineke-ids}) implies that the generating series, as well as DT invariants of $Q^s$ and $Q^w$ must agree, and thus DT invariants of their diagonalizations $Q^s_{\infty}$ and $Q^w_{\infty}$ must be equal too. This equality implies the infinite set of relations (\ref{eq:3d4dDTrelation}), which then imposes constraints on $c^{a,b}_{k}$ or equivalently $\Omega^{4d}_k(\gamma)_{i_p}$. We show that these constraints can be effectively solved. 

In what follows we present details of the above program. Diagonalization of $Q^w$ and $Q^s$ is discussed respectively in subsection \ref{ssec-diagonalizeQw} and appendix \ref{app-diagonalizeQs}. In subsection \ref{ssec-trees} we determine the structure of trees of unlinkins, and in subsection \ref{sec:symmetricwallcrossingformula} we derive the identity (\ref{eq:detailed3d4dDTrelation}) and write is in a simplified form as (\ref{eq:simplified3d4drelation}), see also (\ref{eq:3d4dDTrelation}). Using this identity, in appendices \ref{app-m3}, \ref{app-m4} and \ref{app-m6} we determine several invariants $c^{a,b}_{k}$ of $m$-Kronecker quivers for $m=3,4,6$.

\subsection{Diagonalizing infinite quiver $Q^w$}   \label{ssec-diagonalizeQw}

Rather than following the full diagonalization procedure, we look at all the ways one can find a particular identification for a loop inside the cone. To start, we look at identifications that already exist within the quiver. We first look at a particular state inside this dense cone, for example with one of the identifications in (\ref{ident5})
\begin{align} \label{identification}
q^{ \frac{a^{2} +  b^{2}-mab-a-b +k}{2} } x^{a}_{2}  x_{1}^{b},
\end{align}
and check how it can be matched after unlinking of some pairs of nodes inside the cone, i.e. after the adjacency matrix transformation (\ref{eq:unlinking}) combined with the identification (\ref{unlinking-identification}). %In unlinking a new node $n$ is produced and $a$ and $b$ are the nodes that are being unlinked:
%\begin{align}
%& C^{(unlinked)}_{ab} = C_{ab} -1, \   \    \  C^{(unlinked)}_{nn} = C_{aa}+2C_{ab}+ C_{bb}  -1, \\ \nonumber
%\\
%& C^{(unlinked)}_{in} = C_{ai}+C_{bi}- \delta_{ai}  - \delta_{bi}, \   \    \   \   \   \text{otherwise}   \   \   \ C^{(unlinked)}_{ij} = C_{ij}
%\end{align}

\subsubsection*{Contributions from a single unlinking}

In the first step  
% in any diagonalization that determines the DT invariants of a wild Kronecker quiver is to 
we consider all pairs of nodes in the dense cone that 
% choose a set of pairs of nodes inside the dense cone and look at all possible pairs that can be unlinked within the two paired sets of DT invariants such that they 
give a contribution to the final identification of the form (\ref{identification}). This means that we initially look at all contributions from pairs of nodes with the same charges, which we call $c_{1}$ and $d_{1}$ and spin $k_{1}$, paired with $c_{2}$ and $d_{2}$ and spin $k_{2}$. We should find all single possible unlinkings between these pairs of nodes that can generate the final identification within the infinite quiver. We recall that the identification of variables from unlinking is generated by $x_{ij} = q^{-\frac{1}{2}}x_{i}x_{j}$, so we must sum over all contributions that match this. 

We also define the shift in the identification for a node in the infinite quiver of the form $S_{c_{i},d_{i},k_{i}}$. This comes from the shift in the identification in the case of a negative power of the dilogarithm, see (\ref{ident5}). If the power of the dilogarithm is negative this is 1 (and the number of loops also jumps by one), otherwise this is 0.

First, we consider other states inside the cone:
\begin{align}
\begin{split}
x'_{1} &=q^{ \frac{c_{1}^{2} +  d_{1}^{2}-mc_{1}d_{1}-c_{1}-d_{1}-S_{c_{1},d_{1},k_{1}} +k_{1}}{2} } x^{c_{1}}_{2}  x_{1}^{d_{1}}, \\ 
x'_{2} &= q^{ \frac{c_{2}^{2} +  d_{2}^{2}-mc_{2}d_{2}-c_{2}-d_{2} -S_{c_{2}, d_{2}, k_{2}}+k_{2}}{2} } x^{c_{2}}_{2}  x_{1}^{d_{2}},
\end{split}
\end{align}
for which the unlinking gives the identification
\begin{align} \label{unlinkident}
x'_{12} =q^{ \frac{c_{1}^{2} + c_{2}^{2}  +  d_{1}^{2}+ d_{2}^{2}  -m(c_{1}d_{1}+ c_{2}d_{2} )-c_{1}-d_{1} -c_{2}-d_{2}-S_{c_{1},d_{1},k_{1}}-S_{c_{2}, d_{2}, k_{2}}+k_{1}+k_{2}-1}{2} } x^{c_{1}+c_{2}}_{2}  x_{1}^{d_{1}+d_{2}}.
\end{align}
%For this to contribute to a Donaldson-Thomas invariant there are a set of conditions that must be satisfied: the identification must match the final identification of the state that we are calculating the contributions for. 
Comparing this with (\ref{identification}) we get a set of conditions that we label by $\Box$
\begin{empheq}[box=\mymathbox]{align} \label{conditionssingleunlink}
\begin{split}
& \Box : \\ 
& c_{1}+c_{2} = a, \   \  \ d_{1}+d_{2} = b,  \\  \\ 
& a^{2} +  b^{2}-mab-a-b +k \  = \  c_{1}^{2} + c_{2}^{2}  +  d_{1}^{2}+ d_{2}^{2}  -m(c_{1}d_{1}+ c_{2}d_{2} ) \ + \\
& \qquad -c_{1}-d_{1} -c_{2}-d_{2} -S_{c_{1}, d_{1}, k_{1}}-S_{c_{2}, d_{2}, k_{2}}+k_{1}+k_{2}-1   
\end{split}
\end{empheq} 
for all pairs one can choose in this first unlinking. %So this means one must sum over all these contributions together with their unlinking. 
We will introduce a diagram that represents all the possible unlinkings that lead to this final identification.

\begin{figure}[h!]
	
	\label{fig-2-sets-of-nodes}
	
	\begin{center}

 \vspace{-0.0cm}
 
		\begin{tikzpicture}

  \hspace{-0.55cm}

   \node at (-0.9, 0.4){\scalebox{0.9}{$c^{a_{1},b_{1}}_{k}$}};

   %\node at (5.05, 0.4){\scalebox{0.9}{$c^{a_{1},b_{1}}_{k+1}$}};

   %\node at (11.05, 0.4){\scalebox{0.9}{$c^{a_{1},b_{1}}_{k+2}$}};

   %\node at (-6.9, 0.4){\scalebox{0.9}{$c^{a_{1},b_{1}}_{k-1}$}};

   %\node at (-4.0, -0.1){\scalebox{0.7}{$ma_{1}b_{1}$}};

   %\node at (2.0, -0.1){\scalebox{0.7}{$ma_{1}b_{1}$}};

   %\node at (8.0, -0.1){\scalebox{0.7}{$ma_{1}b_{1}$}};

   %\node at (-4.0, -4.7){\scalebox{0.7}{$ma_{2}b_{2}$}};

   \node at (0.35, -2.5){\scalebox{0.7}{$ma_{1}b_{2}$}};

   \node at (-0.9, -1.0){\scalebox{0.7}{$ma_{1}b_{1}$}};

   \node at (-0.9, -3.7){\scalebox{0.7}{$ma_{2}b_{2}$}};

   %\node at (8.0, -4.7){\scalebox{0.7}{$ma_{2}b_{2}$}};

   %\node at (-5.6, -2.3){\scalebox{0.7}{$ma_{1}b_{2}$}}; 

   %\node at (9.6, -2.3){\scalebox{0.7}{$ma_{1}b_{2}$}};

   %\node at (9.0, 1.63){\scalebox{0.7}{$ma_{1}b_{1}$}};

   \node at (1.0, 1.63){\scalebox{0.7}{$ma_{1}b_{1}$}};

   %\node at (3.0, 1.63){\scalebox{0.7}{$ma_{1}b_{1}+1$}};

   \node at (-3.0, 1.63){\scalebox{0.7}{$ma_{1}b_{1}$}};

   %\node at (-5.0, 1.63){\scalebox{0.7}{$ma_{1}b_{1}+1$}};

    %\node at (9.0, -6.57){\scalebox{0.7}{$ma_{2}b_{2}$}};

   %\node at (7.0, -6.57){\scalebox{0.7}{$ma_{2}b_{2}+1$}};

   \node at (1.0, -6.57){\scalebox{0.7}{$ma_{2}b_{2}$}};

   %\node at (3.0, -6.57){\scalebox{0.7}{$ma_{2}b_{2}+1$}};

   \node at (-3.0, -6.57){\scalebox{0.7}{$ma_{2}b_{2}$}};

   %\node at (-5.0, -6.57){\scalebox{0.7}{$ma_{2}b_{2}+1$}};     

   \node at (-0.9, -4.7){\scalebox{0.9}{$c^{a_{2},b_{2}}_{\tilde{k}}$}};

   %\node at (5.05, -4.7){\scalebox{0.9}{$c^{a_{2},b_{2}}_{\tilde{k}+1}$}};

   %\node at (11.05, -4.7){\scalebox{0.9}{$c^{a_{2},b_{2}}_{\tilde{k}+2}$}};

   %\node at (-6.9, -4.7){\scalebox{0.9}{$c^{a_{2},b_{2}}_{\tilde{k}-1}$}};

\path [draw= red, dashed, line width=0.6mm] (-3, -0.15) [bend right= 0]  to [bend left =  -0] (1, -4.7);

%\path [draw=black, dashed, line width=0.2mm] (-3, -0.15) [bend right= 0]  to [bend left =  -0] (3, -4.7);

%\path [draw=black, dashed, line width=0.2mm] (-3, -0.15) [bend right= 0]  to [bend left =  -0] (7, -4.7);

%\path [draw=black, dashed, line width=0.2mm] (-3, -0.15) [bend right= 0]  to [bend left =  -0] (9, -4.7);

\path [draw=black, dashed, line width=0.3mm] (1, -0.15) [bend right= 0]  to [bend left =  -0] (1, -4.7);

\path [draw=black, dashed, line width=0.3mm] (-3, -0.15) [bend right= 0]  to [bend left =  -0] (-3, -4.7);

\path [draw=black, dashed, line width=0.1mm] (-3, -4.7) [bend right= -35]  to [bend left =  35] (1, -4.7);

%\path [draw=black, dashed, line width=0.1mm] (-3, -4.7) [bend right= -35]  to [bend left =  35] (3, -4.7);

%\path [draw=black, dashed, line width=0.1mm] (-3, -4.7) [bend right= -35]  to [bend left =  35] (7, -4.7);

%\path [draw=black, dashed, line width=0.1mm] (-3, -4.7) [bend right= -35]  to [bend left =  35] (9, -4.7);

%\path [draw=black, dashed, line width=0.1mm] (-5, -4.7) [bend right= -35]  to [bend left =  35] (1, -4.7);

%\path [draw=black, dashed, line width=0.1mm] (-5, -4.7) [bend right= -35]  to [bend left =  35] (3, -4.7);

%\path [draw=black, dashed, line width=0.1mm] (-5, -4.7) [bend right= -35]  to [bend left =  35] (7, -4.7);

\path [draw=black, dashed, line width=0.3mm] (-3, -4.7) [bend right= 0]  to [bend left =  -0] (1, -0.15);

%\path [draw=black, dashed, line width=0.3mm] (-3, -4.7) [bend right= 0]  to [bend left =  -0] (3, -0.15);

\path [draw=black, dashed, line width=0.1mm] (-3, -0.15) [bend right= 30]  to [bend left =  -30] (1, -0.15);

  \path [draw= gray ,line width=0.8mm] (-3, 0.74) circle (0.5cm);

   \path [draw= black,fill=yellow,line width=0.4mm] (-3, 0) circle (0.3cm);

   \path [draw= gray ,line width=0.8mm] (1, 0.74) circle (0.5cm);

   \path [draw=black,fill=yellow,line width=0.4mm] (1, 0) circle (0.3cm);

     \path [draw=gray,line width=0.8mm] (-3, -5.74) circle (0.5cm);

   \path [draw=black,fill=yellow,line width=0.4mm] (-3, -5) circle (0.3cm);

\path [draw=gray,line width=0.8mm] (1, -5.74) circle (0.5cm);

   \path [draw=black,fill=yellow,line width=0.4mm] (1, -5) circle (0.3cm);

    \path [draw= blue ,fill=black,line width=0.4mm] (-1.8, -5.3) circle (0.02cm);

    \path [draw= blue ,fill=black,line width=0.4mm] (-1.5, -5.3) circle (0.02cm);

    \path [draw= blue ,fill=black,line width=0.4mm] (-1.2, -5.3) circle (0.02cm);

    \path [draw= blue ,fill=black,line width=0.4mm] (-0.9, -5.3) circle (0.02cm);

    \path [draw= blue ,fill=black,line width=0.4mm] (-0.6, -5.3) circle (0.02cm); 

    \path [draw= blue ,fill=black,line width=0.4mm] (-0.3, -5.3) circle (0.02cm);

\path [draw= blue ,fill=black,line width=0.4mm] (5.73, -5.3) circle (0.02cm);

    \path [draw= blue ,fill=black,line width=0.4mm] (5.43, -5.3) circle (0.02cm);

    \path [draw= blue ,fill=black,line width=0.4mm] (5.13, -5.3) circle (0.02cm);

    \path [draw= blue ,fill=black,line width=0.4mm] (4.83, -5.3) circle (0.02cm);

    \path [draw= blue ,fill=black,line width=0.4mm] (4.53, -5.3) circle (0.02cm); 

    \path [draw= blue ,fill=black,line width=0.4mm] (4.23, -5.3) circle (0.02cm);

\path [draw= blue ,fill=black,line width=0.4mm] (5.73, -0.2) circle (0.02cm);

    \path [draw= blue ,fill=black,line width=0.4mm] (5.43, -0.2) circle (0.02cm);

    \path [draw= blue ,fill=black,line width=0.4mm] (5.13, -0.2) circle (0.02cm);

    \path [draw= blue ,fill=black,line width=0.4mm] (4.83, -0.2) circle (0.02cm);

    \path [draw= blue ,fill=black,line width=0.4mm] (4.53, -0.2) circle (0.02cm); 

    \path [draw= blue ,fill=black,line width=0.4mm] (4.23, -0.2) circle (0.02cm);

\path [draw= blue ,fill=black,line width=0.4mm] (-1.8, -0.2) circle (0.02cm);

    \path [draw= blue ,fill=black,line width=0.4mm] (-1.5, -0.2) circle (0.02cm);

    \path [draw= blue ,fill=black,line width=0.4mm] (-1.2, -0.2) circle (0.02cm);

    \path [draw= blue ,fill=black,line width=0.4mm] (-0.9, -0.2) circle (0.02cm);

    \path [draw= blue ,fill=black,line width=0.4mm] (-0.6, -0.2) circle (0.02cm); 

    \path [draw= blue ,fill=black,line width=0.4mm] (-0.3, -0.2) circle (0.02cm);

\path [draw= blue ,fill=black,line width=0.4mm] (-7.83, -5.3) circle (0.02cm);

    \path [draw= blue ,fill=black,line width=0.4mm] (-7.43, -5.3) circle (0.02cm);

    \path [draw= blue ,fill=black,line width=0.4mm] (-7.13, -5.3) circle (0.02cm);

    \path [draw= blue ,fill=black,line width=0.4mm] (-6.83, -5.3) circle (0.02cm);

    \path [draw= blue ,fill=black,line width=0.4mm] (-6.53, -5.3) circle (0.02cm); 

    \path [draw= blue ,fill=black,line width=0.4mm] (-6.23, -5.3) circle (0.02cm);

    \path [draw= blue ,fill=black,line width=0.4mm] (-7.83, -0.2) circle (0.02cm);

    \path [draw= blue ,fill=black,line width=0.4mm] (-7.43, -0.2) circle (0.02cm);

    \path [draw= blue ,fill=black,line width=0.4mm] (-7.13, -0.2) circle (0.02cm);

    \path [draw= blue ,fill=black,line width=0.4mm] (-6.83, -0.2) circle (0.02cm);

    \path [draw= blue ,fill=black,line width=0.4mm] (-6.53, -0.2) circle (0.02cm); 

    \path [draw= blue ,fill=black,line width=0.4mm] (-6.23, -0.2) circle (0.02cm);

		\end{tikzpicture}
		
	\end{center}

\caption{This diagram shows 2 sets of nodes with the same charge and spin (here depicted next to each other) including all possible linkings. The red line shows one of the pairs of nodes that must be unlinked.}

\end{figure}
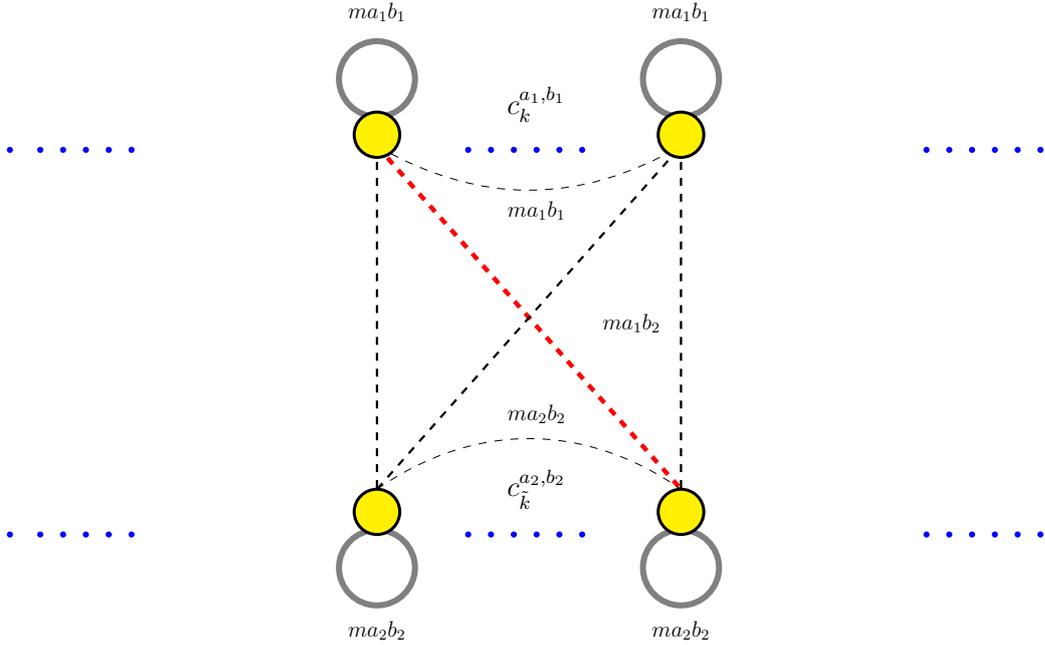

We can now organize this into a sum over all the unlinkings that must be done. This in practice is a product over the DT invariants for the pair of sets of nodes we are unlinking 
\begin{align}
%\sum_{\Box  %c_{1}+c_{2} = a,  \ d_{1}+d_{2} = b,  \   c_{1}^{2} + c_{2}^{2}  +  d_{1}^{2}+ d_{2}^{2}  -m(c_{1}d_{1}+ c_{2}d_{2} )-c_{1}-d_{1} -c_{2}-d_{2}-S_{k_{1}}-S_{k_{2}}+k_{1}+k_{2}-1   =   a^{2} +  b^{2}-mab-a-b +k   
%}      
c^{c_{1}, d_{1}}_{k_{1}}  c^{c_{2}, d_{2}}  _{k_{2}} 
\end{align}
for a contribution to the count of the final identification. However, we should also note that the final identification can be more general because we can also have identifications that are not present in the original infinite quiver but are present in the diagonalised quiver. In this case we generalize
\begin{align*}
& a^{2} + b^{2}-mab-a-b +k \quad \rightarrow \quad \Delta 
\end{align*} 
such that $\Delta = a^{2} +  b^{2}-mab-a-b +k$ only holds when the final identification exists in the original quiver. %In this more general case the sum becomes

%\begin{align}
%\sum_{c_{1}+c_{2} = a,  \ d_{1}+d_{2} = b,  \   c_{1}^{2} + c_{2}^{2}  +  d_{1}^{2}+ d_{2}^{2}  -m(c_{1}d_{1}+ c_{2}d_{2} )-c_{1}-d_{1} -c_{2}-d_{2}-S_{k_{1}}-S_{k_{2}}+k_{1}+k_{2}-1   = \Delta }  \\     c^{c_{1}, d_{1}}_{k_{1}}  c^{c_{2}, d_{2}}  _{k_{2}} 
%\end{align}

To compute the contribution for a particular open Donaldson-Thomas invariant we must determine how many loops exist on the new node for the particular unlinking. This can be counted using the standard formula for unlinking:
\begin{align}
C_{vv} = C_{aa} +C_{bb} +2C_{ab}-1,
\end{align}
where $v$ labels the new node generated by the unlinking. 
%We can use the linking and loop numbers to determine 
We see that the number of loops on this new entry, which we denote $C_{vv,l}$, equals
\begin{align}
\begin{split}
C_{vv,l} & = C_{aa} +C_{bb} +2C_{ab}-1 = \\
& = (mc_{1}d_{1}+S_{c_{1}, d_{1}, k_{1}})+ (mc_{2}d_{2}+S_{c_{2}, d_{2}, k_{2}} )+2(mc_{1}d_{2}-l) -1
\end{split}
\end{align}
where, as before, $ S_{c_{1}, d_{1}, k_{1}},  S_{c_{2}, d_{2}, k_{2}} $ are either $0$ or $1$ depending on the power of the dilogarithm. 
% and we let $C_{vv,l}$ be the number of loops on the quiver. 

At this point we note that we must sum over all contributions that come from the full unlinking of any pairs of nodes we may choose that satisfy the conditions (\ref{conditionssingleunlink}). This means that for a pair of nodes we must unlink it over and over again until all pairs of arrows are removed and sum over the contributions. At every step the linking between the nodes reduces by one. For this we can introduce $l$ as the number of linkings already removed from a pair of nodes from the previous unlinkings. From the formula for unlinking we see that each step of unlinking therefore produces a different number of loops on the final node, labeled by $l$, produced by the unlinking. Now there is a DT invariant of the form
\begin{align}
\Omega^{C_{vv,l}}_{r,s}(q^{\frac{\Delta}{2}} x^{a}_{2}  x_{1}^{b})^{r}q^{\frac{s}{2}} =  \Omega^{C_{vv,l}}_{r,s} x^{ar}_{2}  x_{1}^{br}  q^{ (\frac{\Delta r+s}{2}) }.
\end{align}
Consequently, we are looking for all contributions to this DT invariant of the symmetric quiver (which should be the same before and after wall-crossing). Then we have to sum over all the possible loops generated by the unlinking of a pair of nodes. 
%We define $\Box$ as the set of charges satisfying this relation
%\begin{align}
%& \Box:  c_{1}+c_{2} = a,  \ d_{1}+d_{2} = b,  \   c_{1}^{2} + c_{2}^{2}  +  d_{1}^{2}+ d_{2}^{2}  -m(c_{1}d_{1}+ c_{2}d_{2} )-c_{1}-d_{1} -c_{2}-d_{2}-S_{k_{1}}-S_{k_{2}} \\ \nonumber & +k_{1}+k_{2}-1   =    \Delta
%\end{align}
We must therefore define a sum by including the contributions from all unlinkings depending on the number of arrows between the nodes one is unlinking
\begin{align}
\sum^{mc_{1}d_{2}-1}_{ l = 0 }   c^{c_{1}, d_{1}}_{k_{1}}  c^{c_{2}, d_{2}}  _{k_{2}} \Omega^{C_{vv, l}}_{r,s} x^{ar}_{2}  x_{1}^{br}  q^{( \frac{\Delta r+s}{2}) }.
\end{align}
Here $l$ is the number of pairs of arrows removed after unlinking and one must sum over these arrows for every pair of nodes, as we also sum over the nodes, until every pair of arrows is unlinked.

\subsubsection*{Contributions from multiple pairs of unlinkings} \label{sec:multipairs}

Now we have to try all possible sequences of unlinkings, so we have to use all the unlinking relations  to unlink the quiver step by step. This means that we unlink pairs of nodes in the original infinite quiver $Q^{w}$ to produce new nodes that result from this unlinking. After this, these new nodes will have some linking to each other. Then we proceed by unlinking these new nodes from each other which again results in further nodes being created which can again be unlinked both from each other and the previous nodes created by the unlinking or originally existing within the infinite quiver. These sequences of unlinkings result in a specific identification for the final node created by the full sequence. As before we must sum over all these sequences that result in a particular final identification which we choose in the diagonalization. 

%For this the relation
%\begin{align}
%C^{(unlinked)}_{iv} = C_{ai}+C_{bi}- \delta_{ai}  - \delta_{bi},
%\end{align}
For this we must consider a sequence of unlinkings starting with nodes labeled by charges of the form $c_{i},d_{i}$, where we choose $i= 1, \ldots , n$ as a set of nodes originally in $Q^{w}$ that are unlinked in this process,\footnote{At the moment $i$ just labels the nodes in the set being unlinked to give the final identification. This does not correspond to the node number given in the adjacency matrix. Later we will introduce $i_{p}$ as this node number on $Q^{w}$ where $p$ will play the role that $i$ does here.} and we now generalize (\ref{unlinkident}) such that the identifications must be related in the form
\begin{align} \label{eq:productofidentifications1}
 \prod^{n}_{i=1} %(-1)^{s_{k_{i}}}
 q^{ \frac{c_{i}^{2} +  d_{i}^{2}-mc_{i}d_{i}-c_{i}-d_{i}-S_{c_{i}, d_{i}, k_{i}}+k_{i}-(n-1)}{2} } x^{c_{i}}_{2}  x_{1}^{d_{i}} = %(-1)^{s_{k_{a,b,\Delta}}u}
 q^{ \frac{\Delta}{2} } x^{a}_{2}  x_{1}^{b}. 
\end{align}

This means the conditions (\ref{conditionssingleunlink}) generalize to these new conditions for multiple pairs of unlinkings $\Box_{n}$  that must now be satisfied 
\begin{empheq}[box=\mymathbox]{align}
\begin{split} \label{conditionsmultiplepairunlink}
& \Box_{n}: \  \   \  \sum^{n}_{i= 1} c_{i} =a, \  \   \sum^{n}_{i=1} d_{i} =b, \ \ \\  
& \sum^{n}_{i=1}( c_{i}^{2} +  d_{i}^{2}-mc_{i}d_{i}-c_{i}-d_{i}-S_{c_{i}, d_{i}, k_{i}} +k_{i})-(n-1) = \Delta, \\ 
\\ 
%& \prod^{n}_{i=1} (-1)^{s_{k_{i}}} = (-1)^{s_{k_{a,b,\Delta}}}, \\ \nonumber
&  s_{c_{i}, d_{i}, k_{i}} = c_{i}^{2} +  d_{i}^{2}-mc_{i}d_{i}-1 +k_{i}, \\   
&  \\
& S_{c_{i}, d_{i}, k_{i}} = 1  \  \ \forall s_{c_{i}, d_{i}, k_{i}} \in 2 \mathbb{Z}+1, \  \ S_{c_{i}, d_{i}, k_{i}} = 0 \  \  \ \forall s_{c_{i},d_{i}, k_{i}} \in 2 \mathbb{Z} 
\end{split}
\end{empheq} 
where we remember that
\begin{align}
\Delta = a^{2} +  b^{2}-mab-a-b +k  \label{Delta}
\end{align}
only if the final identification is contained in the original infinite quiver. Now we have to sum over all possible “trees” of unlinkings, which is a convenient way to describe ways to combine the identifications of all the nodes involved. 

\subsubsection*{First contribution: ways to unlink two nodes to obtain a single final identification}

The first contribution is simply a sum of all the ways to combine two nodes as done above, but the sum is over all nodes in the tree and we are using the new conditions (\ref{conditionsmultiplepairunlink}). Therefore, we must first generalize the previous expression to a sum over all pairs of nodes $i$ and $j$, that we can label as $i,j \in 1, 2 $ and 
$i \neq j$
\begin{align}
\sum^{mc_{i}d_{j}-1}_{ \Box, l = 0 }   c^{c_{i}, d_{i}}_{k_{i}}  c^{c_{j}, d_{j}}  _{k_{j}} \Omega^{C_{vv, l}}_{u,s} %(-1)^{s_{k_{a,b, \Delta}}u}
x^{au}_{2}  x_{1}^{bu}  q^{( \frac{\Delta u+s}{2}) },
\end{align}

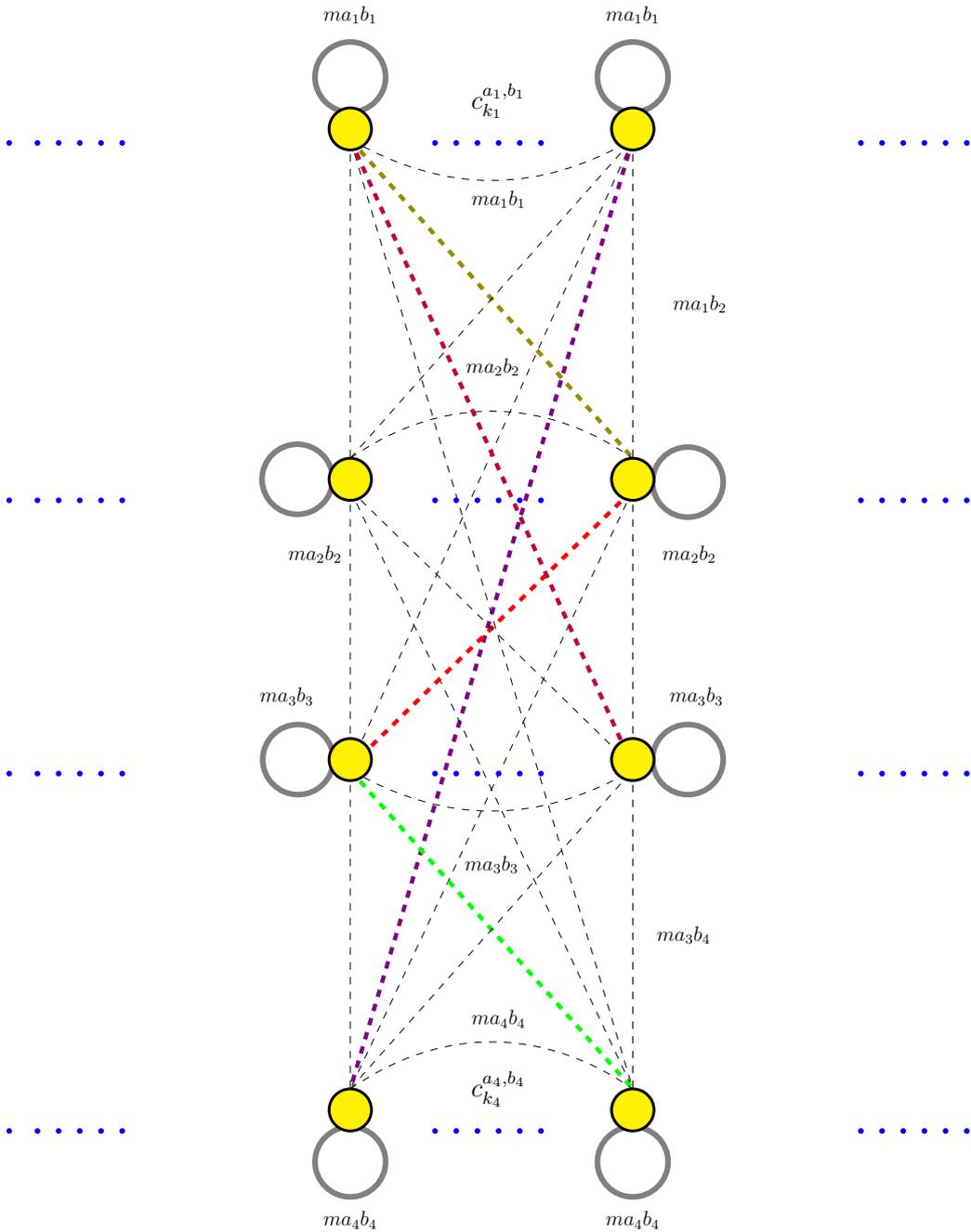
\begin{figure}[h!]
	
	\label{fig-many-sets-of-nodes}
	
	\begin{center}

 \vspace{-0.0cm}
 
		\begin{tikzpicture}

  \hspace{-0.55cm}

   \node at (-0.9, 0.4){\scalebox{0.9}{$c^{a_{1},b_{1}}_{k_{1}}$}};

   %\node at (5.05, 0.4){\scalebox{0.9}{$c^{a_{1},b_{1}}_{k+1}$}};

   %\node at (11.05, 0.4){\scalebox{0.9}{$c^{a_{1},b_{1}}_{k+2}$}};

   %\node at (-6.9, 0.4){\scalebox{0.9}{$c^{a_{1},b_{1}}_{k-1}$}};

   %\node at (-4.0, -0.1){\scalebox{0.7}{$ma_{1}b_{1}$}};

   %\node at (2.0, -0.1){\scalebox{0.7}{$ma_{1}b_{1}$}};

   %\node at (8.0, -0.1){\scalebox{0.7}{$ma_{1}b_{1}$}};

   %\node at (-4.0, -4.7){\scalebox{0.7}{$ma_{2}b_{2}$}};

   \node at (1.95, -2.5){\scalebox{0.7}{$ma_{1}b_{2}$}};

   \node at (-0.9, -1.0){\scalebox{0.7}{$ma_{1}b_{1}$}};

   \node at (-0.98, -3.4){\scalebox{0.7}{$ma_{2}b_{2}$}};

   %\node at (8.0, -4.7){\scalebox{0.7}{$ma_{2}b_{2}$}};

   %\node at (-5.6, -2.3){\scalebox{0.7}{$ma_{1}b_{2}$}}; 

   %\node at (9.6, -2.3){\scalebox{0.7}{$ma_{1}b_{2}$}};

   %\node at (9.0, 1.63){\scalebox{0.7}{$ma_{1}b_{1}$}};

   \node at (1.0, 1.63){\scalebox{0.7}{$ma_{1}b_{1}$}};

   %\node at (3.0, 1.63){\scalebox{0.7}{$ma_{1}b_{1}+1$}};

   \node at (-3.0, 1.63){\scalebox{0.7}{$ma_{1}b_{1}$}};

   %\node at (-5.0, 1.63){\scalebox{0.7}{$ma_{1}b_{1}+1$}};

    %\node at (9.0, -6.57){\scalebox{0.7}{$ma_{2}b_{2}$}};

   %\node at (7.0, -6.57){\scalebox{0.7}{$ma_{2}b_{2}+1$}};

   \node at (1.8, -6.07){\scalebox{0.7}{$ma_{2}b_{2}$}};

   %\node at (3.0, -6.57){\scalebox{0.7}{$ma_{2}b_{2}+1$}};

   \node at (-3.5, -6.07){\scalebox{0.7}{$ma_{2}b_{2}$}};

   %\node at (-5.0, -6.57){\scalebox{0.7}{$ma_{2}b_{2}+1$}};     

   %\node at (-0.9, -4.7){\scalebox{0.9}{$c^{a_{2},b_{2}}_{\tilde{k}}$}};

   %\node at (5.05, -4.7){\scalebox{0.9}{$c^{a_{2},b_{2}}_{\tilde{k}+1}$}};

   %\node at (11.05, -4.7){\scalebox{0.9}{$c^{a_{2},b_{2}}_{\tilde{k}+2}$}};

   %\node at (-6.9, -4.7){\scalebox{0.9}{$c^{a_{2},b_{2}}_{\tilde{k}-1}$}};

 \path [draw= violet, dashed, line width=0.6mm] (1, -0.15) [bend right= 0]  to [bend left =  -0] (-3, -13.7);

  \path [draw=black, dashed, line width=0.01mm] (1, -5.15) [bend right= 0]  to [bend left =  -0] (-3, -13.7);

  \path [draw=black, dashed, line width=0.01mm] (-3, -0.15) [bend right= 0]  to [bend left =  -0] (1, -13.7);

  \path [draw=black, dashed, line width=0.01mm] (-3, -5.15) [bend right= 0]  to [bend left =  -0] (1, -13.7);

  \path [draw=black, dashed, line width=0.01mm] (-3, -5.15) [bend right= 0]  to [bend left =  -0] (-3, -9.1);

  \path [draw= black, dashed, line width=0.01mm] (1, -5.15) [bend right= 0]  to [bend left =  -0] (1, -9.1);

\path [draw= red, dashed, line width=0.6mm] (-3, -9.1) [bend right= 0]  to [bend left =  -0] (1, -5.15);

\path [draw=black, dashed, line width=0.01mm] (-3, -9.1) [bend right= 0]  to [bend left =  -0] (1, -0.15);

\path [draw=purple, dashed, line width=0.6mm] (-3, -0.15) [bend right= 0]  to [bend left =  -0] (1, -9.1);

\path [draw= black, dashed, line width=0.01mm] (-3, -5.15) [bend right= 0]  to [bend left =  -0] (1, -9.1);

\path [draw= olive, dashed, line width=0.6mm] (-3, -0.15) [bend right= 0]  to [bend left =  -0] (1, -4.7);

%\path [draw=black, dashed, line width=0.2mm] (-3, -0.15) [bend right= 0]  to [bend left =  -0] (3, -4.7);

%\path [draw=black, dashed, line width=0.2mm] (-3, -0.15) [bend right= 0]  to [bend left =  -0] (7, -4.7);

%\path [draw=black, dashed, line width=0.2mm] (-3, -0.15) [bend right= 0]  to [bend left =  -0] (9, -4.7);

\path [draw=black, dashed, line width=0.01mm] (1, -0.15) [bend right= 0]  to [bend left =  -0] (1, -4.7);

\path [draw=black, dashed, line width=0.01mm] (-3, -0.15) [bend right= 0]  to [bend left =  -0] (-3, -4.7);

\path [draw=black, dashed, line width=0.01mm] (-3, -4.7) [bend right= -35]  to [bend left =  35] (1, -4.7);

%\path [draw=black, dashed, line width=0.1mm] (-3, -4.7) [bend right= -35]  to [bend left =  35] (3, -4.7);

%\path [draw=black, dashed, line width=0.1mm] (-3, -4.7) [bend right= -35]  to [bend left =  35] (7, -4.7);

%\path [draw=black, dashed, line width=0.1mm] (-3, -4.7) [bend right= -35]  to [bend left =  35] (9, -4.7);

%\path [draw=black, dashed, line width=0.1mm] (-5, -4.7) [bend right= -35]  to [bend left =  35] (1, -4.7);

%\path [draw=black, dashed, line width=0.1mm] (-5, -4.7) [bend right= -35]  to [bend left =  35] (3, -4.7);

%\path [draw=black, dashed, line width=0.1mm] (-5, -4.7) [bend right= -35]  to [bend left =  35] (7, -4.7);

\path [draw=black, dashed, line width=0.01mm] (-3, -4.7) [bend right= 0]  to [bend left =  -0] (1, -0.15);

%\path [draw=black, dashed, line width=0.3mm] (-3, -4.7) [bend right= 0]  to [bend left =  -0] (3, -0.15);

\path [draw=black, dashed, line width=0.01mm] (-3, -0.15) [bend right= 30]  to [bend left =  -30] (1, -0.15);

  \path [draw= gray ,line width=0.8mm] (-3, 0.74) circle (0.5cm);

   \path [draw= black,fill=yellow,line width=0.4mm] (-3, 0) circle (0.3cm);

   \path [draw= gray ,line width=0.8mm] (1, 0.74) circle (0.5cm);

   \path [draw=black,fill=yellow,line width=0.4mm] (1, 0) circle (0.3cm);

     \path [draw=gray,line width=0.8mm] (-3.75, -5.0) circle (0.5cm);

   \path [draw=black,fill=yellow,line width=0.4mm] (-3, -5) circle (0.3cm);

\path [draw=gray,line width=0.8mm] (1.78, -5.04) circle (0.5cm);

   \path [draw=black,fill=yellow,line width=0.4mm] (1, -5) circle (0.3cm);

    \path [draw= blue ,fill=black,line width=0.4mm] (-1.8, -5.3) circle (0.02cm);

    \path [draw= blue ,fill=black,line width=0.4mm] (-1.5, -5.3) circle (0.02cm);

    \path [draw= blue ,fill=black,line width=0.4mm] (-1.2, -5.3) circle (0.02cm);

    \path [draw= blue ,fill=black,line width=0.4mm] (-0.9, -5.3) circle (0.02cm);

    \path [draw= blue ,fill=black,line width=0.4mm] (-0.6, -5.3) circle (0.02cm); 

    \path [draw= blue ,fill=black,line width=0.4mm] (-0.3, -5.3) circle (0.02cm);

\path [draw= blue ,fill=black,line width=0.4mm] (5.73, -5.3) circle (0.02cm);

    \path [draw= blue ,fill=black,line width=0.4mm] (5.43, -5.3) circle (0.02cm);

    \path [draw= blue ,fill=black,line width=0.4mm] (5.13, -5.3) circle (0.02cm);

    \path [draw= blue ,fill=black,line width=0.4mm] (4.83, -5.3) circle (0.02cm);

    \path [draw= blue ,fill=black,line width=0.4mm] (4.53, -5.3) circle (0.02cm); 

    \path [draw= blue ,fill=black,line width=0.4mm] (4.23, -5.3) circle (0.02cm);

\path [draw= blue ,fill=black,line width=0.4mm] (5.73, -0.2) circle (0.02cm);

    \path [draw= blue ,fill=black,line width=0.4mm] (5.43, -0.2) circle (0.02cm);

    \path [draw= blue ,fill=black,line width=0.4mm] (5.13, -0.2) circle (0.02cm);

    \path [draw= blue ,fill=black,line width=0.4mm] (4.83, -0.2) circle (0.02cm);

    \path [draw= blue ,fill=black,line width=0.4mm] (4.53, -0.2) circle (0.02cm); 

    \path [draw= blue ,fill=black,line width=0.4mm] (4.23, -0.2) circle (0.02cm);

\path [draw= blue ,fill=black,line width=0.4mm] (-1.8, -0.2) circle (0.02cm);

    \path [draw= blue ,fill=black,line width=0.4mm] (-1.5, -0.2) circle (0.02cm);

    \path [draw= blue ,fill=black,line width=0.4mm] (-1.2, -0.2) circle (0.02cm);

    \path [draw= blue ,fill=black,line width=0.4mm] (-0.9, -0.2) circle (0.02cm);

    \path [draw= blue ,fill=black,line width=0.4mm] (-0.6, -0.2) circle (0.02cm); 

    \path [draw= blue ,fill=black,line width=0.4mm] (-0.3, -0.2) circle (0.02cm);

\path [draw= blue ,fill=black,line width=0.4mm] (-7.83, -5.3) circle (0.02cm);

    \path [draw= blue ,fill=black,line width=0.4mm] (-7.43, -5.3) circle (0.02cm);

    \path [draw= blue ,fill=black,line width=0.4mm] (-7.13, -5.3) circle (0.02cm);

    \path [draw= blue ,fill=black,line width=0.4mm] (-6.83, -5.3) circle (0.02cm);

    \path [draw= blue ,fill=black,line width=0.4mm] (-6.53, -5.3) circle (0.02cm); 

    \path [draw= blue ,fill=black,line width=0.4mm] (-6.23, -5.3) circle (0.02cm);

    \path [draw= blue ,fill=black,line width=0.4mm] (-7.83, -0.2) circle (0.02cm);

    \path [draw= blue ,fill=black,line width=0.4mm] (-7.43, -0.2) circle (0.02cm);

    \path [draw= blue ,fill=black,line width=0.4mm] (-7.13, -0.2) circle (0.02cm);

    \path [draw= blue ,fill=black,line width=0.4mm] (-6.83, -0.2) circle (0.02cm);

    \path [draw= blue ,fill=black,line width=0.4mm] (-6.53, -0.2) circle (0.02cm); 

    \path [draw= blue ,fill=black,line width=0.4mm] (-6.23, -0.2) circle (0.02cm);

   %\node at (-0.9, -8.7){\scalebox{0.9}{$c^{a_{1},b_{1}}_{k}$}};

   %\node at (5.05, 0.4){\scalebox{0.9}{$c^{a_{1},b_{1}}_{k+1}$}};

   %\node at (11.05, 0.4){\scalebox{0.9}{$c^{a_{1},b_{1}}_{k+2}$}};

   %\node at (-6.9, 0.4){\scalebox{0.9}{$c^{a_{1},b_{1}}_{k-1}$}};

   %\node at (-4.0, -0.1){\scalebox{0.7}{$ma_{1}b_{1}$}};

   %\node at (2.0, -0.1){\scalebox{0.7}{$ma_{1}b_{1}$}};

   %\node at (8.0, -0.1){\scalebox{0.7}{$ma_{1}b_{1}$}};

   %\node at (-4.0, -4.7){\scalebox{0.7}{$ma_{2}b_{2}$}};

   \node at (1.72, -11.5){\scalebox{0.7}{$ma_{3}b_{4}$}};

   \node at (-1.0, -10.5){\scalebox{0.7}{$ma_{3}b_{3}$}};

   \node at (-0.9, -12.7){\scalebox{0.7}{$ma_{4}b_{4}$}};

   %\node at (8.0, -4.7){\scalebox{0.7}{$ma_{2}b_{2}$}};

   %\node at (-5.6, -2.3){\scalebox{0.7}{$ma_{1}b_{2}$}}; 

   %\node at (9.6, -2.3){\scalebox{0.7}{$ma_{1}b_{2}$}};

   %\node at (9.0, 1.63){\scalebox{0.7}{$ma_{1}b_{1}$}};

   \node at (1.9, -8.1){\scalebox{0.7}{$ma_{3}b_{3}$}};

   %\node at (3.0, 1.63){\scalebox{0.7}{$ma_{1}b_{1}+1$}};

   \node at (-3.9, -8.1){\scalebox{0.7}{$ma_{3}b_{3}$}};

   %\node at (-5.0, 1.63){\scalebox{0.7}{$ma_{1}b_{1}+1$}};

    %\node at (9.0, -6.57){\scalebox{0.7}{$ma_{2}b_{2}$}};

   %\node at (7.0, -6.57){\scalebox{0.7}{$ma_{2}b_{2}+1$}};

   \node at (1.0, -15.57){\scalebox{0.7}{$ma_{4}b_{4}$}};

   %\node at (3.0, -6.57){\scalebox{0.7}{$ma_{2}b_{2}+1$}};

   \node at (-3.0, -15.57){\scalebox{0.7}{$ma_{4}b_{4}$}};

   %\node at (-5.0, -6.57){\scalebox{0.7}{$ma_{2}b_{2}+1$}};     

   \node at (-0.9, -13.7){\scalebox{0.9}{$c^{a_{4},b_{4}}_{k_{4}}$}};

   %\node at (5.05, -4.7){\scalebox{0.9}{$c^{a_{2},b_{2}}_{\tilde{k}+1}$}};

   %\node at (11.05, -4.7){\scalebox{0.9}{$c^{a_{2},b_{2}}_{\tilde{k}+2}$}};

   %\node at (-6.9, -4.7){\scalebox{0.9}{$c^{a_{2},b_{2}}_{\tilde{k}-1}$}};

\path [draw= green, dashed, line width=0.6mm] (-3, -9.15) [bend right= 0]  to [bend left =  -0] (1, -13.7);

%\path [draw=black, dashed, line width=0.2mm] (-3, -0.15) [bend right= 0]  to [bend left =  -0] (3, -4.7);

%\path [draw=black, dashed, line width=0.2mm] (-3, -0.15) [bend right= 0]  to [bend left =  -0] (7, -4.7);

%\path [draw=black, dashed, line width=0.2mm] (-3, -0.15) [bend right= 0]  to [bend left =  -0] (9, -4.7);

\path [draw= black, dashed, line width=0.01mm] (1, -9.15) [bend right= 0]  to [bend left =  -0] (1, -13.7);

\path [draw=black, dashed, line width=0.01mm] (-3, -9.15) [bend right= 0]  to [bend left =  -0] (-3, -13.7);

\path [draw=black, dashed, line width=0.01mm] (-3, -13.7) [bend right= -35]  to [bend left =  35] (1,-13.7);

%\path [draw=black, dashed, line width=0.1mm] (-3, -4.7) [bend right= -35]  to [bend left =  35] (3, -4.7);

%\path [draw=black, dashed, line width=0.1mm] (-3, -4.7) [bend right= -35]  to [bend left =  35] (7, -4.7);

%\path [draw=black, dashed, line width=0.1mm] (-3, -4.7) [bend right= -35]  to [bend left =  35] (9, -4.7);

%\path [draw=black, dashed, line width=0.1mm] (-5, -4.7) [bend right= -35]  to [bend left =  35] (1, -4.7);

%\path [draw=black, dashed, line width=0.1mm] (-5, -4.7) [bend right= -35]  to [bend left =  35] (3, -4.7);

%\path [draw=black, dashed, line width=0.1mm] (-5, -4.7) [bend right= -35]  to [bend left =  35] (7, -4.7);

\path [draw=black, dashed, line width=0.01mm] (-3, -13.7) [bend right= 0]  to [bend left =  -0] (1, -9.15);

%\path [draw=black, dashed, line width=0.3mm] (-3, -4.7) [bend right= 0]  to [bend left =  -0] (3, -0.15);

\path [draw=black, dashed, line width=0.01mm] (-3, -9.15) [bend right= 30]  to [bend left =  -30] (1, -9.15);

  \path [draw= gray ,line width=0.8mm] (-3.74, -9.0) circle (0.5cm);

   \path [draw= black,fill=yellow,line width=0.4mm] (-3, -9) circle (0.3cm);

   \path [draw= gray ,line width=0.8mm] (1.78, -9.0) circle (0.5cm);

   \path [draw=black,fill=yellow,line width=0.4mm] (1, -9) circle (0.3cm);

     \path [draw=gray,line width=0.8mm] (-3, -14.74) circle (0.5cm);

   \path [draw=black,fill=yellow,line width=0.4mm] (-3, -14) circle (0.3cm);

\path [draw=gray,line width=0.8mm] (1, -14.74) circle (0.5cm);

   \path [draw=black,fill=yellow,line width=0.4mm] (1, -14) circle (0.3cm);

    \path [draw= blue ,fill=black,line width=0.4mm] (-1.8, -14.3) circle (0.02cm);

    \path [draw= blue ,fill=black,line width=0.4mm] (-1.5, -14.3) circle (0.02cm);

    \path [draw= blue ,fill=black,line width=0.4mm] (-1.2, -14.3) circle (0.02cm);

    \path [draw= blue ,fill=black,line width=0.4mm] (-0.9, -14.3) circle (0.02cm);

    \path [draw= blue ,fill=black,line width=0.4mm] (-0.6, -14.3) circle (0.02cm); 

    \path [draw= blue ,fill=black,line width=0.4mm] (-0.3, -14.3) circle (0.02cm);

\path [draw= blue ,fill=black,line width=0.4mm] (5.73, -14.3) circle (0.02cm);

    \path [draw= blue ,fill=black,line width=0.4mm] (5.43, -14.3) circle (0.02cm);

    \path [draw= blue ,fill=black,line width=0.4mm] (5.13, -14.3) circle (0.02cm);

    \path [draw= blue ,fill=black,line width=0.4mm] (4.83, -14.3) circle (0.02cm);

    \path [draw= blue ,fill=black,line width=0.4mm] (4.53, -14.3) circle (0.02cm); 

    \path [draw= blue ,fill=black,line width=0.4mm] (4.23, -14.3) circle (0.02cm);

\path [draw= blue ,fill=black,line width=0.4mm] (5.73, -9.2) circle (0.02cm);

    \path [draw= blue ,fill=black,line width=0.4mm] (5.43, -9.2) circle (0.02cm);

    \path [draw= blue ,fill=black,line width=0.4mm] (5.13, -9.2) circle (0.02cm);

    \path [draw= blue ,fill=black,line width=0.4mm] (4.83, -9.2) circle (0.02cm);

    \path [draw= blue ,fill=black,line width=0.4mm] (4.53, -9.2) circle (0.02cm); 

    \path [draw= blue ,fill=black,line width=0.4mm] (4.23, -9.2) circle (0.02cm);

\path [draw= blue ,fill=black,line width=0.4mm] (-1.8, -9.2) circle (0.02cm);

    \path [draw= blue ,fill=black,line width=0.4mm] (-1.5, -9.2) circle (0.02cm);

    \path [draw= blue ,fill=black,line width=0.4mm] (-1.2, -9.2) circle (0.02cm);

    \path [draw= blue ,fill=black,line width=0.4mm] (-0.9, -9.2) circle (0.02cm);

    \path [draw= blue ,fill=black,line width=0.4mm] (-0.6, -9.2) circle (0.02cm); 

    \path [draw= blue ,fill=black,line width=0.4mm] (-0.3, -9.2) circle (0.02cm);

\path [draw= blue ,fill=black,line width=0.4mm] (-7.83, -14.3) circle (0.02cm);

    \path [draw= blue ,fill=black,line width=0.4mm] (-7.43, -14.3) circle (0.02cm);

    \path [draw= blue ,fill=black,line width=0.4mm] (-7.13, -14.3) circle (0.02cm);

    \path [draw= blue ,fill=black,line width=0.4mm] (-6.83, -14.3) circle (0.02cm);

    \path [draw= blue ,fill=black,line width=0.4mm] (-6.53, -14.3) circle (0.02cm); 

    \path [draw= blue ,fill=black,line width=0.4mm] (-6.23, -14.3) circle (0.02cm);

    \path [draw= blue ,fill=black,line width=0.4mm] (-7.83, -9.2) circle (0.02cm);

    \path [draw= blue ,fill=black,line width=0.4mm] (-7.43, -9.2) circle (0.02cm);

    \path [draw= blue ,fill=black,line width=0.4mm] (-7.13, -9.2) circle (0.02cm);

    \path [draw= blue ,fill=black,line width=0.4mm] (-6.83, -9.2) circle (0.02cm);

    \path [draw= blue ,fill=black,line width=0.4mm] (-6.53, -9.2) circle (0.02cm); 

    \path [draw= blue ,fill=black,line width=0.4mm] (-6.23, -9.2) circle (0.02cm);

		\end{tikzpicture}
		
	\end{center}

\caption{This diagram shows many sets of nodes with the same spin and charges in parallel. The colored linkings represent the different ways one must unlink the pairs of nodes to get a contribution to the final identification. This then results in a product of the Donaldson-Thomas invariants.}

\end{figure}

where we sum over all possible combinations of $i,j$ and now have:
\begin{align}
\begin{split}
C_{vv,l} & = C_{ii} +C_{jj} +2C_{ij}-1 = \\
& = (mc_{i}d_{i}+S_{c_{i}, d_{i}, k_{i}})+ (mc_{j}d_{j}+S_{c_{j}, d_{j}, k_{j}} )+2(mc_{i}d_{j}-l) -1.
\end{split}
\end{align}

However, in a sense this can be considered as a tree where the only way to obtain the end node is to take a product of only two sets of nodes. In general we should be taking products of all $n$ sets of nodes that are represented as end points of the tree. 

%**********
%**********

\subsection{Derivation of wall-crossing formula from trees of unlinkings}  \label{ssec-trees}

Our aim in this section is to derive a general wall-crossing formula that computes $c^{a,b}_{k}$. We do this by diagonalizing both symmetric quivers $Q^s$ and $Q^w$. This requires unlinking of more than two nodes and analysis of all possible combinations in which this can be done. This naturally forms the structure of trees, which we label as $\mathcal{T}$, where one unlinks different pairs of nodes and then subsequently unlinks the resulting nodes from each other. The idea is to sum over all possible trees of unlinkings that result in a particular open DT invariant of the symmetric quiver on both sides of the wall and compare them to find equations for the DT invariants. This means summing over trees that give either the same final identification for the quiver generating parameters or an identification (usually a factor) that still results in the same charges and spin of the open DT for both sides.

\subsubsection*{Contributions from multiple successive unlinkings} \label{subsec:multiplesucessiveunlinkings}

We start looking at these trees by considering that in full generality we must use the unlinkings to unlink three or more nodes successively so that the identification matches that for the final node created that we are using to get the Donaldson-Thomas invariants (later in (\ref{firstwallcrossingformula}) 
%sec. \ref{subsec:differentfinalident} 
we will also consider how nodes with a different final identification can contribute to the same open DT invariant). For this we label all initial nodes (which can be chosen within a set of possible nodes with the same identification) that are being unlinked in this process by $i_{1}, \ldots , i_{n}$ (that are contained at the endpoints of the unlinking tree). Hereby we can also label the endnodes within a tree, denoted as $i_{p}$, from 
 %i_{1}, ... , i_{n}  
 $p=  1, \ldots , n$ (the first until the last node in the tree) and  $i_{p} \neq i_{q}, \ \forall p,q $. Here it is important to clearly distinguish the full $i_{p}$ labeling a particular node on $Q^{w}$ from the label $p$ which tracks and orders the nodes within the much smaller set of initial nodes used in a particular sequence of unlinkings or tree $\mathcal{T}$ that gives a chosen final identification. We must again take products over DT invariants but this time over all possible contributions because at each endpoint of the tree labeled by $i_{p}$, there are $c^{c_{i_{p}}, d_{i_{p}}}_{k_{i_{p}}}$ possible nodes one must unlink in a diagonalization. For this we must take the sum over all such possible products\footnote{Later we will show that we must take this sum over all contributions that give the same fixed values for $au,bu$ and $\Delta u+s$.}
\begin{align} \label{eq:sumovertrees1}
\sum^{\eta_{\mu v}}_{n, \Box_{n}, \ \mathcal{T}, \ g_{\mathcal{T}}, \  l_{\mu v} = 0 } \prod^{n}_{p=1}  c^{c_{i_{p}}, d_{i_{p}}}_{k_{i_{p}}} \Omega^{C^{loop}_{ \mathcal{T}  }}_{u,s} %(-1)^{s_{k_{a,b}}u}
x^{au}_{2}  x_{1}^{bu}  q^{( \frac{\Delta u+s}{2}) }.
\end{align}
At this point, to define the sum there are several new definitions we must introduce to define sets of sequences of unlinkings we must sum over. First, we consider $S$ and define this as the set of all trees $\mathcal{T}$ of unlinkings that produces the final identification which can be labeled as $(a,b, \Delta)$. 
One can consider these trees $\mathcal{T}$ with $n$ end nodes as a series of steps representing the partition of the $n$ end nodes. For example, the first step in the partition is just
\begin{align} \label{firstpartition}
a(1)+b(1) = n.
\end{align}
Now we continue this process started in (\ref{firstpartition}) in partitioning the tree so that for every split in the tree one can write
\begin{align} \label{treesplit}
a(m+1)[y_{m}(m)]+b(m+1)[y_{m}(m)] = [y(m)]
\end{align}
where we now have $a(m+1)$ and $b(m+1)$ that represent the number of end nodes on each side of the split after this further partition of the tree. One should continue this process until one is at the endpoint of the tree so that there is only a single split into 2 nodes left in the tree at which the tree then terminates. We note that there are $n$ nodes at the end of the tree and therefore also $n$ distinct paths. We can label each path within the tree to its end nodes as $p$ and call $m_{p}$ the $m$-th split, as shown in (\ref{treesplit}), in the tree along the path $p$.

We define $\mathcal{T} \in S$ as a single tree or full sequence of unlinkings that obtains the final identification. We must be careful to remember that we are summing over all possible unlinkings that can give us the final identification and we can split this sum into a sum over trees which we must define accordingly. This means that in our definition of $\mathcal{T}$ several sequences of unlinkings with distinct end nodes with the same charge $(a,b)$ and spin $k$ can be combined into the same sequence and defined as the same tree $\mathcal{T}$. This will be explained in more detail for subtrees in (\ref{subtree-1})-(\ref{subtreeset}) 
% sec. \ref{sec:subtrees} 
and for trees containing the same end nodes multiple times in (\ref{conditionsonfinalidentification-1})-(\ref{conditionsonfinalidentification}).
%sec.  \ref{sec:startingnodeseveraltimes}. 
However, as we must still distinguish sequences of unlinkings with endnodes with different charge and spin, these must be considered different trees. 

One can also extend the set $S$ to a larger set $T$ such that $S \subset T$. This larger set now also contains all the intermediate steps in the unlinking of all the trees rather than just the final trees.
Once this is done one can call $S_{\mathcal{T}}$ the set of all the unlinkings (including the intermediate steps) that have to be done to obtain the tree $\mathcal{T}$ and now consider elements in this set $g_{\mathcal{T}} \in S_{\mathcal{T}} \subset T$.  %to be an element in the set of all possible compositions of unlinkings that can be used to obtain the final identification of the form.

Below we write down the sets explicitly in terms of unions of compositions of unlinkings. We first note that to define the set $S$ we must consider the recursive sequences of unlinkings within a tree and then take the union over all trees. The first step, considering (\ref{firstpartition}), is the division of two subtrees that link to provide the final identification\footnote{At every fork in the tree we can choose which branch we label as $a$ and which we call $b$.}  
\begin{align}   \label{2subtrees}
 \bigcup_{i_{a(1)},\ i_{b(1)}, \ a(1)+b(1) = n} \  \  [ i_{a(1)},i_{b(1)}], \  \  \ i_{a(1)} = \ \   [ i_{a(2)a(1)},i_{b(2)a(1)}]_{\ a(2)[a(1)]+b(2)[a(1)] = a(1)} \  \ \\ \nonumber 
i_{b(1)} = \ \   [ i_{a(2)b(1)},i_{b(2)b(1)}]_{\ a(2)[b(1)]+b(2)[b(1)] = b(1)} \  \
\end{align}
This can be iterated by repeatedly applying the steps in (\ref{treesplit}) until the endpoint of the tree $\mathcal{T}$ such that we can now define the set
\begin{align} \label{setfortrees}
& S: \  \  \ \  \  \bigcup_{ \mathcal{T} } \  \\ \nonumber & \text{ \scalebox{0.7}{$\mathcal{T} =   y_{i} \in \{a,b\}, \ p, \ m_{p}: \ 1, ...,m_{pmax},\  \  \ (a(m_{pmax}+1), b(m_{p}+1)) =(1_{a}, b(m_{p}+1)), \  \ (a(m_{p}+1), b(m_{pmax}+1)) =(a_{m_{p}+1}, 1_{b})$}} \\ \nonumber & \text{\scalebox{0.7}{$a(m+1)[y_{m}(m)]+b(m+1)[y_{m}(m)] = [y(m)]$}}  \\ \nonumber \\ \nonumber  \     
& [[ ... [[ i_{{1_{a}[a(m) y_{m-1}(m-1)... ,y_{1}(1)}]},i_{1_{b}[a(m) y_{m-1}(m-1)... ,y_{1}(1)]}],  [ i_{1_{a}[b(m)y_{m-1}(m-1)... ,y_{1}(1)]}, \\ \nonumber & i_{1_{b}[b(m)y_{m-1}(m-1) ... ,y_{1}(1)]}]], ... , [i_{[1_{a}[y_{m}(m), ... , y_{1}(1)]},i_{1_{b}[y_{m}(m), ... , y_{1}(1)]}], \ ... \ ] ... \ ]],  %[[ i_{p_{a(1)a(2)}},i_{p_{a(1)b(2)}}],[ i_{p_{a(1)a(2)}},i_{p_{a(1)b(2)}}]]] \  \ 
\end{align}
%
%
%\begin{align*}
%S:  \bigcup^{n}_{ s(j),t, p_{ij} = 1} \  \  [[ [ i_{p_{11}},i_{p_{21}}],i_{p_{31}}, \ ... \ , i_{p_{s(1)1}} ], \ ... \ ,[ i_{p_{1t}},i_{p_{2t}}],i_{p_{3t}}, \ ... \ , i_{p_{s(t)t}}]],
%\end{align*}
%
where we have split the full tree into all the possible paths to each endpoint of the tree where we use the labels $m_{p} = {1, \ldots ,m_{pmax}}$. 
In the definition above we have labeled 
\begin{align}
a(m_{pmax}+1), b(m_{pmax}+1)=1, \qquad  a(m_{qmax}+1), b(m_{qmax}+1)=1,  
\end{align}
for two distinct paths in the unlinking tree labeled by $p$ and $q$. In fact we should draw these trees to illustrate the point:\footnote{Note that in our definition we consider the sequence of unlinkings with any exchange of $a$ and $b$ at any fork but with the same nodes being unlinked as the same tree $\mathcal{T}$. One should therefore not sum over such permutations as they represent the same unlinkings.}

\begin{figure}[h!]
	
	\label{fig-tree}
	
	\begin{center}

 \vspace{0cm}
 
		\begin{tikzpicture}

  \hspace{0cm}

\path [draw= blue ,fill=black,line width=0.4mm] (11.73, -5.3) circle (0.02cm);

    \path [draw= blue ,fill=black,line width=0.4mm] (11.43, -5.3) circle (0.02cm);

    \path [draw= blue ,fill=black,line width=0.4mm] (11.13, -5.3) circle (0.02cm);

    \path [draw= blue ,fill=black,line width=0.4mm] (10.83, -5.3) circle (0.02cm);

    \path [draw= blue ,fill=black,line width=0.4mm] (14.03, -5.3) circle (0.02cm);

    \path [draw= blue ,fill=black,line width=0.4mm] (13.73, -5.3) circle (0.02cm);

    \path [draw= blue ,fill=black,line width=0.4mm] (13.43, -5.3) circle (0.02cm); 

    \path [draw= blue ,fill=black,line width=0.4mm] (13.13, -5.3) circle (0.02cm);

    \path [draw= blue ,fill=black,line width=0.4mm] (14.59, -4.3) circle (0.02cm);

    \path [draw= blue ,fill=black,line width=0.4mm] (14.29, -4.3) circle (0.02cm);

    \path [draw= blue ,fill=black,line width=0.4mm] (13.99, -4.3) circle (0.02cm); 

    \path [draw= blue ,fill=black,line width=0.4mm] (13.69, -4.3) circle (0.02cm);

    \path [draw= blue ,fill=black,line width=0.4mm] (18.09, -5.3) circle (0.02cm);

    \path [draw= blue ,fill=black,line width=0.4mm] (17.79, -5.3) circle (0.02cm);

    \path [draw= blue ,fill=black,line width=0.4mm] (17.49, -5.3) circle (0.02cm); 

    \path [draw= blue ,fill=black,line width=0.4mm] (17.19, -5.3) circle (0.02cm);

    \path [draw= blue ,fill=black,line width=0.4mm] (21.59, -5.3) circle (0.02cm);

    \path [draw= blue ,fill=black,line width=0.4mm] (21.29, -5.3) circle (0.02cm);

    \path [draw= blue ,fill=black,line width=0.4mm] (20.99, -5.3) circle (0.02cm); 

    \path [draw= blue ,fill=black,line width=0.4mm] (20.69, -5.3) circle (0.02cm);

    \path [draw= blue ,fill=black,line width=0.4mm] (22.59, -2.25) circle (0.02cm);

    \path [draw= blue ,fill=black,line width=0.4mm] (22.29, -2.25) circle (0.02cm);

    \path [draw= blue ,fill=black,line width=0.4mm] (21.99, -2.25) circle (0.02cm); 

    \path [draw= blue ,fill=black,line width=0.4mm] (21.69, -2.25) circle (0.02cm);

    \path [draw= blue ,fill=black,line width=0.4mm] (22.89, 0.75) circle (0.02cm);

    \path [draw= blue ,fill=black,line width=0.4mm] (22.59, 0.75) circle (0.02cm);

    \path [draw= blue ,fill=black,line width=0.4mm] (22.29, 0.75) circle (0.02cm); 

    \path [draw= blue ,fill=black,line width=0.4mm] (21.99, 0.75) circle (0.02cm);

    \path [draw= blue ,fill=black,line width=0.4mm] (15.13, -2.9) circle (0.02cm);

    \path [draw= blue ,fill=black,line width=0.4mm] (14.83, -2.9) circle (0.02cm);

    \path [draw= blue ,fill=black,line width=0.4mm] (14.53, -2.9) circle (0.02cm); 

    \path [draw= blue ,fill=black,line width=0.4mm] (14.23, -2.9) circle (0.02cm);

    \path [draw=black, line width=0.4mm] (11.31, -5.11) [bend right= 0]  to [bend left =  -0] (13.8, 1.2);

    \path [draw=black, line width=0.4mm] (14.1, -4.07) [bend right= 0]  to [bend left =  -0] (12.1, -3.2);

    \path [draw=black, line width=0.4mm] (13.7, -5.07) [bend right= 0]  to [bend left =  -0] (11.7, -4.2);

    \path [draw=black, line width=0.4mm] (14.65, -2.57) [bend right= 0]  to [bend left =  -0] (12.65, -1.7); 

    \path [draw=black, line width=0.4mm] (13.6, 0.6) [bend right= 0]  to [bend left =  -0] (17, -1);

    \path [draw=black, line width=0.4mm] (17, -1) [bend right= 0]  to [bend left =  -0] (19, 0);

    \path [draw=black, line width=0.4mm] (17, -1) [bend right= 0]  to [bend left =  -0] (19, -3);

    \path [draw=black, line width=0.4mm] (19, -3) [bend right= 0]  to [bend left =  -0] (21, -5);

    \path [draw=black, line width=0.4mm] (19, -3) [bend right= 0]  to [bend left =  -0] (17.6, -5.1);

    \path [draw=black, line width=0.4mm] (19, 0) [bend right= 0]  to [bend left =  -0] (22, 1.0);

    \path [draw=black, line width=0.4mm] (19, 0) [bend right= 0]  to [bend left =  -0] (22, -2);

\node at (15, -3.3){\scalebox{0.9}{$c^{a_{1},b_{1}}_{\tilde{k}}$}};

\node at (22.3, -2.8){\scalebox{0.9}{$c^{a_{2},b_{2}}_{k}$}};

\node at (12.7, 0.3){\scalebox{0.9}{$a(1)$}};

\node at (14.7, 0.7){\scalebox{0.9}{$b(1)$}};

\node at (11.8, -1.8){\scalebox{0.9}{$a(m_{1})$}};

\node at (13.8, -1.7){\scalebox{0.9}{$1_{b}$}};

\node at (12.8, -5.0){\scalebox{0.9}{$1_{b}$}};

\node at (10.8, -4.9){\scalebox{0.9}{$1_{a}$}};

\node at (12.5, -0.5){\scalebox{0.9}{$.$}};

\node at (12.5, -0.7){\scalebox{0.9}{$.$}};

\node at (12.5, -0.9){\scalebox{0.9}{$.$}};

\node at (12.5, -0.5){\scalebox{0.9}{$.$}};

\node at (12.5, -0.7){\scalebox{0.9}{$.$}};

\node at (12.5, -0.9){\scalebox{0.9}{$.$}};

\node at (14.5, -0.5){\scalebox{1.0}{$.$}};

\node at (14.8, -0.5){\scalebox{1.0}{$.$}};

\node at (15.1, -0.5){\scalebox{1.0}{$.$}};

\node at (16.9, -1.8){\scalebox{0.9}{$a(m_{p})$}};

\node at (17.6, 0.0){\scalebox{0.9}{$b(m_{p})$}};

\end{tikzpicture}
		
	\end{center}
	
\end{figure}

In this set $m_{p}$ must take the full range from $1, \ldots ,m_{pmax}$ where we choose the $m_{pmax}$ to be the value at the end of the tree. 

Note that we have $m_{p} = m_{q}$ if up until this point $m_{p}$ and $m_{q}$ lie along the same path on the tree. This is the end result obtained after unlinking all nodes. However, we can also define a larger set $T$ such that $S \subset T$. This set includes all intermediate unlinking steps and composition of trees:
\begin{align}
\hspace{-1.5cm}
& T: 
\\ \nonumber & \  \  \  \   \bigcup_{ \mathcal{T} } \\ \nonumber
& \text{ \scalebox{0.7}{ $\mathcal{T} =  y_{i} \in \{a,b\}, \ p, \ m_{p}: \  1, ... , m_{p_{max}}\ (a(m_{p_{max}}+1), b(m_{p_{max}}+1)) \in \ [(a(1),b(1)), ... ,(1,1)]$}} 
\\ \nonumber
&  \text{ \scalebox{0.7}{ $ a(m+1)[y_{m}(m)]+b(m+1)[y_{m}(m)] = [y(m)]$ }}
\\ \nonumber
\\ \nonumber  \     
& [[ ... [[ i_{{a(m+1)[a(m) y_{m-1}(m-1)... ,y_{1}(1)}]},i_{b(m+1)[a(m) y_{m-1}(m-1)... ,y_{1}(1)]}],  [ i_{a(m+1)[b(m)y_{m-1}(m-1)... ,y_{1}(1)]}, \\ \nonumber & i_{b(m+1)[b(m)y_{m-1}(m-1) ... ,y_{1}(1)]}]], ... , [i_{[a(m+1)[y_{m}(m), ... , y_{1}(1)]},i_{b(m+1)[y_{m}(m), ... , y_{1}(1)]}], \ ... \ ] ... \ ]], 
%T:  \bigcup^{n'}_{ s(j),t, n', p_{ij} = 1} \  \  [[ [ i_{p_{11}},i_{p_{21}}],i_{p_{31}}, \ ... \ , i_{p_{s(1)1}} ], \ ... \ ,[ i_{p_{1t}},i_{p_{2t}}],i_{p_{3t}}, \ ... \ , i_{p_{s(t)t}}]],
\end{align}
Finally, there is one more set that we must define. That is, for every $\mathcal{T} \in S$, we look at the set of all possible unlinkings that have to be completed to obtain $\mathcal{T}$. We call this set $S_{\mathcal{T}}$ 
\begin{align} \label{treeset}
S_{\mathcal{T}}:  
\\ \nonumber & \  \  \  \   \bigcup \\ \nonumber
& \text{ \scalebox{0.7}{ $  y_{i} \in \{a,b\}, \ m_{p}: \  1, ... , m_{p_{max}}\ (a(m_{p_{max}}+1), b(m_{p_{max}}+1)) \in \ [(a(1),b(1)), ... ,(1,1)]$}} 
\\ \nonumber
&  \text{ \scalebox{0.7}{ $ a(m+1)[y_{m}(m)]+b(m+1)[y_{m}(m)] = [y(m)]$ }}
\\ \nonumber
\\ \nonumber  \     
& [[ ... [[ i_{{a(m+1)[a(m) y_{m-1}(m-1)... ,y_{1}(1)}]},i_{b(m+1)[a(m) y_{m-1}(m-1)... ,y_{1}(1)]}],  [ i_{a(m+1)[b(m)y_{m-1}(m-1)... ,y_{1}(1)]}, \\ \nonumber & i_{b(m+1)[b(m)y_{m-1}(m-1) ... ,y_{1}(1)]}]], ... , [i_{[a(m+1)[y_{m}(m), ... , y_{1}(1)]},i_{b(m+1)[y_{m}(m), ... , y_{1}(1)]}], \ ... \ ] ... \ ]], \\ \nonumber
\\ \nonumber
\longrightarrow 
& \  \  \ [[ ... [[ i_{{1_{a}[a(m) y_{m-1}(m-1)... ,y_{1}(1)}]},i_{1_{b}[a(m) y_{m-1}(m-1)... ,y_{1}(1)]}],  [ i_{1_{a}[b(m)y_{m-1}(m-1)... ,y_{1}(1)]}, \\ \nonumber & i_{1_{b}[b(m)y_{m-1}(m-1) ... ,y_{1}(1)]}]], ... , [i_{[1_{a}[y_{m}(m), ... , y_{1}(1)]},i_{1_{b}[y_{m}(m), ... , y_{1}(1)]}], \ ... \ ] ... \ ]]. 
%\bigcup^{n'}_{ s(j),t, q_{ij} = 1} \  \  [[ [ i_{q_{11}},i_{q_{21}}],i_{q_{31}}, \ ... \ , i_{q_{s(1)1}} ], \ ... \ ,[ i_{q_{1t}},i_{q_{2t}}],i_{q_{3t}}, \ ... \ , i_{q_{s(t)t}}]], \\ \hookrightarrow [[ [ i_{p_{11}},i_{p_{21}}],i_{p_{31}}, \ ... \ , i_{p_{s_{h}(1)1}} ], \ ... \ ,[ i_{p_{1t_{h}}},i_{p_{2t_{h}}}],i_{p_{3t_{h}}}, \ ... \ , i_{p_{s_{h}(t_{h})t_{h}}}]]_{h}
\end{align}
To sum over all possible combinations we need to use the unlinking formula (\ref{eq:unlinking}) for the final number of loops at the end after all the unlinkings have been done. However, for this we must also sum over all the possible linkings between the nodes that should be unlinked to get the final identification. 

\subsubsection*{Subtrees of unlinkings} \label{sec:subtrees}

We remember that at each step $g_{\mathcal{T}}$ in (\ref{treeset}) we have to unlink the nodes over and over again until they are fully unlinked.  However, remembering the formula for the change in the adjacency matrix under unlinking, we see that the initial linking between two nodes at step $g_{\mathcal{T}}$ depends on data of previous steps that were taken before $g_{\mathcal{T}}$. This means one must also consider the data of the subtrees used to obtain $g_{\mathcal{T}}$. As, when one unlinks 2 nodes and then unlinks the resulting node from a third node, the number of linkings between the resulting node and this third node depends on the number of linkings between the original 2 nodes that were being unlinked and this third node. We take these original 2 nodes to be part of a tree resulting from other unlinkings that occurred before the step $g_{\mathcal{T}}$. We must therefore find a way to define subtrees of steps unlinkings or one could say substrees of subtrees. 

For this we introduce the set of all possible subtrees within the tree $\mathcal{T}$, existing as an element in the set (\ref{setfortrees}), which include the subtrees $g_{\mathcal{T} \mu}$ that are generated by the steps of unlinkings one must go through before one can take the step $g_{\mathcal{T}}$. We call this set $S_{\mathcal{T} \mu}$ so that $g_{\mathcal{T} \mu} \in S_{\mathcal{T} \mu}$. Equivalently, one can define an ordering on the steps of unlinking $g_{\mathcal{T}}$ so that $g_{\mathcal{T}} = g_{\mathcal{T}0}$, and we can have an ordering of the form
\begin{align}   \label{subtree-1}
 g_{\mathcal{T}0} \rightarrow g_{\mathcal{T}1} \rightarrow g_{\mathcal{T} 2} \rightarrow g_{\mathcal{T} 3 }   \quad  \cdots
 \end{align}
along a branch of the subtree. Alternatively, one can define multiple branches by writing
\begin{align}
\begin{split}
 & g_{\mathcal{T}0} \rightarrow g_{\mathcal{T}1} \rightarrow g_{\mathcal{T} 2} \rightarrow  g_{\mathcal{T} 3 } \  \rightarrow \  \ldots \ \rightarrow g_{\mathcal{T}n}\\ 
 & \  \  \ \  \  \   \   \   \   \  \  \   \   \ \  \  \  \  \  \  \  \  \ \  \  \  \   g_{\mathcal{T} n+1 } \rightarrow \ldots \rightarrow  g_{\mathcal{T} m }
 \end{split}
 \end{align}

Now we must sum over all the possible linking numbers that contribute to the final identification (and should in general give a different number of loops at this final step in the tree of unlinkings). These include linking numbers between all the subtrees that contribute. For example, we can define the maximum number of linkings between the subtrees $g_{ \mathcal{T} \mu}$ and $g_{\mathcal{T} v}$ as $\zeta_{\mu v}$.
%This linking number we denote by $\zeta_{g_{h\mu}}$. 
Here we must sum over both $\mu$ and $v$. Another way to interpret this is to identify a subtree with the final node generated by all the possible unlinkings within this subtree. Then one can understand $\zeta_{\mu v}$ as the largest element in the adjacency  matrix between subtrees $\zeta_{\mu v} =C_{uv}$ after part of the diagonalisation has been done. One can also include the end nodes of the tree in this definition --- or conversely the starting nodes of a sequence of unlinkings.
We can say that $u=i, v=j$ such that $\zeta_{\mu v} = C_{ij}$.\footnote{We note that for the first step of unlinking the previous result is recovered $\zeta_{ij}  %= C_{ij}= 
= mc_{i}d_{j}$ . $i$ and $j$ are now just nodes in the original infinite quiver chosen from any of the first two starting nodes respectively defined for the particular tree of unlinkings.} In practice when one carries out the full diagonalization we remember that we are removing one pair of arrows at a time. 

This means that $\zeta_{\mu v}$ is the maximum number of arrows between two nodes generated by subtrees in some step in the diagonalization and one must sum over all the possible linking numbers including the contributions arising from one pair of arrows being removed from $\zeta_{\mu v}$ at a time. 

For this we define 
\begin{equation}
l_{\mu v} = (\textrm{the number of arrows that have already been removed}) \in \mathbb{N}  \label{lmunu}
\end{equation}
so that the number of arrows still linking two nodes is $C_{\mu v} =\zeta_{\mu v} - l_{\mu v}$ and $l_{\mu v}: 0, \ldots , \eta_{\mu v}$. When $l_{\mu v} = \eta_{\mu v}$ the last arrows defined between these subtrees are removed and the number of linkings between the nodes $\mu$ and $v$ at this step is  $\zeta_{s t}+1 = \zeta_{\mu v} - \eta_{\mu v} =C_{st}+1$. Now $\mu, v$ and $s,t$ represent the same pair of nodes that are repeatedly unlinked in the diagonalization but we have now redefined them as part of different branches or subtrees. We can repeat this process until this pair of nodes is completely unlinked. This division of the linking numbers into different subtrees allows one to subsequently combine trees of unlinkings arising from initially unlinking the same pairs of nodes.    

These subtrees contained within $g_{\mathcal{T}}$ in (\ref{treeset}) can also be represented diagramatically as inclusions into the larger sequence of unlinkings represented by $g_{\mathcal{T}}$. This definition is introduced as it affects the contributions to the sum over the linking numbers at step $g_{\mathcal{T}}$. These inclusions of subtrees take the following form:
\begin{align} \label{subtreeset}
S_{\mathcal{T} \mu}:  
\\ \nonumber & \  \  \  \   \bigcup  \\ \nonumber
& \text{ \scalebox{0.7}{ $  y_{i} \in \{a,b\}, \ m_{p}: \ m_{p_{min}}, ... , m_{p_{max}}\ (a(m_{p_{max}}+1), b(m_{p_{max}}+1)) \in \ [(a(1),b(1)), ... ,(1,1)]$}} 
\\ \nonumber
&  \text{ \scalebox{0.7}{ $ a(m+1)[y_{m}(m)]+b(m+1)[y_{m}(m)] = [y(m)]$ }}
\\ \nonumber
\\ \nonumber  \     
& [[ i_{{a(m+1)[a(m) y_{m-1}(m-1)... ,y_{1}(1)}]},i_{b(m+1)[a(m) y_{m-1}(m-1)... ,y_{1}(1)]}],  [ i_{a(m+1)[b(m)y_{m-1}(m-1)... ,y_{1}(1)]}, \\ \nonumber & i_{b(m+1)[b(m)y_{m-1}(m-1) ... ,y_{1}(1)]}]] \\ \nonumber
\\ \nonumber
\hookrightarrow 
& [[ ... [[ i_{{a(m+1)[a(m) y_{m-1}(m-1)... ,y_{1}(1)}]},i_{b(m+1)[a(m) y_{m-1}(m-1)... ,y_{1}(1)]}],  [ i_{a(m+1)[b(m)y_{m-1}(m-1)... ,y_{1}(1)]}, \\ \nonumber & i_{b(m+1)[b(m)y_{m-1}(m-1) ... ,y_{1}(1)]}]], ... , [i_{[a(m+1)[y_{m}(m), ... , y_{1}(1)]},i_{b(n+1)[y_{m}(m), ... , y_{1}(1)]}], \ ... \ ] ... \ ]]
%\bigcup^{n'}_{ s(j),t, q_{ij} = 1} \  \  [[ [ i_{q_{11}},i_{q_{21}}],i_{q_{31}}, \ ... \ , i_{q_{s(1)1}} ], \ ... \ ,[ i_{q_{1t}},i_{q_{2t}}],i_{q_{3t}}, \ ... \ , i_{q_{s(t)t}}]], \\ \hookrightarrow [[ [ i_{p_{11}},i_{p_{21}}],i_{p_{31}}, \ ... \ , i_{p_{s_{h}(1)1}} ], \ ... \ ,[ i_{p_{1t_{h}}},i_{p_{2t_{h}}}],i_{p_{3t_{h}}}, \ ... \ , i_{p_{s_{h}(t_{h})t_{h}}}]]_{h}
\end{align}

%We remember that we have $n$ nodes (multiplied by the DT invariant).

\begin{figure}[h!]
	
	\label{fig-sets-of-many-nodes-2}
	
	\begin{center}

 \vspace{-0.0cm}
 
		\begin{tikzpicture}

  \hspace{-0.55cm}

   \node at (-0.9, 0.4){\scalebox{0.9}{$c^{a_{1},b_{1}}_{k_{1}}$}};

   %\node at (5.05, 0.4){\scalebox{0.9}{$c^{a_{1},b_{1}}_{k+1}$}};

   %\node at (11.05, 0.4){\scalebox{0.9}{$c^{a_{1},b_{1}}_{k+2}$}};

   %\node at (-6.9, 0.4){\scalebox{0.9}{$c^{a_{1},b_{1}}_{k-1}$}};

   %\node at (-4.0, -0.1){\scalebox{0.7}{$ma_{1}b_{1}$}};

   %\node at (2.0, -0.1){\scalebox{0.7}{$ma_{1}b_{1}$}};

   %\node at (8.0, -0.1){\scalebox{0.7}{$ma_{1}b_{1}$}};

   %\node at (-4.0, -4.7){\scalebox{0.7}{$ma_{2}b_{2}$}};

   \node at (1.95, -2.5){\scalebox{0.7}{$ma_{1}b_{2}$}};

   \node at (-0.9, -1.0){\scalebox{0.7}{$ma_{1}b_{1}$}};

   \node at (-0.98, -3.4){\scalebox{0.7}{$ma_{2}b_{2}$}};

   %\node at (8.0, -4.7){\scalebox{0.7}{$ma_{2}b_{2}$}};

   %\node at (-5.6, -2.3){\scalebox{0.7}{$ma_{1}b_{2}$}}; 

   %\node at (9.6, -2.3){\scalebox{0.7}{$ma_{1}b_{2}$}};

   %\node at (9.0, 1.63){\scalebox{0.7}{$ma_{1}b_{1}$}};

   \node at (1.0, 1.63){\scalebox{0.7}{$ma_{1}b_{1}$}};

   %\node at (3.0, 1.63){\scalebox{0.7}{$ma_{1}b_{1}+1$}};

   \node at (-3.0, 1.63){\scalebox{0.7}{$ma_{1}b_{1}$}};

   %\node at (-5.0, 1.63){\scalebox{0.7}{$ma_{1}b_{1}+1$}};

    %\node at (9.0, -6.57){\scalebox{0.7}{$ma_{2}b_{2}$}};

   %\node at (7.0, -6.57){\scalebox{0.7}{$ma_{2}b_{2}+1$}};

   \node at (3.0, -5.5){\scalebox{0.7}{$ma_{2}b_{2}$}};

   %\node at (3.0, -6.57){\scalebox{0.7}{$ma_{2}b_{2}+1$}};

   \node at (-3.5, -6.07){\scalebox{0.7}{$ma_{2}b_{2}$}};

   %\node at (-5.0, -6.57){\scalebox{0.7}{$ma_{2}b_{2}+1$}};     

   %\node at (-0.9, -4.7){\scalebox{0.9}{$c^{a_{2},b_{2}}_{\tilde{k}}$}};

   %\node at (5.05, -4.7){\scalebox{0.9}{$c^{a_{2},b_{2}}_{\tilde{k}+1}$}};

   %\node at (11.05, -4.7){\scalebox{0.9}{$c^{a_{2},b_{2}}_{\tilde{k}+2}$}};

   %\node at (-6.9, -4.7){\scalebox{0.9}{$c^{a_{2},b_{2}}_{\tilde{k}-1}$}};

\path [draw= black, dashed, line width=0.01mm] (8, -5.3) [bend right= 0]  to [bend left =  -0] (-3, -13.7);

\path [draw= red, dashed, line width=0.6mm] (8, -5.3) [bend right= 0]  to [bend left =  -0] (1, -13.7);

\path [draw=black, thick, line width=0.6mm] (0.75, -4.9)   to   (4, -2.5);

\path [draw=black, thick, line width=0.6mm] (0.8, 0.1)   to   (4, -2.5);

\path [draw=black, thick, line width=0.6mm] (0.7, -8.9)   to   (4, -7.0);

\path [draw=black, thick, line width=0.6mm] (0.8, -5.19)   to   (4, -7.0);

\path [draw=black, thick, line width=0.6mm] (4, -7)   to   (8, -5.3);

\path [draw=black, thick, line width=0.6mm] (4, -2.5)   to   (8, -5.3);

\path [draw=gray,line width=0.8mm] (4, -6.3) circle (0.5cm);

 \path [draw=black,fill= green,line width=0.4mm] (4, -7) circle (0.3cm);

\path [draw=gray,line width=0.8mm] (4, -1.75) circle (0.5cm);

 \path [draw=black,fill= green,line width=0.4mm] (4, -2.5) circle (0.3cm);

\path [draw=gray,line width=0.8mm] (8, -4.55) circle (0.5cm);

 \path [draw=black,fill= green,line width=0.4mm] (8, -5.3) circle (0.3cm);

 \path [draw= black, dashed, line width=0.01mm] (1, -0.15) [bend right= 0]  to [bend left =  -0] (-3, -13.7);

  \path [draw=black, dashed, line width=0.01mm] (1, -5.15) [bend right= 0]  to [bend left =  -0] (-3, -13.7);

  \path [draw=black, dashed, line width=0.01mm] (-3, -0.15) [bend right= 0]  to [bend left =  -0] (1, -13.7);

  \path [draw=black, dashed, line width=0.01mm] (-3, -5.15) [bend right= 0]  to [bend left =  -0] (1, -13.7);

  \path [draw=black, dashed, line width=0.01mm] (-3, -5.15) [bend right= 0]  to [bend left =  -0] (-3, -9.1);

  \path [draw= black, dashed, line width=0.01mm] (1, -5.15) [bend right= 0]  to [bend left =  -0] (1, -9.1);

\path [draw=  black, dashed, line width=0.01mm] (-3, -9.1) [bend right= 0]  to [bend left =  -0] (1, -5.15);

\path [draw=black, dashed, line width=0.01mm] (-3, -9.1) [bend right= 0]  to [bend left =  -0] (1, -0.15);

\path [draw=black, dashed, line width=0.01mm] (-3, -0.15) [bend right= 0]  to [bend left =  -0] (1, -9.1);

\path [draw= black, dashed, line width=0.01mm] (-3, -5.15) [bend right= 0]  to [bend left =  -0] (1, -9.1);

\path [draw= black, dashed, line width=0.01mm] (-3, -0.15) [bend right= 0]  to [bend left =  -0] (1, -4.7);

%\path [draw=black, dashed, line width=0.2mm] (-3, -0.15) [bend right= 0]  to [bend left =  -0] (3, -4.7);

%\path [draw=black, dashed, line width=0.2mm] (-3, -0.15) [bend right= 0]  to [bend left =  -0] (7, -4.7);

%\path [draw=black, dashed, line width=0.2mm] (-3, -0.15) [bend right= 0]  to [bend left =  -0] (9, -4.7);

\path [draw=black, dashed, line width=0.01mm] (1, -0.15) [bend right= 0]  to [bend left =  -0] (1, -4.7);

\path [draw=black, dashed, line width=0.01mm] (-3, -0.15) [bend right= 0]  to [bend left =  -0] (-3, -4.7);

\path [draw=black, dashed, line width=0.01mm] (-3, -4.7) [bend right= -35]  to [bend left =  35] (1, -4.7);

%\path [draw=black, dashed, line width=0.1mm] (-3, -4.7) [bend right= -35]  to [bend left =  35] (3, -4.7);

%\path [draw=black, dashed, line width=0.1mm] (-3, -4.7) [bend right= -35]  to [bend left =  35] (7, -4.7);

%\path [draw=black, dashed, line width=0.1mm] (-3, -4.7) [bend right= -35]  to [bend left =  35] (9, -4.7);

%\path [draw=black, dashed, line width=0.1mm] (-5, -4.7) [bend right= -35]  to [bend left =  35] (1, -4.7);

%\path [draw=black, dashed, line width=0.1mm] (-5, -4.7) [bend right= -35]  to [bend left =  35] (3, -4.7);

%\path [draw=black, dashed, line width=0.1mm] (-5, -4.7) [bend right= -35]  to [bend left =  35] (7, -4.7);

\path [draw=black, dashed, line width=0.01mm] (-3, -4.7) [bend right= 0]  to [bend left =  -0] (1, -0.15);

%\path [draw=black, dashed, line width=0.3mm] (-3, -4.7) [bend right= 0]  to [bend left =  -0] (3, -0.15);

\path [draw=black, dashed, line width=0.01mm] (-3, -0.15) [bend right= 30]  to [bend left =  -30] (1, -0.15);

  \path [draw= gray ,line width=0.8mm] (-3, 0.74) circle (0.5cm);

   \path [draw= black,fill=yellow,line width=0.4mm] (-3, 0) circle (0.3cm);

   \path [draw= gray ,line width=0.8mm] (1, 0.74) circle (0.5cm);

   \path [draw=black,fill=yellow,line width=0.4mm] (1, 0) circle (0.3cm);

     \path [draw=gray,line width=0.8mm] (-3.75, -5.0) circle (0.5cm);

   \path [draw=black,fill=yellow,line width=0.4mm] (-3, -5) circle (0.3cm);

\path [draw=gray,line width=0.8mm] (1.78, -5.04) circle (0.5cm);

   \path [draw=black,fill=yellow,line width=0.4mm] (1, -5) circle (0.3cm);

    \path [draw= blue ,fill=black,line width=0.4mm] (-1.8, -5.3) circle (0.02cm);

    \path [draw= blue ,fill=black,line width=0.4mm] (-1.5, -5.3) circle (0.02cm);

    \path [draw= blue ,fill=black,line width=0.4mm] (-1.2, -5.3) circle (0.02cm);

    \path [draw= blue ,fill=black,line width=0.4mm] (-0.9, -5.3) circle (0.02cm);

    \path [draw= blue ,fill=black,line width=0.4mm] (-0.6, -5.3) circle (0.02cm); 

    \path [draw= blue ,fill=black,line width=0.4mm] (-0.3, -5.3) circle (0.02cm);

\path [draw= blue ,fill=black,line width=0.4mm] (5.73, -5.3) circle (0.02cm);

    \path [draw= blue ,fill=black,line width=0.4mm] (5.43, -5.3) circle (0.02cm);

    \path [draw= blue ,fill=black,line width=0.4mm] (5.13, -5.3) circle (0.02cm);

    \path [draw= blue ,fill=black,line width=0.4mm] (4.83, -5.3) circle (0.02cm);

    \path [draw= blue ,fill=black,line width=0.4mm] (4.53, -5.3) circle (0.02cm); 

    \path [draw= blue ,fill=black,line width=0.4mm] (4.23, -5.3) circle (0.02cm);

\path [draw= blue ,fill=black,line width=0.4mm] (5.73, -0.2) circle (0.02cm);

    \path [draw= blue ,fill=black,line width=0.4mm] (5.43, -0.2) circle (0.02cm);

    \path [draw= blue ,fill=black,line width=0.4mm] (5.13, -0.2) circle (0.02cm);

    \path [draw= blue ,fill=black,line width=0.4mm] (4.83, -0.2) circle (0.02cm);

    \path [draw= blue ,fill=black,line width=0.4mm] (4.53, -0.2) circle (0.02cm); 

    \path [draw= blue ,fill=black,line width=0.4mm] (4.23, -0.2) circle (0.02cm);

\path [draw= blue ,fill=black,line width=0.4mm] (-1.8, -0.2) circle (0.02cm);

    \path [draw= blue ,fill=black,line width=0.4mm] (-1.5, -0.2) circle (0.02cm);

    \path [draw= blue ,fill=black,line width=0.4mm] (-1.2, -0.2) circle (0.02cm);

    \path [draw= blue ,fill=black,line width=0.4mm] (-0.9, -0.2) circle (0.02cm);

    \path [draw= blue ,fill=black,line width=0.4mm] (-0.6, -0.2) circle (0.02cm); 

    \path [draw= blue ,fill=black,line width=0.4mm] (-0.3, -0.2) circle (0.02cm);

\path [draw= blue ,fill=black,line width=0.4mm] (-7.83, -5.3) circle (0.02cm);

    \path [draw= blue ,fill=black,line width=0.4mm] (-7.43, -5.3) circle (0.02cm);

    \path [draw= blue ,fill=black,line width=0.4mm] (-7.13, -5.3) circle (0.02cm);

    \path [draw= blue ,fill=black,line width=0.4mm] (-6.83, -5.3) circle (0.02cm);

    \path [draw= blue ,fill=black,line width=0.4mm] (-6.53, -5.3) circle (0.02cm); 

    \path [draw= blue ,fill=black,line width=0.4mm] (-6.23, -5.3) circle (0.02cm);

    \path [draw= blue ,fill=black,line width=0.4mm] (-7.83, -0.2) circle (0.02cm);

    \path [draw= blue ,fill=black,line width=0.4mm] (-7.43, -0.2) circle (0.02cm);

    \path [draw= blue ,fill=black,line width=0.4mm] (-7.13, -0.2) circle (0.02cm);

    \path [draw= blue ,fill=black,line width=0.4mm] (-6.83, -0.2) circle (0.02cm);

    \path [draw= blue ,fill=black,line width=0.4mm] (-6.53, -0.2) circle (0.02cm); 

    \path [draw= blue ,fill=black,line width=0.4mm] (-6.23, -0.2) circle (0.02cm);

   %\node at (-0.9, -8.7){\scalebox{0.9}{$c^{a_{1},b_{1}}_{k}$}};

   %\node at (5.05, 0.4){\scalebox{0.9}{$c^{a_{1},b_{1}}_{k+1}$}};

   %\node at (11.05, 0.4){\scalebox{0.9}{$c^{a_{1},b_{1}}_{k+2}$}};

   %\node at (-6.9, 0.4){\scalebox{0.9}{$c^{a_{1},b_{1}}_{k-1}$}};

   %\node at (-4.0, -0.1){\scalebox{0.7}{$ma_{1}b_{1}$}};

   %\node at (2.0, -0.1){\scalebox{0.7}{$ma_{1}b_{1}$}};

   %\node at (8.0, -0.1){\scalebox{0.7}{$ma_{1}b_{1}$}};

   %\node at (-4.0, -4.7){\scalebox{0.7}{$ma_{2}b_{2}$}};

   \node at (1.72, -11.5){\scalebox{0.7}{$ma_{3}b_{4}$}};

   \node at (-1.0, -10.5){\scalebox{0.7}{$ma_{3}b_{3}$}};

   \node at (-0.9, -12.7){\scalebox{0.7}{$ma_{4}b_{4}$}};

   %\node at (8.0, -4.7){\scalebox{0.7}{$ma_{2}b_{2}$}};

   %\node at (-5.6, -2.3){\scalebox{0.7}{$ma_{1}b_{2}$}}; 

   %\node at (9.6, -2.3){\scalebox{0.7}{$ma_{1}b_{2}$}};

   %\node at (9.0, 1.63){\scalebox{0.7}{$ma_{1}b_{1}$}};

   \node at (1.9, -10.5){\scalebox{0.7}{$ma_{3}b_{3}$}};

   %\node at (3.0, 1.63){\scalebox{0.7}{$ma_{1}b_{1}+1$}};

   \node at (-3.9, -10.1){\scalebox{0.7}{$ma_{3}b_{3}$}};

   %\node at (-5.0, 1.63){\scalebox{0.7}{$ma_{1}b_{1}+1$}};

    %\node at (9.0, -6.57){\scalebox{0.7}{$ma_{2}b_{2}$}};

   %\node at (7.0, -6.57){\scalebox{0.7}{$ma_{2}b_{2}+1$}};

   \node at (1.0, -15.57){\scalebox{0.7}{$ma_{4}b_{4}$}};

   %\node at (3.0, -6.57){\scalebox{0.7}{$ma_{2}b_{2}+1$}};

   \node at (-3.0, -15.57){\scalebox{0.7}{$ma_{4}b_{4}$}};

   %\node at (-5.0, -6.57){\scalebox{0.7}{$ma_{2}b_{2}+1$}};     

   \node at (-0.9, -13.7){\scalebox{0.9}{$c^{a_{4},b_{4}}_{k_{4}}$}};

   %\node at (5.05, -4.7){\scalebox{0.9}{$c^{a_{2},b_{2}}_{\tilde{k}+1}$}};

   %\node at (11.05, -4.7){\scalebox{0.9}{$c^{a_{2},b_{2}}_{\tilde{k}+2}$}};

   %\node at (-6.9, -4.7){\scalebox{0.9}{$c^{a_{2},b_{2}}_{\tilde{k}-1}$}};

\path [draw= black, dashed, line width=0.01mm] (-3, -9.15) [bend right= 0]  to [bend left =  -0] (1, -13.7);

%\path [draw=black, dashed, line width=0.2mm] (-3, -0.15) [bend right= 0]  to [bend left =  -0] (3, -4.7);

%\path [draw=black, dashed, line width=0.2mm] (-3, -0.15) [bend right= 0]  to [bend left =  -0] (7, -4.7);

%\path [draw=black, dashed, line width=0.2mm] (-3, -0.15) [bend right= 0]  to [bend left =  -0] (9, -4.7);

\path [draw= black, dashed, line width=0.01mm] (1, -9.15) [bend right= 0]  to [bend left =  -0] (1, -13.7);

\path [draw=black, dashed, line width=0.01mm] (-3, -9.15) [bend right= 0]  to [bend left =  -0] (-3, -13.7);

\path [draw=black, dashed, line width=0.01mm] (-3, -13.7) [bend right= -35]  to [bend left =  35] (1,-13.7);

%\path [draw=black, dashed, line width=0.1mm] (-3, -4.7) [bend right= -35]  to [bend left =  35] (3, -4.7);

%\path [draw=black, dashed, line width=0.1mm] (-3, -4.7) [bend right= -35]  to [bend left =  35] (7, -4.7);

%\path [draw=black, dashed, line width=0.1mm] (-3, -4.7) [bend right= -35]  to [bend left =  35] (9, -4.7);

%\path [draw=black, dashed, line width=0.1mm] (-5, -4.7) [bend right= -35]  to [bend left =  35] (1, -4.7);

%\path [draw=black, dashed, line width=0.1mm] (-5, -4.7) [bend right= -35]  to [bend left =  35] (3, -4.7);

%\path [draw=black, dashed, line width=0.1mm] (-5, -4.7) [bend right= -35]  to [bend left =  35] (7, -4.7);

\path [draw=black, dashed, line width=0.01mm] (-3, -13.7) [bend right= 0]  to [bend left =  -0] (1, -9.15);

%\path [draw=black, dashed, line width=0.3mm] (-3, -4.7) [bend right= 0]  to [bend left =  -0] (3, -0.15);

\path [draw=black, dashed, line width=0.01mm] (-3, -9.15) [bend right= 30]  to [bend left =  -30] (1, -9.15);

  \path [draw= gray ,line width=0.8mm] (-3.74, -9.0) circle (0.5cm);

   \path [draw= black,fill=yellow,line width=0.4mm] (-3, -9) circle (0.3cm);

   \path [draw= gray ,line width=0.8mm] (1.78, -9.0) circle (0.5cm);

   \path [draw=black,fill=yellow,line width=0.4mm] (1, -9) circle (0.3cm);

     \path [draw=gray,line width=0.8mm] (-3, -14.74) circle (0.5cm);

   \path [draw=black,fill=yellow,line width=0.4mm] (-3, -14) circle (0.3cm);

\path [draw=gray,line width=0.8mm] (1, -14.74) circle (0.5cm);

   \path [draw=black,fill=yellow,line width=0.4mm] (1, -14) circle (0.3cm);

    \path [draw= blue ,fill=black,line width=0.4mm] (-1.8, -14.3) circle (0.02cm);

    \path [draw= blue ,fill=black,line width=0.4mm] (-1.5, -14.3) circle (0.02cm);

    \path [draw= blue ,fill=black,line width=0.4mm] (-1.2, -14.3) circle (0.02cm);

    \path [draw= blue ,fill=black,line width=0.4mm] (-0.9, -14.3) circle (0.02cm);

    \path [draw= blue ,fill=black,line width=0.4mm] (-0.6, -14.3) circle (0.02cm); 

    \path [draw= blue ,fill=black,line width=0.4mm] (-0.3, -14.3) circle (0.02cm);

\path [draw= blue ,fill=black,line width=0.4mm] (5.73, -14.3) circle (0.02cm);

    \path [draw= blue ,fill=black,line width=0.4mm] (5.43, -14.3) circle (0.02cm);

    \path [draw= blue ,fill=black,line width=0.4mm] (5.13, -14.3) circle (0.02cm);

    \path [draw= blue ,fill=black,line width=0.4mm] (4.83, -14.3) circle (0.02cm);

    \path [draw= blue ,fill=black,line width=0.4mm] (4.53, -14.3) circle (0.02cm); 

    \path [draw= blue ,fill=black,line width=0.4mm] (4.23, -14.3) circle (0.02cm);

\path [draw= blue ,fill=black,line width=0.4mm] (5.73, -9.2) circle (0.02cm);

    \path [draw= blue ,fill=black,line width=0.4mm] (5.43, -9.2) circle (0.02cm);

    \path [draw= blue ,fill=black,line width=0.4mm] (5.13, -9.2) circle (0.02cm);

    \path [draw= blue ,fill=black,line width=0.4mm] (4.83, -9.2) circle (0.02cm);

    \path [draw= blue ,fill=black,line width=0.4mm] (4.53, -9.2) circle (0.02cm); 

    \path [draw= blue ,fill=black,line width=0.4mm] (4.23, -9.2) circle (0.02cm);

\path [draw= blue ,fill=black,line width=0.4mm] (-1.8, -9.2) circle (0.02cm);

    \path [draw= blue ,fill=black,line width=0.4mm] (-1.5, -9.2) circle (0.02cm);

    \path [draw= blue ,fill=black,line width=0.4mm] (-1.2, -9.2) circle (0.02cm);

    \path [draw= blue ,fill=black,line width=0.4mm] (-0.9, -9.2) circle (0.02cm);

    \path [draw= blue ,fill=black,line width=0.4mm] (-0.6, -9.2) circle (0.02cm); 

    \path [draw= blue ,fill=black,line width=0.4mm] (-0.3, -9.2) circle (0.02cm);

\path [draw= blue ,fill=black,line width=0.4mm] (-7.83, -14.3) circle (0.02cm);

    \path [draw= blue ,fill=black,line width=0.4mm] (-7.43, -14.3) circle (0.02cm);

    \path [draw= blue ,fill=black,line width=0.4mm] (-7.13, -14.3) circle (0.02cm);

    \path [draw= blue ,fill=black,line width=0.4mm] (-6.83, -14.3) circle (0.02cm);

    \path [draw= blue ,fill=black,line width=0.4mm] (-6.53, -14.3) circle (0.02cm); 

    \path [draw= blue ,fill=black,line width=0.4mm] (-6.23, -14.3) circle (0.02cm);

    \path [draw= blue ,fill=black,line width=0.4mm] (-7.83, -9.2) circle (0.02cm);

    \path [draw= blue ,fill=black,line width=0.4mm] (-7.43, -9.2) circle (0.02cm);

    \path [draw= blue ,fill=black,line width=0.4mm] (-7.13, -9.2) circle (0.02cm);

    \path [draw= blue ,fill=black,line width=0.4mm] (-6.83, -9.2) circle (0.02cm);

    \path [draw= blue ,fill=black,line width=0.4mm] (-6.53, -9.2) circle (0.02cm); 

    \path [draw= blue ,fill=black,line width=0.4mm] (-6.23, -9.2) circle (0.02cm);

		\end{tikzpicture}
		
	\end{center}

\caption{As in the previous figure, the diagram shows many sets of nodes in parallel with the same charges and spin that represent the same DT invariants. However, we now show a tree of sucessive unlinkings (shown by the green nodes and the black lines connecting them). We also show how this tree can connect back to a starting node at a later stage in the unlinking steps. This is shown for example by the red line which can be unlinked in the next step.}

\end{figure}
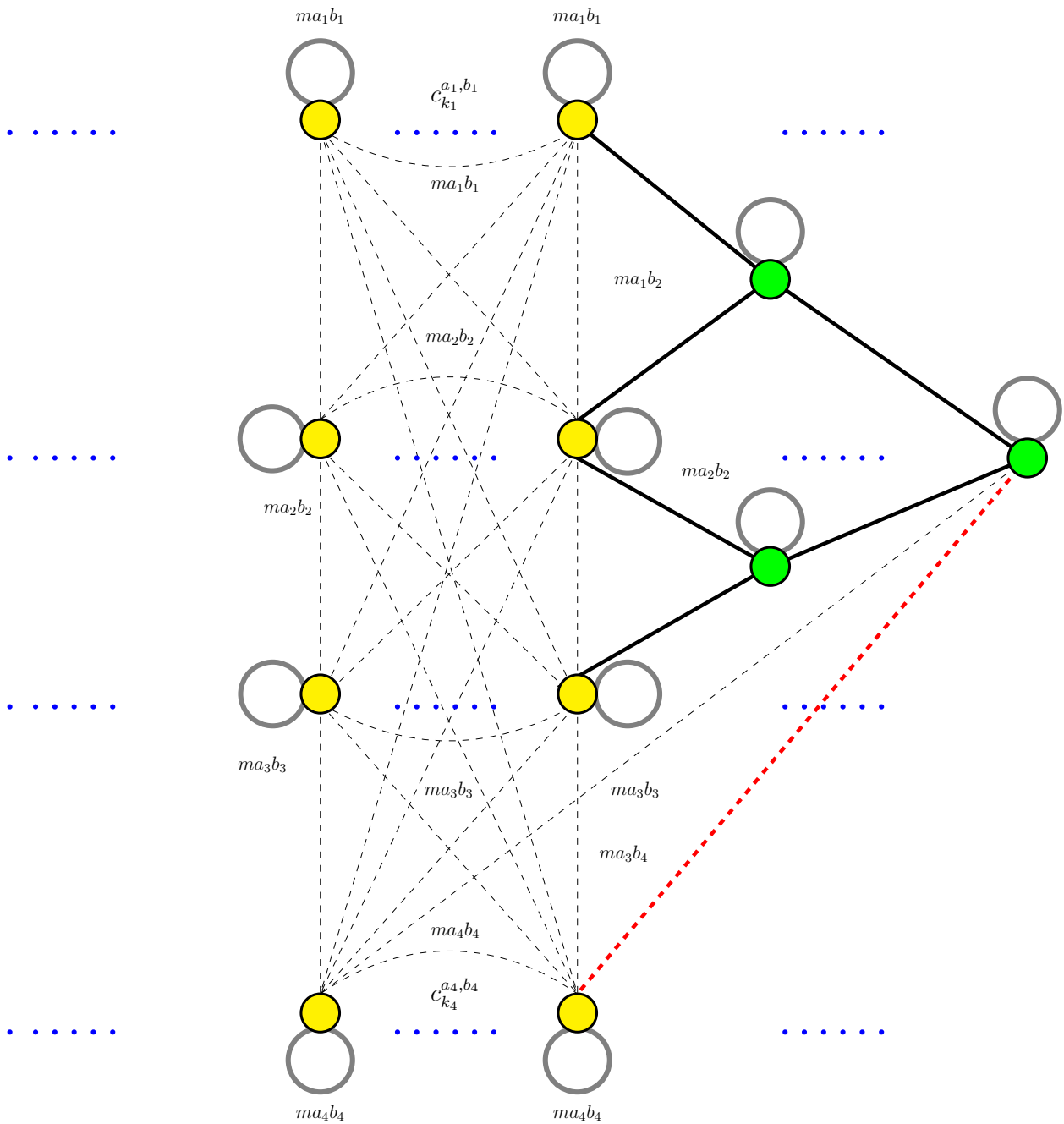

\subsubsection*{Computing the number of loops for the node with the final identification}

Now we have to determine the number of loops on the node for the final identification after all steps of the unlinking have been completed. For this we have to recursively apply the standard formula (\ref{eq:unlinking}) for unlinking %from sec. \ref{sec:linkingandunlinking}
until one obtains the number of loops on the node with the final identification. As this formula for the loops contains the linking numbers we must also apply this part of the formula. At first we recall the notation $i_{1}, \ldots ,i_{n}$ for the end nodes of the tree, or equivalently the starting nodes for the unlinking, and look at the first unlinking in the diagonalization. We use the notation (introduced in (\ref{2subtrees}), 
%from the previous subsection \ref{subsec:multiplesucessiveunlinkings} 
for steps in the tree of unlinkings) $[i_{j},i_{k}]$ to denote the new node (usually called $n$) that arises from the first unlinking step that unlinks two end nodes  $i_{j}$ and $i_{k}$ in the tree or equivalently starting nodes in the diagonalization. These end nodes can be considered the final decompositions of the unlinkings in the trees, where one successively lists the number of nodes at each split of the tree until one obtains $1$.    

We proceed by stating that the number of loops one gets after the first step of unlinking is given by the standard unlinking formula (\ref{eq:unlinking}) written as:    
\begin{align}
\begin{split}
C_{nn} &= C^{loop}_{[i_{j},i_{k}]} = C_{i_{j},i_{j}} +C_{i_{k},i_{k}} +2C_{i_{j},i_{k}}-1 = \\ 
& = (mc_{i_{j}}d_{i_{j}}+s_{a_{i_{j}},b_{i_{j}},k_{i_{j}}}
)+ (mc_{i_{k}}d_{i_{k}}+s_{a_{i_{k}},b_{i_{k}},k_{i_{k}}})+2(mc_{i_{j}}d_{i_{k}}-l_{jk}) -1.
\end{split}
\end{align}
This is written in terms of the loops and linking already existing in the infinite quiver, and the $l_{jk}$, as mentioned in the previous subsection, % \ref{sec:subtrees}, 
represents the removal of one link at a time where one increases $l_{jk}$ until all pairs of arrows have been removed.\footnote{We recall from the previous subsection %sec.\phantom{.}\ref{sec:subtrees} 
that this can also mean all arrows removed between particular subtrees rather than the absolute number of arrows between 2 nodes. This will later arise for multiple subtrees ending on the same nodes.} The next step is to generalize this to all the intermediate trees and the corresponding unlinkings one must do to get the final identification. At this point we must switch notation and use the notations for the elements of the set $S_{\mathcal{T} \mu}$ from (\ref{subtreeset}) which includes all possible intermediate steps as subtrees in the sequences in unlinkings that can produce the final identification. This means we can start by writing the number of loops of a node that resulted from an unlinking in terms of this number for the nodes that were unlinked, but furthermore with these nodes written as being the result of a previous step of unlinking. 
This can be computed as: 
\begin{align} \label{eq:generalisedloops}
&  C^{loop}_{[i_{{a(m+1)[a(m) y_{m-1}(m-1)... ,y_{1}(1)}]},i_{b(m+1)[a(m) y_{m-1}(m-1)... ,y_{1}(1)]}]} = \\ \nonumber \\ \nonumber
& C^{loop}_{[i_{{a(m+2)[a(m+1) y_{m}(m)... ,y_{1}(1)}]},i_{b(m+2)[a(m+1) y_{m}(m)... ,y_{1}(1)]}]}+ C^{loop}_{[i_{{a(m+2)[b(m+1) y_{m}(m)... ,y_{1}(1)}]},i_{b(m+2)[a(m+1) y_{m}(m)... ,y_{1}(1)]}]} \\ \nonumber \\ \nonumber
& + 2(C^{link}_{[i_{{a(m+1)[a(m) y_{m-1}(m-1)... ,y_{1}(1)}]},i_{b(m+1)[a(m) y_{m-1}(m-1)... ,y_{1}(1)]}]} \\ \nonumber \\ \nonumber & -l_{[i_{{a(m+1)[a(m) y_{m-1}(m-1)... ,y_{1}(1)}]},i_{b(m+1)[a(m) y_{m-1}(m-1)... ,y_{1}(1)]}]})-1,
\end{align}

In general $C^{loop}$ should exist for every possible intermediate unlinking tree, but to allow us to write this here we label these elements by the compositions of unlinkings that one can do successively.  

Now we define again
\begin{align}
l_{[i_{{a(m+1)[a(m) y_{m-1}(m-1)... ,y_{1}(1)}]},i_{b(m+1)[a(m) y_{m-1}(m-1)... ,y_{1}(1)]}]}
\end{align}
 as the number of linkings that have already been removed using unlinking. This can run until there is no more linkings between the nodes. It can be thought of as a generalization of the $l_{jk}$ used when unlinking the endnodes of the tree or the starting nodes in the sequence of unlinkings. In fact, in general, this is precisely $l_{\mu v}$ introduced in (\ref{lmunu}).
 %from the previous subsection \ref{sec:subtrees}.
 
We have now written down a recursion relation for the number of loops for a particular node that is the result of some subtree of unlinkings. However, we can clearly see that this equation contains the linking numbers between the loops we are unlinking. This means that we must also find a recursion relation for the linking numbers between the nodes in the quiver at a general intermediate unlinking step at some point in the diagonalization. As we did for the formula for the loops we can start by considering the nodes at the end of the tree (the first to be unlinked).

In general when two starting nodes $i_{j}$ and $i_{k}$ in the original infinite quiver $Q^{w}$ are unlinked, we follow the equation for unlinking (\ref{eq:unlinking}) such that the linking of the new node created by the unlinking (where the unlinking step can be described as $[i_{j},i_{k}]$) to the other nodes in the quiver, denoted by $i_{l}$, now becomes: 
\begin{align}
C^{link}_{[[i_{j},i_{k}],i_{l}]} = C_{i_{j}i_{l}}+C_{i_{k}i_{l}}- \delta_{i_{j}i_{l}}  - \delta_{i_{k}i_{l}}.
\end{align}
As with the number of loops in (\ref{eq:generalisedloops}) we should generalize this to all steps of the unlinking in the subtrees described by the set $S_{\mathcal{T} \mu}$ within the diagonalization by writing this in terms of the combinatorial model  presented as we did for the number of loops
\begin{align} \label{eq:linkequations}
& C^{link}_{[i_{{a(m+1)[a(m) y_{m-1}(m-1)... ,y_{1}(1)}]},i_{b(m+1)[a(m) y_{m-1}(m-1)... ,y_{1}(1)]}]} \\ \nonumber \\ \nonumber = &  \ C^{link}_{[[i_{{a(m+2)[a(m+1) y_{m}(m)... ,y_{1}(1)]}},i_{{b(m+2)[a(m+1) y_{m}(m)... ,y_{1}(1)]}}],i_{b(m+1)[a(m) y_{m-1}(m-1)... ,y_{1}(1)]}]} \\ \nonumber \\ \nonumber 
& \hspace{-1cm} =  (C^{link}_{[i_{{a(m+2)[a(m+1) y_{m}(m)... ,y_{1}(1)]}},i_{b(m+1)[a(m) y_{m-1}(m-1)... ,y_{1}(1)]}]} - l_{[i_{{a(m+2)[a(m+1) y_{m}(m)... ,y_{1}(1)]}},i_{b(m+1)[a(m) y_{m-1}(m-1)... ,y_{1}(1)]}]})  \\ \nonumber \\ \nonumber & \hspace{-1cm}+ (C^{link}_{[i_{{b(m+2)[a(m+1) y_{m}(m)... ,y_{1}(1)]}},i_{b(m+1)[a(m) y_{m-1}(m-1)... ,y_{1}(1)]}]} - l_{[i_{{b(m+2)[a(m+1) y_{m}(m)... ,y_{1}(1)]}},i_{b(m+1)[a(m) y_{m-1}(m-1)... ,y_{1}(1)]}]}) \\ \nonumber \\ \nonumber
& \hspace{-1cm}- \delta_{i_{{a(m+2)[a(m+1) y_{m}(m)... ,y_{1}(1)]}},i_{b(m+1)[a(m) y_{m-1}(m-1)... ,y_{1}(1)]}}  - \delta_{i_{{b(m+2)[a(m+1) y_{m}(m)... ,y_{1}(1)]}},i_{b(m+1)[a(m) y_{m-1}(m-1)... ,y_{1}(1)]}}.
\end{align}

We note that here we are using notation in a slightly nonstandard way. The square brackets containing two entries underneath $C^{link}$ or $C^{loop}$ are not indices of the adjacency matrix but now in the case of $C^{loop}$ represent the two nodes that are unlinked to obtain the one we are reading off the number of loops from. This node can be considered the endpoint of a single subtree in $S_{\mathcal{T} \mu}$. In the case of $C^{link}$ the two entries simply represent the two nodes that are linked. However, it should be noted that in terms of subtrees each node is now the final node of a distinct subtree so that $C_{\mu v}^{link} = \zeta_{\mu v}$.

At this point all the nodes in (\ref{eq:linkequations}) are distinct and we do not yet consider contributions of trees that return to the same nodes. If this becomes the case\footnote{This means that $\delta_{i_{{a(m+2)[a(m+1) y_{m}(m)... ,y_{1}(1)]}},i_{b(m+1)[a(m) y_{m-1}(m-1)... ,y_{1}(1)]}} = 1$.} then one must proceed by defining the self linking of the trees as the number of loops (as with a node on a quiver) and continue the recursion with the loop equations (\ref{eq:generalisedloops}) for that subtree, i.e. 
\begin{align}
C^{loop}_{[i_{a(m+2)[b(m+1)a(m) y_{m-1}(m-1)... ,y_{1}(1)]}, i_{b(m+2)[b(m+1)a(m) y_{m-1}(m-1)... ,y_{1}(1)]}]}.
\end{align}
Later we will also take these into account. However, this means that, at the moment, the following equations must hold: 
\begin{align}
\begin{split}
 \delta_{i_{{a(m+2)[a(m+1) y_{m}(m)... ,y_{1}(1)]}},i_{b(m+1)[a(m) y_{m-1}(m-1)... ,y_{1}(1)]}} = 0,
 \\  
 \delta_{i_{{b(m+2) [a(m+1) y_{m}(m)... ,y_{1}(1)]}},i_{b(m+1)[a(m) y_{m-1}(m-1)... ,y_{1}(1)]}} = 0.
 \end{split}
\end{align}
The linking numbers are then just
\begin{align}
& C^{link}_{[i_{{a(m+1)[a(m) y_{m-1}(m-1)... ,y_{1}(1)]}]},i_{b(m+1)[a(m) y_{m-1}(m-1)... ,y_{1}(1)]}]} \\ \nonumber \\ \nonumber = &  \ C^{link}_{[[i_{{a(m+2)[a(m+1) y_{m}(m)... ,y_{1}(1)]}},i_{{b(m+2)[a(m+1) y_{m}(m)... ,y_{1}(1)]}}],i_{b(m+1)[a(m) y_{m-1}(m-1)... ,y_{1}(1)]}]} \\ \nonumber \\ \nonumber 
& \hspace{-1cm} =  C^{link}_{[i_{{a(m+2)[a(m+1) y_{m}(m)... ,y_{1}(1)]}},i_{b(m+1)[a(m) y_{m-1}(m-1)... ,y_{1}(1)]}]}- l_{[i_{{a(m+2)[a(m+1) y_{m}(m)... ,y_{1}(1)]}},i_{b(m+1)[a(m) y_{m-1}(m-1)... ,y_{1}(1)]}]} \\ \nonumber \\ \nonumber & \hspace{-1cm} + C^{link}_{[i_{{b(m+2)[a(m+1) y_{m}(m)... ,y_{1}(1)]}},i_{b(m+1)[a(m) y_{m-1}(m-1)... ,y_{1}(1)]}]}- l_{[i_{{b(m+2)[a(m+1) y_{m}(m)... ,y_{1}(1)]}},i_{b(m+1)[a(m) y_{m-1}(m-1)... ,y_{1}(1)]}]}.
\end{align}

%********
%********

\subsubsection*{Unlinking a given starting node several times} \label{sec:startingnodeseveraltimes}

Now we note that it is possible, at the start of the diagonalization process, to unlink several nodes and then at some point unlink the new nodes that result from this process back from the nodes that were already unlinked. This should now give a degeneracy of the nodes that were unlinked several times. Now we proceed with this next generalization to the examples where the starting nodes are unlinked multiple times. This means we can on a tree define an equivalence class of the endpoints of the trees, defined in (\ref{setfortrees}),
%sec. \ref{subsec:multiplesucessiveunlinkings}, 
that arise by unlinking the same starting node that originally exists in the infinite quiver $Q^{w}$. Now we only have $j$ distinct end nodes of the tree such that the previously used set of endnodes is now equivalent to a smaller set $i_{1}, \ldots , i_{n} \sim i_{1}, \ldots , i_{j}$, for $j \leq n $. Therefore we must identify certain endpoints of the tree into an equivalence class when inserted into the tree 
\begin{align}
i_{p_{1}} \sim i_{p_{2}} \sim \ldots \sim i_{p_{d_{p}}} \sim i_{p}  \label{conditionsonfinalidentification-1}
\end{align}
 with multiplicity $d_{p}$. Alternatively, we now still have the endnodes labeled as $i_{p}$ but with $p \in 1, \ldots , j$ with a multiplicity rather than the $p \in 1, \ldots , n$ that we were previously using. This multiplicity is now the number of times a starting node is unlinked as part of a tree. 
In the language of the unlinking tree we can remember that each path can be labeled by its endpoint (or starting node), and that $m_{q}$ is the number of splittings in the tree from the final node to each starting node labeled by $q$ (before the equivalence class is used). This means that, for example, as mentioned before, the tree splits into subtrees at each fork until there is only one node left at the last fork:  $a,b(m_{qmax}+1) = 1_{a,b}$.

\begin{figure}[h!]
	
	\label{fig-tree-2}
	
	\begin{center}

 \vspace{0cm}
 
		\begin{tikzpicture}

  \hspace{0cm}

\path [draw= blue ,fill=black,line width=0.4mm] (11.73, -5.3) circle (0.02cm);

    \path [draw= blue ,fill=black,line width=0.4mm] (11.43, -5.3) circle (0.02cm);

    \path [draw= blue ,fill=black,line width=0.4mm] (11.13, -5.3) circle (0.02cm);

    \path [draw= blue ,fill=black,line width=0.4mm] (10.83, -5.3) circle (0.02cm);

    \path [draw= blue ,fill=black,line width=0.4mm] (20.43, -3.3) circle (0.02cm);

    \path [draw= blue ,fill=black,line width=0.4mm] (20.13, -3.3) circle (0.02cm);

    \path [draw= blue ,fill=black,line width=0.4mm] (19.83, -3.3) circle (0.02cm); 

    \path [draw= blue ,fill=black,line width=0.4mm] (19.53, -3.3) circle (0.02cm);

    \path [draw= blue ,fill=black,line width=0.4mm] (14.59, -4.3) circle (0.02cm);

    \path [draw= blue ,fill=black,line width=0.4mm] (14.29, -4.3) circle (0.02cm);

    \path [draw= blue ,fill=black,line width=0.4mm] (13.99, -4.3) circle (0.02cm); 

    \path [draw= blue ,fill=black,line width=0.4mm] (13.69, -4.3) circle (0.02cm);

    \path [draw= blue ,fill=black,line width=0.4mm] (16.13, -3.3) circle (0.02cm);

    \path [draw= blue ,fill=black,line width=0.4mm] (15.83, -3.3) circle (0.02cm);

    \path [draw= blue ,fill=black,line width=0.4mm] (15.43, -3.3) circle (0.02cm); 

    \path [draw= blue ,fill=black,line width=0.4mm] (15.13, -3.3) circle (0.02cm);

    \path [draw=black, line width=0.4mm] (11.31, -5.11) [bend right= 0]  to [bend left =  -0] (13.8, 1.2);

    \path [draw=black, line width=0.4mm] (14.1, -4.07) [bend right= 0]  to [bend left =  -0] (12.1, -3.2);

    \path [draw=black, line width=0.4mm] (13.6, 0.6) [bend right= 0]  to [bend left =  -0] (20, -3);

\path [draw=black, line width=0.4mm] (15.65, -2.98) [bend right= 0]  to [bend left =  -0] (17.65, -1.7);

\node at (11.4, -5.95){\scalebox{0.9}{$c^{a_{1},b_{1}}_{\tilde{k}}$}};

\node at (20.2, -3.95){\scalebox{0.9}{$c^{a_{1},b_{1}}_{\tilde{k}}$}};

\end{tikzpicture}
		
	\end{center}
	
\end{figure}

Therefore we can say that we have in general,  with both $p,q \in 1, \ldots , n$ as before,  
\begin{align}
i_{1_{a,b}[y_{m_{pmax}}(m_{pmax}), \ldots , y_{1}(1)]} \neq i_{1_{a,b}[y_{m_{qmax}}(m_{qmax}), \ldots , y_{1}(1)]}  
\end{align}
However, we can now have endpoints of the tree being in the same equivalence class when inserted into the unlinking tree in particular cases so that 
 \begin{align}
i_{1_{a,b}[y_{m_{p_{i}max}}(m_{pmax}), \ldots , y_{1}(1)]} \sim i_{1_{a,b}[y_{m_{p_{j}max}}(m_{q_{i}max}), \ldots , y_{1}(1)]}    
 \end{align}
 where $i_{p},j_{p} \in 1, \ldots , d_{p}$. The unlinking tree and the number of loops on the final node can be computed as before, however, we must make some modifications to the product over DT invariants as well as the equation for the final identification. As we are now taking the product over a particular starting node $p$ with an additional multiplicity of $d_{p}$ this must enter the product formula (\ref{eq:sumovertrees1}), so that we return to: 
\begin{align} \label{eq:productDT2}
\sum^{\eta_{\mu v}}_{j, n \  \Box_{j,n}, \ \mathcal{T} , \ g_{\mathcal{T}}, \  l_{\mu v} = 0 } \prod^{j}_{p=1} \Big(  c^{c_{i_{p}}, d_{i_{p}}}_{k_{i_{p}}} \Big)^{d_{p}}  \Omega^{C^{loop}_{\mathcal{T}}}_{u,s}%(-1)^{s_{k_{a,b,\Delta}}u} 
x^{au}_{2}  x_{1}^{bu}  q^{( \frac{\Delta u+s}{2}) }.
\end{align}
We should also write
\begin{align}
  \   \sum^{j}_{p=1} d_{p} =n.
\end{align}
Now we also have to modify $\Delta$ and $\Box_{n}$ from (\ref{conditionsmultiplepairunlink})-(\ref{Delta})
%sec. \ref{sec:multipairs} 
to generalize to the case of unlinking the node multiple times. This means we have to rewrite the product in equation (\ref{eq:productofidentifications1}) as 
\begin{align} \label{eq:productofidentifications2}
 \prod^{j}_{p=1} %(-1)^{s_{k_{p}}d_{p}}
 q^{ \frac{(e_{p}^{2} +  f_{p}^{2}-me_{p}f_{p}-e_{p}-f_{p}-S_{e_{p},f_{p}, k_{p}}+k_{p})d_{p}-(n-1)}{2} } x^{e_{p}d_{p}}_{2}  x_{1}^{f_{p}d_{p}} = %(-1)^{s_{k_{a,b,\Delta}}}
 q^{ \frac{\Delta}{2} } x^{a}_{2}  x_{1}^{b}. 
\end{align}
This means the conditions for $\Box_{n}$ from (\ref{conditionsmultiplepairunlink}), which we now call $\Box_{j,n}$, must be satisfied 
\begin{empheq}[box=\mymathbox]{align} 
\begin{split} \label{conditionsonfinalidentification}
& \Box_{j,n}: \  \   \  \sum^{j}_{p= 1} e_{p}d_{p} =a, \  \   \sum^{j}_{p=1} f_{p}d_{p} =b, \ \ \\  
& \sum^{j}_{p =1} (e_{p}^{2} +  f_{p}^{2}-me_{p}f_{p}-e_{p}-f_{p} - S_{e_{p}, f_{p}, k_{p}}+k_{p})d_{p}-(n-1) = \Delta, \\  
& \\  
% &  \prod^{m}_{p=1} (-1)^{s_{k_{p}}d_{p}} = (-1)^{s_{k_{a,b,\Delta}}} \\ \nonumber
&   s_{e_{p}, f_{p}, k_{p}} = e_{p}^{2} +  f_{p}^{2}-me_{p}f_{p}-1 +k_{p},
\\   \\  
& S_{e_{p}, f_{p}, k_{p}} = 1  \  \ \forall s_{e_{p}, f_{p}, k_{p}} \in 2 \mathbb{Z}+1, \  \ S_{e_{p}, f_{p}, k_{p}} = 0 \  \  \ \forall s_{e_{p}, f_{p}, k_{p}} \in 2 \mathbb{Z}
\end{split}
\end{empheq} 

% \vspace{1cm}

\subsubsection*{Unlinking a starting node multiple times and then unlinking the resulting nodes from each other}

The result must be modified if we unlink multiple starting nodes with the same charge and spin and then unlink the resulting nodes from each other. If one then unlinks the nodes arising from this back from the first nodes that were unlinked (the endnodes of the tree) the combinatorics of the DT invariants then also works differently. At first one can see that there are  $c^{c_{i_{p}}, d_{i_{p}}}_{k_{i_{p}}}$ endnodes that must be unlinked in the diagonalization which forms the product. %However, then we must node that this unlinking has already been done and there are only $c^{c_{i_{p}}, d_{i_{p}}}_{k_{i_{p}}}-1$ nodes left to unlink in the same way. 

In trees of unlinkings where at no later stage one returns to the same node, or alternatively unlinks nodes within the set representing the same charge and spin, one must unlink all the nodes with charge and spin $(c,d,k)$. However, for example, if at a later stage of the tree one returns to this set of nodes with this particular charge and spin the combinatorics becomes more complicated. 
In these cases one can also define the trees such that one has to unlink just a smaller number of nodes which we can denote by $ ( c^{c_{i_{p}}, d_{i_{p}}}_{k_{i_{p}}}-r^{c_{i_{p}}, d_{i_{p}}}_{k_{i_{p}},q} ) $ for each charge and spin for this DT invariant. Here, we have here introduced $q$ to to track a particular end node of the tree within the number of times this tree $ \mathcal{T} $ returns to the same node. This means that $q:1, \ldots , q_{max}$. After that we can also simplify by calling the nodes removed from the DT invariant that determines how many unlinkings must be done  $r^{c_{i_{p}}, d_{i_{p}}}_{k_{i_{p}},q} = r^{\mathcal{T}}_{pq}$. 

%In general, however, we might not start with $c^{c_{i_{p}}, d_{i_{p}}}_{k_{i_{p}}}$ ways that the first unlinking for a starting node originally in the quiver can be done. This is because if for example one unlinks a pair of nodes in the infinte quiver, but not all with a particular charge and spin, and then returns to unlink this starting node at some later stage of the diagonalisation the number of arrows between the new node from the diagonalisation and the original node might be different even within the same set of charges and spins. 

%This means that initially the number of ways one can start the unlinking is actually $c^{c_{i_{p}}, d_{i_{p}}}_{k_{i_{p}}}-r^{min}_{pq}$. One then proceeds with the unlinking (for this specific tree with fixed final identification and number of loops on the final node) until one can finally only unlink $c^{c_{i_{p}}, d_{i_{p}}}_{k_{i_{p}}}-r^{max}_{pq}$ nodes.

In this case the product within the sum (\ref{eq:productDT2}) must be modified such that
\begin{align} \label{eq:DTproductdifference}
\sum^{\eta_{\mu v}}_{j, n \  \Box_{j,n}, \ \mathcal{T}, \ g_{\mathcal{T}}, \  l_{\mu v} = 0 } \prod^{j, \ q_{max}}_{p,q =1} %\prod^{r^{max}_{pq}}_{r_{pq} = r^{min}_{pq}} 
\Big(c^{c_{i_{p}}, d_{i_{p}}}_{k_{i_{p}}}-r^{\mathcal{T}}_{pq}\Big)  \ \Omega^{C^{loop}_{\mathcal{T}}}_{u,s} %(-1)^{s_{k_{a,b,\Delta}u}} 
x^{au}_{2}  x_{1}^{bu}  q^{( \frac{\Delta u+s}{2})}.
\end{align}
We must here redefine some variables to complete this generalization. We are again using the same equivalence class for the endpoints of the unlinking tree such that 
%\begin{align}
%i_{[1_{a,b}[y_{m_{pimax}}(m_{pmax}), ... , y_{1}(1)]} \sim i_{[1_{a,b}[y_{m_{p_{j}max}}(m_{q_{i}max}), ... , y_{1}(1)]}
%\end{align}
  $i_{p},j_{p} \in 1, \ldots , d_{p}$. This means that for every starting node $p$ with a particular charge and spin identification there are $q_{max} =d_{p}$ factors corresponding to all the possible times a tree contains the starting node in the same equivalence class.

 %The next step is to find $\Box_{n,m}$ and $\Delta$. It is a good idea to keep the original definition of $d_{p}$ as the number of end nodes in the tree with particular identification. This means we can let $q_{max} = d_{p}$. 
 
 %We can see here that there is a contribution to the final identification for every factor in the product as this product is counting the number of ways to include another identification in the sequence of unlinkings. This means that there are $r^{max}_{pq} -r^{min}_{pq}+1$ contributions from each such factor and we must sum over $q$ to include every possible product:

%\begin{align}
   %d_{p} =  \sum^{q_{max}}_{q =1} (r^{max}_{pq} - r^{min}_{pq}+1).
%\end{align}

%Now we must also note that of course $r^{max}_{pq} < c^{c_{i_{p}}, d_{i_{p}}}_{k_{i_{p}}}$ and we can define $t_{pq} = c^{c_{i_{p}}, d_{i_{p}}}_{k_{i_{p}}} - (r^{max}_{pq} - r^{min}_{pq})$ such that $r^{max}_{pq} -
%r^{min}_{pq}=c^{c_{i_{p}}, d_{i_{p}}}_{k_{i_{p}}} - t_{pq} $. Now we can let:

%\begin{align}
  % d_{p} =  \sum^{q_{max}}_{q =1} (c^{c_{i_{p}}, d_{i_{p}}}_{k_{i_{p}}} - t_{pq} +1 ).
%\end{align}

We should note that the $r^{\mathcal{T}}_{pq}$ depend on the details of the specific tree $ \mathcal{T}$, for example, on how many times the tree returns to a particular end node within the same equivalence class and which specific end nodes have previously been unlinked. This is because the linking between new nodes in the diagonalization and the original nodes depends on what nodes were previously unlinked.  Now we can define a new variable  
\begin{align}
t^{\mathcal{T}}_{pq} = \Big(c^{c_{i_{p}}, d_{i_{p}}}_{k_{i_{p}}}-r^{\mathcal{T}}_{pq}\Big).
\end{align}
We must of course only allow solutions with $t^{ \mathcal{T}}_{pq} > 0$. There are several interesting sequences of unlinkings that often occur in a tree that should be considered separately: 

\begin{itemize}
    \item The first case entails unlinking $c^{c_{i_{p}}, d_{i_{p}}}_{k_{i_{p}}}$ starting nodes with several different charges and spin from each other and then combining the nodes in later unlinkings. Specifically this works by combining the nodes resulting from the unlinking of pairs of sets representing two different charges and spin in the next unlinking steps. This is done by dividing the end nodes into segments representing separate subtrees. When we unlink only $\Big(c^{c_{i_{p}}, d_{i_{p}}}_{k_{i_{p}}}-r^{\mathcal{T}}_{pq}\Big) $ nodes with the same charge and spin we quickly obtain the product in the previous equation (\ref{eq:DTproductdifference}) which we write as
    %Initially one has $c^{c_{i_{p}}, d_{i_{p}}}_{k_{i_{p}}}$ ways to choose 2 nodes to unlink, then $c^{c_{i_{p}}, d_{i_{p}}}_{k_{i_{p}}}-1, \ ... \ , c^{c_{i_{p}}, d_{i_{p}}}_{k_{i_{p}}}-r^{max}_{pq}$. Therefore one picks up the factor
\begin{align}
\prod^{j, \ q_{max}}_{p,q =1} %\prod^{r^{max}_{pq}}_{r_{pq} = r^{min}_{pq}} 
\Big(c^{c_{i_{p}}, d_{i_{p}}}_{k_{i_{p}}}-r^{\mathcal{T}}_{pq}\Big).   
\end{align}

\item When we unlink several nodes with the same charges and spin from another node, as shown in fig. \ref{unlinkfromothernode}, and then combine the results in later stages of unlinking the combinatorics changes and we have to look at how many ways one can choose starting points of the unlinkings from the set of nodes with the same DT invariant. To do this one can divide up the nodes with the same DT invariant into several pieces with lengths of the form
$\Big(c^{c_{i_{p}}, d_{i_{p}}}_{k_{i_{p}}}-r^{\mathcal{T}}_{pq}\Big)$ and apply the same formula as before. However, we should note that we may have to define multiple subtrees with the same nodes at their endpoints\footnote{These will then start with a different initial linking number as described in detail in sec. \ref{sec:subtrees}.} (but which can have different factors in general) to capture every possible sequence of unlinkings.

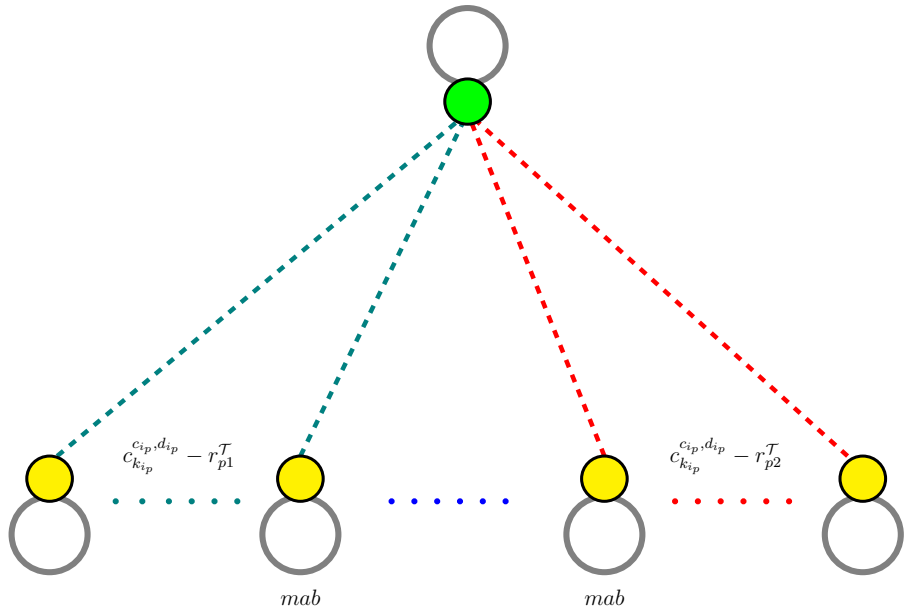
\begin{figure}[h!]	
	
	\begin{center}

 \vspace{-0.0cm}
 
		\begin{tikzpicture}

  \hspace{-0.55cm}

   %\node at (-0.9, 0.4){\scalebox{0.9}{$c^{a_{1},b_{1}}_{k}$}};

   %\node at (5.05, 0.4){\scalebox{0.9}{$c^{a_{1},b_{1}}_{k+1}$}};

   %\node at (11.05, 0.4){\scalebox{0.9}{$c^{a_{1},b_{1}}_{k+2}$}};

   %\node at (-6.9, 0.4){\scalebox{0.9}{$c^{a_{1},b_{1}}_{k-1}$}};

   %\node at (-4.0, -0.1){\scalebox{0.7}{$ma_{1}b_{1}$}};

   %\node at (2.0, -0.1){\scalebox{0.7}{$ma_{1}b_{1}$}};

   %\node at (8.0, -0.1){\scalebox{0.7}{$ma_{1}b_{1}$}};

   %\node at (-4.0, -4.7){\scalebox{0.7}{$ma_{2}b_{2}$}};

   %\node at (8.0, -4.7){\scalebox{0.7}{$ma_{2}b_{2}$}};

   %\node at (-5.6, -2.3){\scalebox{0.7}{$ma_{1}b_{2}$}}; 

   %\node at (9.6, -2.3){\scalebox{0.7}{$ma_{1}b_{2}$}};

   %\node at (9.0, 1.63){\scalebox{0.7}{$ma_{1}b_{1}$}};

   %\node at (1.0, 1.63){\scalebox{0.7}{$ma_{1}b_{1}$}};

   %\node at (3.0, 1.63){\scalebox{0.7}{$ma_{1}b_{1}+1$}};

   %\node at (-3.0, 1.63){\scalebox{0.7}{$ma_{1}b_{1}$}};

   %\node at (-5.0, 1.63){\scalebox{0.7}{$ma_{1}b_{1}+1$}};

    %\node at (9.0, -6.57){\scalebox{0.7}{$ma_{2}b_{2}$}};

   %\node at (7.0, -6.57){\scalebox{0.7}{$ma_{2}b_{2}+1$}};

   \node at (1.0, -6.57){\scalebox{0.7}{$mab$}};

   %\node at (3.0, -6.57){\scalebox{0.7}{$mab+1$}};

   \node at (-3.0, -6.57){\scalebox{0.7}{$mab$}};

   \node at (-4.6, -4.7){\scalebox{0.7}{$c^{c_{i_{p}}, d_{i_{p}}}_{k_{i_{p}}}-r^{ \mathcal{T}  }_{p1} $}};     

   \node at (2.6, -4.7){\scalebox{0.7}{$c^{c_{i_{p}}, d_{i_{p}}}_{k_{i_{p}}}-r^{ \mathcal{T}  }_{p2} $}};

\path [draw= red, dashed, line width=0.6mm] (4.2, -4.7) [bend right= 0]  to [bend left =  -0] (-0.8, -0.15);

\path [draw= red, dashed, line width=0.6mm] (1, -4.7) [bend right= 0]  to [bend left =  -0] (-0.8, -0.15);

\path [draw= teal, dashed, line width=0.6mm] (-6.2, -4.7) [bend right= 0]  to [bend left =  -0] (-0.8, -0.15);

\path [draw= teal, dashed, line width=0.6mm] (-3, -4.7) [bend right= 0]  to [bend left =  -0] (-0.8, -0.15);

%\path [draw=black, dashed, line width=0.3mm] (-3, -4.7) [bend right= 0]  to [bend left =  -0] (3, -0.15);

%\path [draw=black, dashed, line width=0.1mm] (-3, -0.15) [bend right= 30]  to [bend left =  -30] (1, -0.15);

   \path [draw= gray ,line width=0.8mm] (-0.8, 0.74) circle (0.5cm);

   \path [draw=black,fill=green,line width=0.4mm] (-0.8, 0) circle (0.3cm);

     \path [draw=gray,line width=0.8mm] (-3, -5.74) circle (0.5cm);

   \path [draw=black,fill=yellow,line width=0.4mm] (-3, -5) circle (0.3cm);

\path [draw=gray,line width=0.8mm] (-6.3, -5.74) circle (0.5cm);

   \path [draw=black,fill=yellow,line width=0.4mm] (-6.3, -5) circle (0.3cm);

\path [draw=gray,line width=0.8mm] (4.4, -5.74) circle (0.5cm);

\path [draw=black,fill=yellow,line width=0.4mm] (4.4, -5) circle (0.3cm);

\path [draw=gray,line width=0.8mm] (1, -5.74) circle (0.5cm);

   \path [draw=black,fill=yellow,line width=0.4mm] (1, -5) circle (0.3cm);

    \path [draw= blue ,fill=black,line width=0.4mm] (-1.8, -5.3) circle (0.02cm);

    \path [draw= blue ,fill=black,line width=0.4mm] (-1.5, -5.3) circle (0.02cm);

    \path [draw= blue ,fill=black,line width=0.4mm] (-1.2, -5.3) circle (0.02cm);

    \path [draw= blue ,fill=black,line width=0.4mm] (-0.9, -5.3) circle (0.02cm);

    \path [draw= blue ,fill=black,line width=0.4mm] (-0.6, -5.3) circle (0.02cm); 

    \path [draw= blue ,fill=black,line width=0.4mm] (-0.3, -5.3) circle (0.02cm);

\path [draw= red ,fill=black,line width=0.4mm] (3.43, -5.3) circle (0.02cm);

    \path [draw= red ,fill=black,line width=0.4mm] (3.13, -5.3) circle (0.02cm);

    \path [draw= red ,fill=black,line width=0.4mm] (2.83, -5.3) circle (0.02cm);

    \path [draw= red ,fill=black,line width=0.4mm] (2.53, -5.3) circle (0.02cm);

    \path [draw= red ,fill=black,line width=0.4mm] (2.23, -5.3) circle (0.02cm); 

    \path [draw= red ,fill=black,line width=0.4mm] (1.93, -5.3) circle (0.02cm);

   \path [draw= teal ,fill=black,line width=0.4mm] (-5.43, -5.3) circle (0.02cm);

    \path [draw= teal ,fill=black,line width=0.4mm] (-5.08, -5.3) circle (0.02cm);

    \path [draw= teal ,fill=black,line width=0.4mm] (-4.73, -5.3) circle (0.02cm);

    \path [draw= teal ,fill=black,line width=0.4mm] (-4.43, -5.3) circle (0.02cm);

    \path [draw= teal ,fill=black,line width=0.4mm] (-4.13, -5.3) circle (0.02cm); 

    \path [draw= teal ,fill=black,line width=0.4mm] (-3.83, -5.3) circle (0.02cm);

		\end{tikzpicture}
		
	\end{center}

\caption{Here the tree is generated by unlinking the green node once from the range of nodes labeled by the red dots (and lines on the right) and once from the range labelled by the green dots. The tree then proceeds by unlinking the nodes resulting from this from each other.     }

\label{unlinkfromothernode}

\end{figure}

\item 

Now we must consider what happens when we start the tree of unlinkings by unlinking two nodes for the same state, as shown in fig. \ref{fig:unlinkingwithinthesameDT} -- that is within the same set of nodes for a particular DT invariant $c^{c_{i_{p}}, d_{i_{p}}}_{k_{i_{p}}}$. There is then also the possibility of combining these in later steps in the unlinking sequence. As with the previous step we must partition the nodes within the set of DT invariants into non-overlapping pieces and define separate trees if necessary to sum over all possible sequences of unlinkings.

It is interesting to note that for the last case one could choose to define the trees in a way that one end of the tree is restricted to one node whereas the other (that this node is being unlinked from) sweeps the remaining nodes. This can be continued such that in the next tree a second node is chosen and then the other end sweeps all the nodes that must be unlinked from it. This process can be continued until all the nodes have been unlinked from each other (and contribute to the same final identification). In this case one obtains a product over DT invariants, of the form  $(c^{c_{i_{p}}, d_{i_{p}}}_{k_{i_{p}}} -1 )(c^{c_{i_{p}}, d_{i_{p}}}_{k_{i_{p}}} -2)(c^{c_{i_{p}}, d_{i_{p}}}_{k_{i_{p}}} -3)\, \cdots \, 1$, as a prefactor in the wall-crossing formula in front of the number of loops for the node with the final identification. It would be interesting to see if and in what cases this can be arranged into binomial coefficients, i.e., the number of ways one can choose $r$ nodes to combine in some later stage of the tree of unlinkings from $c^{c_{i_{p}}, d_{i_{p}}}_{k_{i_{p}}}$ nodes for a particular charge and spin.

\end{itemize}

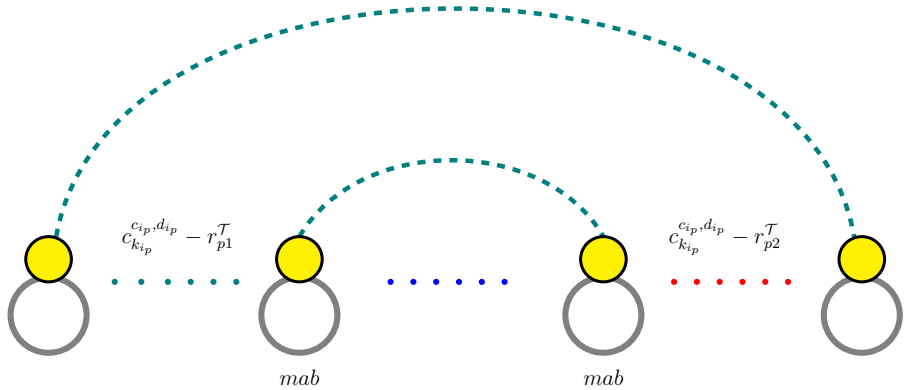
\begin{figure}[h!]	
	
	\begin{center}

 \vspace{-0.0cm}
 
		\begin{tikzpicture}

  \hspace{-0.55cm}

   %\node at (-0.9, 0.4){\scalebox{0.9}{$c^{a_{1},b_{1}}_{k}$}};

   %\node at (5.05, 0.4){\scalebox{0.9}{$c^{a_{1},b_{1}}_{k+1}$}};

   %\node at (11.05, 0.4){\scalebox{0.9}{$c^{a_{1},b_{1}}_{k+2}$}};

   %\node at (-6.9, 0.4){\scalebox{0.9}{$c^{a_{1},b_{1}}_{k-1}$}};

   %\node at (-4.0, -0.1){\scalebox{0.7}{$ma_{1}b_{1}$}};

   %\node at (2.0, -0.1){\scalebox{0.7}{$ma_{1}b_{1}$}};

   %\node at (8.0, -0.1){\scalebox{0.7}{$ma_{1}b_{1}$}};

   %\node at (-4.0, -4.7){\scalebox{0.7}{$ma_{2}b_{2}$}};

   %\node at (8.0, -4.7){\scalebox{0.7}{$ma_{2}b_{2}$}};

   %\node at (-5.6, -2.3){\scalebox{0.7}{$ma_{1}b_{2}$}}; 

   %\node at (9.6, -2.3){\scalebox{0.7}{$ma_{1}b_{2}$}};

   %\node at (9.0, 1.63){\scalebox{0.7}{$ma_{1}b_{1}$}};

   %\node at (1.0, 1.63){\scalebox{0.7}{$ma_{1}b_{1}$}};

   %\node at (3.0, 1.63){\scalebox{0.7}{$ma_{1}b_{1}+1$}};

   %\node at (-3.0, 1.63){\scalebox{0.7}{$ma_{1}b_{1}$}};

   %\node at (-5.0, 1.63){\scalebox{0.7}{$ma_{1}b_{1}+1$}};

    %\node at (9.0, -6.57){\scalebox{0.7}{$ma_{2}b_{2}$}};

   %\node at (7.0, -6.57){\scalebox{0.7}{$ma_{2}b_{2}+1$}};

   \node at (1.0, -6.57){\scalebox{0.7}{$mab$}};

   %\node at (3.0, -6.57){\scalebox{0.7}{$mab+1$}};

   \node at (-3.0, -6.57){\scalebox{0.7}{$mab$}};

   \node at (-4.6, -4.7){\scalebox{0.7}{$c^{c_{i_{p}}, d_{i_{p}}}_{k_{i_{p}}}-r^{ \mathcal{T} }_{p1} $}};     

   \node at (2.6, -4.7){\scalebox{0.7}{$c^{c_{i_{p}}, d_{i_{p}}}_{k_{i_{p}}}-r^{\mathcal{T} }_{p2} $}};

\path [draw= teal, dashed, line width=0.6mm] (-6.2, -4.7) [bend right= -80]  to [bend left =  80] (4.3, -4.7);

\path [draw= teal, dashed, line width=0.6mm] (-3, -4.7) [bend right= -60]  to [bend left =  60] (1, -4.7);

%\path [draw=black, dashed, line width=0.3mm] (-3, -4.7) [bend right= 0]  to [bend left =  -0] (3, -0.15);

%\path [draw=black, dashed, line width=0.1mm] (-3, -0.15) [bend right= 30]  to [bend left =  -30] (1, -0.15);

     \path [draw=gray,line width=0.8mm] (-3, -5.74) circle (0.5cm);

   \path [draw=black,fill=yellow,line width=0.4mm] (-3, -5) circle (0.3cm);

\path [draw=gray,line width=0.8mm] (-6.3, -5.74) circle (0.5cm);

   \path [draw=black,fill=yellow,line width=0.4mm] (-6.3, -5) circle (0.3cm);

\path [draw=gray,line width=0.8mm] (4.4, -5.74) circle (0.5cm);

\path [draw=black,fill=yellow,line width=0.4mm] (4.4, -5) circle (0.3cm);

\path [draw=gray,line width=0.8mm] (1, -5.74) circle (0.5cm);

   \path [draw=black,fill=yellow,line width=0.4mm] (1, -5) circle (0.3cm);

    \path [draw= blue ,fill=black,line width=0.4mm] (-1.8, -5.3) circle (0.02cm);

    \path [draw= blue ,fill=black,line width=0.4mm] (-1.5, -5.3) circle (0.02cm);

    \path [draw= blue ,fill=black,line width=0.4mm] (-1.2, -5.3) circle (0.02cm);

    \path [draw= blue ,fill=black,line width=0.4mm] (-0.9, -5.3) circle (0.02cm);

    \path [draw= blue ,fill=black,line width=0.4mm] (-0.6, -5.3) circle (0.02cm); 

    \path [draw= blue ,fill=black,line width=0.4mm] (-0.3, -5.3) circle (0.02cm);

\path [draw= red ,fill=black,line width=0.4mm] (3.43, -5.3) circle (0.02cm);

    \path [draw= red ,fill=black,line width=0.4mm] (3.13, -5.3) circle (0.02cm);

    \path [draw= red ,fill=black,line width=0.4mm] (2.83, -5.3) circle (0.02cm);

    \path [draw= red ,fill=black,line width=0.4mm] (2.53, -5.3) circle (0.02cm);

    \path [draw= red ,fill=black,line width=0.4mm] (2.23, -5.3) circle (0.02cm); 

    \path [draw= red ,fill=black,line width=0.4mm] (1.93, -5.3) circle (0.02cm);

   \path [draw= teal ,fill=black,line width=0.4mm] (-5.43, -5.3) circle (0.02cm);

    \path [draw= teal ,fill=black,line width=0.4mm] (-5.08, -5.3) circle (0.02cm);

    \path [draw= teal ,fill=black,line width=0.4mm] (-4.73, -5.3) circle (0.02cm);

    \path [draw= teal ,fill=black,line width=0.4mm] (-4.43, -5.3) circle (0.02cm);

    \path [draw= teal ,fill=black,line width=0.4mm] (-4.13, -5.3) circle (0.02cm); 

    \path [draw= teal ,fill=black,line width=0.4mm] (-3.83, -5.3) circle (0.02cm);

		\end{tikzpicture}
		
	\end{center}

\caption{We are now unlinking nodes within the set for the same DT invariant for $c_{i_{p}},d_{i_{p}}, k_{i_{p}}$. We are looking at a tree $\mathcal{T}$ defined by unlinking all  nodes within the range of nodes taken to lie within the green dashed lines.  }    

\label{fig:unlinkingwithinthesameDT}

\end{figure}

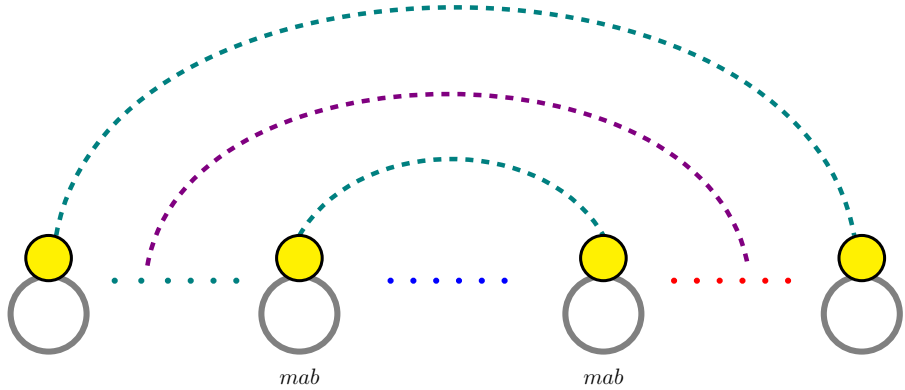
\begin{figure}[h!]

	\begin{center}

 \vspace{-0.0cm}
 
		\begin{tikzpicture}

  \hspace{-0.55cm}

   %\node at (-0.9, 0.4){\scalebox{0.9}{$c^{a_{1},b_{1}}_{k}$}};

   %\node at (5.05, 0.4){\scalebox{0.9}{$c^{a_{1},b_{1}}_{k+1}$}};

   %\node at (11.05, 0.4){\scalebox{0.9}{$c^{a_{1},b_{1}}_{k+2}$}};

   %\node at (-6.9, 0.4){\scalebox{0.9}{$c^{a_{1},b_{1}}_{k-1}$}};

   %\node at (-4.0, -0.1){\scalebox{0.7}{$ma_{1}b_{1}$}};

   %\node at (2.0, -0.1){\scalebox{0.7}{$ma_{1}b_{1}$}};

   %\node at (8.0, -0.1){\scalebox{0.7}{$ma_{1}b_{1}$}};

   %\node at (-4.0, -4.7){\scalebox{0.7}{$ma_{2}b_{2}$}};

   %\node at (8.0, -4.7){\scalebox{0.7}{$ma_{2}b_{2}$}};

   %\node at (-5.6, -2.3){\scalebox{0.7}{$ma_{1}b_{2}$}}; 

   %\node at (9.6, -2.3){\scalebox{0.7}{$ma_{1}b_{2}$}};

   %\node at (9.0, 1.63){\scalebox{0.7}{$ma_{1}b_{1}$}};

   %\node at (1.0, 1.63){\scalebox{0.7}{$ma_{1}b_{1}$}};

   %\node at (3.0, 1.63){\scalebox{0.7}{$ma_{1}b_{1}+1$}};

   %\node at (-3.0, 1.63){\scalebox{0.7}{$ma_{1}b_{1}$}};

   %\node at (-5.0, 1.63){\scalebox{0.7}{$ma_{1}b_{1}+1$}};

    %\node at (9.0, -6.57){\scalebox{0.7}{$ma_{2}b_{2}$}};

   %\node at (7.0, -6.57){\scalebox{0.7}{$ma_{2}b_{2}+1$}};

   \node at (1.0, -6.57){\scalebox{0.7}{$mab$}};

   %\node at (3.0, -6.57){\scalebox{0.7}{$mab+1$}};

   \node at (-3.0, -6.57){\scalebox{0.7}{$mab$}};

\path [draw= teal, dashed, line width=0.6mm] (-6.2, -4.7) [bend right= -80]  to [bend left =  80] (4.3, -4.7);

\path [draw= teal, dashed, line width=0.6mm] (-3, -4.7) [bend right= -60]  to [bend left =  60] (1, -4.7);

\path [draw= violet, dashed, line width=0.6mm] (-5, -5.1) [bend right= -80]  to [bend left =  80] (2.9, -5.1);

%\path [draw=black, dashed, line width=0.3mm] (-3, -4.7) [bend right= 0]  to [bend left =  -0] (3, -0.15);

%\path [draw=black, dashed, line width=0.1mm] (-3, -0.15) [bend right= 30]  to [bend left =  -30] (1, -0.15);

     \path [draw=gray,line width=0.8mm] (-3, -5.74) circle (0.5cm);

   \path [draw=black,fill=yellow,line width=0.4mm] (-3, -5) circle (0.3cm);

\path [draw=gray,line width=0.8mm] (-6.3, -5.74) circle (0.5cm);

   \path [draw=black,fill=yellow,line width=0.4mm] (-6.3, -5) circle (0.3cm);

\path [draw=gray,line width=0.8mm] (4.4, -5.74) circle (0.5cm);

\path [draw=black,fill=yellow,line width=0.4mm] (4.4, -5) circle (0.3cm);

\path [draw=gray,line width=0.8mm] (1, -5.74) circle (0.5cm);

   \path [draw=black,fill=yellow,line width=0.4mm] (1, -5) circle (0.3cm);

    \path [draw= blue ,fill=black,line width=0.4mm] (-1.8, -5.3) circle (0.02cm);

    \path [draw= blue ,fill=black,line width=0.4mm] (-1.5, -5.3) circle (0.02cm);

    \path [draw= blue ,fill=black,line width=0.4mm] (-1.2, -5.3) circle (0.02cm);

    \path [draw= blue ,fill=black,line width=0.4mm] (-0.9, -5.3) circle (0.02cm);

    \path [draw= blue ,fill=black,line width=0.4mm] (-0.6, -5.3) circle (0.02cm); 

    \path [draw= blue ,fill=black,line width=0.4mm] (-0.3, -5.3) circle (0.02cm);

\path [draw= red ,fill=black,line width=0.4mm] (3.43, -5.3) circle (0.02cm);

    \path [draw= red ,fill=black,line width=0.4mm] (3.13, -5.3) circle (0.02cm);

    \path [draw= red ,fill=black,line width=0.4mm] (2.83, -5.3) circle (0.02cm);

    \path [draw= red ,fill=black,line width=0.4mm] (2.53, -5.3) circle (0.02cm);

    \path [draw= red ,fill=black,line width=0.4mm] (2.23, -5.3) circle (0.02cm); 

    \path [draw= red ,fill=black,line width=0.4mm] (1.93, -5.3) circle (0.02cm);

   \path [draw= teal ,fill=black,line width=0.4mm] (-5.43, -5.3) circle (0.02cm);

    \path [draw= teal ,fill=black,line width=0.4mm] (-5.08, -5.3) circle (0.02cm);

    \path [draw= teal ,fill=black,line width=0.4mm] (-4.73, -5.3) circle (0.02cm);

    \path [draw= teal ,fill=black,line width=0.4mm] (-4.43, -5.3) circle (0.02cm);

    \path [draw= teal ,fill=black,line width=0.4mm] (-4.13, -5.3) circle (0.02cm); 

    \path [draw= teal ,fill=black,line width=0.4mm] (-3.83, -5.3) circle (0.02cm);

		\end{tikzpicture}
		
	\end{center}

\caption{To include the contributions from  unlinkings of nodes (unlinking within the green and red dots) inside the 2 green lines one can define additional trees by further dividing up the set of nodes at the endpoints of the tree.}

\label{fig:unlinkingwithinthesameDT2}

\end{figure}

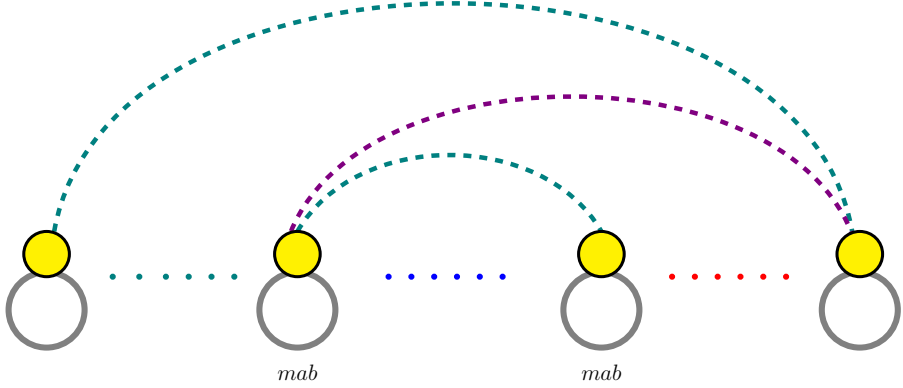
\begin{figure}[h!]	
	
	\begin{center}

 \vspace{-0.0cm}
 
		\begin{tikzpicture}

  \hspace{-0.55cm}

   %\node at (-0.9, 0.4){\scalebox{0.9}{$c^{a_{1},b_{1}}_{k}$}};

   %\node at (5.05, 0.4){\scalebox{0.9}{$c^{a_{1},b_{1}}_{k+1}$}};

   %\node at (11.05, 0.4){\scalebox{0.9}{$c^{a_{1},b_{1}}_{k+2}$}};

   %\node at (-6.9, 0.4){\scalebox{0.9}{$c^{a_{1},b_{1}}_{k-1}$}};

   %\node at (-4.0, -0.1){\scalebox{0.7}{$ma_{1}b_{1}$}};

   %\node at (2.0, -0.1){\scalebox{0.7}{$ma_{1}b_{1}$}};

   %\node at (8.0, -0.1){\scalebox{0.7}{$ma_{1}b_{1}$}};

   %\node at (-4.0, -4.7){\scalebox{0.7}{$ma_{2}b_{2}$}};

   %\node at (8.0, -4.7){\scalebox{0.7}{$ma_{2}b_{2}$}};

   %\node at (-5.6, -2.3){\scalebox{0.7}{$ma_{1}b_{2}$}}; 

   %\node at (9.6, -2.3){\scalebox{0.7}{$ma_{1}b_{2}$}};

   %\node at (9.0, 1.63){\scalebox{0.7}{$ma_{1}b_{1}$}};

   %\node at (1.0, 1.63){\scalebox{0.7}{$ma_{1}b_{1}$}};

   %\node at (3.0, 1.63){\scalebox{0.7}{$ma_{1}b_{1}+1$}};

   %\node at (-3.0, 1.63){\scalebox{0.7}{$ma_{1}b_{1}$}};

   %\node at (-5.0, 1.63){\scalebox{0.7}{$ma_{1}b_{1}+1$}};

    %\node at (9.0, -6.57){\scalebox{0.7}{$ma_{2}b_{2}$}};

   %\node at (7.0, -6.57){\scalebox{0.7}{$ma_{2}b_{2}+1$}};

   \node at (1.0, -6.57){\scalebox{0.7}{$mab$}};

   %\node at (3.0, -6.57){\scalebox{0.7}{$mab+1$}};

   \node at (-3.0, -6.57){\scalebox{0.7}{$mab$}};

\path [draw= teal, dashed, line width=0.6mm] (-6.2, -4.7) [bend right= -80]  to [bend left =  80] (4.3, -4.7);

\path [draw= teal, dashed, line width=0.6mm] (-3, -4.7) [bend right= -60]  to [bend left =  60] (1, -4.7);

\path [draw= violet, dashed, line width=0.6mm] (-3.2, -5.1) [bend right= -80]  to [bend left =  80] (4.4, -5.1);

%\path [draw=black, dashed, line width=0.3mm] (-3, -4.7) [bend right= 0]  to [bend left =  -0] (3, -0.15);

%\path [draw=black, dashed, line width=0.1mm] (-3, -0.15) [bend right= 30]  to [bend left =  -30] (1, -0.15);

     \path [draw=gray,line width=0.8mm] (-3, -5.74) circle (0.5cm);

   \path [draw=black,fill=yellow,line width=0.4mm] (-3, -5) circle (0.3cm);

\path [draw=gray,line width=0.8mm] (-6.3, -5.74) circle (0.5cm);

   \path [draw=black,fill=yellow,line width=0.4mm] (-6.3, -5) circle (0.3cm);

\path [draw=gray,line width=0.8mm] (4.4, -5.74) circle (0.5cm);

\path [draw=black,fill=yellow,line width=0.4mm] (4.4, -5) circle (0.3cm);

\path [draw=gray,line width=0.8mm] (1, -5.74) circle (0.5cm);

   \path [draw=black,fill=yellow,line width=0.4mm] (1, -5) circle (0.3cm);

    \path [draw= blue ,fill=black,line width=0.4mm] (-1.8, -5.3) circle (0.02cm);

    \path [draw= blue ,fill=black,line width=0.4mm] (-1.5, -5.3) circle (0.02cm);

    \path [draw= blue ,fill=black,line width=0.4mm] (-1.2, -5.3) circle (0.02cm);

    \path [draw= blue ,fill=black,line width=0.4mm] (-0.9, -5.3) circle (0.02cm);

    \path [draw= blue ,fill=black,line width=0.4mm] (-0.6, -5.3) circle (0.02cm); 

    \path [draw= blue ,fill=black,line width=0.4mm] (-0.3, -5.3) circle (0.02cm);

\path [draw= red ,fill=black,line width=0.4mm] (3.43, -5.3) circle (0.02cm);

    \path [draw= red ,fill=black,line width=0.4mm] (3.13, -5.3) circle (0.02cm);

    \path [draw= red ,fill=black,line width=0.4mm] (2.83, -5.3) circle (0.02cm);

    \path [draw= red ,fill=black,line width=0.4mm] (2.53, -5.3) circle (0.02cm);

    \path [draw= red ,fill=black,line width=0.4mm] (2.23, -5.3) circle (0.02cm); 

    \path [draw= red ,fill=black,line width=0.4mm] (1.93, -5.3) circle (0.02cm);

   \path [draw= teal ,fill=black,line width=0.4mm] (-5.43, -5.3) circle (0.02cm);

    \path [draw= teal ,fill=black,line width=0.4mm] (-5.08, -5.3) circle (0.02cm);

    \path [draw= teal ,fill=black,line width=0.4mm] (-4.73, -5.3) circle (0.02cm);

    \path [draw= teal ,fill=black,line width=0.4mm] (-4.43, -5.3) circle (0.02cm);

    \path [draw= teal ,fill=black,line width=0.4mm] (-4.13, -5.3) circle (0.02cm); 

    \path [draw= teal ,fill=black,line width=0.4mm] (-3.83, -5.3) circle (0.02cm);

		\end{tikzpicture}
		
	\end{center}

\caption{One can define an additional tree to include the contributions from unlinking the central nodes (labeled) by blue dots from the set of nodes on the right labeled by red dots. Again the two sets of nodes are now at endpoints of this tree.}

\label{fig:unlinkingswithinthesameDTinvariant3}

\end{figure}

%This is still within the cases of unlinking the same starting node more than once but now this step is the very first unlinking one must do. In this case there are $ c^{c_{i_{p}}, d_{i_{p}}}_{k_{i_{p}}} (c^{c_{i_{p}}, d_{i_{p}}}_{k_{i_{p}}}-1)$ ways to take this combination after that there are $ (c^{c_{i_{p}} d_{i_{p}}}_{k_{i_{p}}} -1 ) (c^{c_{i_{p}}, d_{i_{p}}}_{k_{i_{p}}}-2)$ remaining combinations. So we again pick up the product above.

%\item Now we look at unlinking 2 nodes from 2 seperate sets of states with particular spin and charge and then repeating this process for these nodes initially we get the product $c^{c_{i_{p}}, d_{i_{p}}}_{k_{i_{p}}}c^{c_{i_{q}}, d_{i_{q}}}_{k_{i_{q}}}$ then the product $(c^{c_{i_{p}}, d_{i_{p}}}_{k_{i_{p}}}-1) (c^{c_{i_{q}}, d_{i_{q}}}_{k_{i_{q}}}-1)$ and $(c^{c_{i_{p}}, d_{i_{p}}}_{k_{i_{p}}}-2) (c^{c_{i_{q}}, d_{i_{q}}}_{k_{i_{q}}}-2)$. This means that now in the flow tree there is a pair of products which one can label by  $r^{max}_{pq}$ and $r^{max}_{pq'}$.

% \newpage

\subsubsection*{Contributions to DT from nodes with different final identification} \label{subsec:differentfinalident}

In the tree formula above we were looking at all the ways one can obtain the same final identification as described in the product formulae (\ref{eq:productofidentifications1}) and (\ref{eq:productofidentifications2}). Now we want to explicitly look at the contribution to the open DT invariants of the original symmetric quiver. They, and contributions to them, take the form:
\begin{align}
A x^{B}_{2}  x_{1}^{C}  q^{\frac{D}{2}}.
\end{align}
%So we now aim to find $A$ by diagonalising both sides - by comparing the results one should get equations for the $c^{a,b}_{k}$. 
What we should notice now is that different final identifications can contribute to the same open DT invariant, in particular, if the final identification can be multiplied by some factor to give an open DT invariant with the same set of charges.
To find the contributions one must return to the sum of the form shown in equations (\ref{eq:sumovertrees1}) and (\ref{eq:DTproductdifference}). We can make this sum clearer by letting $ \mathcal{T} = \mathcal{T}_{a,b, \Delta}$ be the trees that give the same final identification labeled by $(a,b, \Delta)$. We must generalize this sum by now also summing over these different final identifications that contribute to the same open DT invariant\footnote{Here we should note that the sum over the $g_{\tau}$ implies that at every distinct step $g_{\tau}$ we sum over all the linkings $l_{\mu v}$ between all its subtrees defined in sec.\phantom{.}\ref{sec:subtrees}. However, we  don't count the linkings twice between subtrees associated to multiple $g_{\tau}$ steps.}
\begin{align} \label{firstwallcrossingformula}
Ax^{B}_{2}  x_{1}^{C}  q^{\frac{D}{2}} \  = \  \sum_{ \Box_{a,b, \Delta}} \sum^{\eta_{\mu v}}_{j, n \  \Box_{j,n}, \ \mathcal{T}_{a,b, \Delta}, \ g_{ \mathcal{T}  }, \  l_{\mu v} = 0 } \prod^{j, \ q_{max}}_{p,q =1} %\prod^{r^{max}_{pq}}_{r_{pq} = r^{min}_{pq}} 
\Big(c^{c_{i_{p}}, d_{i_{p}}}_{k_{i_{p}}}-r^{ \mathcal{T}  }_{pq}\Big)  \Omega^{C^{loop}_{ \mathcal{T}   }}_{u,s}
%(-1)^{s_{k_{a,b,\Delta}u}}
x^{au}_{2}  x_{1}^{bu}  q^{( \frac{\Delta u+s}{2}) },
\end{align}
where $ \Box_{a,b, \Delta}: au=B, bu =C, \Delta u +s =D$. Therefore this takes into account contributions with different final identifications using the formula for the DT invariants of $m$-loop quivers so that this different identification contributes to the same open DT of the original symmetric quiver.

\subsubsection*{Contributions to DT from nodes outside the dense cone}

We now recall our definition from sec. \ref{sec:wallcrossing} of the cone $\mathcal{C}$ of infinitely dense BPS rays in which the closed Kronecker DT invariants $c^{a,b}_{k}$ are unknown. Thus far our analysis was general and was considering states within this dense region. 
To complete the formula for the contributions to the DT invariants we must also consider trees of unlinkings with endnodes that are outside this dense cone $\mathcal{C}$ with unknown DT invariants. Outside this cone for $p \notin \mathcal{C}$ we always have $c^{c_{i_{p}}, d_{i_{p}}}_{k_{i_{p}}} =1, \ r^{\mathcal{T}}_{pq} = 0$. For this we must redefine the conditions on the endnodes of a tree with some endnodes being inside and some outside the dense cone $\mathcal{C}$. So for inside the cone $\mathcal{C}$ we have the same conditions (\ref{conditionsonfinalidentification}) as before
\begin{align}
 \prod^{j}_{p=1} %(-1)^{s_{k_{p}}d_{p}}
 q^{ \frac{(e_{p}^{2} +  f_{p}^{2}-me_{p}f_{p}-e_{p}-f_{p}-S_{e_{p}, f_{p}, k_{p}}+k_{p})d_{p}-(n-1)}{2} } x^{e_{p}d_{p}}_{2}  x_{1}^{f_{p}d_{p}} = %(-1)^{s_{k_{a,b,\Delta}}}
 q^{ \frac{\Delta}{2} } x^{a}_{2}  x_{1}^{b}, 
\end{align}

%\begin{align}
%& \Box_{n,m}: \  \   \  \sum^{m}_{p= 1} e_{p}d_{p} =a, \  \   \sum^{m}_{p=1} f_{p}d_{p} =b, \ \ \\ \nonumber & \sum^{m}_{p =1} (e_{p}^{2} +  f_{p}^{2}-me_{p}f_{p}-e_{p}-f_{p} - S_{e_{p}, f_{p}, k_{p}}+k_{p})d_{p}-n-1 = \Delta, \\ \nonumber
%&  \prod^{m}_{p=1} (-1)^{s_{k_{p}}d_{p}} = (-1)^{s_{k_{a,b,\Delta}}}
%\\ \nonumber
%&   s_{e_{p}, f_{p}, k_{p}} = e_{p}^{2} +  f_{p}^{2}-me_{p}f_{p}-e_{p}-f_{p} +k_{p},
%\\ \nonumber \\ \nonumber
%& S_{e_{p}, f_{p}, k_{p}} = 1  \  \ \forall s_{e_{p}, f_{p}, k_{p}} \in 2 \mathbb{Z}+1, \  \ S_{e_{p}, f_{p}, k_{p}} = 0 \  \  \ \forall s_{k_{p}} \in 2 \mathbb{Z},
%\end{align}

\begin{empheq}[box=\mymathbox]{align}
\begin{split}
& \Box_{j,n,}: \  \   \  \sum^{j}_{p= 1} e_{p}d_{p} =a, \  \   \sum^{j}_{p=1} f_{p}d_{p} =b, \ \ \\ 
& \sum^{j}_{p =1} (e_{p}^{2} +  f_{p}^{2}-me_{p}f_{p}-e_{p}-f_{p} - S_{e_{p}, f_{p}, k_{p}}+k_{p})d_{p}-(n-1) = \Delta, \\ 
%&  \prod^{m}_{p=1} (-1)^{s_{k_{p}}d_{p}} = (-1)^{s_{k_{a,b,\Delta}}}
\\ 
&   s_{e_{p}, f_{p}, k_{p}} = e_{p}^{2} +  f_{p}^{2}-me_{p}f_{p}-1 +k_{p},
\\ 
\\ 
& S_{e_{p}, f_{p}, k_{p}} = 1  \  \ \forall s_{e_{p}, f_{p}, k_{p}} \in 2 \mathbb{Z}+1, \  \ S_{e_{p}, f_{p}, k_{p}} = 0 \  \  \ \forall s_{k_{p}} \in 2 \mathbb{Z}
\end{split}
\end{empheq}

%\vspace{0.5cm}

For outside the cone we must modify these conditions. We note that here $k_{p} = 0$ such that
\begin{align}
s_{e_{p},f_{p},k_{p}} = e_{p}^{2} +  f_{p}^{2}-me_{p}f_{p}-1, \qquad  k_{p} =0, 
\end{align}
and we remember that from the dihedral symmetry we have $s_{e_{p},f_{p},k_{p}} = 0$ outside the cone such that the dilogarithm powers are positive. 
To redefine $e_{p}$ and $f_{p}$ we must now define two regions on either side of the cone in the dilogarithm identity. For the left side of the identity (all still in the infinite chamber at weak coupling) we recall the definition of the other regions from sec. \ref{sec:wallcrossing} and call this $\mathcal{C}^{-}$. For $p \in \mathcal{C}^{-}$ we must have\footnote{We have now denoted $n_{p}$ and $n'_{p}$ as the endnode $p$ of the tree outside of $\mathcal{C}$ that coincides with the nodes $n$ or $n'$ in the infinite quiver.}
\begin{align}
e_{p} = \sigma^{n_{p},m}_{1},  \qquad  f_{p} = \sigma^{n_{p},m}_{2}.
\end{align}
On the other side of the infinite product for  $p \in \mathcal{C}^{+}$ we must have:
\begin{align}
e_{p} = \sigma^{n'_{p},m}_{2},  \qquad  f_{p} = \sigma^{n'_{p},m}_{1}.
\end{align}

%**********
%**********

\subsection{The symmetric wall-crossing formula} \label{sec:symmetricwallcrossingformula}

Now we have completed the formula (\ref{firstwallcrossingformula}) for the general contribution to a particular open DT invariant on the right hand side of the full wall-crossing identity, that is for the infinite symmetric quiver $Q^{w}$ in the weak coupling region. We must then repeat this on this right hand side at strong coupling for $Q^{s}$ where there are only two 4d basis BPS states represented by the nodes of the doubled $m$-Kronecker quiver. Once we have derived this we can then choose specific open DT invariants on both sides and compare them to obtain an equation for the general $m$-Kronecker DT invariants. We should then have an equation for the closed invariants for every open invariant. The idea is that these are then solvable and give a new formulation of the $m$-Kronecker DT even in the dense or wild region. 

\subsubsection*{Comparison with finite chamber on other side of the wall}

We here proceed with the general diagonalization of the doubled (symmetric) quiver $Q^{s}$ (with $m$ pairs of arrows). This should work in analogous, but simpler way as for the infinite quiver $Q^{w}$. Firstly, this means that all endpoints of the trees always have identifications of just $x_{2}$ and $x_{1}$. Also we have $c^{1, 0} = c^{0, 1} = 1$. This greatly simplifies the formula for the diagonalization on this side as we can now remove the products of combinations of DT invariants. We look at the same open DT invariant as on the other side:
\begin{align}
A x^{B}_{2}  x_{1}^{C}  q^{\frac{D}{2}}.
\end{align}
The formula for $A$ on this side now reads
\begin{align}
A x^{B}_{2}  x_{1}^{C}  q^{\frac{D}{2}} \  = \  \sum_{ \Box_{c,d, \tilde{\Delta}}} \sum^{\eta_{ \tilde{\mu} \tilde{v}}}_{j,n \  \Box_{j,n}, \ \tilde{ \mathcal{T}}_{c,d, \tilde{\Delta}}, \ g_{ \tilde{\mathcal{T}}}, \  l_{\tilde{\mu} \tilde{ v}} = 0 }   \Omega^{C^{loop}_{\tilde{\mathcal{T}}}}_{w, \tilde{s}} \
%(-1)^{s_{k_{a,b,\Delta}u}}
x^{cw}_{2}  x_{1}^{dw}  q^{( \frac{\tilde{\Delta} w + \tilde{s}}{2}) }.
\end{align}
We now let the trees and subtrees $ \mathcal{T}, \mu $ be $ \tilde{\mathcal{T}}, \tilde{\mu} $ respectively in this chamber on the other side of the wall. We have also redefined $s$ as $\tilde{s}$, $u$ as $w$ and $a,b$ as $c,d$. The $q$ exponent in the final identification is also redefined as $\tilde{\Delta}$. As with the other side the other relations for the contributions should simplify\footnote{Note that we can set $j=2$ for the non-trivial trees as there are only two nodes in this quiver.}
\begin{align}
  %(-1)^{s_{k_{p}}d_{p}}
 q^{\frac{-(n-1)}{2}} x^{d_{2}}_{2}  x_{1}^{d_{1}} = %(-1)^{s_{k_{a,b,\Delta}}}
 q^{ \frac{\tilde{\Delta}}{2} } x^{c}_{2}  x_{1}^{d}, \qquad  d_{1}+d_{2} = n 
\end{align}
so that 
\begin{empheq}[box=\mymathbox]{align}
\Box_{j,n}: \  \   \  d_{2} =c, \  \  d_{1} =d, \ \ -(n-1) = \tilde{\Delta}. 
\end{empheq}
Now we can write down the final equation (or set of equations for the DT invariants) of the non-symmetric Kronecker quivers by directly equating the open DT invariants on both sides of the equation.

\subsubsection*{Equation for the DT invariants of $m$-Kronecker quivers}

Now we have all the information we need to write down an analog of the wall-crossing formula from the perspective of symmetric quivers. We simply compare the diagonalization of the two dual symmetric quivers, i.e. the symmetrized Kronecker quiver $Q^{s}$ and the infinite quiver on the other side of the wall $Q^{w}$. This yields
%
%\begin{align}
 %& \  \sum_{ \Box_{a,b, \Delta}} \sum^{\zeta_{\tilde{\mu} \tilde{v}}}_{n,m \  \Box_{n,m}, \ \tilde{h}_{a,b, \Delta}, \ g_{ \tilde{h}}, \  l_{\tilde{\mu} \tilde{v}} = 1 }   \Omega^{C^{loop}_{\tilde{h}}}_{au,bu,s}
%(-1)^{s_{k_{a,b,\Delta}u}}
%x^{au}_{2}  x_{1}^{bu}  q^{( \frac{\Delta u+s}{2}) }  = \\ \nonumber & \  \sum_{ \Box_{a,b, \Delta}} \sum^{\zeta_{ \mu v}}_{n,m \  \Box_{n,m}, \ h_{a,b, \Delta}, \ g_{h}, \  l_{ \mu v} = 1 } \prod^{m, \ q_{max}}_{p,q =1} % \prod^{r^{max}_{pq}}_{r_{pq} = r^{min}_{pq}} 
%\Big(c^{c_{i_{p}}, d_{i_{p}}}_{k_{i_{p}}}-r^{h}_{pq}\Big)  \Omega^{C^{loop}_{h}}_{au,bu,s}
%(-1)^{s_{k_{a,b,\Delta}u}}
%x^{au}_{2}  x_{1}^{bu}  q^{( \frac{\Delta u+s}{2}) },
%\end{align}
%
\begin{empheq}[box=\tcbhighmath]{align}    
\begin{split} \label{eq:detailed3d4dDTrelation} 
 & \  \sum_{ \Box_{c,d, \tilde{\Delta}}} \sum^{\eta_{\tilde{\mu} \tilde{v}}}_{j,n \  \Box_{j,n}, \ \tilde{\mathcal{T}}_{c,d, \tilde{\Delta}}, \ g_{ \tilde{\mathcal{T}}}, \  l_{\tilde{\mu} \tilde{v}} = 0 }   \Omega^{C^{loop}_{\tilde{\mathcal{T}}}}_{w, \tilde{s}} \
%(-1)^{s_{k_{a,b,\Delta}u}}
x^{cw}_{2}  x_{1}^{dw}  q^{( \frac{\tilde{\Delta} w+ \tilde{s}}{2}) }  = \\ 
\\ 
& \ = \sum_{ \Box_{a,b, \Delta}} \sum^{\eta_{ \mu v}}_{j,n \  \Box_{j, n}, \ \mathcal{T}_{a,b, \Delta}, \ g_{\mathcal{T}}, \  l_{ \mu v} = 0 } \prod^{j, \ q_{max}}_{p,q =1} % \prod^{r^{max}_{pq}}_{r_{pq} = r^{min}_{pq}} 
\Big(c^{c_{i_{p}}, d_{i_{p}}}_{k_{i_{p}}}-r^{\mathcal{T}}_{pq}\Big) \  \Omega^{C^{loop}_{\mathcal{T}}}_{u,s} \
%(-1)^{s_{k_{a,b,\Delta}u}}
x^{au}_{2}  x_{1}^{bu}  q^{( \frac{\Delta u+s}{2}) } = \\ 
\\ 
& \phantom{aaaaaaaaaaaaaaaaaaaaaaa} = A x^{B}_{2}  x_{1}^{C}  q^{\frac{D}{2}}
\end{split}
\end{empheq}
This gives the equations for the DT invariants of $m$-Kronecker quivers in terms of those for $m$-loop quivers. This also takes the form of a wall-crossing formula involving products over DT invariants which is reminiscent of the wall-crossing formulas derived by Alexandrov, Manschot and Pioline \cite{AlexandrovPiolinehttps://doi.org/10.48550/arxiv.1804.06928,Manschot:2014fua,BoltzmannblackholesManschot_2011} using split attractor flow methods first discussed in \cite{DenefSeibergWittenemptyhole_2000, DenefGreeneRaugas_2001}.

\subsubsection*{Simplified relation between 3d and 4d BPS degeneracies}

We now note that this equation describes a relation between the counts of (wild) BPS dyons in 4d class $\mathcal{S}$ theories and 3d BPS vorticies on the boundary. We expect that this is generalizable to all 4d BPS states described by a Kontsevich-Soibelman type wall-crossing formula. To explicitly write down the relation between the 4d and 3d theories we recall that $\Omega^{4d}_{k}(\gamma)_{i_{p}} = c^{c_{i_{p}}, d_{i_{p}}}_{k_{i_{p}}}$ and we also note that in the above equation the degeneracy of the 3d BPS vorticies can be written as $A= \Omega^{3d}_{\mathbf{d},\tilde{k}}$ where $ \mathbf{d} = (B,C)$ is the vortex charge vector and $\tilde{k} = D$ is the 3d spin. Now we write the above equation in a simplified form. We collect all the trees of unlinkings $\mathcal{T}_{a,b, \Delta}$ such that the conditions on the final identification,  $\Box_{j, n}$  and $\Box_{a,b, \Delta}$, produced by the tree result in the charge $\mathbf{d}$ and spin $\tilde{k}$ of the open 3d BPS state. We then call them $\mathbf{T}_{\mathbf{d}, \tilde{k}}$ and must now sum over them. This allows us to simplify the sum in a  way where we also sum over the contributions of the different $m$-loop quivers, arising at the final identifications, to $\Omega^{3d}_{\mathbf{d},\tilde{k}}$ in the diagonalization. This simplified equation takes form
\begin{empheq}[box=\tcbhighmath]{align} \label{eq:simplified3d4drelation}
& \ \Omega^{3d}_{\mathbf{d},\tilde{k}} = \sum_{\mathbf{T}_{\mathbf{d}, \tilde{k}}, \ C^{loop}_{\mathbf{T}}} \prod^{j, \ q_{max}}_{p,q =1} % \prod^{r^{max}_{pq}}_{r_{pq} = r^{min}_{pq}} 
\Big(\Omega^{4d}_{k}(\gamma)_{i_{p}} - r^{\mathbf{T}}_{pq}\Big) \  \Omega^{3d}_{C^{loop}_{\mathbf{T}}} 
\end{empheq}

where $\Omega^{3d}_{\mathbf{d},\tilde{k}}$ is the BPS degeneracy of the 3d theory described by the symmetrised $m$-Kronecker quiver and $\Omega^{3d}_{C^{loop}_{\mathbf{T}}}$ is that of the 3d theory simply described by the $m$-loop quiver where $m = C^{loop}_{\mathbf{T}}$. This is one of the nodes in the infinite diagonal quiver $Q^{w}_{\infty}$ obtained after the diagonalization is complete.

To test the identity (\ref{eq:simplified3d4drelation}) and to illustrate its power, in appendices \ref{app-m3}, \ref{app-m4} and \ref{app-m6} we use it to determine several invariants $c^{a,b}_{k}$ of $m$-Kronecker quivers for $m=3,4,6$ and compare them with known results obtained by other means.

%**********
%**********
%**********

\section{Summary and future directions}   \label{sec-summary}

In summary, we have reformulated Kontsevich and Soibelmans motivic wall-crossing formula (\ref{eq:motivicwallcrossingformula}) as a duality between two 3d $\mathcal{N}=2$ theories. This is analogous to the construction in \cite{Cecotti:2009ufwallcrossingtopologicalstrings} but now the duality manifests itself as an equality between two symmetric quiver generating series, encoding vortex partition functions of the theory \cite{multiskeinsEkholm:2019lmb}, on either side of the wall. In the strong coupling chamber this is given by the symmetrized BPS quiver \cite{alim2011n2new} and in the weak it is given by a new infinite quiver for which we find the adjacency matrix in appendix \ref{Appendix:normal ordering}. In the wild examples where the DT invariants of the $m$-Kronecker quivers are unknown \cite{reineke2023wildquantumdilogarithmidentities} this translates to an unknown number of nodes on the infinite quiver. We then find a new way to determine these refined BPS degeneracies by diagonalizing, as in \cite{Jankowski:2022qdpnew}, the symmetric quivers on both sides of the wall and comparing the open DT invariants of the symmetric quivers. As these must be the same this gives equations that can be solved for the original closed $m$-Kronecker DT. In the appendix \ref{appendix:diagonalize} we use diagonalization to find some low order invariants. Higher order DT invariants can be found using a wall-crossing formula (\ref{eq:detailed3d4dDTrelation}) or (\ref{eq:simplified3d4drelation}) involving trees of unlinkings. This is also equivalent to recursion relations that can determine the invariants order by order starting from those computed in the appendices. Possible future directions for this research include: 
\begin{itemize}
    \item We expect our construction of wall-crossing of the symmetrized BPS quivers to be widely generalizable beyond the $m$-Kronecker quivers to any model of wall-crossing described by a motivic wall-crossing formula. In particular, $m$-Kronecker quivers can be embedded in quivers characterizing theories with higher rank gauge groups such as $SU(3)$, and generalizations should exist to other 4d $\mathcal{N}=2$ models, e.g. trinion theories or theories with added flavors  \cite{Alim:2011ae,alim2011n2new}.

    \item We have derived a wall-crossing formula for Kronecker DT invariants in terms of trees of unlinkings. There currently exists an analogous formula derived from attractor flow trees. These are trees consisting of flow lines of central charges of BPS states, derived from periods of CY 3-folds,  under supergravity equations \cite{AlexandrovPiolinehttps://doi.org/10.48550/arxiv.1804.06928,Manschot:2014fua,BoltzmannblackholesManschot_2011, DenefSeibergWittenemptyhole_2000, DenefGreeneRaugas_2001, Alexandrov:2025sig}. It would be interesting to relate the attractor flow trees to the trees of unlinkings in this paper and we expect that ultimately the wall-crossing formulae for the DT can be related after some rewriting in terms of the open $m$-loop DT.

    \item An important part of our construction of the symmetrization of the BPS quivers involves the generators of the quantum torus algebra used in the arguments of the motivic wall-crossing formula. To symmetrize the quiver they must written in terms of another quantum torus algebra defining quantum quiver A-polynomials annihilating the 3d vortex partition function. For $A_n$ series, this relation between BPS quivers and symmetric quivers is explained geometrically in \cite{kllnps}, and it would be desirable to generalize it to other classes of quivers too. Further, given that in the context of Schur quantization \cite{Gaiotto:2024osrschurquant,Ambrosino:2025qpy} the algebra can be interpreted as the IR fusion algebra of $\frac{1}{2}$-BPS line defects it would be interesting to find such an interpretation of our 3d algebra.         

    \item  There has been recent work relating 4d and 3d $\mathcal{N}=2$ as well as 3d $\mathcal{N}=4$ theories through characters of chiral algebras \cite{Dedushenko:2023cvd, ArabiArdehali:2024ysy, Gaiotto:2024ioj4d3d, ArabiArdehali:2024vli, Kim:2024dxu, Go:2025ixu4d3d}. Importantly the Schur index can be written as a Nahm sum or symmetric quiver generating series \cite{kllnps,Gang:2024loaNahmsum3d4d} containing the symmetrized BPS quiver. This suggests that this index can also be rewritten as the equivalent generating series of an infinite quiver.

\end{itemize}

\newpage 

\appendix

%**************
%**************
%**************

\section{Normal ordering for general wall-crossing identity} \label{Appendix:normal ordering}

Here we present the explicit computations involving the normal ordering of the quantum torus algebra for the general Kontsevich-Soibelman motivic wall-crossing formulae (\ref{eq:motivicwallcrossingformula}) for $m$-Kronecker quivers (including all wild cases), which were given in this form in \cite{reineke2023wildquantumdilogarithmidentities}. Hereby we compute the adjacency matrices of the infinite symmetric quivers $Q^{w}$ discussed in sec. \ref{sec:infinitequivers}. We proceed by normal ordering all possible products of dilogarithms representing nodes in the infinite symmetric quiver to compute their linking numbers. We compute the number of loops on each node by normal ordering the operators at each dilogarithm in the infinite product.

\subsection{Normal ordering of infinite product}

Recall the variables $x$ and $y$ expressed in terms of the quantum torus algebra (\ref{eq:quantumtorusident}). We aim to find the infinite quiver (up to the number of nodes in the cone where the DT invariants are unknown) by substituting the right variables into the quantum dilogarithms on the side of $MS$ with infinitely many states and normal ordering. We start by considering a part of the quiver where the number of nodes are known, that is in the region $\mathcal{C}^{-}$. We therefore first consider the normal ordering on the LHS of the infinite product:
\begin{align} \label{lefthandside}
  : \Phi_{(0,1)}   \Phi_{\sigma (0,1)}  \Phi_{\sigma^{2} (0,1)}  \Phi_{\sigma^{3} (0,1)}  \cdots\,  :.
\end{align}
We recall the form of the dilogarithm on this side 
\begin{align}
\Phi_{a,b} = \Phi( (-1) ^{mab}q^{ \frac{a^{2} +  b^{2} -2mab-1 }{2} } x^{a}y^{b})^{(-1)^{a^{2}+b^{2}-mab-1}}. 
\\ \nonumber
\end{align}
On this LHS we define 
\begin{align}
a =  \sigma^{n} (0,1)_{1} =  \sigma^{n,m} _{1}, \  \  \  b =  \sigma^{n} (0,1)_{2} =\sigma^{n,m}_{2},
\end{align}
 where $1$ and $2$  represent the first and second component of the vector $  \sigma^{n} (0,1)$. 
We now proceed by substituting this into the expression
\begin{align}
&\Phi_{ ( \sigma^{n,m} _{1}, \  \sigma^{n,m} _{2} ) } = \\ \nonumber \\  \nonumber
& = \Phi( (-1) ^{m  \sigma^{n,m} _{1}  \sigma^{n,m} _{2} }q^{ \frac{ (\sigma^{n,m} _{1}) ^{2} +  (\sigma^{n,m} _{2})^{2} -2m \sigma^{n,m} _{1} \sigma^{n,m} _{2}  -1 }{2} } x^{\sigma^{n,m} _{1}  }y^{ \sigma^{n,m} _{2} })^{(-1)^{ (\sigma^{n,m} _{1})^{2}+  (\sigma^{n,m} _{2}) ^{2}-m \sigma^{n,m} _{1} \sigma^{n,m} _{2}   -1}}. \\ \nonumber
\end{align}
At this moment we are assuming that  $  (\sigma^{n,m} _{1})^{2}+  (\sigma^{n,m} _{2}) ^{2}-m \sigma^{n,m} _{1} \sigma^{n,m} _{2}   -1   $ is even so that the dilogarithm for this node takes the form
\begin{align}
&\Phi_{( \sigma^{n,m} _{1},   \sigma^{n,m} _{2} )} = 
 \Phi( (-1) ^{m  \sigma^{n,m} _{1}  \sigma^{n,m} _{2} }q^{ \frac{ (\sigma^{n,m} _{1}) ^{2} +  (\sigma^{n,m} _{2})^{2} -2m \sigma^{n,m} _{1} \sigma^{n,m} _{2}  -1 }{2} } x^{\sigma^{n,m} _{1}  }y^{ \sigma^{n,m} _{2} }).
\end{align}
We can now substitute in the non-commutative generators from the quantum torus algebra
\begin{align}
&\Phi_{ ( \sigma^{n,m} _{1}, \   \sigma^{n,m} _{2} ) } = 
\Phi( (-1) ^{m  \sigma^{n,m} _{1}  \sigma^{n,m} _{2} }q^{ \frac{ (\sigma^{n,m} _{1}) ^{2} +  (\sigma^{n,m} _{2})^{2} -2m \sigma^{n,m} _{1} \sigma^{n,m} _{2}  -1 }{2} } (  (q^{\frac{1}{2}})^{-1} \hat{x}_{2} \hat{y}_{1}^{m} )^{\sigma^{n,m} _{1}  }( (q^{\frac{1}{2}})^{-1} \hat{x}_{1})^{ \sigma^{n,m} _{2} }) 
\\ \nonumber
= & \  \Phi( (-1) ^{m  \sigma^{n,m} _{1}  \sigma^{n,m} _{2} }q^{ \frac{ (\sigma^{n,m} _{1}) ^{2} +  (\sigma^{n,m} _{2})^{2} -2m \sigma^{n,m} _{1} \sigma^{n,m} _{2} -\sigma^{n,m} _{1} -  \sigma^{n,m} _{2}    -1 }{2} } (   \hat{x}_{2} \hat{y}_{1}^{m} )^{\sigma^{n,m} _{1}  }( \hat{x}_{1})^{ \sigma^{n,m} _{2} }) \\ \nonumber
\\ \nonumber
= & \  \Phi( (-1) ^{m  \sigma^{n,m} _{1}  \sigma^{n,m} _{2} }q^{ \frac{ (\sigma^{n,m} _{1}) ^{2} +  (\sigma^{n,m} _{2})^{2}  -\sigma^{n,m} _{1} -  \sigma^{n,m} _{2}    -1 }{2} }  \hat{x}_{2}^{\sigma^{n,m} _{1}  }\hat{x}_{1}^{ \sigma^{n,m} _{2} }   \hat{y}_{1}^{m  \sigma^{n,m} _{1} }  ) 
\end{align}
where we have normal ordered 
\begin{align}
(\hat{x}_{2} \hat{y}_{1}^{m} )^{\sigma^{n,m} _{1}  }( \hat{x}_{1})^{ \sigma^{n,m} _{2} } =   q^{m  \sigma^{n,m} _{1}  \sigma^{n,m} _{2} }   \hat{x}_{2}^{\sigma^{n,m} _{1}  }\hat{x}_{1}^{ \sigma^{n,m} _{2} }   \hat{y}_{1}^{m  \sigma^{n,m} _{1}}, 
\end{align}
using the relations from the quantum torus algebra. The next step is to expand the quantum dilogarithm and normal order the resulting operators to get the number of loops on the particular node. Denoting
\begin{align}
\Phi_{q^{\frac{1}{2}}}(x) = \sum^{\infty}_{k=0} \frac{q^{\frac{k}{2}}x^{k}}{(q;q)_{k}}  = \prod^{\infty} _{i = 0}( 1 - q^{i+\frac{1}{2}}x)^{-1} , \qquad  \Phi_{q^{\frac{1}{2}}}(x)^{-1}
= \sum^{\infty}_{k=0} \frac{(-q^{\frac12})^{k^2} x^{k}}{(q;q)_{k}}
= \prod^{\infty} _{i = 0}( 1 - q^{i+\frac{1}{2}}x),  \label{qdilogs}
\end{align}
we now write 
\begin{align}
\begin{split}
& \Phi_{q^{\frac{1}{2}}}(  (-1) ^{m  \sigma^{n,m} _{1}  \sigma^{n,m} _{2} }q^{ \frac{ (\sigma^{n,m} _{1}) ^{2} +  (\sigma^{n,m} _{2})^{2}  -\sigma^{n,m} _{1} -  \sigma^{n,m} _{2}    -1 }{2} }  \hat{x}_{2}^{\sigma^{n,m} _{1}  }\hat{x}_{1}^{ \sigma^{n,m} _{2} }   \hat{y}_{1}^{m  \sigma^{n,m} _{1}   } ) =  \\  
& = \sum^{\infty}_{k=0} \frac{q^{\frac{k}{2}}}{(q;q)_{k}} (   (-1) ^{m  \sigma^{n,m} _{1}  \sigma^{n,m} _{2} }q^{ \frac{ (\sigma^{n,m} _{1}) ^{2} +  (\sigma^{n,m} _{2})^{2}  -\sigma^{n,m} _{1} -  \sigma^{n,m} _{2}    -1 }{2} }  )^{k}  ( \hat{x}_{2}^{\sigma^{n,m} _{1}  }\hat{x}_{1}^{ \sigma^{n,m} _{2} }   \hat{y}_{1}^{m  \sigma^{n,m} _{1} })^{k} , 
\end{split}
\end{align}
and the factors of $ q^{\frac{k}{2}}$ now cancel,
\begin{align}
\begin{split}
& \Phi_{q^{\frac{1}{2}}}(  (-1) ^{m  \sigma^{n,m} _{1}  \sigma^{n,m} _{2} }q^{ \frac{ (\sigma^{n,m} _{1}) ^{2} +  (\sigma^{n,m} _{2})^{2}  -\sigma^{n,m} _{1} -  \sigma^{n,m} _{2}    -1 }{2} }  \hat{x}_{2}^{\sigma^{n,m} _{1}  }\hat{x}_{1}^{ \sigma^{n,m} _{2} }   \hat{y}_{1}^{m  \sigma^{n,m} _{1}  } ) =  \\ 
& = \sum^{\infty}_{k=0} \frac{1}{(q;q)_{k}} (   (-1) ^{m  \sigma^{n,m} _{1}\sigma^{n,m} _{2}}q^{ \frac{ (\sigma^{n,m} _{1}) ^{2} +  (\sigma^{n,m} _{2})^{2}  -\sigma^{n,m} _{1} -  \sigma^{n,m} _{2} }{2} }  )^{k}  ( \hat{x}_{2}^{\sigma^{n,m} _{1}  }\hat{x}_{1}^{ \sigma^{n,m} _{2} }   \hat{y}_{1}^{m  \sigma^{n,m} _{1}  })^{k}.  
\end{split}
\end{align}
We must normal order the operators in the expansion of the dilogarithm
\begin{align}
( \hat{x}_{2}^{\sigma^{n,m} _{1}  }\hat{x}_{1}^{ \sigma^{n,m} _{2} }   \hat{y}_{1}^{m  \sigma^{n,m} _{1}  })^{k} =  q^{\frac{ (k^{2}-k)  m  \sigma^{n,m} _{1}  \sigma^{n,m} _{2} }{2}} \hat{x}_{2}^{\sigma^{n,m} _{1} k  }\hat{x}_{1}^{ \sigma^{n,m} _{2} k }   \hat{y}_{1}^{m  \sigma^{n,m} _{1}  k },    
\end{align}
so we are left with 
\begin{align}
\begin{split}
& \Phi_{q^{\frac{1}{2}}}(  (-1) ^{m  \sigma^{n,m} _{1}  \sigma^{n,m} _{2} }q^{ \frac{ (\sigma^{n,m} _{1}) ^{2} +  (\sigma^{n,m} _{2})^{2}  -\sigma^{n,m} _{1} -  \sigma^{n,m} _{2}    -1 }{2} }  \hat{x}_{2}^{\sigma^{n,m} _{1}  }\hat{x}_{1}^{ \sigma^{n,m} _{2} }   \hat{y}_{1}^{m  \sigma^{n,m} _{1} } ) =  \\ 
& = \sum^{\infty}_{k=0} \frac{(-q^{\frac{1}{2}})^{ k^{2} m  \sigma^{n,m} _{1} \sigma^{n,m} _{2} }   }{(q;q)_{k}} (q^{ \frac{ (\sigma^{n,m} _{1}) ^{2} +  (\sigma^{n,m} _{2})^{2}  - m  \sigma^{n,m} _{1}  \sigma^{n,m} _{2}    -\sigma^{n,m} _{1} -  \sigma^{n,m} _{2} }{2} }  )^{k}  \hat{x}_{2}^{\sigma^{n,m} _{1} k  }\hat{x}_{1}^{ \sigma^{n,m} _{2} k }   \hat{y}_{1}^{m  \sigma^{n,m} _{1}  k }       ,  
\end{split}
\end{align}
where we have used the fact that $q^{\frac{k^{2}-k}{2}} =  (-q^{\frac{1}{2}})^{k^{2}-k} $ as $k(k-1)$ is even, and we have moved the linear term into the brackets. This already looks like a symmetric quiver generating series and we have also been able to correctly put the negative sign in front of the $q^{\frac{1}{2}}$.  We can immediately see that the number of loops is given by $m  \sigma^{n,m} _{1}  \sigma^{n,m} _{2} $. One can also generate linkings between two nodes of this form. This is done by normal ordering 
\begin{align}
 (\hat{x}_{2}^{\sigma^{n_{1},m} _{1} k  }\hat{x}_{1}^{ \sigma^{n_{1},m} _{2} k }   \hat{y}_{1}^{m  \sigma^{n_{1},m} _{1}  k })   (\hat{x}_{2}^{\sigma^{n_{2},m} _{1} l  }\hat{x}_{1}^{ \sigma^{n_{2},m} _{2} l }   \hat{y}_{1}^{m  \sigma^{n_{2},m} _{1}  l})       ,
\end{align}
which means that by carrying out the normal ordering and moving all the $\hat{y}_{1}$ variables to the right we generate prefactors of the form $ q^{ m  \sigma^{n_{1},m} _{1}   \sigma^{n_{2},m} _{2} kl } =  (-q^{\frac{1}{2}})^{ 2m  \sigma^{n_{1},m} _{1}  \sigma^{n_{2},m} _{2} kl } $. This gives rise to the linking numbers $ m  \sigma^{n_{1},m} _{1}  \sigma^{n_{2},m} _{2} $.
The identifications take the following general form for the nodes labelled by $n$:
\begin{align} \label{eq:ident1}
(q^{ \frac{ (\sigma^{n,m} _{1}) ^{2} +  (\sigma^{n,m} _{2})^{2}  - m  \sigma^{n,m} _{1}  \sigma^{n,m} _{2}    -\sigma^{n,m} _{1} -  \sigma^{n,m} _{2} }{2} } x_{2}^{\sigma^{n,m} _{1}   } x_{1}^{ \sigma^{n,m} _{2}  }  ). 
\end{align}

%**********
%**********

\subsubsection*{Negative power in the dilogarithm} \label{sec:negativedilog}

So far we have assumed that the power of the dilogarithm is always positive which is true outside the dense cone $\mathcal{C}$. However, we will now perform the computation using the general formulation as if the power could be negative. This will become important later when we need to use the same techniques inside $\mathcal{C}$ where the power actually can become negative.  

If 
$ (\sigma^{n,m} _{1})^{2}+  (\sigma^{n,m} _{2}) ^{2}-m \sigma^{n,m} _{1} \sigma^{n,m} _{2}   -1  $ were to be odd then the power of the dilogarithm becomes negative, meaning that we pick up extra loops on the quiver. We must now look at
\begin{align} \label{eq:negativedilog}
 \Phi( (-1) ^{m  \sigma^{n,m} _{1}  \sigma^{n,m} _{2} }q^{ \frac{ (\sigma^{n,m} _{1}) ^{2} +  (\sigma^{n,m} _{2})^{2} -2m \sigma^{n,m} _{1} \sigma^{n,m} _{2}  -1 }{2} } x^{\sigma^{n,m} _{1}  }y^{ \sigma^{n,m} _{2} })^{-1},
\end{align}
and using (\ref{qdilogs}) we have
\begin{align}
P_{1-loop}(x,q) = (q^{\frac{1}{2}}x;q)_{\infty} =   % \prod^{\infty} _{i = 0}( 1-(q^{\frac{1}{2}}x) q^{i}) = 
%\prod^{\infty} _{i = 0}( 1-xq^{i+\frac{1}{2}}) =  
\sum^{\infty}_{d=0} \frac{(-q^{\frac{1}{2}})^{d^{2} }}{(q, q)_{d}}x^{d}
= \Phi_{q^{\frac{1}{2}}}(x)^{-1}.
\end{align}
Finally, we can substitute the variables associated to the charges into this inverse dilogarithm:
\begin{align}
\begin{split}
 & \Phi_{q^{\frac{1}{2}}}( (-1) ^{m  \sigma^{n,m} _{1}  \sigma^{n,m} _{2} }q^{ \frac{ (\sigma^{n,m} _{1}) ^{2} +  (\sigma^{n,m} _{2})^{2} -2m \sigma^{n,m} _{1} \sigma^{n,m} _{2}  -1 }{2} } x^{\sigma^{n,m} _{1}  }y^{ \sigma^{n,m} _{2} })^{-1} =   \\  
 & = P_{1-loop}((-1) ^{m  \sigma^{n,m} _{1}  \sigma^{n,m} _{2}  }q^{ \frac{ (\sigma^{n,m} _{1}) ^{2} +  (\sigma^{n,m} _{2})^{2} -2m \sigma^{n,m} _{1} \sigma^{n,m} _{2}  -1 }{2} } x^{\sigma^{n,m} _{1}  }y^{ \sigma^{n,m} _{2} } ) = \\ 
 & = \sum^{\infty}_{d=0} \frac{(-q^{\frac{1}{2}})^{d^{2} }}{(q, q)_{d}}((-1) ^{m  \sigma^{n,m} _{1}   \sigma^{n,m} _{2} }q^{ \frac{ (\sigma^{n,m} _{1}) ^{2} +  (\sigma^{n,m} _{2})^{2} -2m \sigma^{n,m} _{1} \sigma^{n,m} _{2}  -1 }{2} } x^{\sigma^{n,m} _{1}  }y^{ \sigma^{n,m} _{2} }    )^{d}. 
 \end{split}
\end{align}
Now we must again substitute in the relations between the generators of the quantum torus algebra:
\begin{align}
x = X_{2} =(q^{\frac{1}{2}})^{-1} \hat{x}_{2} \hat{y}_{1}^{m}, \qquad y  = X_{1} =(q^{\frac{1}{2}})^{-1} \hat{x}_{1},
\end{align}
so that we have
\begin{align}
\begin{split}
& \Phi_{q^{\frac{1}{2}}}( (-1) ^{m  \sigma^{n,m} _{1}  \sigma^{n,m} _{2} }q^{ \frac{ (\sigma^{n,m} _{1}) ^{2} +  (\sigma^{n,m} _{2})^{2} -2m \sigma^{n,m} _{1} \sigma^{n,m} _{2}  -1 }{2} } x^{\sigma^{n,m} _{1}  }y^{ \sigma^{n,m} _{2} })^{-1} =   \\  
& = P_{1-loop}((-1) ^{m  \sigma^{n,m} _{1}  \sigma^{n,m} _{2} }q^{ \frac{ (\sigma^{n,m} _{1}) ^{2} +  (\sigma^{n,m} _{2})^{2} -2m \sigma^{n,m} _{1} \sigma^{n,m} _{2}  -1 }{2} } x^{\sigma^{n,m} _{1}  }y^{ \sigma^{n,m} _{2} } ) = \\ 
& = \sum^{\infty}_{d=0} \frac{(-q^{\frac{1}{2}})^{d^{2} }}{(q, q)_{d}}((-1) ^{m  \sigma^{n,m} _{1}  \sigma^{n,m} _{2} }q^{ \frac{ (\sigma^{n,m} _{1}) ^{2} +  (\sigma^{n,m} _{2})^{2} -2m \sigma^{n,m} _{1} \sigma^{n,m} _{2}  -1 }{2} } ( (q^{\frac{1}{2}})^{-1} \hat{x}_{2} \hat{y}_{1}^{m}  )^{\sigma^{n,m} _{1}  }((q^{\frac{1}{2}})^{-1} \hat{x}_{1} )^{ \sigma^{n,m} _{2} }    )^{d}.  
\end{split}
\end{align}
We must at first normal order the arguments inside the quantum dilogarithm
\begin{align}
(\hat{x}_{2} \hat{y}_{1}^{m})^{ \sigma^{n,m} _{1}} (\hat{x}_{1} )^{ \sigma^{n,m} _{2} } =  q^{m  \sigma^{n,m} _{1}  \sigma^{n,m} _{2}   } \hat{x}_{2}^{ \sigma^{n,m} _{1}}  \hat{x}_{1} ^{ \sigma^{n,m} _{2} }  \hat{y}_{1}^{m  \sigma^{n,m} _{1} },
\end{align}
so that we can now write the expression as
\begin{align}
  \sum^{\infty}_{d=0} \frac{(-q^{\frac{1}{2}})^{d^{2} }}{(q, q)_{d}}((-1) ^{m  \sigma^{n,m} _{1}  \sigma^{n,m} _{2} }q^{ \frac{ (\sigma^{n,m} _{1}) ^{2} +  (\sigma^{n,m} _{2})^{2}- \sigma^{n,m} _{1} -  \sigma^{n,m} _{2}  -1 }{2} }  \hat{x}_{2}^{ \sigma^{n,m} _{1}}  \hat{x}_{1} ^{ \sigma^{n,m} _{2} }  \hat{y}_{1}^{m  \sigma^{n,m} _{1} }    )^{d}.  
\end{align}
To get loops we must normal order
\begin{align}
(  \hat{x}_{2}^{ \sigma^{n,m} _{1}}  \hat{x}_{1} ^{ \sigma^{n,m} _{2} }  \hat{y}_{1}^{m  \sigma^{n,m} _{1} } )^{d}  =   q^{\frac{(d^{2}-d) m \sigma^{n,m} _{2}  \sigma^{n,m} _{1} }{2}}\hat{x}_{2}^{ d \sigma^{n,m} _{1}}  \hat{x}_{1} ^{ d \sigma^{n,m} _{2} }  \hat{y}_{1}^{d m  \sigma^{n,m} _{1} },  
\end{align}
where as before we can write this as
\begin{align}
 (  \hat{x}_{2}^{ \sigma^{n,m} _{1}}  \hat{x}_{1} ^{ \sigma^{n,m} _{2} }  \hat{y}_{1}^{m  \sigma^{n,m} _{1} } )^{d}  =   (-q^{\frac{1}{2}} )^{(d^{2}-d) m \sigma^{n,m} _{2}  \sigma^{n,m} _{1} }\hat{x}_{2}^{ d \sigma^{n,m} _{1}}  \hat{x}_{1} ^{ d \sigma^{n,m} _{2} }  \hat{y}_{1}^{d m  \sigma^{n,m} _{1} }.     
\end{align}
Now we can put this back into the expression to obtain the form of the symmetric quiver generating series
\begin{align}
\sum^{\infty}_{d=0} \frac{(-q^{\frac{1}{2}})^{d^{2}(1+  m \sigma^{n,m} _{2}  \sigma^{n,m} _{1} ) }}{(q, q)_{d}}( q^{ \frac{ (\sigma^{n,m} _{1}) ^{2} +  (\sigma^{n,m} _{2})^{2} - m \sigma^{n,m} _{1}  \sigma^{n,m} _{2}    - \sigma^{n,m} _{1} -  \sigma^{n,m} _{2}  -1 }{2} }   )^{d}   \hat{x}_{2}^{ d \sigma^{n,m} _{1}}  \hat{x}_{1} ^{d \sigma^{n,m} _{2} }  \hat{y}_{1}^{d m  \sigma^{n,m} _{1} }   ,  
\end{align}
so the number of loops at this node is  $ 1+  m \sigma^{n,m} _{2}  \sigma^{n,m} _{1}$. We can write the identification as 
\begin{align} \label{eq::ident2}
 (q^{ \frac{ (\sigma^{n,m} _{1}) ^{2} +  (\sigma^{n,m} _{2})^{2} - m \sigma^{n,m} _{1}  \sigma^{n,m} _{2}    - \sigma^{n,m} _{1} -  \sigma^{n,m} _{2}  -1 }{2} }  x_{2}^{  \sigma^{n,m} _{1}} x_{1} ^{ \sigma^{n,m} _{2} } ).
\end{align}
One can now also generate linking between nodes represented by negative powers of the dilogarithm but they should be of the same form as those with positive powers as the operators we must ultimately normal order between two nodes are the same.

%**********
%**********

\subsubsection*{Other side of infinite product}

Now we will look at the other (right hand side) of the infinite product (\ref{eq:motivicwallcrossingformula}) in the region $\mathcal{C}^{+}$ and compute the loops and linkings on this side of the infinite quiver. For this we must normal order the following expression:
\begin{align} \label{righthandside}
: \cdots \Phi_{\sigma^{-3} (1,0)}   \Phi_{\sigma^{-2} (1,0)}  \Phi_{\sigma^{-1}(1,0)}  \Phi_{(1,0)} \cdots :
\end{align}
We can use the simple relations $  (\sigma^{-n}_{1,0})_{1} =    (\sigma^{n}_{0,1})_{2} $, and   $  (\sigma^{-n}_{1,0})_{2} =    (\sigma^{n}_{0,1})_{1} $. This means we can exchange $\sigma^{n,m} _{2}$ and  $\sigma^{n,m} _{1}$ and we call $n'$ the $n'$th dilogarithm on the right of the identity counted from the outside inwards. Therefore, for example, as an analog for (\ref{eq:negativedilog}) we have on the right for dilogarithms with negative exponentials
\begin{align}
\sum^{\infty}_{d=0} \frac{(-q^{\frac{1}{2}})^{d^{2}(1+  m \sigma^{n',m} _{1}  \sigma^{n',m} _{2} ) }}{(q, q)_{d}}( q^{ \frac{ (\sigma^{n',m} _{1}) ^{2} +  (\sigma^{n',m} _{2})^{2} - m \sigma^{n',m} _{1}  \sigma^{n',m} _{2}    - \sigma^{n',m} _{1} -  \sigma^{n',m} _{2}  -1 }{2} }   )^{d}   \hat{x}_{2}^{ d \sigma^{n',m} _{2}}  \hat{x}_{1} ^{d \sigma^{n',m} _{1} }  \hat{y}_{1}^{d m  \sigma^{n',m} _{2} },  
\end{align}
and we notice that actually everything is symmetric in the powers of $q$ in the identifications as well as in the loops so that one can exchange the labels 1 and 2 and keep these expressions. The linkings can now be determined by exchanging $1$ and $2$ whilst using the same normal ordering computation so that we obtain $ m  \sigma^{n'_{2},m} _{2}  \sigma^{n'_{1},m} _{1} $. (We also note that we are exchanging the order of the dilogarithm as $n'_{2}$ is now the dilogarithm furthest to the left when we choose a pair to compute the linkings of). The identifications for the positive and negative powers of the dilogarithm become
\begin{align} \label{eq:ident3}
(  q^{ \frac{ (\sigma^{n',m} _{1}) ^{2} +  (\sigma^{n',m} _{2})^{2}  - m  \sigma^{n',m} _{1}  \sigma^{n',m} _{2}    -\sigma^{n',m} _{1} -  \sigma^{n',m} _{2} }{2} } x_{2}^{\sigma^{n',m} _{2}   } x_{1}^{ \sigma^{n',m} _{1}  }  )
\end{align}
and 
\begin{align} \label{eq:ident4}
( q^{ \frac{ (\sigma^{n',m} _{1}) ^{2} +  (\sigma^{n',m} _{2})^{2} - m \sigma^{n',m} _{1}  \sigma^{n',m} _{2}    - \sigma^{n',m} _{1} -  \sigma^{n',m} _{2}  -1 }{2} }  x_{2}^{  \sigma^{n',m} _{2}} x_{1} ^{ \sigma^{n',m} _{1} } ).
\end{align}
The identifications in the generating parameters from (\ref{eq:ident1}) and (\ref{eq::ident2}) as well as (\ref{eq:ident3})-(\ref{eq:ident4}) are listed as equations (\ref{ident1})-(\ref{ident4}). 

%**********
%**********

\subsubsection*{Linkings between sides of infinite quiver}

We recall that there are two sides of the infinite quantum dilogarithm product (the spectrum generator) which we have called the LHS for (\ref{lefthandside}) and RHS for (\ref{righthandside}). We have just computed the linking numbers between the nodes within both the LHS and the RHS. The next step is to compute the linking numbers between the LHS and the RHS which is done by normal ordering
\begin{align}
 : \Phi_{(0,1)}   \Phi_{\sigma (0,1)}  \Phi_{\sigma^{2} (0,1)}  \Phi_{\sigma^{3} (0,1)}  \cdots \Phi_{\sigma^{-3} (1,0)}   \Phi_{\sigma^{-2} (1,0)}  \Phi_{\sigma(1,0)}  \Phi_{(1,0)} :
\end{align}
We can now start generating the linkings between these opposite sides of the quiver.
To start with, one must again normal order operators:
\begin{align}
\begin{split}
&( \hat{x}_{2}^{ l \sigma^{n,m} _{1}}  \hat{x}_{1} ^{l \sigma^{n,m} _{2} }  \hat{y}_{1}^{l m  \sigma^{n,m} _{1} } )(\hat{x}_{2}^{ d \sigma^{n',m} _{2}}  \hat{x}_{1} ^{d \sigma^{n',m} _{1} }  \hat{y}_{1}^{d m  \sigma^{n',m} _{2} }) = \\
& \qquad = q^{dl m  \sigma^{n,m} _{1}   \sigma^{n',m} _{1}   } \hat{x}_{2}^{ l \sigma^{n,m} _{1}}  \hat{x}_{1} ^{l \sigma^{n,m} _{2} }     \hat{x}_{2}^{ d \sigma^{n',m} _{2}}  \hat{x}_{1} ^{d \sigma^{n',m} _{1} }  \hat{y}_{1}^{l m  \sigma^{n,m} _{1} }    \hat{y}_{1}^{d m  \sigma^{n',m} _{2} }     
\end{split}
\end{align}
which means we obtain linkings of the form $m  \sigma^{n,m} _{1}   \sigma^{n',m} _{1} $.

%*************
%*************

\subsection{Dilogarithms inside the dense cone}
We have yet to consider the dilogarithms inside the dense cone $\mathcal{C}$ which occur in the central part of the infinite product. This is the region in which the Donaldson-Thomas invariants are unknown, which means we don't know the number of dilogarithms for each charge $(a,b)$ and spin $k$:
\begin{align} \label{insidecone}
  \cdots \prod^{\rightarrow}_{\mu_{-} < a/b < \mu_{+}} \Phi_{a,b}^{P_{a,b}}  \cdots 
\end{align}
For the dilogarithms inside the cone we may not know the number of nodes corresponding to DT invariants that we aim to find, but we can still calculate the loops and linkings. We recall that:
\begin{align}
\Phi^{P_{a,b}}_{a,b} = \prod_{k}\Phi( (-1) ^{mab}q^{ \frac{a^{2} +  b^{2} -2mab-1+k }{2} } x^{a}y^{b})^{(-1)^{(a^{2}+b^{2}-mab-1+k)} c^{a,b}_{k} }.
\end{align}
We start by looking at a single dilogarithm for a given $k$:
\begin{align}
\Phi^{k}_{a,b} = \Phi( (-1) ^{mab}q^{ \frac{a^{2} +  b^{2} -2mab-1+k }{2} } x^{a}y^{b})^{(-1)^{(a^{2}+b^{2}-mab-1+k)}}.
\end{align}
As in previous examples we must substitute the variables for the quantum torus algebra. We first look at the case where $  a^{2}+b^{2}-mab-1+k$ is even so that we look at positive powers of the dilogarithm
\begin{align}
\Phi^{k}_{a,b} = \Phi( (-1) ^{mab}q^{ \frac{a^{2} +  b^{2} -2mab-1+k }{2} } x^{a}y^{b}),
\end{align}
and we substitute in the relations between quantum torus algebras
\begin{align}
x  = (q^{\frac{1}{2}})^{-1} \hat{x}_{2} \hat{y}_{1}^{m}, \qquad
y   = (q^{\frac{1}{2}})^{-1} \hat{x}_{1},
\end{align}
therefore we have 
\begin{align}
\begin{split}
\Phi^{k}_{a,b} &= \Phi( (-1) ^{mab}q^{ \frac{a^{2} +  b^{2} -2mab-1+k }{2} } ( (q^{\frac{1}{2}})^{-1} \hat{x}_{2} \hat{y}_{1}^{m} )^{a}((q^{\frac{1}{2}})^{-1} \hat{x}_{1}   )^{b}) = \\
& = \Phi( (-1) ^{mab}q^{ \frac{a^{2} +  b^{2}-a-b -2mab-1+k }{2} } ( \hat{x}_{2} \hat{y}_{1}^{m} )^{a}((\hat{x}_{1}   )^{b}).
\end{split}
\end{align}
We first normal order $( \hat{x}_{2} \hat{y}_{1}^{m} )^{a}(\hat{x}_{1}   )^{b} =   q^{mab}\hat{x}^{a}_{2}  \hat{x}_{1}^{b}    \hat{y}_{1}^{ma} $, so now the dilogarithm becomes
\begin{align}
\Phi^{k}_{a,b} = \Phi( (-1) ^{mab}q^{ \frac{a^{2} +  b^{2}-a-b -1+k }{2} }  \hat{x}^{a}_{2}  \hat{x}_{1}^{b}    \hat{y}_{1}^{ma}   ).
\end{align}
We must now expand this to get loops, so we substitute this into the expansion of the quantum dilogarithm to get
%\begin{align}
%\Phi_{q^{\frac{1}{2}}}( \zeta) = \sum^{\infty}_{l=0} \frac{q^{\frac{l}{2}}}{(q;q)_{l}} \zeta^{l},
%\end{align}
\begin{align}
%\Phi_{q^{\frac{1}{2}}}( \zeta) = 
\sum^{\infty}_{l=0} \frac{q^{\frac{l}{2}}}{(q;q)_{l}} (   (-1) ^{mab}q^{ \frac{a^{2} +  b^{2}-a-b -1+k }{2} }  \hat{x}^{a}_{2}  \hat{x}_{1}^{b}    \hat{y}_{1}^{ma}  )^{l} = %\\ \nonumber \\ \nonumber
\sum^{\infty}_{l=0} \frac{1}{(q;q)_{l}} (   (-1) ^{mab}q^{ \frac{a^{2} +  b^{2}-a-b +k }{2} }  \hat{x}^{a}_{2}  \hat{x}_{1}^{b}    \hat{y}_{1}^{ma}  )^{l}. 
\end{align}
To get the number of loops we must normal order the following expression 
\begin{align}
( \hat{x}^{a}_{2}  \hat{x}_{1}^{b}    \hat{y}_{1}^{ma}  )^{l} =   q^{\frac{mab(l^{2}-l)}{2}}\hat{x}^{al}_{2}  \hat{x}_{1}^{bl}    \hat{y}_{1}^{mal}  =  (-q^{\frac{1}{2}})^{mab(l^{2}-l)}\hat{x}^{al}_{2}  \hat{x}_{1}^{bl}    \hat{y}_{1}^{mal}     
\end{align}
where we have used the fact that $l(l-1)$ is even, so we can write
\begin{align}
\begin{split}
\Phi^{k}_{a,b} &= \sum^{\infty}_{l=0} \frac{q^{\frac{l}{2}}}{(q;q)_{l}} (   (-1) ^{mab}q^{ \frac{a^{2} +  b^{2}-a-b -1+k }{2} }  \hat{x}^{a}_{2}  \hat{x}_{1}^{b}    \hat{y}_{1}^{ma}  )^{l} = \\ 
&= \sum^{\infty}_{l=0} \frac{(-q^{\frac{1}{2}})^{mabl^{2}} }{(q;q)_{l}} ( q^{ \frac{a^{2} +  b^{2}-mab-a-b +k }{2} } )^{l} \hat{x}^{al}_{2}  \hat{x}_{1}^{bl}    \hat{y}_{1}^{mal}, 
\end{split}
\end{align}
and we can now read off that there are $mab$ loops at this node.

%**********
%**********

\subsubsection*{Linkings between nodes of the same charge, different spin}

To consider linkings we should first note that there is a distinct set of nodes with $c^{a,b}_{k}$ elements for every $k$, so one must compute linkings for nodes within this set, for nodes with the same charge but different spin $k$ and finally with both different charge and different spin. Here we start by computing the linkings between two nodes (which we here call 1 and 2) with the same charge and two values $k_{1}$ and $k_{2}$ of the spin so that $k_{2} = k_{1} \pm 2a, a \in \mathbb{Z}$. This ensures the sign of the dilogarithm stays the same.

The linkings can be generated using the following normal ordering operation 
\begin{align}
(\hat{x}^{al}_{2}  \hat{x}_{1}^{bl}    \hat{y}_{1}^{mal} )  (\hat{x}^{ak}_{2}  \hat{x}_{1}^{bk}    \hat{y}_{1}^{mak} ) = q^{mabkl} \hat{x}^{ak}_{2}  \hat{x}_{1}^{bk}   \hat{x}^{al}_{2}  \hat{x}_{1}^{bl}   \hat{y}_{1}^{mal}  \hat{y}_{1}^{mak}.    
\end{align}
Here we immediately see that the linking is $mab$ and as this normal ordering is independent of the spins this will remain the case when simply choosing nodes within the same set of DT invariants -- that is when we specialize to $k_{1} = k_{2}$.  Therefore we always have both $mab$ loops and linkings between nodes with the same $a,b$ and the same sign of the dilogarithm. We should now write down identifications for these variables for both nodes of the form
\begin{align} \label{eq:ident5}
q^{ \frac{a^{2} +  b^{2}-mab-a-b +k_{1} }{2} } x^{a}_{2}  x_{1}^{b} ,  \qquad   q^{ \frac{a^{2} +  b^{2}-mab-a-b +k_{2} }{2} } x^{a}_{2}  x_{1}^{b}.
\end{align}

%**********
%**********

\subsubsection*{Linkings between nodes of different charges}

In general, we have different charges so that the identifications (\ref{eq:ident5}) for the nodes inside the cone on the infinite quiver generalize to
\begin{align} \label{eq:ident5bis}
q^{ \frac{a_{1}^{2} +  b_{1}^{2}-ma_{1}b_{1}-a_{1}-b_{1} +k_{1} }{2} } x^{a_{1}}_{2}  x_{1}^{b_{1}} ,  \qquad   q^{ \frac{a_{2}^{2} +  b_{2}^{2}-ma_{2}b_{2}-a_{2}-b _{2}+k_{2} }{2} } x^{a_{2}}_{2}  x_{1}^{b_{2}}.
\end{align}
In this case the linkings simply become $ma_{1}b_{2}$, as follows from the normal ordering.

%**********
%**********

\subsubsection*{Dilogarithms with negative exponentials}

Now as in \ref{sec:negativedilog} we must consider the case where the exponentials are negative which now arises when $ a^{2}+b^{2}-mab-1+k$ is odd. If this happens we must consider the analog of (\ref{eq:negativedilog}) inside the cone $\mathcal{C}$. This gives
\begin{align}
\Phi^{k}_{a,b} =\Phi( (-1) ^{mab}q^{ \frac{a^{2} +  b^{2} -2mab-1+k }{2} } x^{a}y^{b})^{-1 }.
\end{align}
Recall from the previous section that
%\begin{align}
%P_{1-loop}(x,q) = (qx;q^{2})_{\infty} =   \prod^{\infty} _{i = 0}( 1-(qx) q^{2i}) = \prod^{\infty} _{i = 0}( 1-xq^{2i+1}) =   \sum^{\infty}_{d=0} \frac{(-q)^{d^{2} }}{(q^{2}, q^{2})_{d}}x^{d},
%\end{align}
%and:
%\begin{align}
%\Phi_{q}(x) = \prod^{\infty} _{i = 0}( 1+ q^{2i+1}x)^{-1} , \    \    \    \    \   \Phi_{q}(x)^{-1}  = \prod^{\infty} _{i = 0}( 1+ q^{2i+1}x),
%\end{align}
%such that  
\begin{align}
\Phi_{q^{\frac{1}{2}}}(x)^{-1}  = P_{1-loop}(x,q). 
\end{align}
We must now use the following substitution
\begin{align}
\begin{split}
\Phi^{k}_{a,b} &=   \Phi( (-1) ^{mab}q^{ \frac{a^{2} +  b^{2}-a-b -1+k }{2} }  \hat{x}^{a}_{2}  \hat{x}_{1}^{b}    \hat{y}_{1}^{ma}   )^{-1 } =  \\ 
& = \sum^{\infty}_{d=0} \frac{(-q^{\frac{1}{2}})^{d^{2} }}{(q, q)_{d}}( (-1) ^{mab}q^{ \frac{a^{2} +  b^{2}-a-b -1+k }{2} }  \hat{x}^{a}_{2}  \hat{x}_{1}^{b}    \hat{y}_{1}^{ma}    )^{d}.   
\end{split}
\end{align}
To determine the number of loops we must normal order the following expression 
\begin{align}
(\hat{x}^{a}_{2}  \hat{x}_{1}^{b}    \hat{y}_{1}^{ma} )^{d} =  q^{\frac{(d^{2}-d) mab}{2}}  \hat{x}^{ad}_{2}  \hat{x}_{1}^{bd}    \hat{y}_{1}^{mad} =  (-q^{\frac{1}{2}})^{(d^{2}-d) mab}  \hat{x}^{ad}_{2}  \hat{x}_{1}^{bd}    \hat{y}_{1}^{mad}. 
\end{align}
Putting this back into the expression for the expansion of the quantum dilogarithm we get 
\begin{align}
 \sum^{\infty}_{d=0} \frac{(-q^{\frac{1}{2}})^{(1+mab)d^{2} }}{(q, q)_{d}}( q^{ \frac{a^{2} +  b^{2}-mab-a-b -1+k }{2} } )^{d}  \hat{x}^{ad}_{2}  \hat{x}_{1}^{bd}    \hat{y}_{1}^{mad},   
\end{align}
and recover $1+mab$ loops. The linking numbers should be the same as those for positive exponents as the normal ordering of operators between two nodes stays the same. However, as with the negative exponent outside the dense cone, the identification has shifted and now reads as
\begin{align} \label{eq:ident6}
q^{ \frac{a^{2} +  b^{2}-mab-a-b -1+k }{2} }x^{a}_{2}  x_{1}^{b}.  
\end{align}
The identifications in (\ref{eq:ident5}) for positive exponents and in (\ref{eq:ident6}) for negative exponents are also listed as (\ref{ident5}).

%**********
%**********

\subsection{Linkings between nodes in the dense cone and nodes outside the cone}

Finally, we must compute the linkings between nodes within the dense cone and nodes outside it. This can be done by considering the appropriate normal ordering of the products of quantum torus operators.

%**********
%**********

\subsubsection*{Left hand side of infinite product}

 We will start with the linking the nodes on the left of the identity, as shown in (\ref{lefthandside}), with those inside the cone from (\ref{insidecone}). For this we normal order the dilogarithm product
\begin{align}
 : \Phi_{(0,1)}   \Phi_{\sigma (0,1)}  \Phi_{\sigma^{2} (0,1)}  \Phi_{\sigma^{3} (0,1)}  \cdots \prod_{\mu_{-} < a/b < \mu_{+}} \Phi_{a,b}^{P_{a,b}} :  
\end{align}
Specifically, following the same computation as with the other nodes, we normal order the following monomials 
\begin{align}
 ( \hat{x}_{2}^{ l \sigma^{n,m} _{1}}  \hat{x}_{1} ^{l \sigma^{n,m} _{2} }  \hat{y}_{1}^{l m  \sigma^{n,m} _{1} } )    (\hat{x}^{ad}_{2}  \hat{x}_{1}^{bd}    \hat{y}_{1}^{mad}) = q^{ ld m  \sigma^{n,m} _{1} b } \hat{x}_{2}^{ l \sigma^{n,m} _{1}}  \hat{x}_{1} ^{l \sigma^{n,m} _{2} } \hat{x}^{ad}_{2}  \hat{x}_{1}^{bd} \hat{y}_{1}^{l m  \sigma^{n,m} _{1} }  \hat{y}_{1}^{mad},
\end{align}
therefore the number of links is given by $ m  \sigma^{n,m} _{1} b  $.

%**********
%**********

\subsubsection*{Right hand side of infinite product}

For the other side of the dilogarithm product, we are looking for the linkings from within the cone to the right side of the identity. We normal order
\begin{align}
 : \prod_{\mu_{-} < a/b < \mu_{+}} \Phi_{a,b}^{P_{a,b}}  \cdots\Phi_{\sigma^{-3} (1,0)}   \Phi_{\sigma^{-2} (1,0)}  \Phi_{\sigma(1,0)}  \Phi_{(1,0)} : 
\end{align}
Using the same procedure as on the LHS we compute the linkings between the right hand side of the identity and the nodes inside the dense cone by normal ordering
\begin{align}
 (\hat{x}^{ac}_{2}  \hat{x}_{1}^{bc}    \hat{y}_{1}^{mac}) (\hat{x}_{2}^{ d \sigma^{n',m} _{2}}  \hat{x}_{1} ^{d \sigma^{n',m} _{1} }  \hat{y}_{1}^{d m  \sigma^{n',m} _{2} }) = q^{ma \sigma^{n',m} _{1} cd} \hat{x}^{ac}_{2}  \hat{x}_{1}^{bc} \hat{x}_{2}^{ d \sigma^{n',m} _{2}}  \hat{x}_{1} ^{d \sigma^{n',m} _{1} }\hat{y}_{1}^{mac}\hat{y}_{1}^{d m  \sigma^{n',m} _{2} },
\end{align}
so the linkings we finally obtain become $ma \sigma^{n',m} _{1} $.

%**************
%**************
%**************

\section{Low order computation of DT invariants}  \label{appendix:diagonalize}

In this appendix we test our symmetric wall-crossing formula (\ref{eq:detailed3d4dDTrelation})-(\ref{eq:simplified3d4drelation}) and illustrate its power in explicit examples. We do this by looking at low order computations for the closed DT invariants $c^{a,b}_{k}$ inside the dense cone. These low order computations do not require the use of the trees of unlinkings (as the trees always generate identifications for the open DT invariants of order higher than the initial identifications in $Q^{w}$ which are not required to compute low order $c^{a,b}_{k}$). This means that one can directly read off the closed low order DT invariants just by diagonalising the doubled $m$-Kronecker quivers. We compare the results we obtain using our diagonalization procedure with the known results given by Reineke \cite{reineke2023wildquantumdilogarithmidentities}. Low order DT invariants are usually taken as those involving rays with small ratios $a/b$ and small charges including the charges along the central ray of the scattering diagram (for which the complete expression is known). Here we focus on the rays with charges $(1,k)$ for which a binomial formula exists. 

%************
%************

\subsection{Start of diagonalization of doubled $m$-Kronecker quivers}   \label{app-diagonalizeQs}

As the computation of the low order DT invariants of the Kronecker quivers involves diagonalizing the symmetrized quiver for specific examples, we briefly write the first few steps arising from just unlinking the first two nodes. We carry this out for the general $m$-arrow case that obtains the identifications we are looking for
\begin{align} \label{startofgeneralm}
C= 
\begin{pmatrix}
0  & m \\
m & 0
\end{pmatrix}, \ \
C= 
\begin{pmatrix}
0  & m-1 & m-1\\
m-1 & 0  & m-1 \\
m-1 & m-1 & 2m-1 
\end{pmatrix}, \ \
C= 
\begin{pmatrix}
0  & m-2 & m-1 &  m-2 \\
m-2 & 0  & m-1 &  m-2 \\
m-1 & m-1 & 2m-1 & 2m-2 \\
 m-2 & m-2 & 2m-2 &  2m-3
\end{pmatrix}
\end{align}
This can be written in general for $n_{1}$ arrows removed as
\begin{align} \label{generalmunlink}
C= 
\begin{pmatrix}
0  & m-n_{1} & m-1 &  m-2 & ... & m-n_{1}-2 & m-n_{1}-1  \\
m-n_{1} & 0  & m-1 &  m-2 & ... & m-n_{1}-2 & m-n_{1}-1 \\
m-1 & m-1 & 2m-1 & 2m-2 & ... & 2m-2 & 2m-2 \\
 m-2 & m-2 & 2m-2 &  2m-3 & ... & 2m-4 & 2m-4 \\
 ...   &  ...   &  ...    &   ...    & ... & ...& ...  \\
m-n_{1}-2 & m-n_{1}-2 & 2m-2 & 2m-4 & ... & 2m -2n_{1}-3 & 2(m-n_{1}-2)  \\
m-n_{1}-1   &  m-n_{1}-1   &  2m-2    &   2m-4    & ... & 2(m-n_{1}-2) &2m -2n_{1}-1   \\
\end{pmatrix}
\end{align}
which can then be continued until $n_{1}=m$ and all arrows connecting the two nodes have been removed. Now the next step is to unlink the nodes created by unlinking the original two nodes from those same original nodes (this generates the low order generating function for the open DT invariants). We choose the first node as an example and call  $n_{2}$ the number of times these nodes have been unlinked. We then obtain a matrix containing entries of the form
\begin{small}
\begin{align*}
\hspace{-2.1cm}
 & \  \  \  \  \  \ \  \  \  \  \  \  \  \   \  \   \   \  C= \\ \nonumber
\hspace{-2.1cm}
& \begin{pmatrix}
0  & m-n_{1} & m-1 &  m-2 & ... & m-n_{1}-2 & m-n_{1}-n_{2}-1 & ... & m-n_{1}-n_{2}-2 & ... \\
m-n_{1} & 0  & m-1 &  m-2 & ... & m-n_{1}-2 & m-n_{1}-1 & ... & 2m-2n_{1}-1 & ... \\
m-1 & m-1 & 2m-1 & 2m-2 & ... & 2m-2 & 2m-2 & & & \\
 m-2 & m-2 & 2m-2 &  2m-3 & ... & 2m-4 & 2m-4 & & & \\
 ...   &  ...   &  ...    &   ...    & ... & ...& ... & & & \\
m-n_{1}-2 & m-n_{1}-2 & 2m-2 & 2m-4 & ... & 2m -2n_{1}-3 & 2(m-n_{1}-2) &  & & \\
m-n_{1}-n_{2}-1   &  m-n_{1}-1   &  2m-2    &   2m-4    & ... & 2(m-n_{1}-2) &2m -2n_{1}-1 &  & & \\
    ...        &     ...         &          &           &     &              &             &...  & & \\
m-n_{1}-n_{2}-2 & 2m-2n_{1}-1 &          &           &     &              &             &  & 4m -4n_{1}-2n_{2}-4 & \\
 ...          &     ...    &          &           &     &              &             &  & & ...           
\end{pmatrix}
\end{align*}
\end{small}
This will generate the higher identifications that define the open DT of the doubled Kronecker quivers. Further identifications representing DT with higher charge vector can be generated by repeating this pattern and unlinking this new node again from the original two nodes until all linkings are removed. One must then define $n_{3}, \ldots , n_{m}$ for every new entry that must be unlinked to obtain all combinations that contribute to the specific open DT one is looking for.     

%**********
%**********

\subsection{Initial example of $m=3$ arrows}    \label{app-m3}

We start with the simplest example of wild wall-crossing, which corresponds to the $m$-Kronecker quiver with $m=3$ arrows. Now we will attempt to determine some low order DT invariants inside the dense cone $\mathcal{C}$. We first briefly demonstrate the condition for the cone in the scattering diagram for the 3-arrow quiver to show that the ratios are indeed inside $\mathcal{C}$
\begin{align}
\mu_{\pm} = \frac{3 \pm \sqrt{3^{2}-4}}{2} =  \frac{3 \pm \sqrt{5}}{2}.
\end{align}
We will now choose a set of DT invariants of charges inside this region so that we can start unlinking the symmetric quivers on both sides of the wall. We begin with the adjacency matrix 
\begin{align}
C= 
\begin{pmatrix}
0  & 3 \\
3 & 0
\end{pmatrix}.
\end{align}

%**********
%**********

\subsubsection*{Diagonalization of $Q^{s}$}

We proceed by starting to unlink this quiver following (\ref{startofgeneralm})-(\ref{generalmunlink}) to get the loops on the diagonal
\begin{align}
C= 
\begin{pmatrix}
0  & 2 & 2\\
2 & 0  & 2 \\
2 & 2 & 5 
\end{pmatrix}
\end{align}
We keep going to obtain
\begin{align}
C= 
\begin{pmatrix}
0  & 1 & 2 &  1 \\
1 & 0  & 2 &  1 \\
2 & 2 & 5 & 4 \\
 1 & 1 & 4 &  3
\end{pmatrix}
\end{align}
and now we unlink the last component between the first two nodes to get:
\begin{align}
C= 
\begin{pmatrix}
0  & 0 & 2 &  1  &  0\\
0 & 0  & 2 &  1 & 0\\
2 & 2 & 5 & 4 &    4\\
 1 & 1 & 4 &  3 &  2 \\
0  &  0 &  4  &  2  & 1
\end{pmatrix}
\end{align}
We therefore have 3 nodes with identification $q^{-\frac{1}{2}}x_{1}x_{2}$ and 5, 3 and 1 loops respectively. This means we can write down the open DT invariants with contribution from this identification. We now write down the open DT invariants \cite{Jankowski:2022qdpnew} as
\begin{align}
\Omega_{m-loop}(x,q) = (-1)^{m-1} q^{\frac{m}{2}}x +O(x^{2}).
\end{align}
We further write down
\begin{align}
\begin{split}
\Omega_{1-loop}(q^{-\frac{1}{2}}x_{1}x_{2},q) &=  x_{1}x_{2},  \\  
\Omega_{3-loop}(q^{-\frac{1}{2}}x_{1}x_{2},q) &=  qx_{1}x_{2}, \\  
\Omega_{5-loop}(q^{-\frac{1}{2}}x_{1}x_{2},q) &=  q^{2}x_{1}x_{2}.
\end{split}
\end{align}

%**********
%**********

\subsubsection*{Comparison with $Q^{w}$}

Now we must look at the other side and check if the computation of open DT matches. We recall that the number of loops is $mab$  or  $mab+1$, depending on whether
$   \   a^{2}+b^{2}-mab-1+k   \ $ is even or odd. So we set $a=1, b=1, m=3$ and let this function just become $-2+k$. Now we must check the identification for even values where we have 3 loop quivers
\begin{align}
\begin{split}
q^{ \frac{a^{2} +  b^{2}-mab-a-b +k }{2} }x^{a}_{2}  x_{1}^{b} &= q^{ \frac{-3 +k }{2} }x_{2}  x_{1}, \\ 
\Omega_{3-loop}(q^{ \frac{-3 +k }{2} }x_{2}  x_{1},q) &=  q^{\frac{k}{2}}x_{1}x_{2},
\end{split}
\end{align}
and for odd values where we have 4 loop quivers
\begin{align}
\begin{split}
q^{ \frac{a^{2} +  b^{2}-mab-a-b-1+k }{2} }x^{a}_{2}  x_{1}^{b}  &=  q^{\frac{-4+k }{2}}x_{2}  x_{1}, \\ 
\Omega_{4-loop}(q^{ \frac{-4 +k }{2} }x_{2}  x_{1},q) &= - q^{\frac{k}{2}}x_{1}x_{2},
\end{split}
\end{align}
so we can read off the closed DT invariants by comparing with the open invariants. Here, the only coefficients with non-vanishing DT invariants are for even $k$ and specifically $k= 0, 2 , 4,$ and we therefore have a count of one for each of the DT invariants of the form $c_{0} = 1, c_{2} = 1, c_{4} = 1$ and $ c_{k} = 0$ otherwise. Now we can compare this to the results in Reineke \cite{reineke2023wildquantumdilogarithmidentities}. Firstly, we should note that the closed DT invariants can be described by a generating function (\ref{eq:generatingfunction}) in the form of a Laurent Polynomial
\begin{align}
 P_{a,b}(q) = \sum_{k} c^{a,b}_{k} (-q^{\frac{1}{2}})^{k} \in \mathbb{Q} [q^{\pm\frac{1}{2}}], 
\end{align}
However, it is shown by Reineke that due to the positivity conditions of the DT invariants the polynomial can be written as: 
\begin{align}
 P_{a,b}(q) = \sum_{\tilde{k}} c^{a,b}_{\tilde{k}} q^{ \tilde{k}} \in \mathbb{N} [q], 
 \end{align}
where we can write $k =2 \tilde{k}$. This means that in our computation we can say that for $c_{\tilde{k}}$ we have $c_{0} = 1, c_{1} = 1, c_{2} = 1$ and $ c_{ \tilde{k}} = 0$ otherwise. We can now compare this Reineke's formula in the simplest possible example. It has been shown that
\begin{align} \label{eq:binom}
 P(1, \tilde{k}) =   \binom{m}{\tilde{k}}_{q}, 
\end{align}
and we look at the values 
\begin{align} \label{eq:binom1}
 P(1,1) =  \binom{3}{1}_{q} = \frac{1-q^{3}}{1-q} = 1 + q + q^{2},
 \end{align}
and so, as expected, the DT invariants match. 
We note that due to the symmetry of the $q$ binomial we also have
\begin{align} \label{eq:binom2}
 P(1,2) =  \binom{3}{2}_{q} = \frac{1-q^{3}}{1-q} = 1 + q + q^{2}.
\end{align}

%**********
%**********

\subsubsection*{Test of symmetry of $q$ -- binomial through quiver diagonalisation}

We aim to derive this binomial relation between (\ref{eq:binom1})-(\ref{eq:binom2}) by continuing to diagonalize the quiver. We look for all possible unlinkings that result in the identifications of the form $x^{2}_{2}x_{1}$, so we go back to the last quiver and keep unlinking it to obtain:
\begin{align}
C= 
\begin{pmatrix}
0  & 0 & 1 &  1  &  0 & \\
0 & 0  & 2 &  1 & 0 &  \\
1 & 2 & 5 & 4 &    4 &  \\
 1 & 1 & 4 &  3 &  2 &   \\
0  &  0 &  4  &  2  & 1 &  \\
   &     &    &     &   &  8  \\
\end{pmatrix},
\  \  \  \
C= 
\begin{pmatrix}
0  & 0 & 0 &  1  &  0 & & \\
0 & 0  & 2 &  1 & 0 & & \\
0 & 2 & 5 & 4 &    4 & & \\
 1 & 1 & 4 &  3 &  2 & &  \\
0  &  0 &  4  &  2  & 1 & & \\
   &     &    &     &   &  8 & \\
   &     &    &     &   &    & 6
\end{pmatrix}
\end{align}
\begin{align*}
C= 
\begin{pmatrix}
0  & 0 & 0 &  0  &  0 & & & \\
0 & 0  & 2 &  1 & 0 & & &\\
0 & 2 & 5 & 4 &    4 & & & \\
 0 & 1 & 4 &  3 &  2 & & & \\
0  &  0 &  4  &  2  & 1 & & & \\
   &     &    &     &   &  8 & & \\
   &     &    &     &   &    & 6 & \\
   &     &    &     &   &    &   & 4
\end{pmatrix}
\end{align*}
We now have to use the expressions for the $m$-loop quivers where we can just use the first order approximation given by:
\begin{align}
\Omega_{m-loop}(q^{-1}x^{2}_{2}x_{1},q) = (-1)^{m-1} q^{\frac{m}{2}}(q^{-1}x^{2}_{2}x_{1}),  
\end{align}
where the $q^{-1}$ arises from the two steps of unlinking, so that we have
\begin{align}
\begin{split}
\Omega_{8-loop}(q^{-1}x^{2}_{2}x_{1},q) &= - q^{3} x^{2}_{2}x_{1},
\\  
\Omega_{6-loop}(q^{-1}x^{2}_{2}x_{1},q) &= - q^{2} x^{2}_{2}x_{1},
\\  
\Omega_{4-loop}(q^{-1}x^{2}_{2}x_{1},q) &= - q x^{2}_{2}x_{1}.
\end{split}
\end{align}
We then look at this from the other side of the wall where $x^{2}_{2}x_{1}$ should have no other contributions other than from the node with it as its identification. Again, we have to check whether the function $   \   a^{2}+b^{2}-mab-1+k   \ $ is even or odd now for $m=3, a=2, b=1$. This is now valued at
\begin{align}
 2^{2}+1^{2}-6-1+k   =  -2 + k
\end{align}
as before. For even values we have $6$ loops and for odd values we have $7$ loops. 
We now use the identification for the even case:
\begin{align}
\begin{split}
q^{ \frac{a^{2} +  b^{2}-mab-a-b +k }{2} }x^{a}_{2}  x_{1}^{b} &= q^{ \frac{-4 +k }{2} }x^{2}_{2}  x_{1} \\  
\Omega_{6-loop}(q^{ \frac{-4 +k }{2} }x^{2}_{2}  x_{1},q) &= - q^{3}(q^{ \frac{-4 +k }{2} }x^{2}_{2}x_{1}) = - q^{3}(q^{ \frac{-4 +k }{2} }x^{2}_{2}x_{1}) = - q^{1+ \frac{k}{2} }x^{2}_{2}x_{1},
\end{split}
\end{align}
so we again recover $k= 0, 2, 4$ or $\tilde{k} = 0,1,2$, so that we again have $c_{0},c_{1},c_{1} = 1$ and $c_{\tilde{k}} = 0$ otherwise. For the odd values of $k$ we have
\begin{align}
\begin{split}
q^{ \frac{a^{2} +  b^{2}-mab-a-b-1 +k }{2} }x^{a}_{2}  x_{1}^{b} &= q^{ \frac{-5 +k }{2} }x^{2}_{2}  x_{1} \\ 
\Omega_{7-loop}(q^{ \frac{-4 +k }{2} }x^{2}_{2}  x_{1},q) &= - q^{\frac{7}{2}}(q^{ \frac{-5 +k }{2} }x^{2}_{2}x_{1}) = - q^{ \frac{2 +k }{2} }x^{2}_{2}x_{1} = - q^{1+ \frac{k}{2} }x^{2}_{2}x_{1},
\end{split}
\end{align}
which gives the same result as for the even case, however, one must always put in odd values of $k$ so this never matches the symmetric DT from the unlinking. Therefore, the values of $c_{\tilde{k}}$ must always be 0 if $\tilde{k}$ is odd.

%********
%********

\subsection{Example for $m=4$ arrows}    \label{app-m4}

We can repeat this for the $m=4$ arrow example and start by unlinking this quiver
\begin{align}
C= 
\begin{pmatrix}
0  & 4 \\
4 & 0
\end{pmatrix}.
\end{align}

%********
%********

\subsubsection*{Diagonalization of doubled quiver}

We start with the same diagonalization procedure as in the last example, so the quiver becomes
\begin{align}
C= 
\begin{pmatrix}
0  & 3 & 3\\
3 & 0  & 3 \\
3 & 3 & 7 
\end{pmatrix},
\  \  \  \  \  \  \  \
C= 
\begin{pmatrix}
0  & 2 & 3 & 2 \\
2 & 0  & 3 &  2 \\
3 & 3 & 7 & 6 \\
2 & 2 & 6 &  5
\end{pmatrix}
\end{align}
\begin{align*}
C= 
\begin{pmatrix}
0  & 1 & 3 & 2 & 1\\
1 & 0  & 3 &  2 & 1\\
3 & 3 & 7 & 6 & 6\\
2 & 2 & 6 &  5 & 4  \\
 1 & 1  &  6  &  4  & 3 \\
\end{pmatrix},
\  \  \  \ \  \  \
C= 
\begin{pmatrix}
0  & 0 & 3 & 2 & 1 & 0 \\
0 & 0  & 3 &  2 & 1 & 0 \\
3 & 3 & 7 & 6 & 6 &  6\\
2 & 2 & 6 &  5 & 4 & 4 \\
 1 & 1  &  6  &  4  & 3 & 2 \\
0 &    0  &  6   &   4  &  2  & 1
\end{pmatrix}
\end{align*}
Now we must go through another sequence of unlinkings to obtain identifications of the form $x^{2}_{2}x_{1}$ by unlinking the first node from the other nodes produced in the previous sequence of unlinkings:
\begin{align}
C= 
\begin{pmatrix}
0  & 0 & 2 & 2 & 1 & 0 & \\
0 & 0  & 3 &  2 & 1 & 0 & \\
2 & 3 & 7 & 6 & 6 &  6 & \\
2 & 2 & 6 &  5 & 4 & 4 & \\
 1 & 1  &  6  &  4  & 3 & 2 & \\
0 &    0  &  6   &   4  &  2  & 1 & \\
 &        &      &      &     &    & 12
\end{pmatrix},
\  \  \ C= 
\begin{pmatrix}
0  & 0 & 1 & 2 & 1 & 0 & & \\
0 & 0  & 3 &  2 & 1 & 0 & & \\
1 & 3 & 7 & 6 & 6 &  6 & & \\
2 & 2 & 6 &  5 & 4 & 4 & & \\
 1 & 1  &  6  &  4  & 3 & 2 & & \\
0 &    0  &  6   &   4  &  2  & 1 & &\\
 &        &      &      &     &    & 12 & \\
 &        &      &      &     &    &    & 10
\end{pmatrix}
\end{align}
\begin{align*}
C= 
\begin{pmatrix}
0  & 0 & 0 & 2 & 1 & 0 & & & \\
0 & 0  & 3 &  2 & 1 & 0 & & & \\
0 & 3 & 7 & 6 & 6 &  6 & & & \\
2 & 2 & 6 &  5 & 4 & 4 & & & \\
 1 & 1  &  6  &  4  & 3 & 2 & & & \\
0 &    0  &  6   &   4  &  2  & 1 & & & \\
 &        &      &      &     &    & 12 & & \\
 &        &      &      &     &    &    & 10 & \\
 &        &      &      &     &    &    &    & 8
\end{pmatrix},
\  \ C= 
\begin{pmatrix}
0  & 0 & 0 & 1 & 1 & 0 & & &  & \\
0 & 0  & 3 &  2 & 1 & 0 & & &  & \\
0 & 3 & 7 & 6 & 6 &  6 & & & & \\
1 & 2 & 6 &  5 & 4 & 4 & & & & \\
 1 & 1  &  6  &  4  & 3 & 2 & & & & \\
0 &    0  &  6   &   4  &  2  & 1 & & & & \\
 &        &      &      &     &    & 12 & & & \\
 &        &      &      &     &    &    & 10 & & \\
 &        &      &      &     &    &    &    & 8 & \\
 &        &      &      &     &    &    &    &   & 8
\end{pmatrix}
\end{align*}
\begin{align*}
 C= 
\begin{pmatrix}{}
0  & 0 & 0 & 0 & 1 & 0 & & &  & & \\
0 & 0  & 3 &  2 & 1 & 0 & & &  & & \\
0 & 3 & 7 & 6 & 6 &  6 & & & & & \\
0 & 2 & 6 &  5 & 4 & 4 & & & & & \\
 1 & 1  &  6  &  4  & 3 & 2 & & & & & \\
0 &    0  &  6   &   4  &  2  & 1 & & & & & \\
 &        &      &      &     &    & 12 & & & & \\
 &        &      &      &     &    &    & 10 & & & \\
 &        &      &      &     &    &    &    & 8 & & \\
 &        &      &      &     &    &    &    &   & 8 & \\
 &        &      &      &     &    &    &    &   &   & 6
\end{pmatrix},
\  \   \  \
C= 
\begin{pmatrix}{}
0  & 0 & 0 & 0 & 0 & 0 & & &  & & &\\
0 & 0  & 3 &  2 & 1 & 0 & & &  & & &\\
0 & 3 & 7 & 6 & 6 &  6 & & & & & & \\
0 & 2 & 6 &  5 & 4 & 4 & & & & & & \\
 0 & 1  &  6  &  4  & 3 & 2 & & & & & & \\
0 &    0  &  6   &   4  &  2  & 1 & & & & & & \\
 &        &      &      &     &    & 12 & & & & & \\
 &        &      &      &     &    &    & 10 & & & & \\
 &        &      &      &     &    &    &    & 8 & & & \\
 &        &      &      &     &    &    &    &   & 8 & & \\
 &        &      &      &     &    &    &    &   &   & 6 & \\
 &        &      &      &     &    &    &     &   &   &   & 4
\end{pmatrix}
\end{align*}
Now, as before, we have to use the formula for the DT of $m$-loop quivers
\begin{align}
\Omega_{m-loop}(q^{-1}x^{2}_{2}x_{1},q) = (-1)^{m-1} q^{\frac{m}{2}}(q^{-1}x^{2}_{2}x_{1}),  
\end{align}
so that we now have (remembering we must add the contributions of the two 8 loop quivers)
\begin{align}
\begin{split}
\Omega_{12-loop}(q^{-1}x^{2}_{2}x_{1},q) &= - q^{6}(q^{-1}x^{2}_{2}x_{1}) = - q^{5} x^{2}_{2}x_{1}, \\  
\Omega_{10-loop}(q^{-1}x^{2}_{2}x_{1},q) &= - q^{5}(q^{-1}x^{2}_{2}x_{1}) = - q^{4} x^{2}_{2}x_{1}, \\  
2 \Omega_{8-loop}(q^{-1}x^{2}_{2}x_{1},q) &= - 2 q^{4}(q^{-1}x^{2}_{2}x_{1}) = - 2 q^{3} x^{2}_{2}x_{1},
\\  
\Omega_{6-loop}(q^{-1}x^{2}_{2}x_{1},q) &= - q^{3}(q^{-1}x^{2}_{2}x_{1}) = - q^{2} x^{2}_{2}x_{1},
\\  
\Omega_{4-loop}(q^{-1}x^{2}_{2}x_{1},q) &= - q^{2}(q^{-1}x^{2}_{2}x_{1}) = - q x^{2}_{2}x_{1}.
\end{split}
\end{align}

%********
%********

\subsubsection*{Comparison with $Q^{w}$ on other side}

We must now look at the contribution on the other side and the cases in which the function
\begin{align}
 2^{2}+1^{2}-4 \times 2 \times 1-1+k = -4+k   
\end{align}
is even or odd, where we have set $m=4$ for the number of arrows.
We start with the even examples where the number of loops on the quiver are given by $mab =8$:
\begin{align}
q^{ \frac{a^{2} +  b^{2}-mab-a-b +k }{2} }x^{a}_{2}  x_{1}^{b} = q^{ \frac{-6 +k }{2} }x^{2}_{2}  x_{1} = q^{ -3 +\frac{k }{2} }x^{2}_{2}  x_{1},
\end{align}
and putting this into the formula for the $m$-loop quiver
\begin{align}
\Omega_{8-loop}(q^{-1}x^{2}_{2}x_{1},q) = - q^{4}(q^{ -3 +\frac{k }{2} }x^{2}_{2}  x_{1}) =  - q^{ 1 +\frac{k }{2} }x^{2}_{2}  x_{1}.
\end{align}
Therefore, we can read off the close DT invariants $c_{k}$ as
\begin{align}
c_{0} = 1, \  \ c_{2} = 1, \  \  c_{4} = 2, \ \  c_{6} = 1, \  \ c_{8} = 1,
\end{align}
which match the previous expression. This means that for $c_{\tilde{k}}$ where $\tilde{k} = \frac{k}{2}$ we have  
\begin{align}
c_{0} = 1, \  c_{1} = 1, \  c_{2} = 2, \ c_{3} = 1, \  c_{4} = 1.
\end{align}
This matches the known result from the binomial expression for $c_{\tilde{k}}q^{\tilde{k}}$
\begin{align*}
 P(1,2) =   \binom{4}{2}_{q} = 1+ q + 2q^{2} + q^{3} + q^{4}.
 \end{align*}
We now also check the odd cases 
\begin{align}
q^{ \frac{a^{2} +  b^{2}-mab-a-b-1 +k }{2} }x^{a}_{2}  x_{1}^{b} = q^{ \frac{-7 +k }{2} }x^{2}_{2}  x_{1}.
\end{align}
We remember that for the odd examples we must increase the number of loops by one. This means that the open DT invariants in this example are
\begin{align}
\Omega_{9-loop}(q^{ \frac{-7 +k }{2} }x^{2}_{2}  x_{1},q) = (-1)^{9-1} q^{\frac{9}{2}}(q^{ \frac{-7 +k }{2} }x^{2}_{2}  x_{1}) =  q^{1 + \frac{k }{2} }x^{2}_{2}  x_{1}.
\end{align}
Here we now have terms that again never occur in the diagonalisation. Therefore, as with the previous examples, the contribution here vanishes for any close DT invariants as expected.

%*********
%*********

\subsection{Example for $m=6$ arrows}   \label{app-m6}

Finally, we consider an example with $m=6$ arrows which might give higher degeneracies for the BPS states. For this, following (\ref{startofgeneralm})-(\ref{generalmunlink}), we start by unlinking:
\begin{align}
C= 
\begin{pmatrix}
0  & 6 \\
6 & 0
\end{pmatrix},
\end{align}
and get
\begin{align}
C= 
\begin{pmatrix}
0  & 5 & 5\\
5 & 0  & 5 \\
5 & 5 & 11 
\end{pmatrix},
 \  \  \ C= 
\begin{pmatrix}
0  & 4 & 5 & 4 \\
4 & 0  & 5 &  4 \\
5 & 5 & 11 & 10 \\
4 & 4 & 10 &  9
\end{pmatrix}
\end{align}
\begin{align*}
C= 
\begin{pmatrix}
0  & 3 & 5 & 4 &  3 \\
3 & 0  & 5 &  4 & 3\\
5 & 5 & 11 & 10 & 10 \\
4 & 4 & 10 &  9 & 8  \\
 3 & 3  &  10  &  8  & 7 \\
\end{pmatrix},
\  \  \  \  \  \
C= 
\begin{pmatrix}
0  & 2 & 5 & 4 & 3 & 2 \\
2 & 0  & 5 &  4 & 3 & 2 \\
5 & 5 & 11 & 10 & 10 &  10\\
4 & 4 & 10 &  9 & 8 & 8 \\
 3 & 3  &  10  &  8  & 7 & 6 \\
2 &    2  &  10   &   8  &  6  & 5
\end{pmatrix}
\end{align*}
\begin{align*}
C= 
\begin{pmatrix}
0  & 1 & 5 & 4 & 3 & 2 & 1 \\
1 & 0  & 5 &  4 & 3 & 2 & 1  \\
5 & 5 & 11 & 10 & 10 &  10 & 10 \\
4 & 4 & 10 &  9 & 8 & 8 &  8 \\
 3 & 3  &  10  &  8  & 7 & 6 & 6 \\
2 &    2  &  10   &   8  &  6  & 5 & 4 \\
 1  &    1    &   10    &   8   &  6   &  4  & 3
\end{pmatrix},
\  \  \  \  \
C= 
\begin{pmatrix}
0  & 0 & 5 & 4 & 3 & 2 & 1 & 0 \\
0 & 0  & 5 &  4 & 3 & 2 & 1 & 0 \\
5 & 5 & 11 & 10 & 10 &  10 & 10 & 10 \\
4 & 4 & 10 &  9 & 8 & 8 &  8 & 8 \\
 3 & 3  &  10  &  8  & 7 & 6 & 6 &  6 \\
2 &    2  &  10   &   8  &  6  & 5 & 4 &   4\\
 1  &    1    &   10    &   8   &  6   &  4  & 3 &  2\\ 
  0  &    0     &   10      &   8    &   6   &  4   & 2   & 1
\end{pmatrix}
\end{align*}
If we are now looking for terms of order $x^{2}_{2}x_{1}$ we must, as before, do another series of unlinkings. As the matrices are too large to write down we will write the number of loops in the resulting node and number of linkings to the node representing $x_{2}$.
\begin{align}
 \text{Number of loops}: \  \  20, 18, 16, 14, 12 ; \  \ 16, 14, 12, 10;  \  \ 12, 10, 8;  \  \ 8, 6, 4.
\end{align}
We can immediately read off that the number 12 appears 3 times, and the other numbers appear either once or twice. This can be nicely mapped onto the degeneracies of the BPS states of the closed $m$-Kronecker quiver. Again, to obtain the terms in the sum over the $m$-loop quivers we have to use
\begin{align}
\begin{split}
\Omega_{4-loop}(q^{-1}x^{2}_{2}x_{1},q) &= (-1)^{4-1} q^{\frac{4}{2}}(q^{-1}x^{2}_{2}x_{1}) = - qx^{2}_{2}x_{1}, \\  
\Omega_{6-loop}(q^{-1}x^{2}_{2}x_{1},q) &= (-1)^{6-1} q^{\frac{6}{2}}(q^{-1}x^{2}_{2}x_{1}) = - q^{2}x^{2}_{2}x_{1}, \\  
2\Omega_{8-loop}(q^{-1}x^{2}_{2}x_{1},q) &= 2(-1)^{8-1} q^{\frac{8}{2}}(q^{-1}x^{2}_{2}x_{1}) = -2 q^{3}x^{2}_{2}x_{1}, \\   
2\Omega_{10-loop}(q^{-1}x^{2}_{2}x_{1},q) &= 2(-1)^{10-1} q^{\frac{10}{2}}(q^{-1}x^{2}_{2}x_{1}) = -2 q^{4}x^{2}_{2}x_{1}, \\  
3\Omega_{12-loop}(q^{-1}x^{2}_{2}x_{1},q) &= 2(-1)^{12-1} q^{\frac{12}{2}}(q^{-1}x^{2}_{2}x_{1}) = -3 q^{5}x^{2}_{2}x_{1}, \\  
2\Omega_{14-loop}(q^{-1}x^{2}_{2}x_{1},q) &= 2(-1)^{14-1} q^{\frac{14}{2}}(q^{-1}x^{2}_{2}x_{1}) = -2 q^{6}x^{2}_{2}x_{1}, \\  
2\Omega_{16-loop}(q^{-1}x^{2}_{2}x_{1},q) &= 2(-1)^{16-1} q^{\frac{16}{2}}(q^{-1}x^{2}_{2}x_{1}) = -2 q^{7}x^{2}_{2}x_{1}, \\  
\Omega_{18-loop}(q^{-1}x^{2}_{2}x_{1},q) &= (-1)^{18-1} q^{\frac{18}{2}}(q^{-1}x^{2}_{2}x_{1}) = - q^{8}x^{2}_{2}x_{1}, \\  
\Omega_{20-loop}(q^{-1}x^{2}_{2}x_{1},q) &= (-1)^{20-1} q^{\frac{20}{2}}(q^{-1}x^{2}_{2}x_{1}) = - q^{9}x^{2}_{2}x_{1}.
\end{split}
\end{align}
We now look at the other side of the wall where the identification becomes: 
\begin{align}
q^{ \frac{a^{2} +  b^{2}-mab-a-b +k }{2} }x^{a}_{2}  x_{1}^{b} = q^{ \frac{-10 +k }{2} }x^{2}_{2}  x_{1} = q^{ -5 +\frac{k }{2} }x^{2}_{2}  x_{1}
\end{align}
for $mab = 12$ loop quiver. Plugging this back into the equation for $m$-loop quivers we get
\begin{align}
\Omega_{12-loop}(q^{ \frac{-10+k }{2} }x^{2}_{2}  x_{1},q) = (-1)^{12-1} q^{\frac{12}{2}}(q^{ \frac{-10 +k }{2} }x^{2}_{2}  x_{1}) =  -q^{1 + \frac{k }{2} }x^{2}_{2}  x_{1}.
\end{align}
This can now be compared with the terms obtained above by diagonalization and letting $\tilde{k} = \frac{k}{2}$. This means that the DT invariants take form
\begin{align}
c_{\tilde{k}}: 1 , 1, 2 , 2 , 3 , 2 , 2 , 1 , 1.
\end{align}
We have therefore recovered the Donaldson-Thomas invariants
\begin{align}
 P(1,2) =  \binom{6}{2}_{q} = 1+ q + 2q^{2} + 2q^{3} + 3q^{4} + 2q^{5} + 2q^{6} + q^{7} + q^{8}.
\end{align}

\section*{Acknowledgments}

We thank Piotr Kucharski  and Markus Reineke for insightful discussions. This work has been supported by the OPUS grant no. 2022/47/B/ST2/03313 ``Quantum geometry and BPS states'' funded by the National Science Centre, Poland.

% \clearpage

\bibliography{generatingfunction}
\bibliographystyle{utphys}

\end{document}